\documentclass[a4paper,12pt,conditionalcopyright]{uq-thesis}
\newcommand{\p}[1]{\phantom{#1}}

\newcommand{\fref}[1]{Fig.~\ref{#1}}

\newcommand{\cref}[1]{Chapter~\ref{#1}}

\newcommand{\rmi}{\mathrm{i}}

\newcommand{\ket} [1] {| #1 \rangle}
\newcommand{\bra} [1] {\langle #1 |}
\newcommand{\braket}[2]{\langle #1 | #2 \rangle}

\newcommand{\fuser}{\Upsilon^{\mbox{\tiny \,fuse}}}
\newcommand{\spliter}{\Upsilon^{\mbox{\tiny \,split}}}

\newcommand{\splitt}[3]{\Upsilon^{\mbox{\tiny \,split}}_{#1\rightarrow #2,#3}}
\newcommand{\fuse}[3]{\Upsilon^{\mbox{\tiny \,fuse}}_{#1, #2 \rightarrow #3}}

\newcommand{\splitspin}[3]{\Upsilon^{\mbox{\tiny \,split}}_{j_{#1}t_{j_{#1}}m_{j_{#1}}\rightarrow j_{#2}t_{j_{#2}}m_{j_{#2}},j_{#3}t_{j_{#3}}m_{j_{#3}}}}
\newcommand{\fusespin}[3]{\Upsilon^{\mbox{\tiny \,fuse}}_{j_{#1}t_{j_{#1}}m_{j_{#1}},j_{#2}t_{j_{#2}}m_{j_{#2}}\rightarrow j_{#3}t_{j_{#3}}m_{j_{#3}}}}

\newcommand{\tfuser}{X^{\mbox{\tiny \,fuse}}}
\newcommand{\tspliter}{X^{\mbox{\tiny \,split}}}

\newcommand{\tsplitt}[3]{X^{\mbox{\tiny \,split}}_{#1\rightarrow #2,#3}}
\newcommand{\tfuse}[3]{X^{\mbox{\tiny \,fuse}}_{#1, #2 \rightarrow #3}}

\newcommand{\tsplitspin}[3]{X^{\mbox{\tiny \,split}}_{j_{#1}t_{j_{#1}}\rightarrow j_{#2}t_{j_{#2}},j_{#3}t_{j_{#3}}}}
\newcommand{\tfusespin}[3]{X^{\mbox{\tiny \,fuse}}_{j_{#1}t_{j_{#1}},j_{#2}t_{j_{#2}}\rightarrow j_{#3}t_{j_{#3}}}}

\newcommand{\csplitt}[3]{C^{\mbox{\tiny \,split}}_{#1\rightarrow #2,#3}}
\newcommand{\cfuse}[3]{C^{\mbox{\tiny \,fuse}}_{#1, #2 \rightarrow #3}}

\newcommand{\cfuser}{C^{\mbox{\tiny \,fuse}}}
\newcommand{\cspliter}{C^{\mbox{\tiny \,split}}}

\newcommand{\csplitspin}[3]{C^{\mbox{\tiny \,split}}_{j_{#1}m_{j_{#1}}\rightarrow j_{#2}m_{j_{#2}},j_{#3}m_{j_{#3}}}}
\newcommand{\cfusespin}[3]{C^{\mbox{\tiny \,fuse}}_{j_{#1}m_{j_{#1}},j_{#2}m_{j_{#2}}\rightarrow j_{#3}m_{j_{#3}}}}

\makeatletter

\newcommand{\Rmnum}[1]{\expandafter\@slowromancap\romannumeral #1@}

\newcommand{\half}{\frac{1}{2}}
\newcommand{\mhalf}{\frac{-1}{2}}

\newcommand{\tree}{\boldsymbol{\tau}}

\newcommand{\markend}{~\rule{0.4em}{1.8ex}} %instead of \blacksquare

\newcommand{\mycup}{\hat{\Omega}^{\mbox{\tiny \,cup}}}
\newcommand{\mycap}{\hat{\Omega}^{\mbox{\tiny \,cap}}}

\newcommand{\braid}[3]{R^{\mbox{\tiny \,swap}}_{j_{#1},j_{#2} \rightarrow j_{#3}}}
\newcommand{\braider}{\hat{R}^{\mbox{\tiny \,swap}}}

\usepackage{amssymb}	% AMS symbols package
\usepackage{amsmath}	% AMS extended maths command set
\usepackage{amsthm}		% Creation of "theorem" environments (e.g. Example 1:... \\ Example 2:...)
\usepackage{bm}			% Bold maths symbols
\usepackage{array}		% More sophisticated table column specifications - see http://en.wikibooks.org/wiki/LaTeX/Tables
\usepackage{epstopdf}	% Automatically convert eps figures to PDF if using PDFTeX (e.g. on Mac)
\usepackage{url}		% Adds URL command, e.g. \url{http://www.google.com/}, and improves 
\usepackage[sort&compress,round]{natbib}		% Better selection of cite commands (like in RevTeX). Lots of options to set up for this...
\usepackage[nottoc]{tocbibind}
\usepackage{bigstrut}	%Allows for manual control of spacing in arrays
\usepackage{booktabs}	% Package for creating better looking tables
\usepackage{titlesec}	% Provides commands used for fancy-looking section headings etc.
\usepackage{color} \definecolor{darkgreen}{rgb}{0,0.55,0}
\usepackage[breaklinks=true,colorlinks=true,citecolor=darkgreen,pagebackref=true]{hyperref}
\usepackage{multirow}	% Allows tabular entries to span multiple rows
\usepackage[margin=15pt,font={small,sf},labelfont=bf,labelsep=endash]{caption}
\usepackage[final]{pdfpages}
\titleformat{\subsubsection}[hang]{\centering\slshape\bfseries}{}{0pt}{}{}

\newtheoremstyle{henrytheoremstyle}
  {\topsep}%      Space above
  {\topsep}%      Space below
  {\itshape}%         Body font
  {}%         Indent amount (empty = no indent, \parindent = para indent)
  {\bfseries}% Thm head font
  {:}%        Punctuation after thm head
  {.5em}%     Space after thm head: " " = normal interword space;
  {}%         Thm head spec (can be left empty, meaning `normal')

\theoremstyle{henrytheoremstyle}

\title{\vspace{1.6em}\Large{{Tensor Network States and Algorithms\\ in the presence of \\ Abelian and non-Abelian Symmetries\vspace{.2em}\\}}}

\author{Sukhbinder Singh}
\prevdegrees{B.E. Comp. Sci. (2003), Thapar University, India.}
\division{I}
\school{School of Mathematics and Physics}
\thesistype{Doctor of Philosophy}

\hypersetup{
  pdfinfo={
    Title={Tensor Network States and Algorithms in the presence of Abelian and non-Abelian Symmetries},
    Author={Sukhbinder Singh},
    Keywords={tensor networks, symmetries, U(1) symmetry, SU(2) symmetry, MERA, spin models, spin networks}
  }
}

\tablespagetrue % Include tables list page
\figurespagetrue % Include figures list page
\begin{document}

% Include parts using \include. Single chapters can then be compiled using \includeonly.
% Note a \included file always begins on a new page. Also, because the thesis class is based on
% the "book" type, new chapters (created using \chapter) always begin on the right hand side of a 
% pair of facing pages.

% --- Title page, preface text, abstract, contents, etc.
%***************************************************************************************************************

% ----------------- Begin TITLE --------------->>

\beforepreface

% ----------------- End TITLE --------------->>

% ----------------- ORIGINALITY --------------->>

%\originalitystatement

% ----------------- ACKNOWLEDGEMENTS --------------->>

%\chapter*{Statements of Contributions}
\textbf{Declaration by author}

This thesis is composed of my original work, and contains no material previously published or written by another person except where due reference has been made in the text. I have clearly stated the contribution by others to jointly-authored works that I have included in my thesis.

I have clearly stated the contribution of others to my thesis as a whole, including statistical assistance, survey design, data analysis, significant technical procedures, professional editorial advice, and any other original research work used or reported in my thesis. The content of my thesis is the result of work I have carried out since the commencement of my research higher degree candidature and does not include a substantial part of work that has been submitted to qualify for the award of any other degree or diploma in any university or other tertiary institution. I have clearly stated which parts of my thesis, if any, have been submitted to qualify for another award.

I acknowledge that an electronic copy of my thesis must be lodged with the University Library and, subject to the General Award Rules of The University of Queensland, immediately made available for research and study in accordance with the \textit{Copyright Act 1968}.

I acknowledge that copyright of all material contained in my thesis resides with the copyright holder(s) of that material.

\textbf{Statement of Contributions to Jointly Authored Works Contained in this Thesis}

This thesis is partly by publication. It contains the following publications, which I have co-authored as the first author, directly as Chapters 2,3 and 4.

\citep{Singh10c} - Incorporated as \textit{Chapter 2}. The details of the SU(2) symmetric iTEBD algorithm were developed by myself and Prof. Guifre Vidal. The implementation of the algorithm in MATLAB and numerical simulations was performed by myself. Most of the manuscript was prepared by Prof. Guifre Vidal. Section 5, which describes the details of the symmetric iTEBD algorithm, and Figs. 3 and 4 were prepared by myself. Prof. Huan-Zhou Qiang was responsible for clarifying several results from the representation theory of SU(2) that were crucial to the development of the algorithm. He also pointed out the exact results that were used in Fig.~4 to draw a comparison with the numerical results.

\citep{Singh10b} - Incorporated as \textit{Chapter 3}. The theoretical formalism presented in this paper was developed mostly by myself and Prof. Guifre Vidal. Robert~N.~C.~Pfeifer joined the effort at an advanced stage of the project and contributed ideas that were important for the final presentation of the formalism. The manuscript was mostly written by Prof. Guifre Vidal, in close correspondence with myself and Robert~N.~C.~Pfeifer. Implementation in MATLAB and numerical simulations were performed by me. However, these results were not included in the final draft of the manuscript.

 \citep{Singh11} - Incorporated as \textit{Chapter 4}. The theoretical ideas presented in this paper were developed mostly by myself and Prof. Guifre Vidal. The MATLAB implementation and numerical results reported in the manuscript were performed by myself. Robert~N.~C.~Pfeifer also independently and simultaneously coded an implementation in MATLAB that was an important support for the validity of the ideas presented in the paper. Myself and Robert~N.~C.~Pfeifer jointly proposed a precomputation scheme that was used in both the implementations. The main structure and content of the paper, including all the figures, was prepared by myself in close supervision of Prof. Guifre Vidal. The final draft was critically revised by Prof. Guifre Vidal. Robert. N.C. Pfeifer also contributed to the corrections of the figures and the manuscript at a pre-submission stage, and assisted in writing the Appendix.

%%%%%%%%%%%%%%%%%%%%%%%%%%%%%%%%%%%%%%%%%%%%%%%%%%%%%%%%%%%%%%%%%%%%%%%%%%%%%%%%%%%%%%%%%%%%%%
% Fill in text as appropriate
%%%%%%%%%%%%%%%%%%%%%%%%%%%%%%%%%%%%%%%%%%%%%%%%%%%%%%%%%%%%%%%%%%%%%%%%%%%%%%%%%%%%%%%%%%%%%%

\textbf{Statement of Contributions by Others to the Thesis as a Whole}

The overall motivation and research direction of this thesis was provided by Prof. Guifre Vidal.

%%%%%%%%%%%%%%%%%%%%%%%%%%%%%%%%%%%%%%%%%%%%%%%%%%%%%%%%%%%%%%%%%%%%%%%%%%%%%%%%%%%%%%%%%%%%%%
% Fill in text as appropriate
%%%%%%%%%%%%%%%%%%%%%%%%%%%%%%%%%%%%%%%%%%%%%%%%%%%%%%%%%%%%%%%%%%%%%%%%%%%%%%%%%%%%%%%%%%%%%%
\newpage
\textbf{Statement of Parts of the Thesis Submitted to Qualify for the Award of Another Degree} 

None.
%%%%%%%%%%%%%%%%%%%%%%%%%%%%%%%%%%%%%%%%%%%%%%%%%%%%%%%%%%%%%%%%%%%%%%%%%%%%%%%%%%%%%%%%%%%%%%
% Fill in text as appropriate
%%%%%%%%%%%%%%%%%%%%%%%%%%%%%%%%%%%%%%%%%%%%%%%%%%%%%%%%%%%%%%%%%%%%%%%%%%%%%%%%%%%%%%%%%%%%%%

\textbf{Published Works by the Author Incorporated into the Thesis}

\citep{Singh10c} - Incorporated as Chapter 2.

\citep{Singh10b} - Incorporated as Chapter 3.

\citep{Singh11} - Incorporated as Chapter 4.

%%%%%%%%%%%%%%%%%%%%%%%%%%%%%%%%%%%%%%%%%%%%%%%%%%%%%%%%%%%%%%%%%%%%%%%%%%%%%%%%%%%%%%%%%%%%%%
% Fill in text as appropriate
%%%%%%%%%%%%%%%%%%%%%%%%%%%%%%%%%%%%%%%%%%%%%%%%%%%%%%%%%%%%%%%%%%%%%%%%%%%%%%%%%%%%%%%%%%%%%%

\textbf{Additional Published Works by the Author Relevant to the Thesis but not Forming Part of it} 

None.

\chapter*{Acknowledgments}
The completion of this thesis would have been impossible without the tireless supervision and wise guidance of Guifre Vidal. From him I have learnt numerous practical aspects of both research and life. For that my gratitude remains well beyond words.

I thank all my fellow research group members: Philippe Corboz, Glen Evenbly, Andy Ferris, Jacob Jordan, Ian McCulloch, Roman Orus, Robert Pfeifer and Luca Tagliacozzo, for providing me with an intellectually rich and competitive research environment. I also acknowledge the support of Huan-Qiang Zhou, whose research group I visited twice during the completion of this work. I thank him for always putting pulpable enthusiam into physics and physics into words. 

My personal perseverance was fueled by my parents who have made me what I am today, and by Meru, my dear friend, who has been like my shadow in the last few years, especially during the strenuous times of this project. 

Finally, I would like to make a special mention of the blessings of my grandsire figure, Sant Baba Pritam Das Ji, who forever nourishes my spirit.

% ----------------- ABSTRACT --------------->>
\
\chapter*{Abstract}
Understanding and classifying phases of matter is a vast and important area of research in modern physics. Of special interest are phases at low temperatures where quantum effects are dominant. Theoretical progress is thwarted by a general lack of analytical solutions for quantum many-body systems. Moreover, perturbation theory is often inadequate in the strongly interacting regime. As a result, numerical approaches have become an indispensable tool to address such problems. In recent times, numerical approaches based on tensor networks have caught widespread attention. Tensor network algorithms draw on insights from Quantum Information theory to take advantage of special entanglement properties of low energy quantum many-body states of lattice models. Examples of popular tensor networks include Matrix Product States, Tree tensor Network, Multi-scale Entanglement Renormalization Ansatz and Projected Entangled Pair States. The main impediment of these methods comes from the fact that they can only represent states with a limited amount of entanglement. On the other hand, exploitation of symmetries, a powerful asset for numerical methods, has remained largely unexplored for a broad class of tensor networks algorithms.

In this thesis we extend the formalism of tensor network algorithms to incorporate global internal symmetries. We describe how to both numerically \textit{protect} the symmetry and \textit{exploit} it for computational gain in tensor network simulations. Our formalism is generic. It can readily be adapted to specific tensor network representations and to a wide spectrum of physical symmetries. The latter includes conservation of total particle number (U(1) symmetry) and of total angular momentum (SU(2) symmetry), and also more exotic symmetries (anyonic systems). The generality of the formalism is due to the fact that the symmetry constraints are imposed at the level of individual tensors, in a way that is independent of the details of the tensor network. As a result, we are led to a framework of symmetric tensors. Such tensors are then used as building blocks for tensor network representations of quantum-many states in the presence of symmetry.

For a long time several physical problems of immense interest have remained elusive to numerical methods mostly owing to extremely high simulation costs. These include systems of frustrated magnets and interacting fermions that are relevant in the context of quantum magnetism and high temperature superconductivity. With symmetry now as a potent ally, tensor network algorithms may finally be used to draw positive insights about such systems.

\newpage
\textbf{Keywords:} tensor networks, symmetry, U(1), SU(2), MERA, MPS, spin networks, anyons % Max. ten words

\textbf{Australian and New Zealand Standard Research Classifications (ANZSRC):}
\\\p{0}020401 Condensed Matter Characterisation Technique Development (50\%)
\\\p{0}020603 Quantum Information, Computation and Communication (50\%).
%%%%%%%%%%%%%%%%%%%%%%%%%%%%%%%%%%%%%%%%%%%%%%%%%%%%%%%%%%%%%%%%%%%%%%%%%%%%%%%%%%%%%%%%%%%%%%
% Fill in text as appropriate
%%%%%%%%%%%%%%%%%%%%%%%%%%%%%%%%%%%%%%%%%%%%%%%%%%%%%%%%%%%%%%%%%%%%%%%%%%%%%%%%%%%%%%%%%%%%%%
%% Only if not generating this automatically using the makenomenclature package
%
%
%%%%%%%%%%%%%%%%%%%%%%%%%%%%%%%%%%%%%%%%%%%%%%%%%%%%%%%%%%%%%%%%%%%%%%%%%%%%%%%%%%%%%%%%%%%%%%%
%% Fill in text as appropriate
%%%%%%%%%%%%%%%%%%%%%%%%%%%%%%%%%%%%%%%%%%%%%%%%%%%%%%%%%%%%%%%%%%%%%%%%%%%%%%%%%%%%%%%%%%%%%%%

% ----------------- TABLES OF CONTENTS ETC. --------------->>

\afterpreface
%****************************************************************************************************

% --- Main body of thesis:
\chapter{Introduction\label{sec:introduction}}

The study of quantum many-body phenomena is of pivotal interest in modern physics. Important areas of research, such as the characterization of exotic phases of quantum matter and of quantum phase transitions\citep{Sachdev}, or even the possible realization\citep{Kitaev03, Nayak08} of a quantum computer, rely on our understanding of collective phenomena in quantum many-body systems. Theoretical progress in these research areas is hindered both by a general lack of analytical results as well as the inadequacy of perturbation theory in the strongly interacting regime where the interesting physics often lies. As a result, development of numerical approaches to probe such systems has become a flourishing research industry. However, numerical methods are limited by staggering computational costs.

The number of parameters required to describe a generic quantum many-body wavefunction on a lattice grows exponentially with the number of sites in the lattice. An immediate consequence is that exact diagonalization can only be applied to small systems. In particular, thermodynamic properties often remain inaccessible by this method. Conventional alternatives include Quantum Monte Carlo sampling techniques. These are well established numerical methods that have been used extensively in several areas of Mathematics and Physics. Of interest here is that these techniques have been applied\citep{Prokofev98, Evertz03, Syljuasen02, Sandvik05} successfully to several quantum lattice models. On the other hand, Quantum Monte Carlo techniques suffer from the notorious sign problem\citep{Loh90, Henelius00} that hinders their application to certain systems of immense interest. Notable examples include systems of frustrated magnets and of interacting fermions that are relevant in the context of quantum magnetism and high temperature superconductivity\citep{Anderson87}.

In recent years, new approaches based on tensor networks have caught widespread attention. Such approaches can be regarded as generalizations of the density matrix renormalization group (DMRG) method\citep{White92, White93, Schollwock05, McCulloch08}, which is highly successful for one-dimensional systems. The potential of tensor network algorithms relies on the fact that, as  DMRG, they can address systems of frustrated spins and interacting fermions but, unlike DMRG, they can also be applied to two dimensional systems, both of large size and of infinite size. The main impediment of such methods comes from the fact that simulation costs increase rapidly with the amount of entanglement in the system. Consequently, tensor networks can only represent states with a limited amount of entanglement. On the other hand, exploitation of symmetries has remained largely unexplored for a broad class of tensor networks algorithms.

Symmetries, of fundamental importance in physics, require special treatment in numerical studies. Unless explicitly preserved at the algorithmic level, they are bound to be destroyed by the accumulation of small errors, in which case significant features of the system might be concealed. On the other hand, when properly handled, the presence of a symmetry can be exploited to reduce simulation costs. 

The goal of this thesis is to extend the tensor network formalism to the presence of symmetries. We develop a generic framework that can be applied to adapt any given tensor network representation and algorithm to both numerically protect symmetries and exploit them for computational gain.

\section{Tensor network states and algorithms}

Tensor networks are an efficient parameterization of low energy quantum many-body states of lattice models. The degrees of freedom of the model are arranged on a lattice $\mathcal{L}$ made of $L$ sites where each site is described by a Hilbert space of dimension $d$. As a result, the Hilbert space dimension of $\mathcal{L}$ grows exponentially with the number of sites $L$. Thus, a generic quantum many-body state on the lattice is parameterized by exponentially many parameters. On the other hand, the dynamics of the system are typically governed by a \textit{local} Hamiltonian $\hat{H}$, that is, $\hat{H}$ decomposes as the sum of terms involving only a small number of sites, and whose strength decays with the distance between the sites. The locality of the dynamics often implies that only a relatively small amount of entanglement is present in the ground state. In such circumstances, tensor networks offer a good description of the ground state. Moreover, the description is efficient, in that the total number of parameters encoded into the tensor networks grows roughly linearly with $L$. 

Examples of tensor network states for one dimensional systems include the matrix product state\citep{Fannes92,Ostlund95,Perez-Garcia07} (MPS), which results naturally from both Wilson's numerical renormalization group\citep{Wilson75} and White's DMRG and is also used as a basis for simulation of time evolution, e.g. with the time evolving block decimation (TEBD)\citep{Vidal03,Vidal04,Vidal07} algorithm and variations thereof, often collectively referred to as time-dependent DMRG\citep{Vidal03,Vidal04,Daley04,White04,Schollwock05b,Vidal07}; the tree tensor network\citep{Shi06} (TTN), which follows from coarse-graining schemes where the spins are blocked hierarchically; and the multi-scale entanglement renormalization ansatz\citep{Vidal07b, Vidal08, Evenbly09, Giovannetti08, Pfeifer09, Vidal10} (MERA), which results from a renormalization group procedure known as entanglement renormalization\citep{Vidal07b,Vidal10}. For two dimensional lattices there are generalizations of these three tensor network states, namely projected entangled pair  states\citep{Verstraete04, Sierra98, Nishino98, Nishio04, Murg07, Jordan08, Gu08, Jiang08, Xie09, Murg09} (PEPS), 2D TTN,\citep{Tagliacozzo09, Murg10} and 2D MERA\citep{Evenbly10, Evenbly10b, Aguado08, Cincio08, Evenbly09b, Konig09} respectively. As variational ans\"atze, PEPS and 2D MERA are particularly interesting since they can be used to address large two-dimensional lattices, including systems of frustrated spins\citep{Murg09, Evenbly10} and interacting fermions,\citep{Corboz09, Kraus10, Pineda10, Corboz10, Barthel09, Shi09, Corboz10b, Pizorn10, Gu10} where Monte Carlo techniques fail due to the sign problem. 

Some popular tensor networks are summarized in table \ref{table:tn}.

\begin{table}
%%%%%%%%%%%%%%%%%%%%%%%%%%%%%%%%%%%%%%%%%%%%%%%%%%%%%%%%%%%%%%%%%%%%%%%%%%%%%%%%%%%%%%%%%%%%%%%%
\centering % used for centering table
\begin{tabular}{|c| c|} % centered columns (4 columns)
\hline %inserts double horizontal lines
\textit{One dimensional} &  \textit{Two dimensional} \\ [0.5ex] % inserts table
\hline\hline
MPS & PEPS \\ \hline
1D TTN & 2D TTN \\ \hline
1D MERA & 2D MERA \\
[1ex]% [1ex] adds vertical space
%heading
\hline % inserts single horizontal line
\end{tabular}
\caption{
%[table:degdist]
Popular tensor networks in one and two spatial dimensions.\label{table:tn}
}
\end{table}

\section{Symmetries}

The presence of symmetries is a universal trait of physical theories. Symmetry has become one of the most powerful tools of theoretical physics, as it has become evident that practically all laws of nature originate in symmetries. The importance of symmetries in physical theories was firmly grounded by the famous \textit{Noether's theorem}, a rigorous result that links the presence of a symmetry to the conservation of a physical quantity.

In this thesis we will be concerned with symmetries exhibited in quantum lattice models. The many-body Hamiltonian $\hat H$ may be invariant under certain transformations, which form a group $\mathcal{G}$ of symmetries\citep{Cornwell97}. Under the action of the symmetry transformation, the Hilbert space of the theory is divided into symmetry sectors labeled by quantum numbers or conserved charges. The symmetry group $\mathcal{G}$ may be \textit{Abelian} or \textit{non-Abelian}, depending on whether or not the total effect of applying two symmetry transformations depends on the order in which the transformations are applied. The symmetry sectors associated with an Abelian symmetry correspond to one-dimensional invariant subspaces. In contrast, the dimension of the symmetry sectors associated with a non-Abelian symmetry may be larger than one.

On the lattice, one can distinguish between \textit{space} symmetries, which correspond to some permutation of the sites of the lattice, and \textit{internal} symmetries, which act on the vector space of each site. An example of space symmetry is invariance under translations by some unit cell, which leads to conservation of quasi-momentum. An example of internal symmetry is SU(2) invariance, e.g. spin isotropy in a quantum spin model. An internal symmetry can in turn be \textit{global}, if it transforms the space of each of the lattice sites according to the same transformation (e.g. a spin independent rotation); or \textit{local}, if each lattice site is transformed according to a different transformation (e.g. a spin-dependent rotation), as it is in the case of lattice gauge models. A global internal SU(2) symmetry gives rise to conservation of total spin. Table \ref{table:symmetry} lists some examples of Abelian and non-Abelian physical symmetries.

\begin{table}
%%%%%%%%%%%%%%%%%%%%%%%%%%%%%%%%%%%%%%%%%%%%%%%%%%%%%%%%%%%%%%%%%%%%%%%%%%%%%%%%%%%%%%%%%%%%%%%%
\centering % used for centering table
\begin{tabular}{|c| c| c|} % centered columns (4 columns)
\hline %inserts double horizontal lines
\textit{Conserved physical quantity} &  \textit{Symmetry group} & \textit{Abelian/non-Abelian}\\ [0.5ex] % inserts table
\hline\hline
Parity of particle number & $Z_2$ & Abelian\\ \hline
Particle number, spin projection & U(1) & Abelian\\ \hline
Total angular momentum, total spin & SU(2)& non-Abelian\\ \hline
Particle number and spin & U(1) $\times$ SU(2)& non-Abelian \\ \hline
Spin and isospin & SU(2) $\times$ SU(2)& non-Abelian \\ \hline
Total anyonic charge & e.g. $\mbox{SU(2)}_k$ & e.g.,non-Abelian\\
[1ex]% [1ex] adds vertical space
%heading
\hline % inserts single horizontal line
\end{tabular}
\caption{
%[table:degdist]
Examples of global internal physical symmetries.\label{table:symmetry}
}
\end{table}

By targeting a specific symmetry sector during a calculation, computational costs can often be significantly reduced while explicitly preserving the symmetry. It is therefore not surprising that symmetries play an important role in numerical approaches.

\section{Incorporating symmetries into tensor network algorithms}

Exploiting symmetries has been of great interest in numerical approaches, since it allows selection of a specific charge sector within the kinematic Hilbert space, and leads to significant reduction of computational costs.  

In the context of tensor network algorithms, benefits of exploiting the symmetry have been extensively demonstrated especially in the context of MPS. Both \textit{space} and \textit{internal} symmetries, Abelian and non-Abelian, have been thoroughly incorporated into DMRG code and have been exploited to obtain computational gains\citep{Ostlund95,White92,Schollwock05b,Ramasesha96,Sierra97,Tatsuaki00,McCulloch02,Bergkvist06,Pittel06,McCulloch07,Perez-Garcia08,Sanz09}. %,Vidal07

Symmetries have also been used in more recent proposals to simulate time evolution with MPS\citep{Vidal04,Daley04,White04,Schollwock05b,Vidal07,Daley05,Danshita07,Muth10,Mishmash09,Singh10c,Cai10}. 

%%%%%%%%%%%%%%%%%%%%%%%%%%%%%%%%%%%%%%%%%%%%%%%%%%%%%%%%%%%%%%%%%%%%%%%%%%%%%%%%%%%%%%%%%%%%%%%%
\begin{figure}[t]
\begin{center}
  \includegraphics[width=12cm]{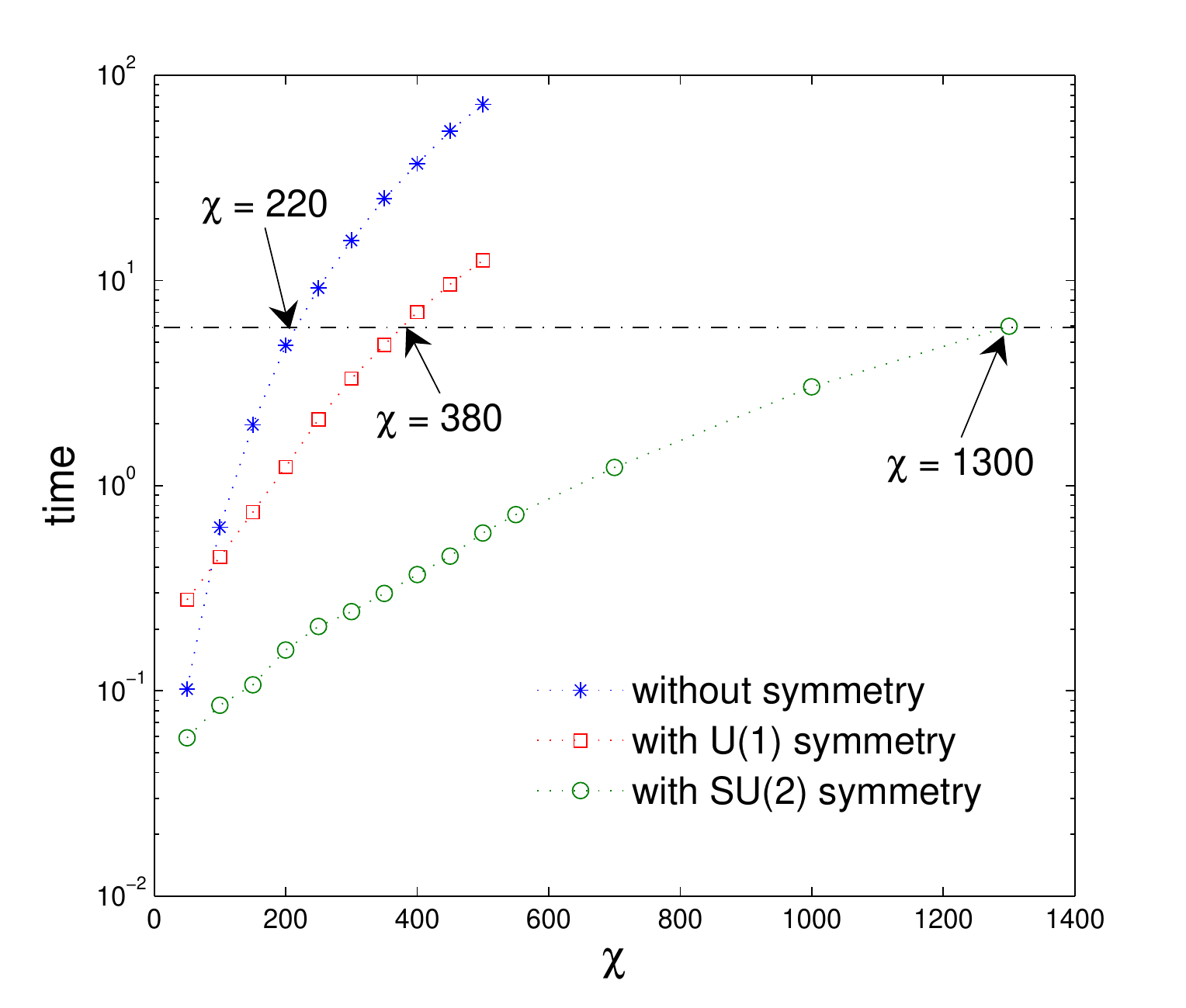}
  \end{center}
\caption{ Computational gain obtained by exploiting the symmetry in an MPS algorithm. Computation time (in seconds) for one iteration of the infinite TEBD algorithm, as a function of the MPS bond dimension $\chi$ is shown. Here $\chi$ is a refinement parameter, a larger $\chi$ leads to a better accuracy of the method. For sufficiently large $\chi$, exploiting symmetry leads to reductions in computation time. The horizontal line on this graph shows that this reduction in computation time equates to the ability to evaluate MPSs with a higher bond dimension $\chi$: For the same cost per iteration incurred when optimizing a regular MPS in MATLAB with bond dimension $\chi=220$, one may choose instead to optimize a U(1)-symmetric MPS with $\chi=380$ or an SU(2)-symmetric MPS with $\chi=1300$. \label{fig:mpscompare}}
\end{figure}
%%%%%%%%%%%%%%%%%%%%%%%%%%%%%%%%%%%%%%%%%%%%%%%%%%%%%%%%%%%%%%%%%%%%%%%%%%%%%%%%%%%%%%%%%%%%%%%%

Figure~\ref{fig:mpscompare} is demonstrative of the colossal computational gain that has been obtained by exploiting the symmetry in the context of the MPS. (In Fig.~\ref{fig:meracompare} we show an analogous comparison for exploiting symmetries in the context of the MERA.)

However, when considering symmetries, it is important to notice that an MPS is a trivalent tensor network. That is, in an MPS each tensor has at most three indices. The Clebsch-Gordan coefficients\citep{Cornwell97} (or coupling coefficients) of a symmetry group are also trivalent, and this makes incorporating the symmetry into an MPS by considering symmetric tensors particularly simple. In contrast, tensor network states with a more elaborate network of tensors, such as MERA or PEPS, consist of tensors having a larger number of indices. In this case a more general formalism is required in order to exploit the symmetry.

In this thesis we will describe how to incorporate a global internal symmetry, given by a compact and reducible group $\mathcal{G}$, into tensor network algorithms. We will develop a generic strategy that is independent of the details of the underlying tensor network. We will do this by imposing the symmetry constraints at the level of individual tensors that constitute the tensor network. We will then also describe how symmetric tensors are manipulated such that the symmetry is both \textit{preserved} and \textit{exploited} for computational gain. Having built a framework of symmetric tensors, we will adapt an arbitrary tensor network to the presence of symmetry by using symmetric tensors as building blocks for the tensor network. The resulting tensor network represents a class of quantum many-body wavefunctions that are invariant (or more generally covariant) under the symmetry transformation. Algorithms based on such symmetric tensor networks will also be adapted to the presence of symmetry. This will be achieved by expressing each step of an algorithm in terms of symmetric manipulations of the tensors. 

As a concrete illustration, we will extensively describe the implementation of U(1) and SU(2) symmetries into the MPS and MERA. With these implementations at hand, we will demonstrate the colossal benefits of incorporating the symmetry into tensor network algorithms. These include addressing specific symmetry sectors of the Hilbert space, compactification of the tensor network representation and computational speedup in numerical simulations. For example, in a lattice spin model endowed with spin isotropy the ground state is constrained to the spin zero or singlet sector of the Hilbert space. Therefore, in a numerical probing for the state, it is sufficient to restrict attention to the singlet subspace. This can, in turn, potentially result in a substantial reduction of computational costs.

\section{Plan of the thesis}

This thesis is comprised of three published papers corresponding to chapters 2,3 and 4 and additional chapters 5 and 6. The material in chapters 5 and 6 was under preparation for publication at the time of submitting this thesis. The following is a brief summary of all the chapters.

In Chapter 2 we dive straight into the core of the problem. We describe the implementation of a non-Abelian symmetry for the case of the simplest tensor network: the MPS. This will serve to illustrate the key points that are required to be considered when implementing symmetries into tensor network algorithms. In addition, this chapter also demonstrates the benefits of exploiting symmetries in the case of MPS algorithms. We adapt the infinite time evolving block decimation (iTEBD) algorithm to the presence of a global SU(2) symmetry. This is of interest in its own right since the iTEBD algorithm has been immensely successful in simulations of infinite 1D quantum many-body systems. This is also the first implementation of a non-Abelian symmetry into the iTEBD algorithm and has resulted in a significant enhancement of this algorithm.

In Chapter 3 we go beyond the class of MPS algorithms. We describe the general strategy to incorporate a wide spectrum of symmetries into more complex tensor network states and algorithms. We consider tensor networks made of symmetric tensors, that is, tensors that are invariant under the action of the symmetry. We develop a formalism to characterize and manipulate symmetric tensors.

In Chapter 4 we implement the general formalism for the case of an Abelian symmetry. The implementation of an Abelian symmetry is simplified by the fact that symmetric tensors are easier to characterize. In a basis labeled by the charges of the symmetry, a symmetric tensor has a sparse block structure. We explain how this block structure can be exploited for computational gain in a practical implementation of the symmetry. The benefits of exploiting the symmetry are numerically demonstrated by exploiting U(1) symmetry in the context of the MERA.

In Chapters 5 and 6 we describe the implementation of a non-Abelian symmetry. The details of implementing the symmetry are more involved, since the structural tensors are highly non-trivial. However, the computational gain that results from exploiting a non-Abelian symmetry is significantly larger than that obtained by exploiting an Abelian symmetry. Moreover, the practical scheme presented to implement a non-Abelian symmetry can be readily extended to incorporate more exotic symmetry constraints such as those corresponding to the presence of anyonic degrees of freedom. The benefits of exploiting a non-Abelian symmetry are numerically demonstrated by means of our implementation of SU(2) symmetry in the context of the MERA. 

Finally, in Chapter 7 we draw conclusions and discuss potential applications and future directions of this work.

%%%%%%%%%%%%%%%%%%%%%%%%%%%%%%%%%%%%%%%%%%%%%%%%%%%%%%%%%%%%%%%%%%%%%%%%%%%%%%%%%%%%%%%%%%%%%%
\textbf{Note on References:} In addition to the references listed at the end of the thesis, chapter wise references appear at the end of chapters 2, 3 and 4 that correspond to published papers.

%%%%%%%%%%%%%%%%%%%%%%%%%%%%%%%%%%%%%%%%%%%%%%%%%%%%%%%%%%%%%%%%%%%%%%%%%%%%%%%%%%%%%%%%%%%%%%
%*************************************************************************************************
\chapter{Exploiting symmetries in MPS algorithms: An Example\label{sec:ch2}}

We kick start the main discussion of the thesis by describing how to incorporate an SU(2) symmetry into a specific MPS algorithm: the iTEBD algorithm. The iTEBD algorithm has been immensely successful in simulations of infinite 1D quantum many-body systems. 

We follow a straightforward implementation of the symmetry into the algorithm. We consider an MPS that is made of trivalent SU(2) invariant tensors. A trivalent SU(2) invariant tensor decomposes into two pieces. One piece contains the degrees of freedom whereas the other corresponds to the Clebsch-Gordan coefficients of SU(2). We describe how the iTEBD algorithm can be enhanced by exploiting this decomposition of the MPS tensors. The resulting symmetric algorithm is obtained to be 300 times faster (See Fig.~\ref{fig:mpscompare}). We use the symmetric algorithm for a numerical study of a critical quantum spin chain.

This chapter serves to illustrate the main conceptual ingredients that are required to incorporate symmetries into tensor network algorithms. In the next chapter, we will generalize these ingredients by going beyond the specific details of the SU(2) symmetry group and the MPS representation. 

\includepdf[pages={2-13}]{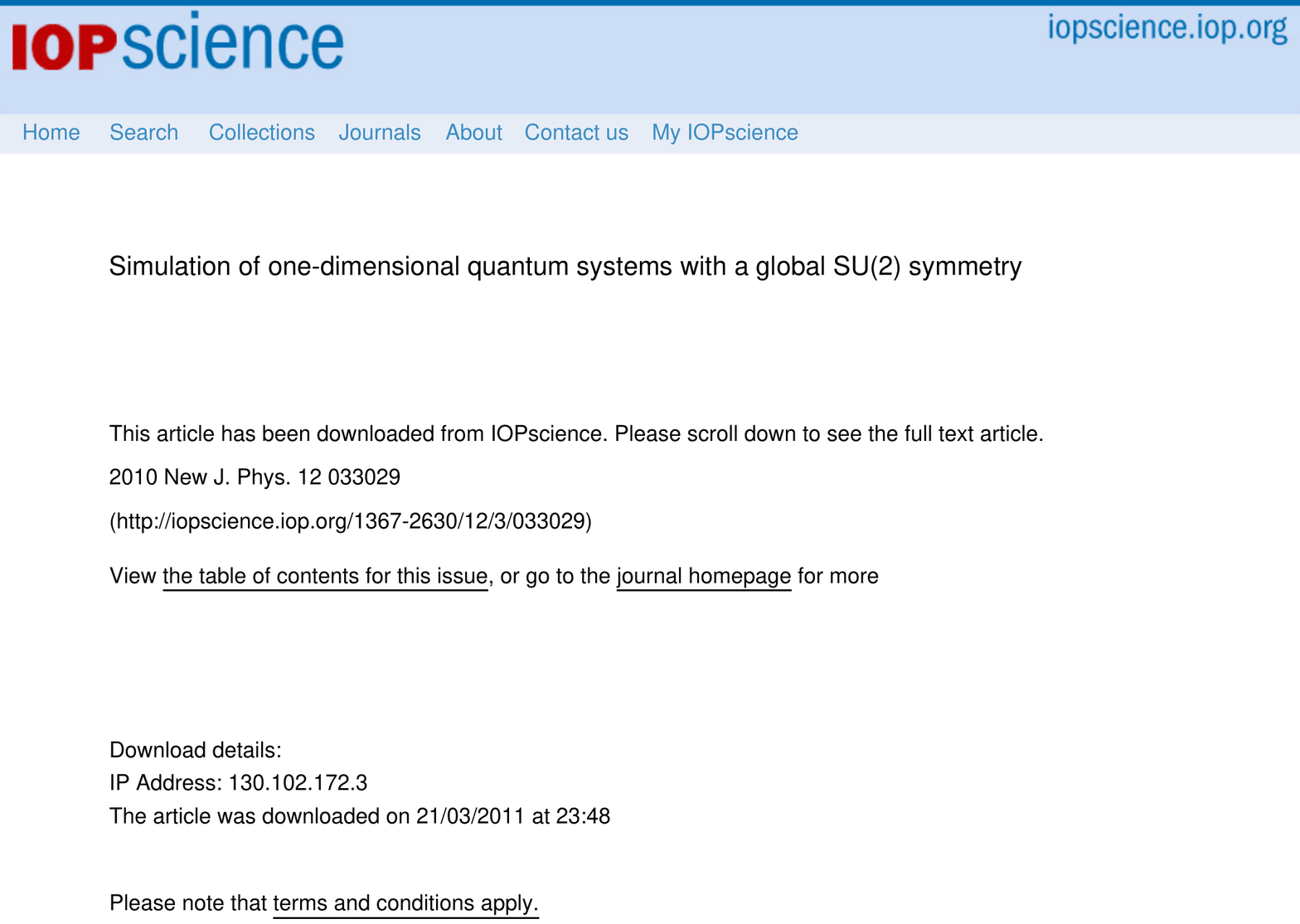}

%*************************************************************************************************

\chapter{Tensor networks and symmetries: Theoretical formalism\label{sec:ch3}}

In this chapter we develop a generic theoretical formalism to incorporate a symmetry into tensor network algorithms. We consider a wide class of symmetries that are described by a compact and reducible group $\mathcal{G}$ that is multiplicity free, that is, the tensor product of two charges of the group does not contain multiple copies of a charge. Our strategy revolves around tensors that are invariant under the action of the symmetry. As a result, we formulate a framework of symmetric tensors. 

A symmetric tensor transforms covariantly (or remains invariant) under the action of the symmetry. In a basis labeled by the symmetry charges, the tensor decomposes into a set of \textit{degeneracy tensors} and \textit{structural tensors}. While the degeneracy tensors contain the degrees of freedom, the components of the structural tensors are generalizations of the coupling coefficients of the group, and are determined completely by the symmetry. Moreover, any symmetry preserving manipulation of the tensor can be performed in parts. For instance, a permutation of the indices of a symmetric tensor breaks into the permutation of the corresponding degeneracy indices and the permutation of the corresponding structural indices. Therefore, this \textit{canonical decomposition} of a symmetric tensor allows for both a compact description of the tensor and a computational speedup in numerical manipulations of it.

We also point out a numerical connection to the formalism of \textit{spin networks}\citep{Penrose71, Major99}. A spin network is a mathematical object that appears, for example, in Loop Quantum Gravity\citep{Rovelli98}, where it is used\citep{Rovelli95} to facilitate a description of quantum spacetime. In our formalism, a tensor network made of symmetric tensors decomposes into a linear superposition of spin networks. Also, manipulating a symmetric tensor network requires \textit{evaluating} a spin network. Thus, our work highlights the importance of spin networks in the context of tensor network algorithms, thus setting the stage for cross-fertilization between these two areas of research.

\includepdf[pages={1-4}]{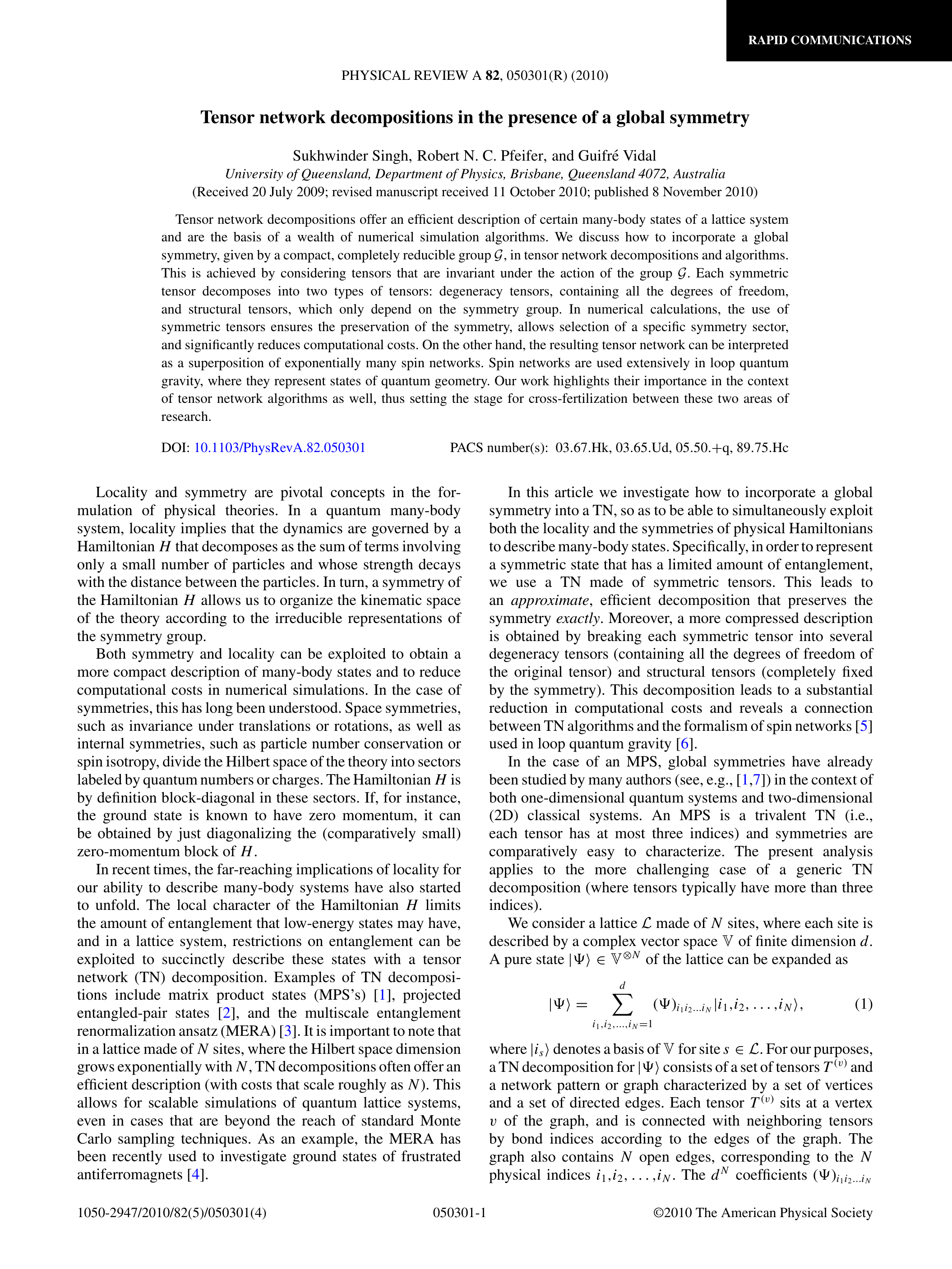}

\section{Errata}

The following equations appear erroneously in the publication. They are to be corrected as follows.

The tensors $P$ and $Q$ that appear in Eqs.11 and 12 do not carry degeneracy indices and spin indices corresponding to the coupled charges $e$ and $f$, since these indices are summed over in the description.

In Eq.~11 the components of $P$ and $Q$ read as $(P^{abcd}_e)_{\alpha_a \beta_b \gamma_c \delta_d}$ and $(Q^{abcd}_e)_{m_a n_b o_c p_d}$ respectively and the sum is only over different values of charge $e$. The corrected equation reads,
\begin{equation}
(T)_{ijkl} = \sum_e (P^{abcd}_e)_{\alpha_a \beta_b \gamma_c \delta_d} (Q^{abcd}_e)_{m_a n_b o_c p_d},\nonumber ~~~~~~~~~~~~~~~~~~(11)
\end{equation}

In Eq.~12 the components of $P$ and $Q$ read as $(\tilde{P}^{abcd}_f)_{\alpha_a \beta_b \gamma_c \delta_d}$ and $(\tilde{Q}^{abcd}_f)_{m_a n_b o_c p_d}$ respectively and the sum is only over different values of charge $f$. The corrected equation reads,
\begin{equation}
(T)_{ijkl} = \sum_f (\tilde{P}^{abcd}_f)_{\alpha_a \beta_b \gamma_c \delta_d} (\tilde{Q}^{abcd}_f)_{m_a n_b o_c p_d},\nonumber ~~~~~~~~~~~~~~~~~~(12)
\end{equation}

A similar correction holds for Eq.~15 which is a generalization of Eqs.11 and 12. The sum is only over different values of the intermediate charges $e_1 \ldots e_{t'}$. The corrected equation reads,
\begin{equation}
(T)_{i_1i_2 \ldots i_t} = \sum_{e_1 \ldots e_{t'}}(P^{a_1 \ldots a_t}_{e_1 \ldots e_{t'}})_{\alpha_{a_1} \ldots \alpha_{a_{t'}}} (Q^{a_1 \ldots a_t}_{e_1 \ldots e_{t'}})_{m{a_1} \ldots m_{a_{t'}}},\nonumber ~~~~~~~~~(15)
\end{equation}

%*************************************************************************************************

\chapter{Implementation of Abelian symmetries\label{sec:ch4}}

In this chapter we specialize the general formalism to the case of Abelian symmetries. Abelian symmetries appear frequently in the context of lattice models with particles (bosons or fermions) as well as those with spins. In the former, they include particle number conservation and parity conservation, whereas in the latter they appear as conservation of spin projection.

The analysis of exploiting an Abelian symmetry is made simple by the fact that the structural tensors in this case are trivial. On the other hand, an implementation of an Abelian symmetry serves to expose the practical difficulties that are encountered when incorporating symmetries into complicated tensor networks. We pay special attention to such implementation level concerns. Certain operations in the algorithm depend only on the symmetry and not on the components of the tensors involved. We exploit this fact to \textit{precompute} the output of such operations and store their result in memory. This is particularly advantageous in an \textit{iterative} algorithm where tensor components are updated or optimized by repeating a set of computations. The runtime cost of the iterative algorithm can be significantly reduced by reusing the precomputed results from memory. By making use of precomputation we obtained a substantial computational gain from exploiting the symmetry in our \textsc{MATLAB} implementation. However, this was achieved at the expense of storing potentially large amounts of precomputed data.

The discussion is conducted in the specific context of U(1) symmetry associated, for example, with conservation of particle number or of spin projection. We describe how to implement elementary tensor manipulations such as permutation and reshape of indices in a U(1) symmetric way. We also present a concrete implementation of the U(1) symmetry in the context of the MERA. We consider a MERA that is made of U(1) symmetric tensors. Then using the U(1) MERA we demonstrate the benefits of including symmetries into tensor networks.

\includepdf[pages={1-22}]{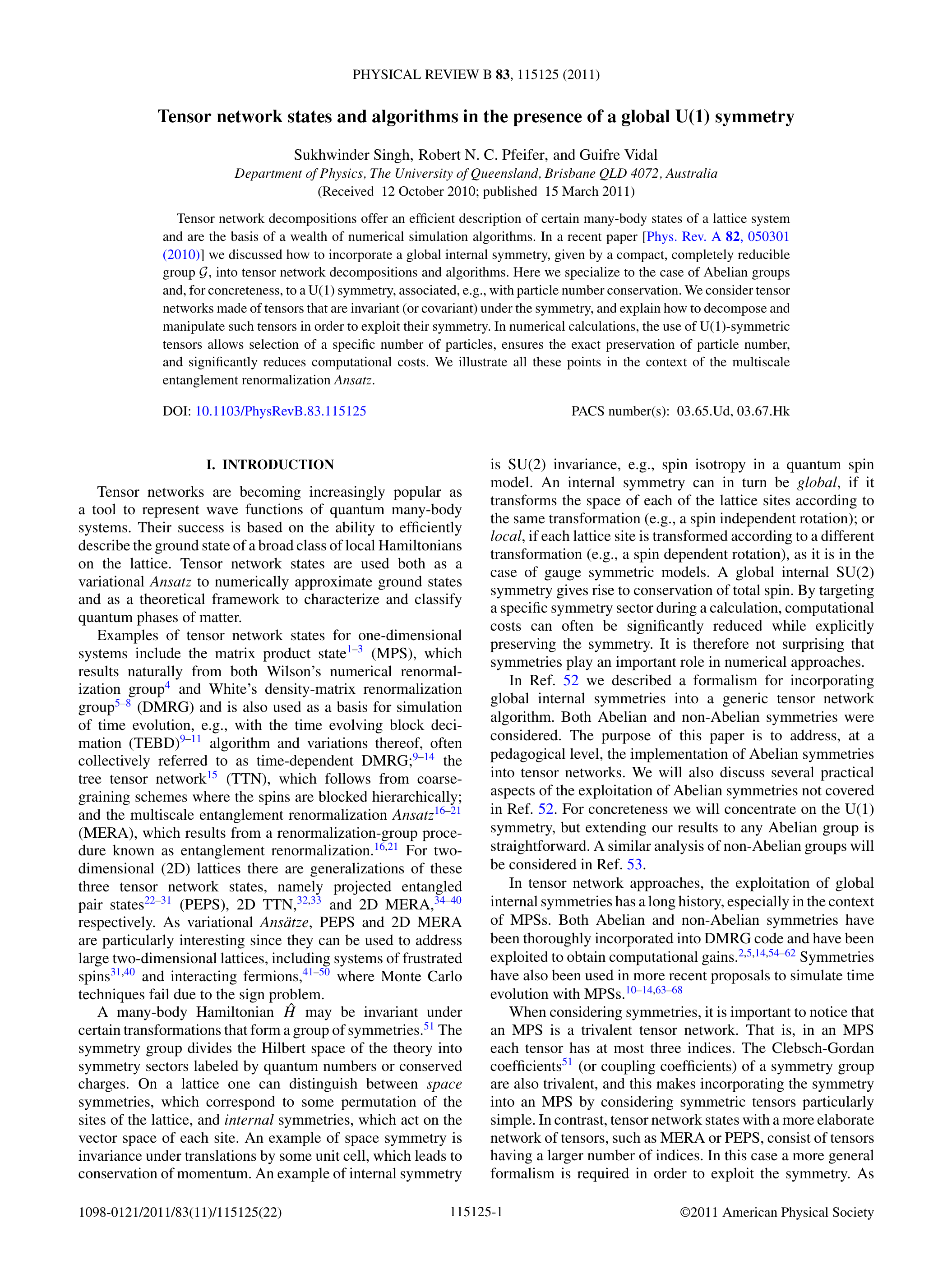}

%*************************************************************************************************

\chapter{Implementation of non-Abelian symmetries \Rmnum{1} \label{sec:ch5}}

In this and the following chapter we will address the implementation of global non-Abelian symmetries into tensor network algorithms. We consider the specific context of an internal SU(2) symmetry, that gives rise to spin isotropy. This is an extremely important symmetry that appears amply in lattice spin models.

In this chapter we will focus on the conceptual aspects of incorporating the symmetry. The theoretic formalism developed in Chapter 3 will be adapted to the specific case of SU(2) symmetry. We consider tensors that are invariant under the action of SU(2). The structural tensors, that are part of the canonical decomposition, are highly non-trivial (and are given in terms of the Clebsch-Gordan coefficients). However, the key advantage of the canonical decomposition is that it allows tensor manipulations, such as reshape or permutation of indices, to be broken into an independent manipulation of degeneracy tensors and of structural tensors. 

In the context of numerical simulations the canonical decomposition leads to a computational gain. Computational cost is incurred only when manipulating degeneracy tensors. One the other hand, structural tensors are manipulated \textit{algebraically} by exploiting properties of the Clebsch-Gordan coefficients. Manipulations of the structural tensors reduce to evaluating a spin network. This process involves integrating over the degrees of freedom associated with spin projection, which are therefore suppressed in the outcome of the tensor manipulation, as is expected in a spin isotropic description.

In Chapter 6 we will describe a specific implementation scheme of elementary manipulations of SU(2)-invariant tensors. We will address several concerns that are of importance in the practical implementations of the symmetry.

\section{Preparatory Review: Tensor network formalism \label{sec:tensornetwork}}

In this section we review the basic formalism of tensors and tensor networks. Even though we do not make any explicit reference to symmetry here, our formalism is directed towards SU(2)-invariant tensors.

We begin by recalling the basic notion of a tensor. A \textit{tensor} $\hat{T}$ is a multidimensional array of complex numbers $\hat{T}_{i_1i_2 \ldots i_k}$. The \textit{rank} of a tensor is the number $k$ of indices. The \textit{size} of an index $i$, denoted $|i|$, is the number of values that the index takes, $i \in {1, 2, \ldots, |i|}$. The \textit{size} of a tensor $\hat{T}$, denoted $|\hat{T}|$, is the number of complex numbers it contains, namely, $|\hat{T}|~=~|i_1|~\times~|i_2|~\times~\ldots~\times~|i_k|$. 

\subsection{Tensors as linear maps \label{sec:tensornetwork:tensors}}

For the purpose of this thesis we regard a rank-$k$ tensor as a linear map. To this end, we first equip each index $i_l,~l=1,2,\ldots,k$, of the tensor with a direction: `in' or `out', that is, either incoming into the tensor or outgoing from the tensor respectively. We denote by $\vec{D}$ the directions associated with the indices of tensor $\hat{T}$, namely, $\vec{D}(l) =$ `in' if $i_l$ is an incoming index and $\vec{D}(l) =$ `out' if $i_l$ is outgoing.

Let us also use index $i$ of the tensor to label a basis $\ket{i}$ of a complex vector space $\mathbb{V}^{[i]} \cong \mathbb{C}^{|i|}$ of dimension $|i|$. Then a rank-one ($k = 1$) tensor with an outgoing index $i$ represents a vector in $\mathbb{V}^{[i]}$, a rank-two ($k = 2$) tensor $\hat{T}$ with one incoming index $a$ and one outgoing index $b$ represents a matrix and so on. 

A tensor can be unambiguously regarded as a linear map from a vector space to complex numbers $\mathbb{C}$. For instance, a vector can be regarded as a linear map from $\mathbb{V}^{[i]}$ to $\mathbb{C}$, a matrix $\hat{T}$ can be regarded as a linear map from $(\mathbb{V}^{[a]})^* \otimes \mathbb{V}^{[b]}$ to $\mathbb{C}$ where $(\mathbb{V}^{[a]})^*$ is the dual of vector space $\mathbb{V}^{[a]}$ etc. More generally, we can use a rank-$k$ tensor $\hat{T}$ to define a linear map from the tensor product of $k$ vector spaces to $\mathbb{C}$ in the following way. Define a set $\mathbb{W}^{[i_l]},~l=1,2,\ldots,k$, of $k$ spaces where
\begin{equation}
 \mathbb{W}^{[i_l]} = \left\{ 
	\begin{array}{cc} \mathbb{V}^{[i_l]} &\mbox{ if } \vec{D}(l)=\mbox{`out'},\\ 
	 									(\mathbb{V}^{[i_l]})^* &\mbox{if } \vec{D}(l)=\mbox{`in'},
	\end{array} \right.
\end{equation}
where the $(\mathbb{V}^{[i_l]})^*$ is the dual of vector space $\mathbb{V}^{[i_l]}$. Then tensor $\hat{T}$ can be regarded as a linear map from the product space $\bigotimes_l  \mathbb{W}^{[i_l]}$ to $\mathbb{C}$,
\begin{equation}
\hat{T} : \bigotimes_l  \mathbb{W}^{[i_l]} \rightarrow \mathbb{C}. \label{eq:tensormap}
\end{equation}
We will find this viewpoint of tensors useful for subsequent generalization to SU(2)-invariant tensors.
%%%%%%%%%%%%%%%%%%%%%%%%%%%%%%%%%%%%%%%%%%%%%%%%%%%%%%%%%%%%%%%%%%%%%%%%%%%%%%%%%%%%%%%%%%%%%%%%
\begin{figure}[t]
\begin{center}
  \includegraphics[width=8cm]{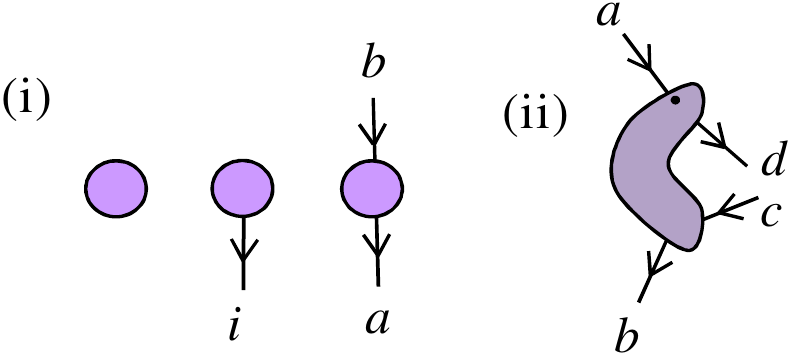}
  \end{center}
\caption{ Graphical representation of tensors by means of a shape e.g. circle or a ``blob'' with emanating lines corresponding to indices of the tensor. Indices may be incoming or outgoing as indicated by arrows. (i) Graphical representations of tensors with rank $0,1$ and $2$, corresponding to a complex number $c \in \mathbb{C}$, a vector $\ket{v} \in \mathbb{C}^{[i]}$ and a matrix $\hat{M} \in \mathbb{C}^{|a|\times|b|}$. (ii) Graphical representation of a tensor $\hat{T}$ with components $\hat{T}_{abcd}$ with directions $\vec{D} = \{\mbox{`in', `out', `in', `out'}\}$. Notice that the indices emerge in a counterclockwise order, perpendicular to the boundary of the blob; the first index is identified by the presence of by a dot within the blob close to the index. All arrows in a tensor diagram point downwards, that may be interpreted as choosing a metaphorical ``time'' that flows downwards.\label{fig:tensor}} 
\end{figure}
%%%%%%%%%%%%%%%%%%%%%%%%%%%%%%%%%%%%%%%%%%%%%%%%%%%%%%%%%%%%%%%%%%%%%%%%%%%%%%%%%%%%%%%%%%%%%%%%
%%%%%%%%%%%%%%%%%%%%%%%%%%%%%%%%%%%%%%%%%%%%%%%%%%%%%%%%%%%%%%%%%%%%%%%%%%%%%%%%%%%%%%%%%%%%%%%%
%\begin{figure}[t]
%\begin{center}
%  \includegraphics[width=8cm]{tensor}
%\end{center}  
%\caption{Graphical representation of tensors by means of a shape e.g. circle or a ``blob'' with emanating lines corresponding to indices of the tensor. Indices may be incoming or outgoing as indicated by arrows. (i) Graphical representations of tensors with rank $0,1$ and $2$, corresponding to a complex number $c \in \mathbb{C}$, a vector $\ket{v} \in \mathbb{C}^{[i]}$ and a matrix $\hat{M} \in \mathbb{C}^{|a|\times|b|}$. (ii) Graphical representation of a tensor $\hat{T}$ with components $\hat{T}_{abcd}$ with directions $\vec{D} = \{\mbox{`in', `out', `in', `out'}\}$. Notice that the indices emerge in a counterclockwise order, perpendicular to the boundary of the blob; the first index is identified as the one emerging closest to a mark (dot) inside the boundary of the blob. We impose that all arrows in a tensor diagram point downwards. \label{fig:tensor}} 
%\end{figure}
%%%%%%%%%%%%%%%%%%%%%%%%%%%%%%%%%%%%%%%%%%%%%%%%%%%%%%%%%%%%%%%%%%%%%%%%%%%%%%%%%%%%%%%%%%%%%%%%

It is convenient to use a graphical representation of tensors, as illustrated in Fig. 1, where a tensor is depicted as a ``blob'' (or by a shape e.g., circle, square etc.) and each of its indices is represented by a line emerging \textit{perpendicular} from the boundary of the blob. In order to specify which index corresponds to which emerging line, we follow the prescription that the lines corresponding to indices $\{i_1, i_2, \ldots , i_k\}$ emerge in counterclockwise order. The first index corresponds to the line emerging closest to a mark (black dot) inside the boundary of the blob (or the first line encountered while proceeding counterclockwise from nine o'clock in case the tensor is depicted as a circle without a mark). The direction of an index is depicted by attaching an arrow to the line corresponding to the index. We follow a convention that all arrows in a diagram point downwards.

\subsection{Elementary manipulations of a tensor\label{sec:tensor:manipulations}}

A tensor can be transformed into another tensor in several elementary ways. These include, \textit{reversing} the direction of one or several of its indices, \textit{permuting} its indices, and/or \textit{reshaping} its indices.

\begin{figure}[t]
\begin{center}
  \includegraphics[width=10cm]{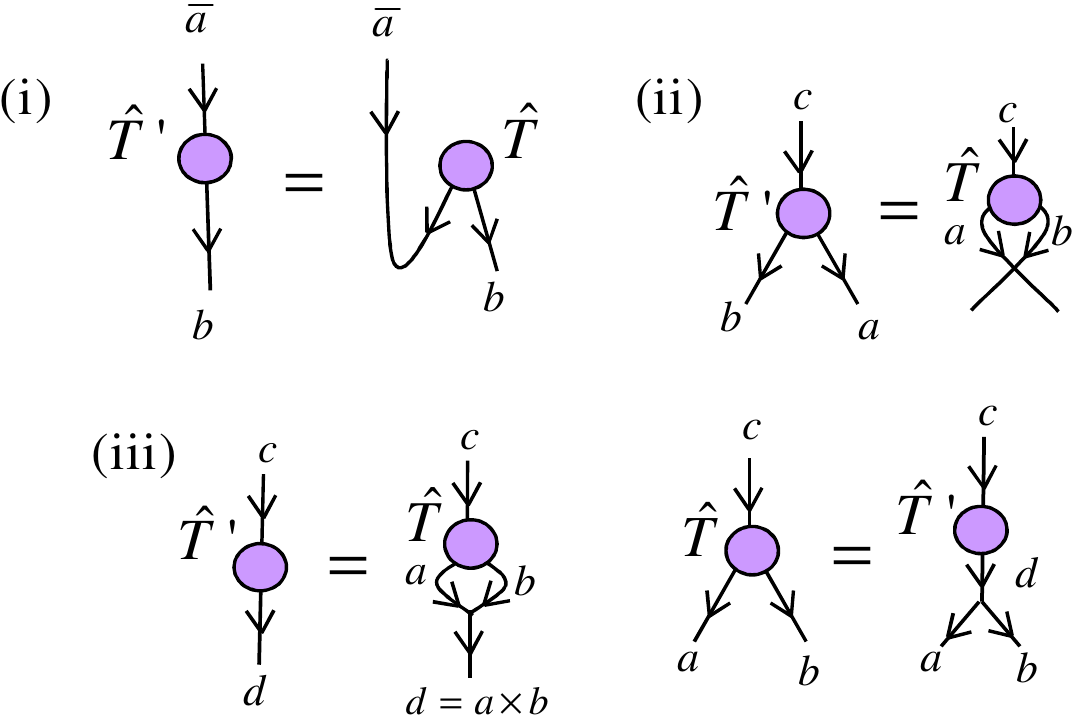}
\end{center}  
\caption{Transformations of a tensor: (i) Reversing direction of an index, Eq.~(\ref{eq:bend}). $(ii)$ Permutation of indices,  Eq.~(\ref{eq:permute}). (ii) Fusion of indices $a$ and $b$ into $d = a \times b$, Eq.~(\ref{eq:fuse}); splitting of index $d=a \times b$ into $a$ and $b$, Eq.~(\ref{eq:split}).\label{fig:tensorman}} 
\end{figure}

\textit{Reversing} the direction of an index corresponds to mapping the vector space that is associated with the index to its dual. For example, in 
\begin{equation}
	(\hat{T}')_{\overline{a}b} = \hat{T}_{ab},
	\label{eq:bend}
\end{equation}
if index $a$ is associated to a vector space $\mathbb{V}^{[a]}$, then index $\overline{a}$ that is obtained by reversing the direction of $a$ is associated with the dual space $(\mathbb{V}^{[a]})^*$. Since all arrows in a diagram point downwards, reversing the direction of an index $i$ is depicted [Fig.~\ref{fig:tensorman}(i)] by `bending' the line corresponding to $i$ upwards if it is an outgoing index or downwards if it is an incoming index. Since tensor $\hat{T}'$ is components wise equal to tensor $\hat{T}$ arrows appear to be irrelevant in the absence of the symmetry. However, arrows will play an important role when we consider SU(2)-invariant tensors since they specify how the group acts on each index of a given tensor.
%%%%%%%%%%%%%%%%%%%%%%%%%%%%%%%%%%%%%%%%%%%%%%%%%%%%%%%%%%%%%%%%%%%%%%%%%%%%%%%%%%%%%%%%%%%%%%%%
%%%%%%%%%%%%%%%%%%%%%%%%%%%%%%%%%%%%%%%%%%%%%%%%%%%%%%%%%%%%%%%%%%%%%%%%%%%%%%%%%%%%%%%%%%%%%%%%

A \textit{permutation} of indices corresponds to creating a new tensor $\hat{T}'$ from $\hat{T}$ by simply changing the order in which the indices appear, e.g.
\begin{equation}
	(\hat{T}')_{bac} = \hat{T}_{abc}.
	\label{eq:permute}
\end{equation}
Permutation of indices is depicted by intercrossing indices, as illustrated in Fig.~\ref{fig:tensorman}(ii). Note that when the permutation involves the first index of the tensor the mark, that indicates the first index, is also shifted to a new location within the blob. It is useful to note that an arbitrary permutation of the indices can be broken into a sequence of \textit{swaps} of adjacent indices wherein the position of two indices are interchanged at a time. 

Last but not the least, a tensor $\hat{T}$ can be \textit{reshaped} into a new tensor $\hat{T}'$ by `fusing' two adjacent indices into a single index and/or `splitting' an index into two indices. For instance, in 
\begin{eqnarray}
	(\hat{T}')_{dc} = \hat{T}_{abc},~~~~~~~d = a\times b,
	\label{eq:fuse}
\end{eqnarray}
tensor $\hat{T}'$ is obtained from tensor $\hat{T}$ by fusing indices $a \in \left\{1, \cdots, |a|\right\}$ and $b \in \left\{1, \cdots, |b|\right\}$ together into a single index $d$ of size $|d| = |a| \cdot |b|$ that runs over all pairs of values of $a$ and $b$, i.e.  $ d \in \left\{ (1,1), (1,2), \cdots, (|a|, |b|-1), (|a|,|b|) \right\}$, whereas in
\begin{eqnarray}
	\hat{T}_{abc} = (\hat{T}')_{dc},~~~~~~~d = a\times b,
	\label{eq:split}
\end{eqnarray}
tensor $\hat{T}$ is recovered from $\hat{T}'$ by splitting index $d$ of $\hat{T}'$ back into indices $a$ and $b$. Reshape of the indices is depicted as shown in Fig.~\ref{fig:tensorman}(iii). 

\subsection{Multiplication of two tensors\label{sec:tensor:multiply}}

\begin{figure}[t]
\begin{center}
  \includegraphics[width=10cm]{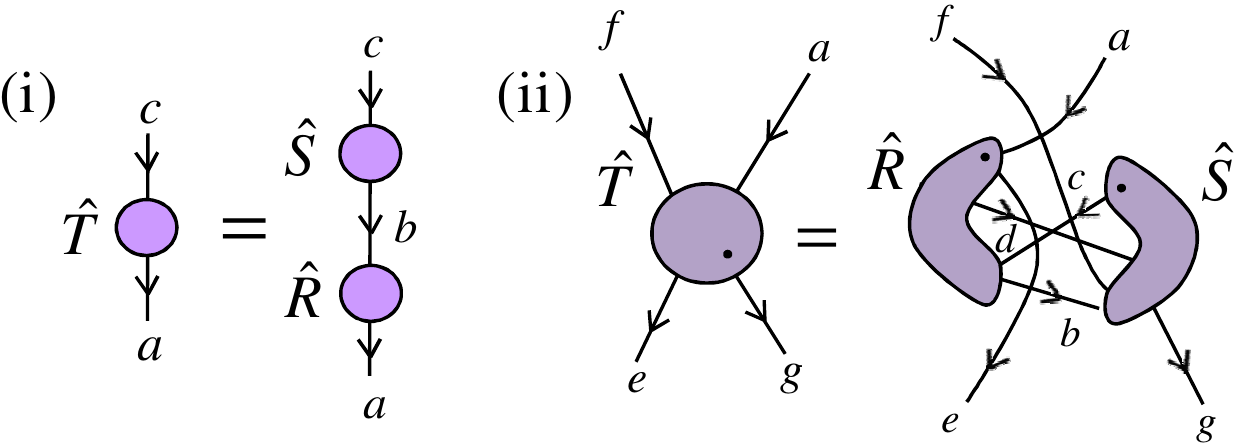}
\end{center}  
\caption{(i) Graphical representation of the matrix multiplication of two matrices $\hat{R}$ and $\hat{S}$ into a new matrix $\hat{T}$, Eq.~(\ref{eq:Mmultiply}). (ii) Graphical representation of an example of the multiplication of two tensors $\hat{R}$ and $\hat{S}$ into a new tensor $\hat{T}$, Eq.~(\ref{eq:tensormult}). \label{fig:multiply1}} 
\end{figure}

Given two matrices $\hat{R}$ and $\hat{S}$ with components $\hat{R}_{ab}$ and $\hat{S}_{bc}$, we can multiply them together to obtain a new matrix $\hat{T}$, $\hat{T} = \hat{R}\cdot \hat{S}$, with components
\begin{equation}
	\hat{T}_{ac} = \sum_{b} \hat{R}_{ab}\hat{S}_{bc},
	\label{eq:Mmultiply}
\end{equation}
by summing over or \textit{contracting} index $b$. The multiplication of matrices $\hat{R}$ and $\hat{S}$ is represented graphically by connecting together the emerging lines of $\hat{R}$ and $\hat{S}$ corresponding to the contracted index, as shown in Fig.~\ref{fig:multiply1}(i).

Matrix multiplication can be generalized to tensors, such that, an incoming index of one tensor is identified and contracted with an outgoing index of another. For instance, given tensor $\hat{R}$ with components $\hat{R}_{abcde}$ and directions $\{\mbox{`in', `out', `in', `out', `out'}\}$, and tensor $\hat{S}$ with components $\hat{S}_{cdfbg}$ and directions $\{\mbox{`out', `in', `in', `in', `out'}\}$, we can define a tensor $\hat{T}$ with components $\hat{T}_{gafe}$ that are given by,
\begin{equation}
	\hat{T}_{afge} = \sum_{bcd} \hat{R}_{abcde}\hat{S}_{cdfbg}.
\label{eq:tensormult}
\end{equation}
Note that each of the indices $b, c$ and $d$, that are contracted, is incoming into one tensor and outgoing from the other. The multiplication is represented graphically by connecting together the lines emerging from $\hat{R}$ and $\hat{S}$ corresponding to each of these indices, as shown in Fig.~\ref{fig:multiply1}(ii).

%%%%%%%%%%%%%%%%%%%%%%%%%%%%%%%%%%%%%%%%%%%%%%%%%%%%%%%%%%%%%%%%%%%%%%%%%%%%%%%%%%%%%%%%%%%%%%%%
%%%%%%%%%%%%%%%%%%%%%%%%%%%%%%%%%%%%%%%%%%%%%%%%%%%%%%%%%%%%%%%%%%%%%%%%%%%%%%%%%%%%%%%%%%%%%%%%

\begin{figure}[t]
\begin{center}
  \includegraphics[width=10cm]{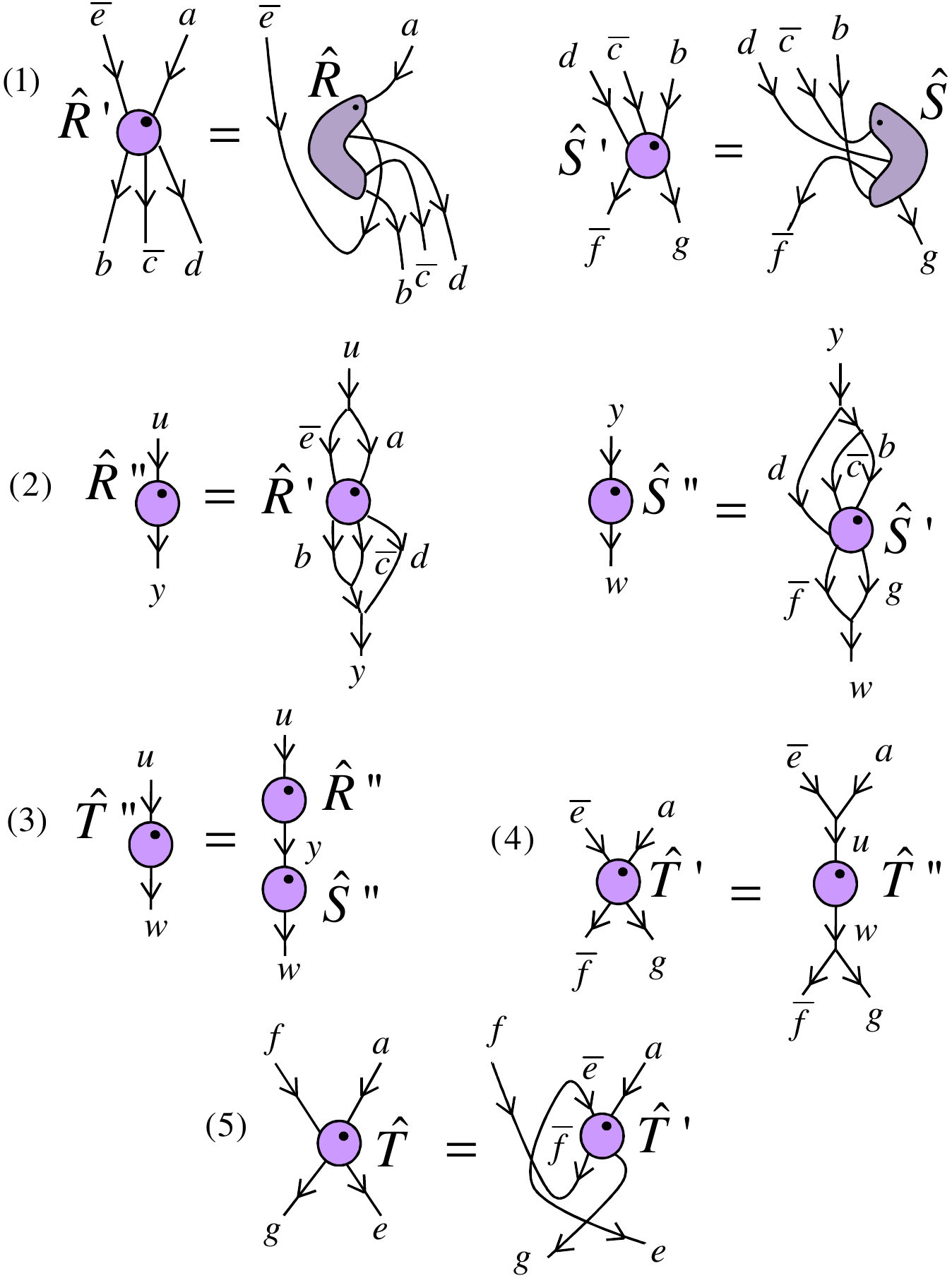}
\end{center}  
\caption{Graphical representations of the five elementary steps 1-5 into which one can decompose the multiplication of the tensors of Eq.~(\ref{eq:tensormult}).\label{fig:multiply2}} 
\end{figure}

Multiplication of two tensors can be broken down into a sequence of elementary steps by transforming the tensors into matrices, multiplying the matrices together, and then transforming the resulting matrix back into a tensor. Next we describe these steps for the contraction given in Eq.~(\ref{eq:tensormult}). They are illustrated in Fig.~\ref{fig:multiply2}. 

%%%%%%%%%%%%%%%%%%%%%%%%%%%%%%%%%%%%%%%%%%%%%%%%%%%%%%%%%%%%%%%%%%%%%%%%%%%%%%%%%%%%%%%%%%%%%%%%
%%%%%%%%%%%%%%%%%%%%%%%%%%%%%%%%%%%%%%%%%%%%%%%%%%%%%%%%%%%%%%%%%%%%%%%%%%%%%%%%%%%%%%%%%%%%%%%%
\begin{enumerate}
 
	\item \textit{Reverse} and \textit{Permute} the indices of tensor $\hat{R}$ in such a way that the indices $b, c$ and $d$ that are contracted appear in the last positions as \textit{outgoing} indices and in a given order, e.g. $b\overline{c}d$, and the remaining indices $a$ and $e$ appear in the first positions as \textit{incoming} indices; similarly reverse and permute the indices of $\hat{S}$ so that the indices $b, c$ and $d$ appear in the first positions as \textit{incoming} indices and in the same order, $\overline{b}c\overline{d}$, and the remaining indices $\overline{f}$ and $g$ appear in the last positions as \textit{outgoing} indices,
	\begin{align}
	(\hat{R}')_{a\overline{e} ~b\overline{c}d} &= (\hat{R})_{abcde},   \nonumber \\
	(\hat{S}')_{\overline{b}c\overline{d} ~\overline{f}g} &= (\hat{S})_{cdfbg}. \label{eq:multi1}
	\end{align}
		
	\item \textit{Reshape} tensor $\hat{R}'$ into a matrix $\hat{R}''$ by fusing into a single index $u$ all the indices that are not contracted, $u = a\times \overline{e}$, and into a single index $y$ all indices that are contracted, $y = b \times \overline{c} \times d$; similarly reshape tensor $\hat{S}'$ into a matrix $\hat{S}''$ with indices $\overline{y} = \overline{b} \times \overline{c} \times d$ and $w = \overline{f}\times g$ (indices $b, c$ and $d$ are required to be fused according to the \textit{same} fusion sequence in the two tensors. A possible fusion sequence may involve, for example, first fusing $b$ and $c$ and then fusing the resulting index with $d$),
		\begin{align}
		(\hat{R}'')_{uy} &= (\hat{R}')_{a\overline{e}b\overline{c}d}, \nonumber \\
		(\hat{S}'')_{yw} &= (\hat{S}')_{b\overline{c}d\overline{f}g}. \label{eq:multi2}
	\end{align}
	
	\item \textit{Multiply} matrices $\hat{R}''$ and $\hat{S}''$ to obtain a matrix $\hat{T}'$ with components
	\begin{equation}
	(\hat{T}'')_{uw} = \sum_{y} (\hat{R}'')_{uy} ~~(\hat{S}'')_{y w}. \label{eq:multi3}
	\end{equation}
	
	\item \textit{Reshape} matrix $\hat{T}'$ into a tensor $\hat{T}$ by splitting indices $u = a\times \overline{e}$ and $w = \overline{f}\times g$, 
		\begin{equation}
	(\hat{T'})_{aefg} = (\hat{T}'')_{uw}.  \label{eq:multi4}
	\end{equation}

	\item \textit{Reverse} and \textit{Permute} indices of tensor $\hat{T}'$ in the order in which they appear in $\hat{T}$,
	\begin{equation}
	\hat{T}_{afge} = (\hat{T}')_{a\overline{e}\overline{f}g}.  \label{eq:multi5}
	\end{equation}
\end{enumerate}
The contraction of Eq.~(\ref{eq:tensormult}) can be implemented at once, without breaking the multiplication down into elementary steps. However, it is often more convenient to compose the above elementary steps since, for instance, in this way one can use existing linear algebra libraries for matrix multiplication. In addition, it can be seen that the leading computational cost in multiplying two large tensors is not changed when decomposing the contraction in the above steps.

\subsection{Factorization of a tensor\label{sec:tensor:factorize}}

\begin{figure}[t]
\begin{center}
  \includegraphics[width=10cm]{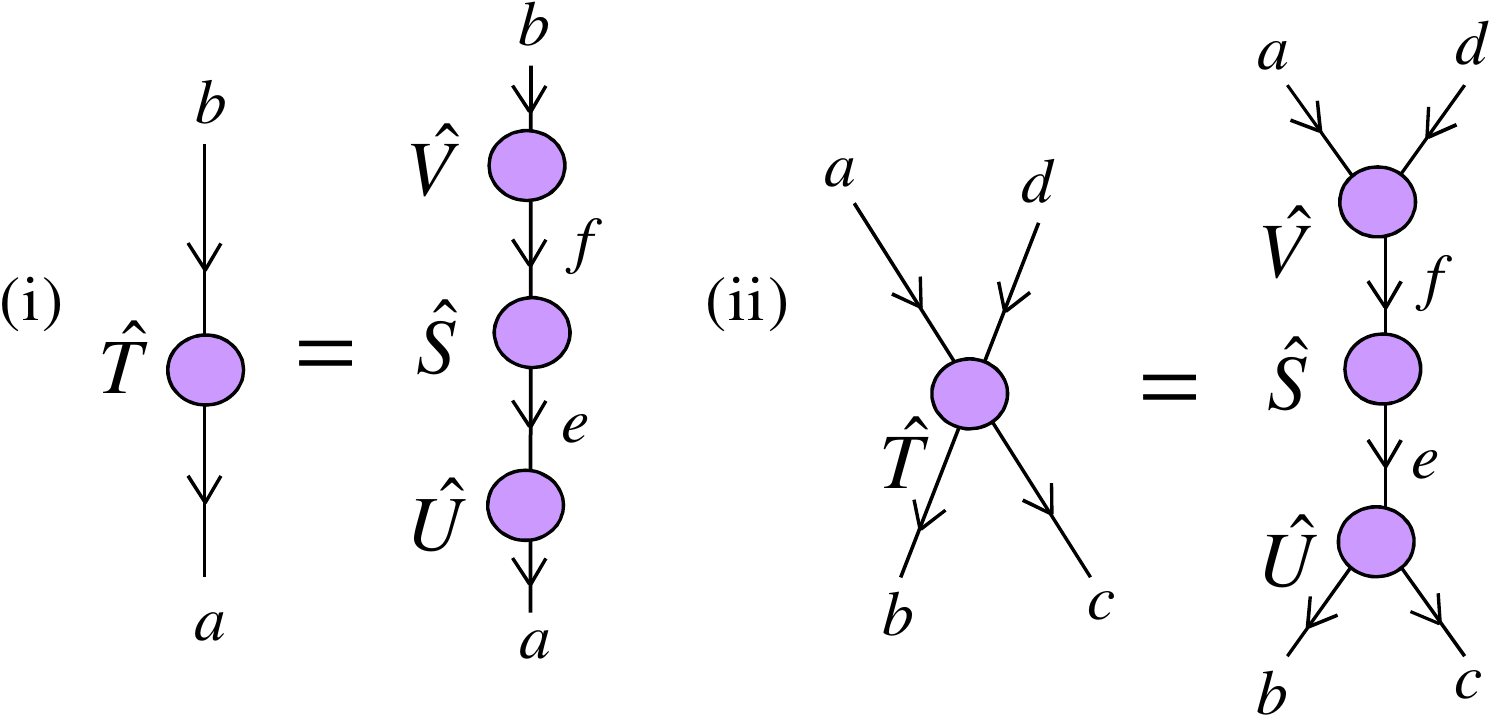}
\end{center}
\caption{(i) Factorization of a matrix $\hat{T}$ according to a singular value decomposition, Eq.~(\ref{eq:singular}). (ii) Factorization of a rank-4 tensor $\hat{T}$ according to one of several possible singular value decompositions. \label{fig:decompose}} 
\end{figure}

A matrix $\hat{T}$ can be factorized into the product of two (or more) matrices in one of several canonical forms. For instance, the \textit{singular value decomposition}
\begin{equation}
	\hat{T}_{ab} = \sum_{c,d} \hat{U}_{ac}\hat{S}_{cd}\hat{V}_{db} 
	= \sum_{c} \hat{U}_{ac}s_{c}\hat{V}_{cb}
	\label{eq:singular}
\end{equation}
factorizes $\hat{T}$ into the product of two unitary matrices $\hat{U}$ and $\hat{V}$, and a diagonal matrix $\hat{S}$ with non-negative diagonal elements $s_c = \hat{S}_{cc}$ known as the \textit{singular values} of $\hat{T}$ [Fig.~\ref{fig:decompose}(i)].
%%%%%%%%%%%%%%%%%%%%%%%%%%%%%%%%%%%%%%%%%%%%%%%%%%%%%%%%%%%%%%%%%%%%%%%%%%%%%%%%%%%%%%%%%%%%%%%%
%%%%%%%%%%%%%%%%%%%%%%%%%%%%%%%%%%%%%%%%%%%%%%%%%%%%%%%%%%%%%%%%%%%%%%%%%%%%%%%%%%%%%%%%%%%%%%%%
 On the other hand, the \textit{eigenvalue} or \textit{spectral decomposition} of a square matrix $\hat{T}$ is of the form
\begin{equation}
	\hat{T}_{ab} = \sum_{c,d} \hat{M}_{ac}D_{cd}(\hat{M}^{-1})_{db} 
	= \sum_{c} \hat{M}_{ac}\lambda_{c}(\hat{M}^{-1})_{cb}
	\label{eq:spectral}
\end{equation}
where $\hat{M}$ is an invertible matrix whose columns encode the eigenvectors $\ket{\lambda_c}$ of $\hat{T}$, 
\begin{equation}
	\hat{T} \ket{\lambda_{c}} = \lambda_c \ket{\lambda_c},
\end{equation}
$\hat{M}^{-1}$ is the inverse of $\hat{M}$, and $\hat{D}$ is a diagonal matrix, with the eigenvalues $\lambda_c=\hat{D}_{cc}$ on its diagonal. Other useful factorizations include the LU decomposition, the QR decomposition, etc. We refer to any such decomposition generically as a \textit{matrix factorization}.

A tensor $\hat{T}$ with more than two indices can be converted into a matrix in several ways by specifying how to join its indices into two subsets. After specifying how tensor $\hat{T}$ is to be regarded as a matrix, we can factorize $\hat{T}$ according to any of the above matrix factorizations, as illustrated in Fig.~\ref{fig:decompose}(ii) for a singular value decomposition. This requires first reversing directions, permuting and reshaping the indices of $\hat{T}$ to form a matrix, then decomposing the latter, and finally restoring the open indices of the resulting matrices into their original form by undoing the reshapes, permutations and reversal of directions.

%%%%%%%%%%%
\subsection{Tensor networks and their manipulation\label{sec:tensor:TN}}

\begin{figure}[t]
\begin{center}
  \includegraphics[width=10cm]{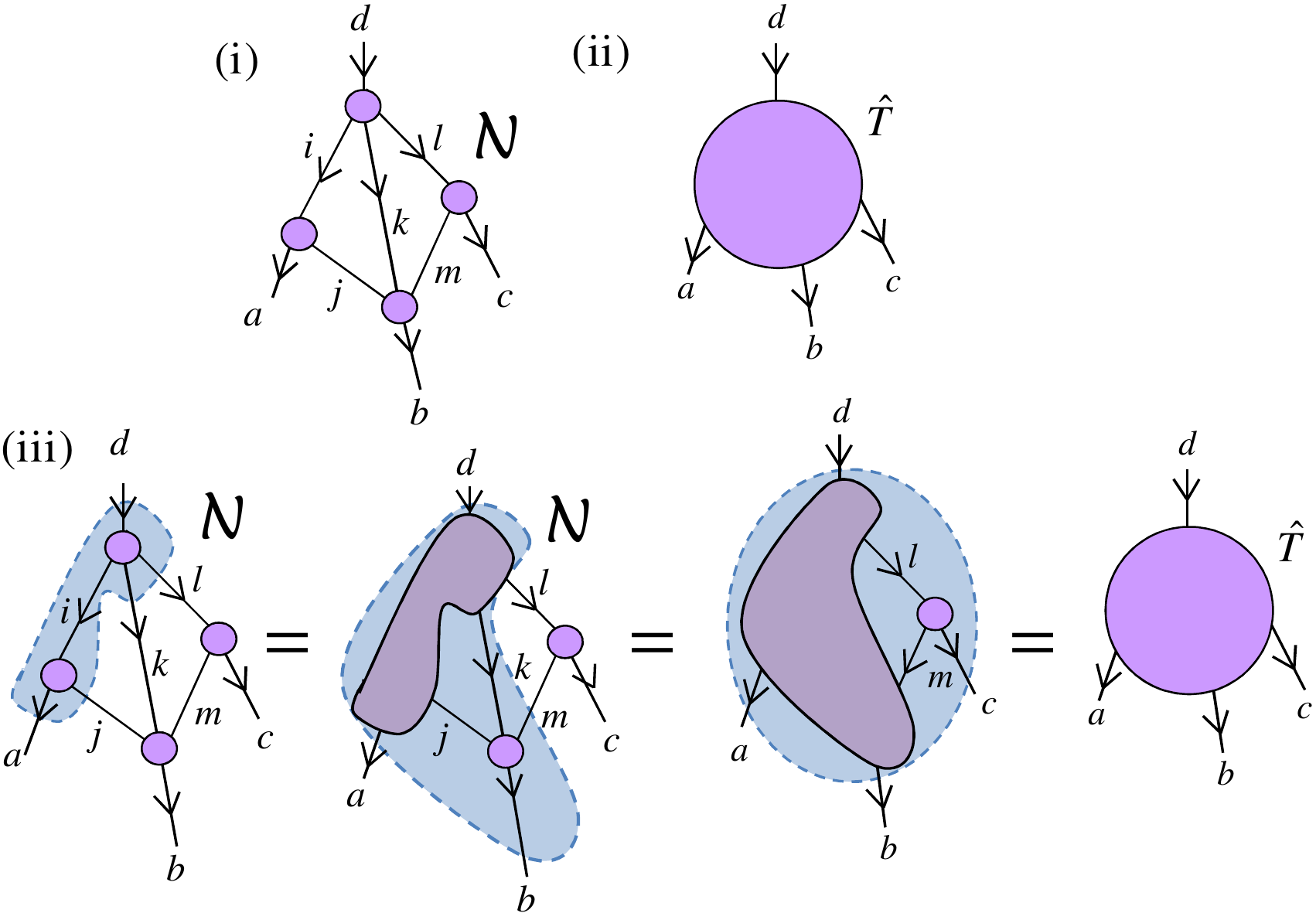}
\end{center}  
\caption{(i) Example of a tensor network $\mathcal{N}$. (ii) Tensor $\hat{T}$ of which the tensor network $\mathcal{N}$ could be a representation. (iii) Tensor $\hat{T}$ can be obtained from $\mathcal{N}$ through a sequence of contractions of pairs of tensors. Shading indicates the two tensors to be multiplied together at each step. The product tensor is depicted by a blob that covers the two tensors that are multiplied.\label{fig:TN}}
\end{figure}

A \textit{tensor network} $\mathcal{N}$ is a set of tensors whose indices are connected according to a network pattern, e.g. Fig.~\ref{fig:TN}. 

%%%%%%%%%%%%%%%%%%%%%%%%%%%%%%%%%%%%%%%%%%%%%%%%%%%%%%%%%%%%%%%%%%%%%%%%%%%%%%%%%%%%%%%%%%%%%%%%
%%%%%%%%%%%%%%%%%%%%%%%%%%%%%%%%%%%%%%%%%%%%%%%%%%%%%%%%%%%%%%%%%%%%%%%%%%%%%%%%%%%%%%%%%%%%%%%%

Given a tensor network $\mathcal{N}$, a single tensor $\hat{T}$ can be obtained by contracting all the indices that connect the tensors in $\mathcal{N}$ [\fref{fig:TN}(ii)]. Here, the indices of tensor $\hat{T}$ correspond to the open indices of the tensor network $\mathcal{N}$. We then say that the tensor network $\mathcal{N}$ is a tensor network decomposition of $\hat{T}$. One way to obtain $\hat{T}$ from $\mathcal{N}$ is through a sequence of contractions involving two tensors at a time [Fig.~\ref{fig:TN}(iii)]. Notice how a tensor that is obtained by contracting a region of a tensor network is conveniently depicted by a blob or shape that covers that region.

From a tensor network decomposition $\mathcal{N}$ for a tensor $\hat{T}$, another tensor network decomposition for the same tensor $\hat{T}$ can be obtained in many ways. One possibility is to replace two tensors in $\mathcal{N}$ with the tensor resulting from contracting them together, as is done in each step of Fig.~\ref{fig:TN}(iii). Another way is to replace a tensor in $\mathcal{N}$ with a decomposition of that tensor (e.g. with a singular value decomposition). In this thesis, we will be concerned with manipulations of a tensor network that, as in the case of multiplying two tensors or decomposing a tensor, can be broken down into a sequence of operations from the following list:
\begin{enumerate}
	\item Reversal of direction of indices of a tensor, Eq.~(\ref{eq:bend}).
	\item Permutation of the indices of a tensor, Eq.~(\ref{eq:permute}).
	\item Reshape of the indices of a tensor, Eqs.~(\ref{eq:fuse})-(\ref{eq:split}).
	\item Multiplication of two matrices, Eq.~(\ref{eq:Mmultiply}).
	\item Factorization of a matrix (e.g. singular value decomposition Eq.~(\ref{eq:singular}) or spectral decomposition Eq.~(\ref{eq:spectral}).
\end{enumerate}
These operations constitute a set $\mathcal{P}$ of \textit{primitive} operations for tensor network manipulations (or, at least, for the type of manipulations we will be concerned with). In Sec.~\ref{sec:blockmoves} (and again in Chapter 6) we discuss how this set $\mathcal{P}$ of primitive operations can be generalized to tensors that are invariant under the action of the group SU(2).

Next we review basic background material concerning the representation theory of the group SU(2) without reference to tensor network states and algorithms. This review is distributed over Sections \ref{sec:symmetry}, \ref{sec:su2lattice} and \ref{sec:fusiontree}. We refer the reader to \citep{Cornwell97} and Chapters 3 and 4 of \citep{Sakurai} for additional supporting material for these sections.

\section {Representations of the group SU(2)}
\label{sec:symmetry}

In this section we consider the action of SU(2) on a vector space that is an irreducible representation of the group, and then more generally on a vector space $\mathbb{V}$ that is a reducible representation, namely, $\mathbb{V}$ decomposes as a direct sum of (possibly degenerate) irreducible representations. We also characterize vectors belonging to $\mathbb{V}$ and linear operators acting on $\mathbb{V}$ that are invariant under the action of SU(2). 

Let $\mathbb{V}$ be a finite dimensional vector space on which SU(2) acts \textit{unitarily} by means of transformations $\hat{W}_{\textbf{r}}~:~\mathbb{V}~\rightarrow~\mathbb{V}$,
\begin{equation}
\hat{W}_{\textbf{r}} \equiv e^{i\textbf{r}\cdot \textbf{J}} = e^{i(r_x\hat{J}_x+r_y\hat{J}_y+r_z\hat{J}_z)}.
\label{eq:exp}
\end{equation}
Here $\textbf{r} \equiv (r_x,r_y,r_z) \in \mathbb{R}^3$ parameterizes the group elements and $\textbf{J} \equiv (\hat{J}_x, \hat{J}_y, \hat{J}_z)$; $\hat{J}_x, \hat{J}_y$ and $\hat{J}_z$ are traceless hermitian operators that are said to \textit{generate} the representation $\hat{W}_{\textbf{r}}$ of SU(2). These operators close the lie algebra su(2), namely,
\begin{equation}
[\hat{J}_{\alpha}, \hat{J}_{\beta}] = i\sum_{\gamma=x,y,z}\epsilon_{\alpha\beta\gamma}\hat{J}_{\gamma}, ~~~~~~~ \alpha,\beta = x,y,z,
\label{eq:algebra}
\end{equation}
where $\epsilon_{\alpha\beta\gamma}$ is the Levi-Civita symbol. The operators $\hat{J}_x, \hat{J}_y$ and $\hat{J}_z$ are associated, for example, with the projection of angular momentum or spin along the three spatial directions $x, y$ and $z$, respectively.

It follows that,
\begin{equation}
[\textbf{J}^2, \hat{J}_{\alpha}] = 0, ~~~~~~ \textbf{J}^2 =\sum_{\alpha=x,y,z}\hat{J}_{\alpha}^2.\label{eq:casimir}
\end{equation}

%%%%%%%%%%%
\subsection{Irreducible representations\label{sec:symmetry:irreps}}

Let vector space $\mathbb{V}$ transform as an irreducible representation (or irrep) of SU(2) with spin $j$. Here $j$ can take values $0,\frac{1}{2},1,\frac{3}{2},2,\ldots$ and $\mathbb{V}$ has dimension $2j+1$. We choose an orthonormal basis $\ket{jm_j}$, the \textit{spin basis}, in $\mathbb{V}$ that is a simultaneous eigenbasis of the operators $\textbf{J}^2$ and $\hat{J}_{z}$,
\begin{align}
\textbf{J}^{2}\ket{j m_j} &= j(j+1)\ket{j m_j}, \nonumber \\
\hat{J}_{z} \ket{jm_j} &= m_j \ket{jm_j}.
\label{eq:irrepz}
\end{align}
Here $m_j$ is the magnitude of the spin projection along the $z$ direction and can assume values in the range $\{-j, -j\!+\!1,\ldots, j\}$. In this basis, the action of the operators $\hat{J}_{x}$ and $\hat{J}_{y}$ on the space $\mathbb{V}$ is conveniently described in terms of the raising operator $\hat{J}_{+} = \hat{J}_{x} + i\hat{J}_{y}$ and the lowering operator $\hat{J}_{-} = \hat{J}_{x} - i\hat{J}_{y}$ as
\begin{equation}
\hat{J}_{\pm} \ket{jm_j} = \sqrt{j(j+1) - m_j (m_j \pm 1)} \ket{j, (m_j \pm 1)}.
\label{eq:irrepladder}
\end{equation}
The operator $\textbf{J}^2$ can be written as
\begin{equation}
\textbf{J}^{2} = j(j+1)\hat{I}_{2j+1},
\end{equation}
where $\hat{I}_{2j+1}$ is the Identity acting on the irrep $j$.

\textbf{Example 1: } Consider that vector space $\mathbb{V}$ is a spin $j=0$ irrep of SU(2). Then $\mathbb{V}$ has dimension one, $\mathbb{V}~\cong~\mathbb{C}$. The operators $\hat{J}_{\alpha}$ are trivial, $\hat{J}_x~=~\hat{J}_y~=~\hat{J}_z~=~(0)$.\markend

\textbf{Example 2: }Consider a two-dimensional vector space $\mathbb{V}$ that transforms as an irrep $j=\half$. Then the orthogonal vectors (in column vector notation)
\begin{align}
\begin{pmatrix} 1 \\ 0 \end{pmatrix}  \equiv \;  \ket{j\!=\!\half, m_{\half}\!=\!-\half},
\begin{pmatrix} 0 \\ 1 \end{pmatrix}  \equiv \;  \ket{j\!=\!\half, m_{\half}\!=\!\half},
\label{eq:basiseg1}
\end{align}
form a basis of $\mathbb{V}$. In this basis the operators $\hat{J}_{x}, \hat{J}_{y}, \hat{J}_{z}$ and $\textbf{J}^2$ read as
\begin{align}
\hat{J}_{x}  \equiv \; \begin{pmatrix} 0 & \half \\ \half & 0 \end{pmatrix},~
\hat{J}_{y}  \equiv \;   \begin{pmatrix} 0 & -\frac{i}{2}  \\ \frac{i}{2} & 0 \end{pmatrix},~
\hat{J}_{z}  \equiv \; \begin{pmatrix} -\half & 0 \\ 0 & \half \end{pmatrix},~\textbf{J}^2 \equiv \;  \begin{pmatrix} \frac{3}{4} & 0 \\ 0 & \frac{3}{4} \end{pmatrix}. \label{eq:eg2c1}
\end{align}
In terms of Pauli matrices $\hat{\sigma}_{\alpha}$ we have
\begin{equation}
\hat{J}_{\alpha} = \frac{\hat{\sigma}_{\alpha}}{2}, ~~~ \alpha = x,y,z.\markend
\end{equation}

\textbf{Example 3: } Also consider a three-dimensional vector space $\mathbb{V}$ that transforms as an irrep $j=1$. The orthogonal vectors
\begin{align}
\begin{pmatrix} 1 \\ 0 \\ 0\end{pmatrix}  \equiv \;  \ket{j\!=\!1, m_{1}\!=\!-1}, ~~
\begin{pmatrix} 0 \\ 1 \\ 0 \end{pmatrix}  \equiv \;  \ket{j\!=\!1, m_{1}\!=\!0}, ~~
\begin{pmatrix} 0 \\ 0 \\ 1 \end{pmatrix}  \equiv \;  \ket{j\!=\!1, m_{1}\!=\!1},
\label{eq:basiseg2}
\end{align}
form a basis of $\mathbb{V}$. In this basis, operators $\hat{J}_x, \hat{J}_y, \hat{J}_z$ and $\textbf{J}^2$ read as
\begin{align}
\hat{J}_x  &\equiv \;  \begin{pmatrix} 0 & \frac{1}{\sqrt{2}} & 0\\\frac{1}{\sqrt{2}} & 0 & \frac{1}{\sqrt{2}}\\0 & \frac{1}{\sqrt{2}} & 0 \end{pmatrix},
\hat{J}_y  \equiv \;  \begin{pmatrix} 0 & \frac{i}{\sqrt{2}} & 0\\-\frac{i}{\sqrt{2}} & 0 & \frac{i}{\sqrt{2}}\\0 & -\frac{i}{\sqrt{2}} & 0 \end{pmatrix}, \nonumber \\
\hat{J}_z  &\equiv \; \begin{pmatrix} -1 & 0 & 0\\0 & 0 & 0\\ 0 & 0 & 1 \end{pmatrix}, ~~~~~\textbf{J}^2 \equiv \begin{pmatrix} 2 & 0 & 0\\0 & 2 & 0\\ 0 & 0 & 2 \end{pmatrix}.\markend
\label{eq:eg2c2}
\end{align}

\subsection{Reducible representations \label{sec:symmetry:rreps}}

More generally, SU(2) can act on the vector space $\mathbb{V}$ reducibly, in that, $\mathbb{V}$ may decompose as the direct sum of irreps of SU(2),
\begin{equation}
\boxed{
\begin{split}\mathbb{V} &\cong \bigoplus_j d_j\mathbb{V}_{j} \\
&\cong \bigoplus_j\left(\mathbb{D}_j \otimes \mathbb{V}_j\right).
\end{split}
}
\label{eq:decoV}
\end{equation}
Here space $\mathbb{V}_{j}$ accommodates a spin $j$ irrep of SU(2) and $d_j$ is the number of times $\mathbb{V}_{j}$ occurs in $\mathbb{V}$. The decomposition can also be re-written in terms of a $d_j-$dimensional space $\mathbb{D}_j$. We say that irrep $j$ is $d_j$-fold degenerate and that $\mathbb{D}_j$ is the degeneracy space. The total dimension of space $\mathbb{V}$ is given by $\sum_j d_j (2j+1)$.

Let $t_j=1,2,\ldots,d_j$ label an orthonormal basis $\ket{jt_j}$ in the space $\mathbb{D}_j$. Then a natural choice of basis of the space $\mathbb{V}$ is the set of orthonormal vectors $\ket{jt_jm_j} \equiv \ket{jt_j}\otimes\ket{jm_j}$, where $j$ assumes various values that occur in the direct sum decomposition, Eq.~(\ref{eq:decoV}).

In this basis the action of SU(2) on $\mathbb{V}$ is given by
\begin{equation}
\hat{W}_{\textbf{r}} \equiv \bigoplus_j \left(\hat{I}_{d_j} \otimes \hat{W}_{\textbf{r},j}\right),
\label{eq:decow}
\end{equation}
as generated by the operators
\begin{equation}
\hat{J}_{\alpha} \equiv \bigoplus_j \left(\hat{I}_{d_j} \otimes \hat{J}_{\alpha, j}\right),~~~\alpha=x,y,z.
\label{eq:decoS}
\end{equation}
Here $\hat{I}_{d_j}$ is a $d_j \times d_j$ Identity and operators $\hat{J}_{\alpha, j}$ generate the irreducible represention $\hat{W}_{\textbf{r},j}$ on space $\mathbb{V}_j$. 

The operator $\textbf{J}^2$ takes the form
\begin{equation}
\textbf{J}^2 \equiv \bigoplus_j j(j+1)\left(\hat{I}_{d_j}\otimes\hat{I}_{2j+1}\right).\label{eq:decoJJ}
\end{equation}

\textbf{Example 4:} Let vector space $\mathbb{V}$ transform as an irrep $j=\half$ with a finite degeneracy $d_{\half} = 3$. The space $\mathbb{V}$ decomposes as $\mathbb{V}~\cong~\mathbb{D}_{\half}~\otimes~\mathbb{V}_{\half}$ where $\mathbb{D}_{\half}$ is a three-dimensional degeneracy space and $\mathbb{V}_{\half}$ corresponds to the space of Example 1. 

The total dimension of space $\mathbb{V}$ is $d_{\half}(2j+1) = 6$. The vectors
\begin{align}
\begin{pmatrix} 1 \\ 0 \\ 0 \end{pmatrix}  \equiv \ket{j=\!\half, t_0\!= 1},~~
\begin{pmatrix} 0 \\ 1 \\ 0 \end{pmatrix}  \equiv   \ket{j=\!\half, t_0\!= 2},~~
\begin{pmatrix} 0 \\ 0 \\ 1 \end{pmatrix}  \equiv \ket{j=\!\half, t_1\!= 3}.
\label{eq:basiseg21}
\end{align}
form a basis of $\mathbb{D}_{\half}$. A basis of $\mathbb{V}$ can be obtained as the product of the the basis (\ref{eq:basiseg21}) of $\mathbb{D}_{\half}$ and the basis (\ref{eq:basiseg1}) of $\mathbb{V}_{\half}$. In this basis of the operators $\hat{J}_{\alpha}$ take the form of Eq.~(\ref{eq:decoS}). For instance,
\begin{equation}
\hat{J}_x  \equiv \begin{pmatrix} 1 & 0 & 0 \\ 0 & 1 & 0 \\ 0 & 0 & 1 \end{pmatrix} \otimes \begin{pmatrix} 0 & \half \\ \half & 0 \end{pmatrix} = 
\begin{pmatrix} 0 & \half & 0 & 0 & 0 & 0 \\ \half & 0 & 0 & 0 & 0 & 0 \\ 0 & 0 & 0 & \half & 0 & 0 \\ 0 & 0 & \half & 0 & 0 & 0 \\ 0 & 0 & 0 & 0 & 0 & \half \\ 0 & 0 & 0 & 0 & \half & 0 \end{pmatrix}
\label{eq:eg2}
\end{equation}
Similarly, operators $\hat{J}_y$ and $\hat{J}_z$ read as
\begin{align}
\hat{J}_y  &\equiv 
\begin{pmatrix} 0 & \frac{i}{2} & 0 & 0 & 0 & 0 \\ -\frac{i}{2} & 0 & 0 & 0 & 0 & 0 \\ 0 & 0 & 0 & \frac{i}{2} & 0 & 0 \\ 0 & 0 & -\frac{i}{2} & 0 & 0 & 0 \\ 0 & 0 & 0 & 0 & 0 & \frac{i}{2} \\ 0 & 0 & 0 & 0 & -\frac{i}{2} & 0 \end{pmatrix}, \nonumber \\
\hat{J}_z  &\equiv
\begin{pmatrix} -\half  & 0 & 0 & 0 & 0 & 0 \\ 0 & \half  & 0 & 0 & 0 & 0 \\ 0 & 0 & -\half  & 0 & 0 & 0 \\ 0 & 0 & 0 & \half  & 0 & 0 \\ 0 & 0 & 0 & 0 & -\half  & 0 \\ 0 & 0 & 0 & 0 & 0 & \half \end{pmatrix}.
\end{align}
The operator $\textbf{J}^2$ reads
\begin{equation}
\textbf{J}^2  \equiv \half(\half+1)\hat{I}_{3}\otimes\hat{I}_{2} \equiv 
 \begin{pmatrix} \frac{3}{4} & 0 & 0 & 0 & 0 & 0 \\ 0 & \frac{3}{4} & 0 & 0 & 0 & 0 \\ 0 & 0 & \frac{3}{4} & 0 & 0 & 0 \\ 0 & 0 & 0 & \frac{3}{4} & 0 & 0 \\ 0 & 0 & 0 & 0 & \frac{3}{4} & 0 \\ 0 & 0 & 0 & 0 & 0 & \frac{3}{4} \end{pmatrix}.\markend
 \label{eq:eg2s2}
\end{equation}

\textbf{Example 5:} Consider a five-dimensional Hilbert space $\mathbb{V}$ that decomposes into two different irreps $j=0$ and $j=1$ with degeneracy dimensions $d_0=2$ and $d_1=1$ so that irrep $j=0$ is two-fold degenerate. The space $\mathbb{V}$ decomposes as $\mathbb{V}~\cong~(\mathbb{D}_0~\otimes~ \mathbb{V}_{0})~\oplus~(\mathbb{D}_1~\otimes~\mathbb{V}_{1})$, where $\mathbb{D}_0$ is the two-dimensional degeneracy space of irrep $j=0$ and $\mathbb{D}_1$ is the one-dimensional degeneracy space of irrep $j=1$. 

The orthogonal vectors
\begin{align}
\begin{pmatrix} 1 \\ 0 \\ 0 \\ 0 \\ 0 \end{pmatrix}  &\equiv \;  \ket{j=0, t_0 = 1, m_0=0},\nonumber \\
\begin{pmatrix} 0 \\ 1 \\ 0 \\ 0 \\ 0 \end{pmatrix}  &\equiv \;  \ket{j=0, t_0 = 2, m_0=0}, \nonumber \\
\begin{pmatrix} 0 \\ 0 \\ 1 \\ 0 \\ 0 \end{pmatrix}  &\equiv \;  \ket{j=1, t_1 = 1, m_1=-1},\nonumber \\
\begin{pmatrix} 0 \\ 0 \\ 0 \\ 1 \\ 0 \end{pmatrix}  &\equiv \;  \ket{j=1, t_1 = 1, m_1=0},\nonumber \\
\begin{pmatrix} 0 \\ 0 \\ 0 \\ 0 \\ 1 \end{pmatrix}  &\equiv \;  \ket{j=1, t_1 = 1, m_1=1}.
\label{eq:basiseg3}
\end{align}
form a basis of $\mathbb{V}$. In this basis, the operators $\hat{J}_{\alpha}$ take the form
\begin{equation}
\hat{J}_{\alpha} = (\hat{I}_{d_0} \otimes \hat{J}_{\alpha, 0}) \oplus (\hat{I}_{d_1} \otimes \hat{J}_{\alpha, 1}),~~~\alpha=x,y,z, 
\end{equation}
where $\hat{J}_{\alpha, 0}$ and $\hat{J}_{\alpha, 1}$ are the generators of irrep $j=0$ (Examples 1) and irrep $j=1$ (Examples 3) respectively. Operators $\hat{J}_{\alpha}$ and $\textbf{J}^2$ read as
\begin{align}
\hat{J}_x  &\equiv \; \begin{pmatrix} 0 & 0 & 0 & 0 & 0 \\ 0 & 0 & 0 & 0 & 0 \\ 0 & 0 & 0 & \frac{1}{\sqrt{2}} & 0 \\0 & 0 & \frac{1}{\sqrt{2}} & 0 & \frac{1}{\sqrt{2}} \\ 0 & 0 & 0 & \frac{1}{\sqrt{2}} & 0 \end{pmatrix},\nonumber \\
\hat{J}_y  &\equiv \;  \begin{pmatrix} 0 & 0 & 0 & 0 & 0 \\ 0 & 0 & 0 & 0 & 0 \\ 0 & 0 & 0 & \frac{i}{\sqrt{2}} & 0 \\0 & 0 & -\frac{i}{\sqrt{2}} & 0 & \frac{i}{\sqrt{2}} \\0 & 0 & 0 & -\frac{i}{\sqrt{2}} & 0 \end{pmatrix}, \nonumber \\
\hat{J}_z  &\equiv \; \begin{pmatrix} 0 & 0 & 0 & 0 & 0 \\ 0 & 0 & 0 & 0 & 0 \\ 0 & 0 & -1 & 0 & 0 \\ 0 & 0 & 0 & 0 & 0 \\0 & 0 & 0 & 0 & 1 \end{pmatrix}, \nonumber \\
\textbf{J}^2 &\equiv \; \begin{pmatrix} 0 & 0 & 0 & 0 & 0 \\ 0 & 0 & 0 & 0 & 0 \\ 0 & 0 & 2 & 0 & 0 \\0 & 0 & 0 & 2 & 0 \\0 & 0 & 0 & 0 & 2 \end{pmatrix}.\markend
\label{eq:eg3}
\end{align}

%%%%%%%%%%%%
\subsection{SU(2)-invariant states and operators \label{sec:symmetry:states}}

We are interested in states and operators that have a simple transformation rule under the action of SU(2). 

A pure state $\ket{\Psi} \in \mathbb{V}$ with a well defined spin $j$ belongs to the subspace $\mathbb{D}_j \otimes \mathbb{V}_j$. In the spin basis $\ket{jt_jm_j}$ the state $\ket{\Psi}$ can be expanded as
\begin{equation}
\ket{\Psi} = \sum_{t_jm_j} (\Psi_j)_{t_jm_j} \ket{jt_jm_j},
\label{eq:welldefj}
\end{equation}
Under the action of SU(2), $\ket{\Psi}$ transforms to another pure state $\ket{\Psi'}$, $\ket{\Psi'}~=~\hat{W}_{\textbf{r}}\ket{\Psi}$, within the same subspace $\mathbb{D}_j~\otimes~\mathbb{V}_j$. The components $(\Psi'_j)_{t'_jm'_j}$ of $\ket{\Psi'}$ are related to those of $\ket{\Psi}$ as
\begin{equation}
(\Psi'_j)_{t'_jm'_j} = \sum_{t_jm_j} (W_{r,j})_{t'_jm'_jt_jm_j}(\Psi_j)_{t_jm_j}.
\label{eq:welldefj1}
\end{equation}
In case of vanishing spin $j=0$, the state $\ket{\Psi}$ transforms trivially under the action of SU(2), $\hat{W}_{r,j=0} \equiv 1$. In this case Eq.~(\ref{eq:welldefj1}) reduces to,
\begin{equation}
\hat{W}_{\textbf{r}} \ket{\Psi} = \ket{\Psi}, ~~~~~ \forall \textbf{r} \in \mathbb{R}^3.
	\label{eq:invPsi1}
\end{equation}
That is, $\ket{\Psi}$ remains \textit{invariant} under the action of SU(2). Equivalently, it is annihilated by the action of generators,
\begin{equation}
\hat{J}_{\alpha}\ket{\Psi} = 0, ~~~~~~~~~~~\alpha=x,y,z.
\label{eq:invPsi2}
\end{equation}
An SU(2)-invariant state $\ket{\Psi}$ can be expanded in the basis $\ket{j~=~0,~t_0~,m_0~=~0}$ of the spin $j=0$ subspace, $\mathbb{D}_{0} \otimes \mathbb{V}_0$, 
\begin{equation}
	\boxed{\ket{\Psi} = \sum_{t_0} (\Psi_0)_{t_0} \ket{j=0,t_0,m_0=0},}
	\label{eq:nPsi3}
\end{equation}
where we have used $(\Psi_0)_{t_0}$ as a shorthand for $(\Psi_{j=0})_{t_0,m_0=0}$.

A linear operator $\hat{T}: \mathbb{V} \rightarrow \mathbb{V}$ is SU(2)-invariant if it commutes with the action of the group,
\begin{equation}
	[\hat{T}, \hat{W}_{\textbf{r}}] = 0, ~~~~~~~~~\forall \textbf{r} \in \mathbb{R}^3,
	\label{eq:invOp2}
\end{equation}
or equivalently, if it commutes with the generators $\hat{J}_{\alpha}$,
\begin{equation}
	[\hat{T}, \hat{J}_{\alpha}] = 0, ~~~~~~~~~~~\alpha=x,y,z.
	\label{eq:invOp1}
\end{equation}
Notice that the operator $\textbf{J}^2$ is SU(2)-invariant, Eq.~(\ref{eq:casimir}).

An SU(2)-invariant operator $\hat{T}$ decomposes as
\begin{align}
&\boxed{\hat{T} = \bigoplus_{j} \left(\hat{T}_{j} \otimes \hat{I}_{2j+1}\right),} \nonumber \\
&\!\!\!\!\!\!\!\!\!\!\!\!\!\!\!\!\!\!\!\!\!\!\!\!\!\!\!\!\!\!\!\!\!\!\!\!\!\!\!\text{(Schur's Lemma)} \label{eq:Schur}
\end{align}
where $\hat{T}_{j}$ is a $d_j\times d_j$ matrix that acts on the degeneracy space $\mathbb{D}_j$. This decomposition implies, for instance, that operator $\hat{T}$ transforms states with a well defined spin $j$ [such as $\ket{\Psi}$ of Eq.~(\ref{eq:welldefj})] into states with the same spin $j$. Thus, SU(2)-invariant operators \textit{conserve} spin. 

\textbf{Example 2 revisited:} A generic state  $\ket{\Psi} \in \mathbb{V}_{\half}$ has the form,
\begin{equation}
\ket{\Psi} = \begin{pmatrix} (\Psi_{j=\half})_{m_{\half}=-\half} \\ (\Psi_{j=\half})_{m_{\half}=\half} \end{pmatrix}~~  \in \mathbb{C}^2.
\end{equation}
and is an eigenstate of $\textbf{J}^2_{\half}$ with eigenvalue $\displaystyle \frac{3}{4}$.

According to Schur's lemma, an SU(2)-invariant operator $\hat{T}$ acting on $\mathbb{V}_{\half}$ must be proportional to the Identity,
\begin{equation}
	\hat{T} = \; (T_{j=\half}) \begin{pmatrix} 1 & 0\\ 0 & 1 \end{pmatrix},~T_{j=\half} \in \mathbb{C}. \label{eq:sparseT1}\markend
\end{equation} 

\textbf{Example 4 revisited:} A generic state $\ket{\Psi}$ in the vector space $\mathbb{V} \cong 3\mathbb{V}_{\half}$ of Example 3 has the form
\begin{equation}
\ket{\Psi} = \begin{pmatrix} (\Psi_{j=\half})_{t_{\half}=1,m_{\half}=-\half} \\ (\Psi_{j=\half})_{t_{\half}=1,m_{\half}=\half} \\ (\Psi_{j=\half})_{t_{\half}=2,m_{\half}=-\half} \\ (\Psi_{j=\half})_{t_{\half}=2,m_{\half}=\half} \\ (\Psi_{j=\half})_{t_{\half}=3,m_{\half}=-\half} \\ (\Psi_{j=\half})_{t_{\half}=3,m_{\half}=\half} \end{pmatrix}~~  \in \mathbb{C}^6.
\end{equation}
Similar to the previous example, $\ket{\Psi}$ is an eigenstate of $\textbf{J}^2_{\half}$ with eigenvalue $\displaystyle \frac{3}{4}$..

An SU(2)-invariant operator $\hat{T} : \mathbb{V} \rightarrow \mathbb{V}$ must be of the form
\begin{align}
	\hat{T} &= \; \begin{pmatrix} T_{11} & T_{12} & T_{13}\\ T_{21} & T_{22} & T_{23} \\ T_{31} & T_{32} & T_{33} \end{pmatrix} \otimes \begin{pmatrix} 1 & 0\\ 0 & 1 \end{pmatrix} \nonumber \\
	&=  \begin{pmatrix} T_{11} & 0 & T_{12} & 0 & T_{13} & 0\\ 0 & T_{11} & 0 & T_{12} & 0 & T_{13} \\ T_{21} & 0 & T_{22} & 0 & T_{23} & 0 \\ 0 & T_{21} & 0 & T_{22} & 0 & T_{23} \\ T_{31} & 0 & T_{32} & 0 & T_{33} & 0 \\ 0 & T_{31} & 0 & T_{32} & 0 & T_{33} \end{pmatrix},\label{eq:sparseT2}
\end{align} 
where we have used $T_{ij}$ as a shorthand for $(T_{\half})_{ij}~\in~\mathbb{C}$. Notice that $\textbf{J}^2$ in Eq.~(\ref{eq:eg2s2}) has this form.\markend

\textbf{Example 5 revisited:} A generic state $\ket{\Psi} \in \mathbb{V},  \mathbb{V} \cong 2\mathbb{V}_0 \oplus \mathbb{V}_1$ has the form
\begin{equation}
	\ket{\Psi} = \begin{pmatrix} (\Psi_0)_{1,0} \\ (\Psi_0)_{2,0} \\ ~~(\Psi_1)_{1,-1} \\ (\Psi_1)_{1,0} \\ (\Psi_1)_{1,1} \end{pmatrix}~~  \in \mathbb{C}^5.
	\label{eq:ex2rev}
\end{equation}
(For simplicity we have omitted explicit labels in the subscripts.) In contrast to the previous two examples, a generic state $\ket{\Psi} \in \mathbb{V}$ is not an eigenstate of $\textbf{J}^2$, that is, $\ket{\Psi}$ is generally not a state with a well defined spin $j$.\\
An SU(2)-invariant vector $\ket{\Psi}$ has the form
\begin{equation}
	\ket{\Psi} = \begin{pmatrix} (\Psi_0)_{1,0} \\ (\Psi_0)_{2,0} \\ 0 \\ 0 \\ 0 \end{pmatrix}, ~~~~~~~~~~~~~
	\label{eq:ex3rev}
\end{equation}
with non-trivial components only in the spin $j=0$ subspace. Notice that this state is annihilated by the action of the operators $\hat{J}_{\alpha}$ [Eq.~(\ref{eq:eg3})] in accordance with Eq.~(\ref{eq:invPsi2}).

A state with a well defined spin $j=1$ must be of the form
\begin{equation}
	\ket{\Psi_1} = \begin{pmatrix} 0 \\ 0 \\ ~~(\Psi_1)_{1,-1} \\ (\Psi_1)_{1,0} \\ (\Psi_1)_{1,1} \end{pmatrix}~~  \in \mathbb{C}^5,
	\label{eq:ex2rev1}
\end{equation}
with non-trivial components only in the spin $j=1$ subspace.

An SU(2)-invariant operator $\hat{T}$ e.g. $\textbf{J}^2$ in Eq.~(\ref{eq:eg3}) has the form
\begin{align}
	\hat{T} &= \; \begin{pmatrix} \left(T_0\right)_{11} & \left(T_0\right)_{12}\\ \left(T_0\right)_{21} &  \left(T_0\right)_{22}\end{pmatrix} \otimes (1) \oplus \begin{pmatrix} \left(T_1\right)_{11} \end{pmatrix} \otimes \begin{pmatrix} 1 & 0 & 0\\ 0 & 1 & 0 \\ 0 & 0 & 1 \end{pmatrix} \nonumber \\
	&= \begin{pmatrix} \left(T_0\right)_{11} & \left(T_0\right)_{12} & 0 & 0 & 0\\ \left(T_0\right)_{21} & \left(T_0\right)_{22} & 0 & 0 & 0 \\ 0 & 0 & \left(T_1\right)_{11} & 0 & 0 \\ 0 & 0 & 0 & \left(T_1\right)_{11} & 0 \\ 0 & 0 & 0 & 0 & \left(T_1\right)_{11} \end{pmatrix},
	\label{eq:sparseT3}
\end{align} 
where $\left(T_0\right)_{11},~\left(T_0\right)_{12},~\left(T_0\right)_{21},~\left(T_0\right)_{22},~\left(T_1\right)_{11}~\in~\mathbb{C}$.\markend

Notice that the SU(2)-invariant vector in Eq.~(\ref{eq:ex3rev}) and the SU(2)-invariant matrices in Eqs.~(\ref{eq:sparseT1}),(\ref{eq:sparseT2}) and (\ref{eq:sparseT3}) have a \textit{sparse} structure. In particular, the non-trivial components of an SU(2)-invariant matrix $\hat{T}$ are organized into blocks $\hat{T}_j$. This block structure can be exploited for computational gain. An SU(2)-invariant matrix can be stored compactly by storing the degeneracy blocks $\hat{T}_j$, while matrix multiplication and matrix factorizations can be performed block-wise [Sec. \ref{sec:blockmoves}] resulting in a significant speedup [Fig.~\ref{fig:multsvdcompare}] for these operations.

%%%%%%%%%%%%%%%%%%%%%%%%%%%%%%
\section{Tensor product of representations \label{sec:su2lattice}}

So far we have described the action of SU(2) on a single vector space. Let us now consider the action of SU(2) on a space $\mathbb{V}$ that is a tensor product of $L$ vector spaces,
\begin{equation}
\mathbb{V} \equiv \bigotimes_{l=1}^{L} \mathbb{V}^{(l)}, \label{eq:latticespace}
\end{equation}
where each vector space $\mathbb{V}^{(l)}, l=1,2,\ldots,L,$ transforms as a finite dimensional representation of SU(2) as generated by spin operators $\hat{J}^{(l)}_{\alpha}, \alpha=x,y,z$. We consider the action of SU(2) on the space $\mathbb{V}$ that is generated by the \textit{total} spin operators,
\begin{equation}
\boxed{\hat{J}_{\alpha} \equiv \sum_{l=1}^{L} \hat{J}^{(l)}_{\alpha}, ~~~ \alpha = x,y,z,} \label{eq:totspinops}
\end{equation}
(each term in the sum acts as $\hat{J}^{(l)}_{\alpha}$ on site $l$ and the Identity on the remaining sites) and which corresponds to the unitary transformations,
\begin{equation}
	\hat{W}_{\textbf{r}} \equiv e^{\rmi\textbf{r}\cdot \textbf{J}} = \bigotimes_{l=1}^L e^{\rmi\textbf{r}\cdot \textbf{J}^{(l)}} = \bigotimes_{l=1}^L \hat{W}_{\textbf{r}}^{(l)}.\label{eq:latticerep}
\end{equation}
As a example consider two vector spaces $\mathbb{V}^{(A)}$ and $\mathbb{V}^{(B)}$ on which the action of SU(2) is generated by spin operators $\hat{J}^{(A)}_{\alpha}$ and $\hat{J}^{(B)}_{\alpha}$ respectively. We can then consider the action of the group on the product space $\mathbb{V}^{(AB)} \cong \mathbb{V}^{(A)} \otimes \mathbb{V}^{(B)}$ as generated by the total spin operators $\hat{J}^{(AB)}_{\alpha},~\alpha=x,y,z,$ that are given by
\begin{equation}
\hat{J}^{(AB)}_{\alpha} \equiv \hat{J}^{(A)}_{\alpha}\otimes \hat{I} + \hat{I} \otimes \hat{J}^{(B)}_{\alpha}.
\end{equation}
Similarly, we can consider the action of SU(2) on the product of three vector spaces, $\mathbb{V}^{(A)}, \mathbb{V}^{(B)}$ and $\mathbb{V}^{(C)}$, that is generated by spin operators $\hat{J}^{(ABC)}_{\alpha},~\alpha=x,y,z$,
\begin{equation}
	\hat{J}^{(ABC)}_{\alpha} \equiv \hat{J}^{(A)}_{\alpha}\otimes \hat{I}\otimes \hat{I} + \hat{I} \otimes \hat{J}^{(B)}_{\alpha}\otimes \hat{I}+\hat{I}\otimes \hat{I}\otimes \hat{J}^{(C)}_{\alpha}, \label{eq:genprodthree}
\end{equation}
where $\hat{J}^{(A)}_{\alpha}, \hat{J}^{(B)}_{\alpha}$ and $\hat{J}^{(C)}_{\alpha}$ are the spin operators that act on the three vector spaces.

A basis of $\mathbb{V}$ can be obtained in terms of the spin basis of each vector space in the product, Eq.~\ref{eq:latticespace}. However, it is convenient to introduce a \textit{coupled} basis: the simultaneous eigenbasis of the \textit{total} spin operators $\hat{J}_z$ and $\textbf{J}^{2}$. In the coupled basis SU(2)-invariant states $\ket{\Psi} \in \mathbb{V}$,
\begin{equation}
\ket{\Psi}=\hat{W}_{\textbf{r}}\ket{\Psi},~~~\forall\textbf{r}\in\mathbb{R}^3, \label{eq:latticeinv}
\end{equation}
and SU(2)-invariant operators $\hat{T}: \mathbb{V} \rightarrow \mathbb{V}$,
\begin{equation}
[\hat{T}, \hat{W}_{\textbf{r}}] = \hat{T},~~~\forall\textbf{r}\in\mathbb{R}^3,
\end{equation}
have a sparse structure, namely, $\ket{\Psi}$ has non-trivial components only in the spin zero sector of $\mathbb{V}$ while $\hat{T}$ is block-diagonal [Eq.~(\ref{eq:Schur})].

In the remainder of the section we focus on the tensor product of \textit{two} representations. We first discuss the case where the two sites transform as irreducible representations and then the more general case of reducible representations. The tensor product of several representations can then be analyzed by considering a sequence of pairwise products. 

\subsection{Tensor product of two irreducible representations\label{sec:symmetry:tp:irrep}}
Let vector spaces $\mathbb{V}^{(A)}$ and $\mathbb{V}^{(B)}$ transform as irreps $j_A$ and $j_B$ respectively. The space $\mathbb{V}^{(AB)}$ is, in general, reducible and decomposes as
\begin{equation}
\mathbb{V}^{(AB)} \cong \bigoplus_{j_{AB}} \mathbb{V}_{j_{AB}}^{(AB)}, 
\label{eq:tensorirrep}
\end{equation}
where the total spin $j_{AB}$ assumes all values in the range 
\begin{equation}
\boxed{
\begin{split}
&j_{AB}:\{|j_A-j_B|, |j_A-j_B|+1, \ldots, j_A+j_B\}. \\
&\mbox{(Fusion rules)}
\end{split}
\label{eq:fusionrules}
}
\end{equation} 
Let $\ket{j_Am_A} \in \mathbb{V}^{(A)}$ and $\ket{j_Bm_B} \in \mathbb{V}^{(B)}$ denote the spin basis of the respective vector spaces. Then the vectors 
\begin{equation}
\ket{j_{A}m_{j_A};j_B m_{j_B}} \equiv \ket{j_{A}m_{j_A}} \otimes \ket{j_B m_{j_B}} \label{eq:prod2}
\end{equation}
form a basis of $\mathbb{V}^{(AB)}$. Introduce a coupled basis $\ket{j_{AB}m_{AB}} \in \mathbb{V}^{(AB)}$ that fulfills
\begin{align}
	{\textbf{J}^2}^{(AB)}\ket{j_{AB} m_{j_{AB}}} &= j_{AB}(j_{AB}+1) \ket{j_{AB} m_{j_{AB}}}, \nonumber \\
	\hat{J}_z^{(AB)}~\ket{j_{AB} m_{j_{AB}}} 
	&= m_{j_{AB}}~\ket{j_{AB} m_{j_{AB}}}.
\end{align}
The coupled basis is related to the product basis (\ref{eq:prod2}) by means of the transformation
\begin{equation}
\boxed{
\begin{split}
\ket{j_{AB} m_{j_{AB}}} = \sum_{m_{j_{A}} m_{j_{B}}} \cfusespin{A}{B}{AB}
\ket{j_{A} m_{j_{A}} ;j_{B} m_{j_{B}}}.
\end{split}
\label{eq:cg}
}
\end{equation}
Here 
\begin{equation}
\cfusespin{A}{B}{AB} \equiv \braket{j_{A}m_{j_A};j_{B}m_{j_B}}{j_{AB}m_{j_{AB}}} 
\end{equation}
are the \textit{Clebsch-Gordan coefficients}, which vanish unless $j_A, j_B$ and $j_{AB}$ are compatible, that is, $j_A, j_B$ and $j_{AB}$ fulfill
\begin{equation}
|j_A-j_B| \leq j_{AB} \leq j_A+j_B, \label{eq:compatiblej}
\end{equation}
and $m_{j_A}, m_{j_B}$ and $m_{j_{AB}}$ fulfill 
\begin{equation}
m_{j_{AB}} = m_{j_A} + m_{j_B}.\label{eq:compatiblem}
\end{equation}
The product basis can be expressed in terms of the coupled basis as
\begin{equation}
\boxed{
\begin{split}
\ket{j_{A} m_{j_{A}};j_{B} m_{j_{B}}} = \sum_{m_{j_{AB}}} \csplitspin{AB}{A}{B} \ket{j_{AB}m_{j_{AB}}},\label{eq:revcg}
\end{split}
}
\end{equation}
where
\begin{equation}
\csplitspin{AB}{A}{B} \equiv \cfusespin{A}{B}{AB}.
\label{eq:splitcg}
\end{equation}

\begin{figure}[t]
\begin{center}
  \includegraphics[width=10cm]{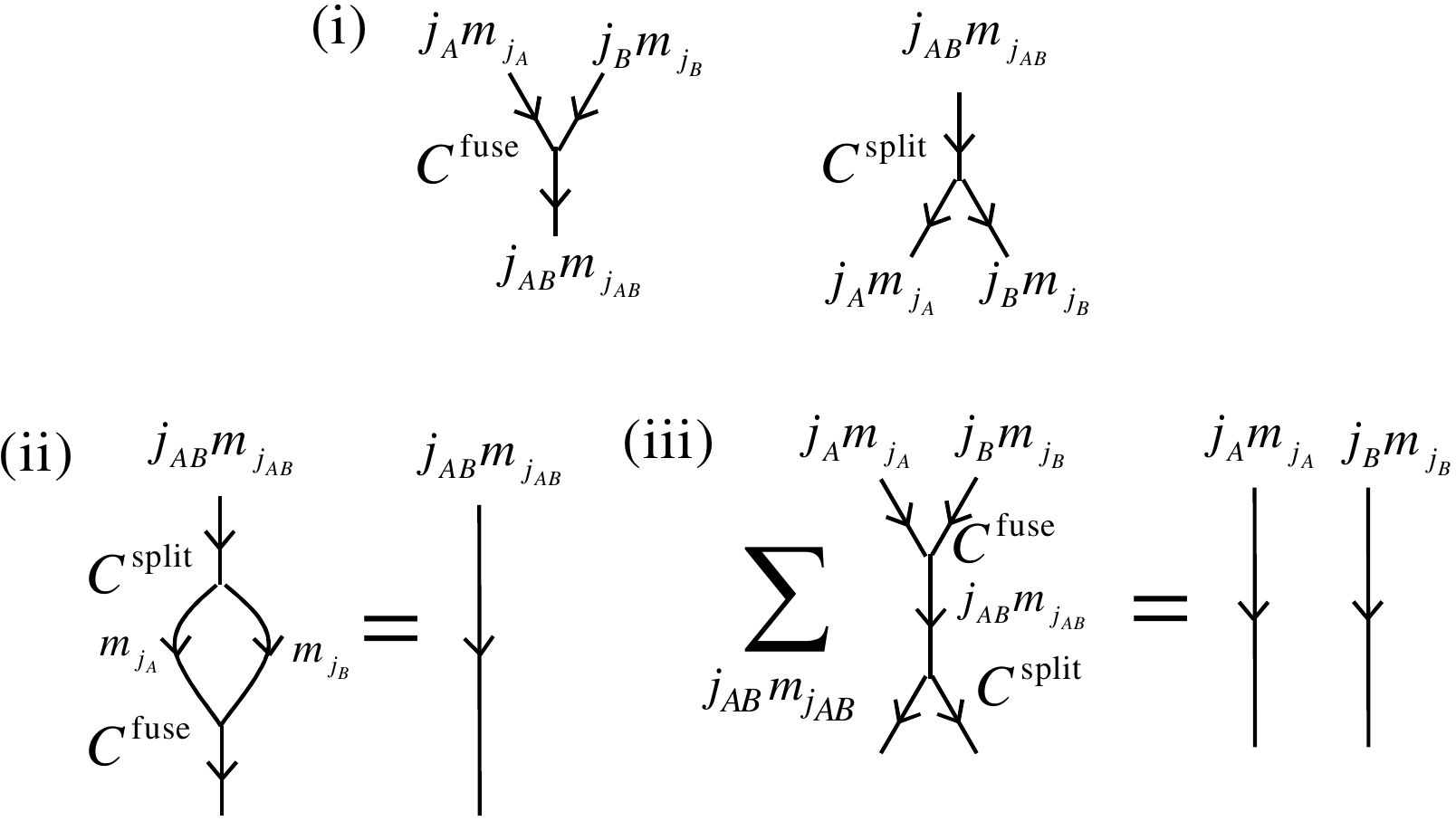}
\end{center}  
\caption{(i) The graphical representation of the Clebsch-Gordan tensors: $\cfuser$ and $\cspliter$. (ii) Tensors $\cfuser$ and $\cspliter$ are isometries and thus yield the Identity when contracted pairwise either as (i) or (ii), Eq.~(\ref{eq:cgunitary}). \label{fig:cg}}
\end{figure}

We graphically represent tensors $\cfuser$ and $\cspliter$ differently from usual tensors, as shown in Fig.~\ref{fig:cg}(i).

Tensor $\cfusespin{A}{B}{AB}$ is depicted by means of two incoming lines and one outgoing line that emerge from a point. The outgoing line corresponds to the spin index $(j_{AB}, m_{j_{AB}})$. The incoming lines that are encountered first and second when proceeding \textit{clockwise} from the outgoing line correspond to the spin indices $(j_{A}, m_{j_{A}})$  and $(j_{B}, m_{j_{B}})$ respectively.
%%%%%%%%%%%%%%%%%%%%%%%%%%%%%%%%%%%%%%%%%%%%%%%%%%%%%%%%%%%%%%%%%%%%%%%%%%%%%%%%%%%%%%%%%%%%%%%%
%%%%%%%%%%%%%%%%%%%%%%%%%%%%%%%%%%%%%%%%%%%%%%%%%%%%%%%%%%%%%%%%%%%%%%%%%%%%%%%%%%%%%%%%%%%%%%%%

Analogously, tensor $\csplitspin{AB}{A}{B}$ is depicted by means of one incoming line and two outgoing lines that emerge from a point. The spin index $(j_{AB}, m_{j_{AB}})$ corresponds to the incoming line in this case while spin indices $(j_{A}, m_{j_{A}})$  and $(j_{B}, m_{j_{B}})$ correspond to the outgoing lines in the order in which they are encountered when proceeding \textit{counterclockwise} from the incoming line.

Tensor $\cfuser$ and tensor $\cspliter$ fulfill the orthogonality identities,
\begin{align}
\sum_{m_{j_A}m_{j_B}}\csplitt{j_{AB}m_{j_{AB}}}{j_{A}m_{j_{A}}}{j_{B}m_{j_{B}}}\cdot\cfuse{j_{A}m_{j_{A}}}{j_{B}m_{j_{B}}}{j'_{AB}m_{j'_{AB}}}
&=\delta_{j_{AB}j'_{AB}}\delta_{m_{j_{AB}}m_{j'_{AB}}} \nonumber \\
\sum_{j_{AB}m_{j_{AB}}}\cfuse{j_{A}m_{j_{A}}}{j_{B}m_{j_{B}}}{j_{AB}m_{j_{AB}}}\cdot\csplitt{j_{AB}m_{j_{AB}}}{j'_{A}m_{j'_{A}}}{j'_{B}m_{j'_{B}}} 
&=\delta_{j_{A}j'_{A}}\delta_{m_{j_{A}}m_{j'_{A}}} \delta_{j_{B}j'_{B}}\delta_{m_{j_{B}}m_{j'_{B}}}
\label{eq:cgunitary}
\end{align}
The special graphical representations for these tensors allows one to depict the above identities in an intuitive way, as shown in Fig.~\ref{fig:cg}.(ii)-(iii).

%%%%%%%%%%%%%%%%%%%%%%%%%% EXAMPLE 4%%%%%%%%%%%%%%%%%%%%%%%%%%%%%%%%
\textbf{Example 6: } Let both vector spaces $\mathbb{V}^{(A)}$ and $\mathbb{V}^{(B)}$ transform as a spin $\frac{1}{2}$ irrep, $j_A=j_B=\half$ (Example 2). The space $\mathbb{V}^{(AB)}$ decomposes into a direct sum of irreps,
\begin{align}
\mathbb{V}^{(AB)} &\cong \mathbb{V}^{(A)} \otimes \mathbb{V}^{(B)} \cong \mathbb{V}^{(AB)}_{0} \oplus \mathbb{V}^{(AB)}_{1}. \label{eq:eg40}
\end{align}
A basis can be introduced in $\mathbb{V}^{(AB)}$ in terms of the spin basis [Eq.~(\ref{eq:basiseg1})] of $\mathbb{V}^{(A)}$ and of $\mathbb{V}^{(B)}$. The coupled basis of $\mathbb{V}^{(AB)}$,
\begin{align}
&\ket{j_{AB}=0, m_{j_{AB}}=0},\nonumber \\
&\ket{j_{AB}=1, m_{j_{AB}}=-1},~\ket{j_{AB}=1, m_{j_{AB}}=0},~\ket{j_{AB}=1, m_{j_{AB}}=1}, \nonumber
\end{align}
is related to the product basis by means of the Clebsch-Gordan coefficients,
\begin{align}
\ket{j_{AB}=0, m_{AB}=0}~~\textbf{=}&~~\cfuse{\half\mhalf}{\half\half}{00} \ket{j_A=\half, m_A=\mhalf;j_B=\half, m_B= \half}\nonumber \\
&+\cfuse{\half\half}{\half\mhalf}{00}\ket{j_A=\half, m_A=\half;j_B=\half, m_B=\mhalf} \label{eq:halfsinglet}\\
\ket{j_{AB}=1, m_{AB}=-1}~~\textbf{=}&~~\cfuse{\half\mhalf}{\half\mhalf}{1-1}\ket{j_A=\half, m_A=\mhalf;j_B=\half, m_B=\mhalf}\\
\ket{j_{AB}=1, m_{AB}=0}~~\textbf{=}&~~\cfuse{\half\half}{\half\mhalf}{10} \ket{j_A=\half, m_A=\half;j_B=\half, m_B=\mhalf}\nonumber\\ 
&+\cfuse{\half\mhalf}{\half\half}{10}\ket{j_A=\half, m_A=\mhalf;j_B=\half, m_B=\half} \\
\ket{j_{AB}=1, m_{AB}=1}~~\textbf{=}&~~\cfuse{\half\half}{\half\half}{11}\ket{j_A=\half, m_A=\half;j_B=\half, m_B=\half}.\label{eq:eg41}
\end{align}
For completeness, we list below the numerical value of the Clebsch-Gordan coefficients that appear in Eqs.~(\ref{eq:halfsinglet})-(\ref{eq:eg41}),
\begin{align}
\cfuse{\half\half}{\half\mhalf}{00} &= \frac{1}{\sqrt{2}}, ~~~ \cfuse{\half\mhalf}{\half\half}{00} = \frac{-1}{\sqrt{2}},~~~\cfuse{\half\mhalf}{\half\mhalf}{1-1} = 1,  \nonumber \\
\cfuse{\half\half}{\half\mhalf}{10} &= \frac{1}{\sqrt{2}},~~~\cfuse{\half\mhalf}{\half\half}{10} = \frac{1}{\sqrt{2}}, ~~~ \cfuse{\half\half}{\half\half}{11} = 1. \nonumber\markend
\end{align}

\subsection{Tensor product of two reducible representations\label{sec:symmetry:tp:general}}

Let us now consider that vector spaces $\mathbb{V}^{(A)}$ and $\mathbb{V}^{(B)}$ transform reducibly under the action of SU(2). We have
\begin{equation}
\mathbb{V}^{(A)} \cong \bigoplus_{j_{A}}\left(\mathbb{D}^{(A)}_{j_A} \otimes \mathbb{V}^{(A)}_{j_A}\right), ~~~\mathbb{V}^{(B)} \cong \bigoplus_{j_{B}}\left(\mathbb{D}^{(B)}_{j_B} \otimes \mathbb{V}^{(B)}_{j_B}\right).
\label{eq:AandB}
\end{equation}
The space $\mathbb{V}^{(AB)}$ decomposes into a direct sum of irreps,
\begin{equation}
\mathbb{V}^{(AB)} \cong \bigoplus_{j_{AB}} d_{j_{AB}} \mathbb{V}^{(AB)}_{j_{AB}} \cong \bigoplus_{j_{AB}}\left(\mathbb{D}^{(AB)}_{j_{AB}} \otimes \mathbb{V}^{(AB)}_{j_{AB}}\right),
\label{eq:decoVAB}
\end{equation}
where the total spin $j_{AB}$ takes all values that are compatible with \textit{any} pair of irreps $j_A$ and $j_B$.

Let $\ket{j_At_{j_A}m_{j_A}}$ and $\ket{j_Bt_{j_B}m_{j_B}}$ denote the spin basis of $\mathbb{V}^{(A)}$ and $\mathbb{V}^{(B)}$ respectively. We introduce a coupled basis $\ket{j_{AB} t_{j_{AB}} m_{j_{AB}}} \in \mathbb{V}^{(AB)}$ that fulfills, 
\begin{align}
	{\textbf{J}^2}^{(AB)}\ket{j_{AB} t_{j_{AB}} m_{j_{AB}}} &= j_{AB}(j_{AB}+1) \ket{j_{AB} t_{j_{AB}} m_{j_{AB}}}, \nonumber \\
	\hat{J}_z^{(AB)}~\ket{j_{AB} t_{j_{AB}} m_{j_{AB}}} 
	&= m_{j_{AB}}~\ket{j_{AB} t_{j_{AB}} m_{j_{AB}}}.
\end{align}
and which is related to the product basis, 
\begin{equation}
\ket{j_{A}t_{j_{A}}m_{j_A};j_B t_{j_{B}} m_{j_B}} \equiv \ket{j_{A}t_{j_{A}}m_{j_A}} \otimes \ket{j_B t_{j_{B}}m_{j_B}}, \nonumber
\end{equation} 
by means of a transformation,
\begin{equation}
\boxed{
\begin{split}
\ket{j_{AB} t_{j_{AB}} m_{j_{AB}}} =
\sum_{t_{j_{A}} t_{j_{B}}}\sum_{m_{j_{A}} m_{j_{B}}} \fusespin{A}{B}{AB}
\ket{j_{A} t_{j_{A}}m_{j_{A}};j_{B} t_{j_{B}} m_{j_{B}}}.
\end{split}
\label{eq:basischange}
}
\end{equation}
The components $\fusespin{A}{B}{AB}$ can be expressed in terms of the Clebsch-Gordan coefficients as
\begin{equation}
\boxed{
\begin{split}
\fusespin{A}{B}{AB} = \tfusespin{A}{B}{AB} \cfusespin{A}{B}{AB}. 
\end{split}
\label{eq:basischange1}
}
\end{equation}
Let us explain how this expression is obtained. From the definition, Eq.~(\ref{eq:basischange}), we have
\begin{align}
\fusespin{A}{B}{AB}\equiv
\braket{j_{AB} t_{j_{AB}} m_{j_{AB}}}{j_{A} t_{j_{A}}m_{j_{A}};j_{B} t_{j_{B}} m_{j_{B}}}.\label{eq:fusedef}
\end{align}
According to the direct sum decomposition, Eq.~(\ref{eq:decoVAB}), each vector $\ket{j_{AB} t_{j_{AB}} m_{j_{AB}}}$ belongs to the subspace $\mathbb{D}^{(AB)}_{j_{AB}} \otimes \mathbb{V}^{(AB)}_{j_{AB}}$ where it factorizes as 
\begin{equation}
\ket{j_{AB} t_{j_{AB}} m_{j_{AB}}} = \ket{j_{AB} t_{j_{AB}}} \otimes \ket{j_{AB} m_{j_{AB}}}.
\end{equation}
Similarly, we can factorize vectors $\ket{j_{A} t_{j_{A}} m_{j_{A}}}$ and $\ket{j_{B} t_{j_{B}} m_{j_{B}}}$. The expression Eq.~(\ref{eq:basischange1}) can then be obtained by substituting these factorizations into Eq.~(\ref{eq:fusedef}) and re-arranging the terms as shown below,
\begin{align}
\fusespin{A}{B}{AB}&=\braket{j_{AB}t_{j_{AB}}}{j_{A}t_{j_A};j_{B}t_{j_B}} \braket{j_{AB}m_{j_{AB}}}{j_{A}m_{j_A};j_{B}m_{j_B}},\nonumber \\
&= \tfusespin{A}{B}{AB} \cfusespin{A}{B}{AB},\nonumber
\end{align}
where $\ket{j_{A}t_{j_A};j_{B}t_{j_B}} \equiv \ket{j_{A}t_{j_A}} \otimes \ket{j_{B}t_{j_B}}$. 

Here $\tfuser$ is a one to one map that relates vectors $\ket{j_{AB} t_{j_{AB}}} \in \mathbb{D}^{(AB)}_{j_{AB}}$ to the vectors $\ket{j_{A}t_{j_A};j_B t_{j_B}} \in \mathbb{D}^{(AB)}_{j_{AB}}$. It can be regarded as a rank-$3$ tensor such that each component of $\tfuser$ is either a zero or a one. We have,
\begin{equation}
\tfusespin{A}{B}{AB} \!\!\!\!\!= \left\{ 
	\begin{array}{cc} 1&\mbox{ if } \ket{j_{AB}t_{j_{AB}}} = \ket{j_{A}t_{j_A};j_{B}t_{j_B}},\\ 
	 									0&\mbox{otherwise }.
	\end{array} \right.
\end{equation}

The product basis can, in turn, be expressed in terms of the coupled basis,
\begin{equation}
\boxed{
\begin{split}
\ket{j_{A} t_{j_{A}}m_{j_{A}};j_{B} t_{j_{B}} m_{j_{B}}} =
\sum_{t_{j_{AB}}}\sum_{m_{j_{AB}}}\splitspin{AB}{A}{B}\ket{j_{AB} t_{j_{AB}} m_{j_{AB}}},\label{eq:revbasischange}
\end{split}
}
\end{equation}
where
\begin{equation}
\boxed{
\begin{split}
\splitspin{AB}{A}{B} =\tsplitspin{AB}{A}{B}\cdot\csplitspin{AB}{A}{B},
\label{eq:u1split1}
\end{split}
}
\end{equation}
and
\begin{equation}
\tsplitspin{AB}{A}{B} \equiv \tfusespin{A}{B}{AB}.
\label{eq:u1split2}
\end{equation}
%%%%%%%%%%%%%%%%%%%%%%%%%%%%%%%%%%%%%%%%%%%%%%%%%%%%%%%%%%%%%%%%%%%%%%%%%%%%%%%%%%%%%%%%%%%%%%%%
\begin{figure}[t]
\begin{center}
  \includegraphics[width=10cm]{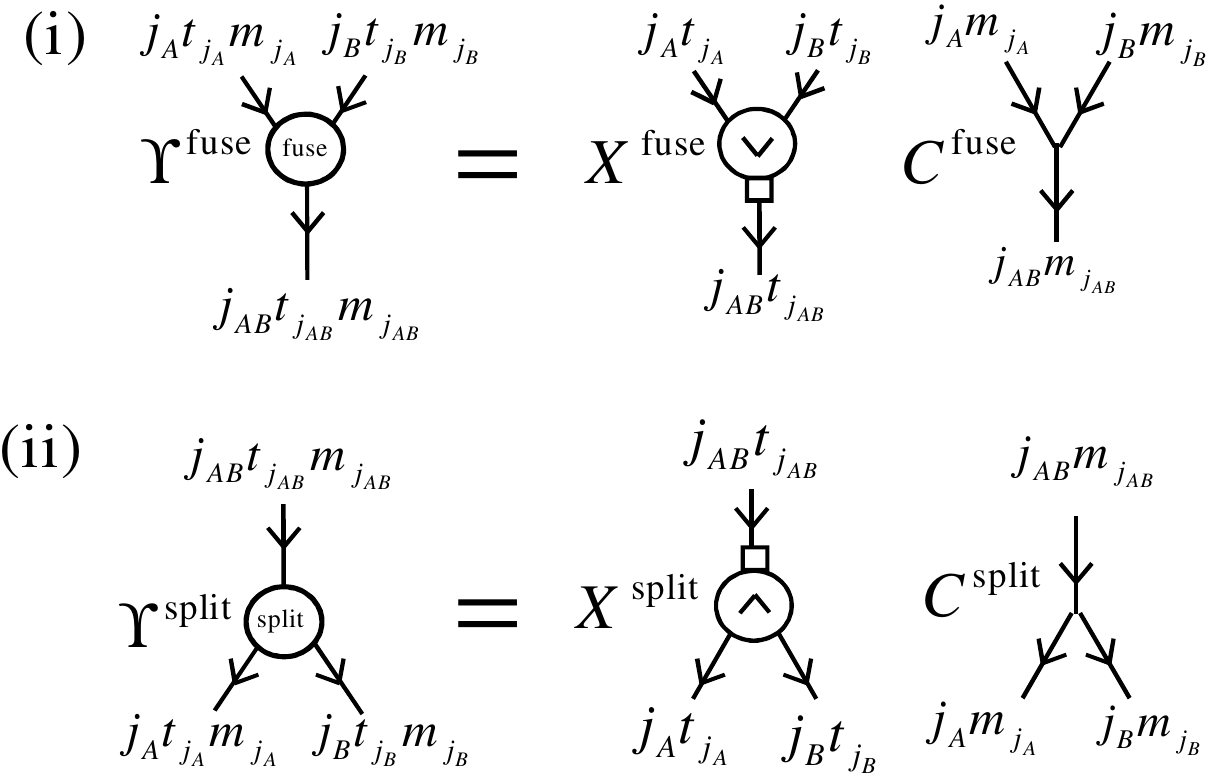}
\end{center}
\caption{The graphical representations of (i) the fusing tensor $\fuser$ and (ii) the splitting tensor. For fixed values of $j_A, j_B$ and $j_{AB}$ each of these tensors  factorizes into a $X$ and a $C$ tensor.\label{fig:su2fuse}}
\end{figure}
%%%%%%%%%%%%%%%%%%%%%%%%%%%%%%%%%%%%%%%%%%%%%%%%%%%%%%%%%%%%%%%%%%%%%%%%%%%%%%%%%%%%%%%%%%%%%%%%

We refer to the tensors $\fuser$ and $\spliter$ as the \textit{fusing} tensor and the \textit{splitting} tensor respectively, since they will play an instrumental role in fusing and splitting indices of an SU(2)-invariant tensor. The special graphical representation of these tensors and their decomposition into $X$ and $C$ tensors is shown in Fig.~\ref{fig:su2fuse}. Tensors $\tfusespin{A}{B}{AB}$ and $\tsplitspin{AB}{A}{B}$ are graphically represented by means of a circle enclosing an arrow head and three lines emerging from the circle corresponding to the three indices of the tensors. The three lines in the diagrams of $\tfuser$ and $\tspliter$ correspond to the degeneracy indices $(j_A, t_{j_A}), (j_B, t_{j_B})$ and $(j_{AB}, t_{j_{AB}})$ by using the same assignment rules that were introduced for tensors $\cfuser$ and $\cspliter$ respectively. Other features of the graphical representation include an arrow head that is placed within the circle to indicate the direction of the fusion and a small rectangle, placed on the line carrying the coupled spins, that represents a permutation of basis elements.

We notice that tensor $\tfuser$ can be decomposed into two pieces. The first piece (depicted as the circle enclosing an arrow head) expresses a basis $\{\ket{j_{A}t_{j_{A}}; j_{B}t_{j_{B}} \equiv \ket{j_{A}t_{j_A}} \otimes \ket{j_{B}t_{j_B}}}\}$ of $\mathbb{D}^{(AB)}$ as the direct product of the basis $\{\ket{j_{A}t_{j_A}}\}$ of $\mathbb{D}^{(A)}$ and the basis $\{\ket{j_{B}t_{j_B}}\}$ of $\mathbb{D}^{(B)}$.  Note that this procedure does not always lead to the set $\{\ket{j_{A}t_{j_{A}}; j_{B}t_{j_{B}}}\}$ being ordered such that states corresponding to the same total spin $j_{AB}$ are adjacent to each other within the set. However, we require that the basis associated to an index be maintained as such (this ensures, for example, that an SU(2)-invariant matrix is block diagonal when expressed in such a basis). This ordering is achieved by means of the second piece (depicted as the small rectangle): a permutation of basis states $\{\ket{j_{A}t_{j_{A}}; j_{B}t_{j_{B}}}\}$ that reorganizes them according to their total spin $j_{AB}$, so that they are identified in an one-to-one correspondence with the coupled states $\{\ket{j_{AB}t_{j_{AB}}}\}$. In particular, this description of the tensors $\tfuser$ and $\tspliter$ can be exploited to multiply together several such tensors, such as in Fig.~\ref{fig:fmove1}(iv), in a fast way.

By construction, a resolution of Identity can be obtained in terms of tensor $\fuser$ and tensor $\spliter$, as shown in Fig.~\ref{fig:su2fuse}(ii)-(iii).

\textbf{Example 7: } Let vector spaces $\mathbb{V}^{(A)}$ and $\mathbb{V}^{(B)}$ correspond to the vector space of Example 4, that is,
\begin{align}
\mathbb{V}^{(A)} &\cong 3\mathbb{V}^{(A)}_{\half} \cong \mathbb{D}_{\half}^{(A)}\otimes\mathbb{V}^{(A)}_{\half}, \nonumber \\
\mathbb{V}^{(B)} &\cong 3\mathbb{V}^{(B)}_{\half} \cong \mathbb{D}_{\half}^{(B)} \otimes \mathbb{V}^{(B)}_{\half}. 
\end{align}
The space $\mathbb{V}^{(AB)}$ decomposes as
\begin{align}
\mathbb{V}^{(AB)} \cong \left(\mathbb{D}^{(AB)}_{0} \otimes \mathbb{V}^{(AB)}_{0}\right) \oplus \left(\mathbb{D}^{(AB)}_{1} \otimes \mathbb{V}^{(AB)}_{1}\right),
  \label{eq:eg50}
\end{align}
where,
\begin{align}
\mathbb{V}_{\half}^{(A)} \otimes \mathbb{V}_{\half}^{(B)} \cong (\mathbb{V}^{(AB)}_{0} \oplus \mathbb{V}^{(AB)}_{1}),  \label{eq:eg71}
\end{align}
and
\begin{align}
\mathbb{D}^{(AB)}_{0} &\cong \mathbb{D}_{\half}^{(A)} \otimes \mathbb{D}_{\half}^{(B)}, \label{eq:degbasis1}\\
\mathbb{D}^{(AB)}_{1} &\cong \mathbb{D}_{\half}^{(A)} \otimes \mathbb{D}_{\half}^{(B)}.\label{eq:eg72}
\end{align}
Recall that the basis of the l.h.s. and r.h.s. of Eq.~(\ref{eq:eg71}) are related by the transformations Eqs.~(\ref{eq:halfsinglet})-(\ref{eq:eg41}). Let us now consider how the basis of the l.h.s. and r.h.s. of Eq.~(\ref{eq:degbasis1}) and of Eq.~(\ref{eq:eg72}) are related. In Eq.~(\ref{eq:degbasis1}), for instance, the vectors 
\begin{equation}
\ket{j_{AB}=0, t_{j_{AB}}} \in \mathbb{D}^{(AB)}_{0},~~t_{j_{AB}}=1,2,\ldots,9, \nonumber
\end{equation}
are related to the vectors 
\begin{equation} 
\ket{j_A = \half, t_{j_A}; j_B = \half, t_{j_B}}, \in \mathbb{D}^{(AB)}_{0} ~~t_{j_A}, t_{j_B} = 1,2,3, \nonumber
\end{equation} 
in straightforward way by associating the vectors in a one to one fashion in the order in which they appear in the respective basis. For example, the change of basis maps the vector
\begin{equation}
\ket{j_{AB}=0, t_{j_{AB}}=1} \mbox{ to } \ket{j_A = \half, t_{j_A}=1; j_B = \half, t_{j_B}=1} \nonumber
\end{equation}
and the vector 
\begin{equation}
\ket{j_{AB}=0, t_{j_{AB}}=2} \mbox{ to } \ket{j_A = \half, t_{j_A}=2; j_B = \half, t_{j_B}=1} \nonumber
\end{equation}
and so on. The basis of the l.h.s. and r.h.s. of Eq.~(\ref{eq:eg72}) are related in a similar way. This one to one mapping can be encoded into $\tfusespin{A}{B}{AB}$ by setting the numerical value of the following components equal to one,
\begin{align}
\tfuse{\half 1}{\half 1}{01} &,~~ \tfuse{\half 1}{\half 2}{02} ,~~ \tfuse{\half 1}{\half 3}{03}, \nonumber \\
\tfuse{\half 2}{\half 1}{01} &,~~ \tfuse{\half 2}{\half 2}{02} ,~~ \tfuse{\half 2}{\half 3}{03}, \nonumber\\
\tfuse{\half 3}{\half 1}{01} &,~~ \tfuse{\half 3}{\half 2}{02} ,~~ \tfuse{\half 3}{\half 3}{03},\nonumber
\end{align}
and
\begin{align}
\tfuse{\half 1}{\half 1}{11} &,~~ \tfuse{\half 1}{\half 2}{12} ,~~ \tfuse{\half 1}{\half 3}{13}, \nonumber\\
\tfuse{\half 2}{\half 1}{11} &,~~ \tfuse{\half 2}{\half 2}{12} ,~~ \tfuse{\half 2}{\half 3}{13}, \nonumber \\
\tfuse{\half 3}{\half 1}{11} &,~~ \tfuse{\half 3}{\half 2}{12} ,~~ \tfuse{\half 3}{\half 3}{13}.\nonumber\markend
\end{align}
%%%%%%%%%%%%%%%%%%%%%%%%%%%%%%%%%%%%%%%%%%%%%%%%%%%%%%%%%%%%%%%%%%%%%%%%%%%%%%%%%%%%%%%%%%%%%%%%
\begin{figure}[t]
\begin{center}
  \includegraphics[width=9cm]{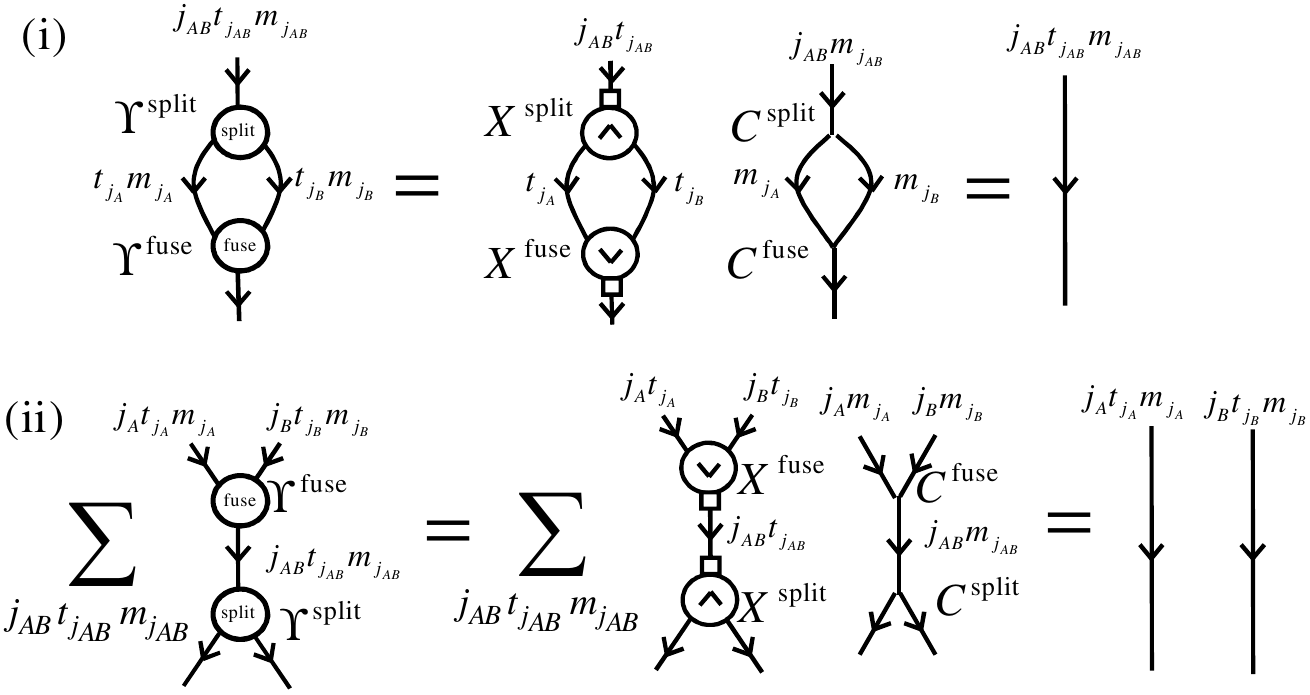}
\end{center}  
\caption{Tensors $\fuser$ and $\spliter$ are unitary and thus yield the Identity when contracted pairwise either as (i) or (ii).
\label{fig:su2fuse1}}
\end{figure}
%%%%%%%%%%%%%%%%%%%%%%%%%%%%%%%%%%%%%%%%%%%%%%%%%%%%%%%%%%%%%%%%%%%%%%%%%%%%%%%%%%%%%%%%%%%%%%%%

\textbf{Example 8: } As another example of the change of basis $\tfuser$, consider that $\mathbb{V}^{(A)}$ and $\mathbb{V}^{(B)}$ correspond to the vector spaces of Example 3 and Example 5 respectively. That is,
\begin{align}
\mathbb{V}^{(A)} &\cong \mathbb{V}^{(A)}_{1} \cong (\mathbb{D}^{(A)}_{1} \otimes \mathbb{V}^{(A)}_{1}), \nonumber \\
\mathbb{V}^{(B)} &\cong 2\mathbb{V}^{(B)}_{0} \oplus \mathbb{V}^{(B)}_{1} \cong (\mathbb{D}_{0}^{(B)}\otimes\mathbb{V}^{(B)}_{0}) \oplus (\mathbb{D}_{1}^{(B)}\otimes\mathbb{V}^{(B)}_{1}).
\end{align}
The space $\mathbb{V}^{(AB)}$ decomposes as
\begin{align}
\mathbb{V}^{(AB)} \cong (\mathbb{D}_{0}^{(AB)} \otimes \mathbb{V}^{(AB)}_{0}) \oplus(\mathbb{D}_{1}^{(AB)} \otimes \mathbb{V}^{(AB)}_{1})\oplus (\mathbb{D}_{2}^{(AB)} \otimes\mathbb{V}^{(AB)}_{2}),
  \label{eq:eg60}
\end{align}
where 
\begin{align}
\mathbb{D}_{0}^{(AB)} &\cong \mathbb{D}_{1}^{(A)} \otimes \mathbb{D}_{1}^{(B)} \label{eq:x1} \\
\mathbb{D}_{1}^{(AB)} &\cong (\mathbb{D}_{0}^{(A)} \otimes \mathbb{D}_{1}^{(B)}) \oplus (\mathbb{D}_{1}^{(A)} \otimes \mathbb{D}_{0}^{(B)}) \label{eq:x2} \\
\mathbb{D}_{2}^{(AB)} &\cong \mathbb{D}_{1}^{(A)} \otimes \mathbb{D}_{1}^{(B)}. \label{eq:x3}
\end{align}
The transformation that relates the bases of l.h.s. and r.h.s. of Eq.~(\ref{eq:x1}) and of Eq.~\ref{eq:x3}) is straightforward, we set
\begin{align}
\tfuse{0 1}{1 1}{0 1} = \tfuse{1 1}{1 1}{2 1} = 1. \nonumber
\end{align}
The basis of the l.h.s. and r.h.s. of Eq.~(\ref{eq:x2}) can be related by mapping the three vectors 
\begin{equation}
\ket{j_{AB}=1, t_{j_{AB}}=1,2,3} \in \mathbb{D}_{1}^{(AB)},\nonumber
\end{equation}
in a one to one manner, to the two vectors 
\begin{equation}
\ket{j_A = 1, t_{j_A}=1; j_B = 0, t_{j_B}=1,2} \in (\mathbb{D}_{0}^{(A)} \otimes \mathbb{D}_{1}^{(B)}),\nonumber
\end{equation}
and the vector 
\begin{equation}
\ket{j_A = 0, t_{j_A}=1; j_B = 1, t_{j_B}=1} \in (\mathbb{D}_{1}^{(A)} \otimes \mathbb{D}_{0}^{(B)}). \nonumber
\end{equation}
This is encoded into $\tfuser$ by setting
\begin{align}
\tfuse{0 2}{1 1}{1 2} = \tfuse{1 1}{1 1}{0 1} = \tfuse{1 1}{1 1}{1 3} = 1.\markend \nonumber
\end{align}

\section{Review: Fusion trees \label{sec:fusiontree}}

When considering the tensor product of more than two representations one can obtain several coupled bases of the product space. These correspond to taking the product of the vector spaces according to different sequences of pairwise products. The spaces are linearly ordered in a given way and we only consider a pairwise product of `adjacent' spaces in this linear ordering. For example, when considering the tensor product of three representations, 
\begin{equation}
\mathbb{V}^{(ABC)} \cong \mathbb{V}^{(A)} \otimes \mathbb{V}^{(B)} \otimes \mathbb{V}^{(C)},
\end{equation}
one can consider either the pairwise products 
\begin{align}
\mathbb{V}^{(D)} &\cong \mathbb{V}^{(A)} \otimes \mathbb{V}^{(B)}, \nonumber \\
\mathbb{V}^{(ABC)} &\cong \mathbb{V}^{(D)} \otimes \mathbb{V}^{(C)}, \label{eq:order1}
\end{align}
or the pairwise products
\begin{align}
\mathbb{V}^{(E)} &\cong \mathbb{V}^{(B)} \otimes \mathbb{V}^{(C)}, \nonumber \\
\mathbb{V}^{(ABC)} &\cong \mathbb{V}^{(A)} \otimes \mathbb{V}^{(E)}. \label{eq:order2}
\end{align}
Considering the tensor product one way or the other leads to two different coupled bases in $\mathbb{V}^{(ABC)}$.

More generally, there exist several choices of a coupled basis in the tensor product of $L$ representations, Eqs.~(\ref{eq:latticespace})-(\ref{eq:latticerep}). A useful way to specify a particular sequence of pairwise products is by means of a \textit{fusion tree}. 

\begin{figure}[t]
\begin{center}
  \includegraphics[width=10cm]{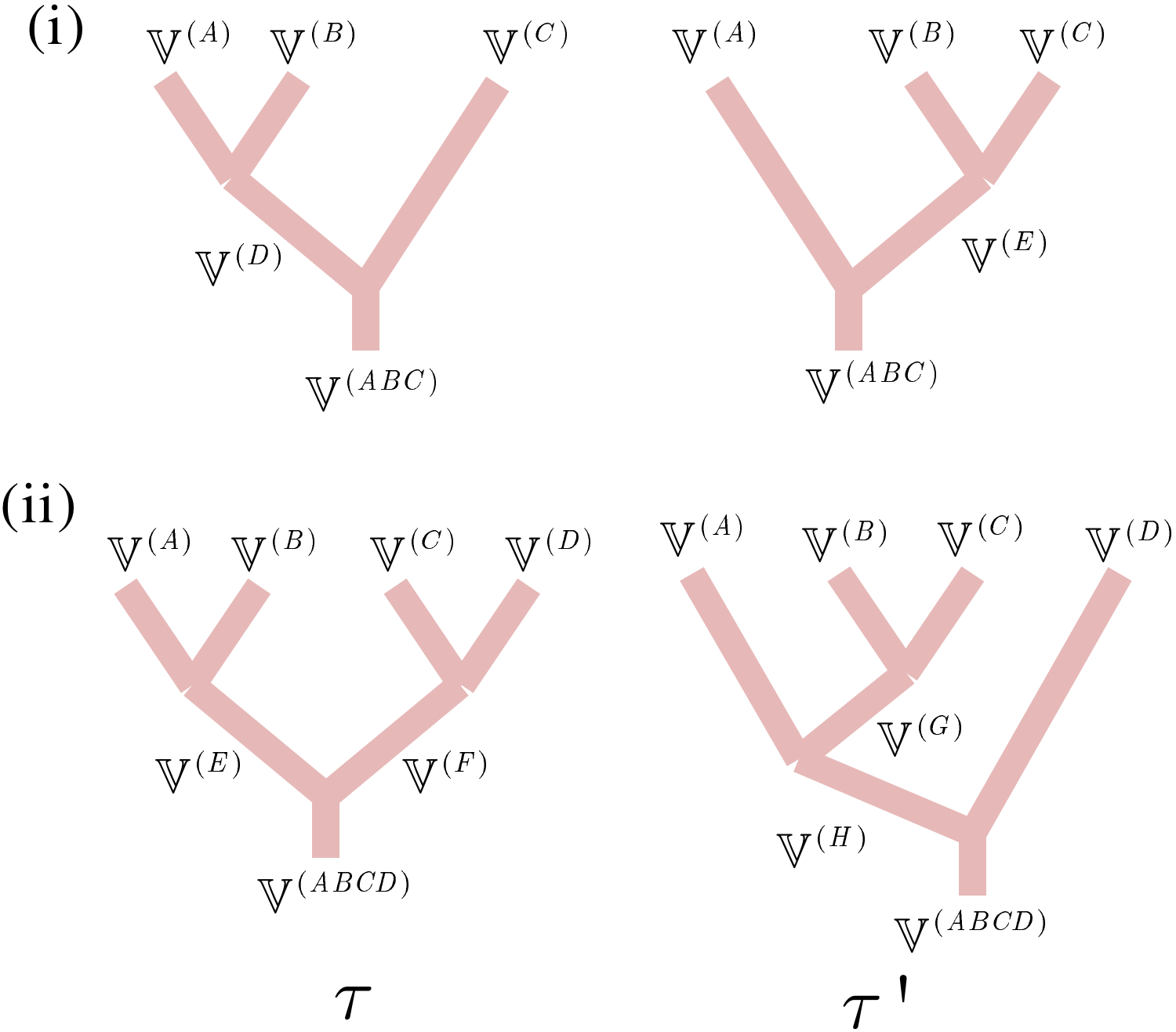}
\end{center}
\caption{Examples of two fusion trees $\tree$ and $\tree'$ that describe two different ways of considering the tensor product of (i) three vector spaces $\mathbb{V}^{(A)}, \mathbb{V}^{(B)}$ and $\mathbb{V}^{(C)}$ and (ii) four vector spaces $\mathbb{V}^{(A)}, \mathbb{V}^{(B)}, \mathbb{V}^{(C)}$ and $\mathbb{V}^{(D)}$. Fusion proceeds from top to bottom.
\label{fig:fusion}}
\end{figure}

A fusion tree, denoted $\tree$, is a \textit{directed} trivalent tree such that each node of the tree represents the tensor product of two incoming spaces into the outgoing space. The tree has a total of $L+1$ open links which correspond to the $L$ vector spaces $\mathbb{V}^{(1)}, \mathbb{V}^{(2)}, \ldots, \mathbb{V}^{(L)}$ and the product space $\mathbb{V}$. The internal links correspond to the intermediate product spaces that appear in a sequence of pairwise products. Figure~\ref{fig:fusion} illustrates two different fusions trees $\tree$ and $\tree'$ that correspond to two different ways of considering the tensor product of three and of four representations. The sequence of fusions proceeds from top to bottom. 
%%%%%%%%%%%%%%%%%%%%%%%%%%%%%%%%%%%%%%%%%%%%%%%%%%%%%%%%%%%%%%%%%%%%%%%%%%%%%%%%%%%%%%%%%%%%%%%%%%%%%%%%%%%%%%%%%%%%%%%%%%
%%%%%%%%%%%%%%%%%%%%%%%%%%%%%%%%%%%%%%%%%%%%%%%%%%%%%%%%%%%%%%%%%%%%%%%%%%%%%%%%%%%%%%%%%%%%%%%%%%%%%%%%%%%%%%%%%%%%%%%%%%
A fusion tree can also be specified as a list of fusions. For example, the fusion trees $\tree$ and $\tree'$ depicted in Fig.~\ref{fig:fusion}(i) can be specified as
\begin{align}
\tree &\equiv \{A,B\rightarrow D; D,C\rightarrow (ABC) \}, \nonumber \\
\tree' &\equiv \{B,C\rightarrow E; A,E\rightarrow (ABC) \}. \nonumber
\end{align}

Fusion trees play an important role in our discussion. In the present context, a fusion tree characterizes a coupled basis of a tensor product space. In Sec.~\ref{sec:symTensor} fusion trees are also used to characterize different canonical decompositions of an SU(2)-invariant tensor.

In the remainder of the section we consider coupled bases that are labeled by different fusion trees. We define the unitary transformation that relates two such coupled bases. This transformation is important since it also relates two different canonical decompositions of an SU(2)-invariant tensor, as discussed in Sec.~\ref{sec:symTensor}. For purpose of illustration, we first characterize the coupled basis in the simple case of the tensor product of three representations before proceeding to the generic case of $L$ representations. 
%The tensor product of \textit{reducible} representations is separately discussed in Appendix A. Despite its limited relevance to the present discussion, the transformation that relates two coupled bases in the general case also relates two different \textit{tree decompositions} of an SU(2)-invariant tensor. This is discussed in Appendix B.

\subsection{Tensor product of three irreps}
Let vector spaces $\mathbb{V}^{(A)}, \mathbb{V}^{(B)}$ and $\mathbb{V}^{(C)}$ transform as irreps $j_A, j_B$ and $j_C$ respectively. The space $\mathbb{V}^{(ABC)}$ is in general reducible, and may contain several copies of an irrep $j_{ABC}$. Let us first consider the sequence (\ref{eq:order1}) of tensor products corresponding to a fusion tree $\tree$. The vector spaces $\mathbb{V}^{(D)}$ and $\mathbb{V}^{(ABC)}$ decompose as
\begin{equation}
\mathbb{V}^{(D)} \cong \bigoplus_{j_D} \mathbb{V}^{(D)}_{j_D},\label{eq:intermediate1}
\end{equation}
and 
\begin{equation}
\mathbb{V}^{(ABC)} \cong \bigoplus_{j_{ABC}, j_D} \mathbb{V}_{j_{ABC}, j_D}^{(ABC)}. \label{eq:tensorirrep3}
\end{equation}
Notice that we can use the values of $j_D$ that appear on the r.h.s. of Eq.~(\ref{eq:intermediate1}) to label different copies of $j_{ABC}$ that appear on the r.h.s. of Eq.~(\ref{eq:tensorirrep3}).  Thus, a coupled basis of $\mathbb{V}^{(ABC)}$ can be labeled as $\ket{j_{ABC}m_{j_{ABC}};\tree;j_D}$. 

Let $(\hat{Q}^{j_D}_{j_A j_B j_C j_{ABC}})_{m_{j_A} m_{j_B} m_{j_C} m_{j_{ABC}}}$ denote the transformation from the product basis\\  $\ket{j_Am_{j_A}}\otimes\ket{j_Bm_{j_B}}\otimes\ket{j_Cm_{j_C}}$ to the coupled basis $\ket{j_{ABC}m_{j_{ABC}};\tree; j_D}$. This change of basis can be expressed in terms of Clebsch-Gordan coefficients as
\begin{align}
&(\hat{Q}^{j_D}_{j_A j_B j_C j_{ABC}})_{m_{j_A} m_{j_B} m_{j_C} m_{j_{ABC}}}\equiv
\sum_{m_{j_D}}\cfusespin{A}{B}{D} \cdot \cfusespin{D}{C}{ABC},
\label{eq:cob1}
\end{align}
where $\cfusespin{A}{B}{D}$ relates the basis $\ket{j_A, m_{j_A}}~\otimes~\ket{j_B,m_{j_B}}$ to the intermediate basis $\ket{j_Dm_{j_D}}$, and $\cfusespin{D}{C}{ABC}$ relates this intermediate basis to the coupled basis $\ket{j_{ABC}m_{j_{ABC}};\tree;j_D}$.

Alternatively, we can first consider the tensor product $\mathbb{V}^{(B)} \otimes \mathbb{V}^{(C)}$ (corresponding to the sequence (\ref{eq:order2}) of tensor products characterized by another fusion tree $\tree'$),
\begin{equation}
\mathbb{V}^{(E)} \cong \mathbb{V}^{(B)} \otimes \mathbb{V}^{(C)} \cong \bigoplus_{j_E} \mathbb{V}^{(E)}_{j_E}, \label{eq:intermediate2}
\end{equation}
and use irrep $j_E$ to label another coupled basis $\ket{j_{ABC}m_{j_{ABC}};\tree';j_E}$ of $\mathbb{V}^{(ABC)}$. Denote by $(\hat{Q}'^{j_E}_{j_A j_B j_Cj_{ABC}})_{m_{j_A} m_{j_B} m_{j_C} m_{j_{ABC}}}$ the change of basis to this new coupled basis. In terms of Clebsch-Gordan coefficients we have
\begin{align}
(\hat{Q}'^{j_E}_{j_A j_B j_C j_{ABC}})_{m_{j_A} m_{j_B} m_{j_C} m_{j_{ABC}}} \equiv
\sum_{m_{j_E}}\cfusespin{B}{C}{E}\cdot\cfusespin{A}{E}{ABC}.
\label{eq:cob2}
\end{align}
%%%%%%%%%%%%%%%%%%%%%%%%%%%%%%%%%%%%%%%%%%%%%%%%%%%%%%%%%%%%%%%%%%%%%%%%%%%%%%%%%%%%%%%%%%%%%%%%%%%%%%%%%%%%%%%%%%%%%%%%%%
\begin{figure}[t]
\begin{center}
  \includegraphics[width=10cm]{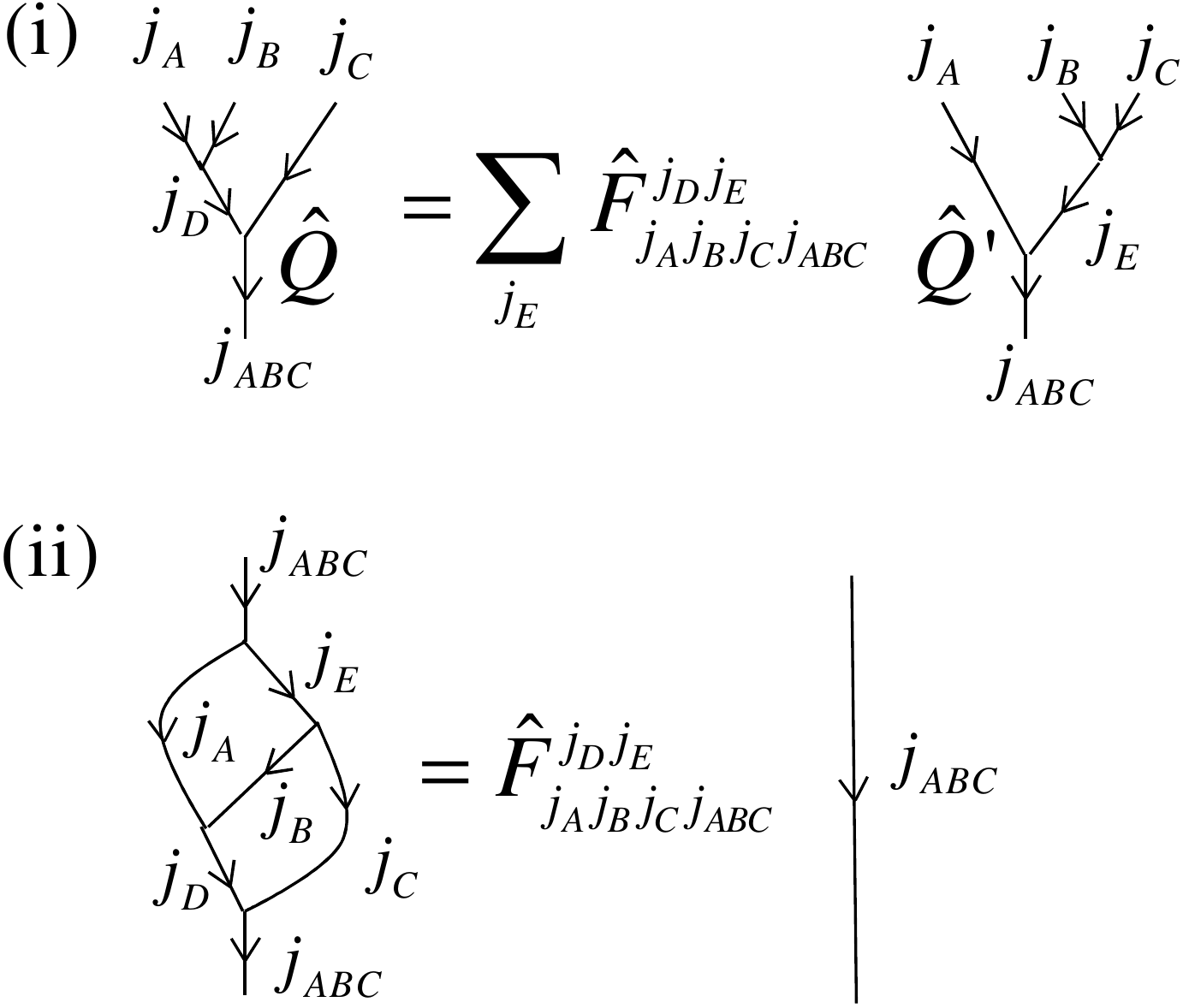}
\end{center}  
\caption{(i) In the tensor product of three irreps, two coupled bases that are labeled by fusion trees $\tree$ (\ref{eq:order1}) and $\tree'$ (\ref{eq:order2}) are related to one another by means of the recoupling coefficients $\hat{F}^{j_D j_E}_{j_A j_B j_C j_{ABC}}$. (ii) A recoupling coefficient is given as the `scalar product' of $\hat{Q}$ and $\hat{Q}'$.
\label{fig:fmove}}
\end{figure}
%%%%%%%%%%%%%%%%%%%%%%%%%%%%%%%%%%%%%%%%%%%%%%%%%%%%%%%%%%%%%%%%%%%%%%%%%%%%%%%%%%%%%%%%%%%%%%%%%%%%%%%%%%%%%%%%%%%%%%%%%%
The two coupled bases $\ket{j_{ABC}m_{j_{ABC}};\tree;j_D}$ and $\ket{j_{ABC}m_{j_{ABC}};\tree';j_E}$ are related by a transformation that is given by a rank-$6$ tensor $\hat{F}$ with components $\hat{F}^{j_Dj_E}_{j_Aj_Bj_Cj_{ABC}}$,
\begin{equation}
\boxed{\hat{Q}'^{j_E}_{j_A j_B j_C j_{ABC}} = \sum_{j_D}\hat{F}^{j_Dj_E}_{j_Aj_Bj_Cj_{ABC}} \hat{Q}^{j_D}_{j_A j_B j_C j_{ABC}}.}
\label{eq:fmove}
\end{equation}
Here $\hat{F}^{j_Dj_E}_{j_Aj_Bj_Cj_{ABC}}$ are the \textit{recoupling coefficients}\citep{Cornwell97} of SU(2). By using Eqs.~(\ref{eq:cob1}) and (\ref{eq:cob2}), the recoupling coefficients can be explicitly expressed in terms of Clebsch-Gordan coefficients,
\begin{align}
\hat{F}^{j_Dj_E}_{j_Aj_Bj_Cj_{ABC}} &\equiv \frac{1}{2j_{ABC}+1} \times \nonumber \\
&\sum_{\textbf{m}} \left(\cfusespin{A}{B}{D} \cfusespin{D}{C}{ABC}\right. \nonumber\\
&~~~\left.\cfusespin{B}{C}{E} \cfusespin{D}{C}{ABC}\right),\label{eq:f}
\end{align}
where $\textbf{m} \equiv \{m_{j_D},m_{j_E},m_{j_A},m_{j_B},m_{j_C},m_{j_{ABC}}\}$. Notice that, since the $m$'s are summed over, the recoupling coefficients depend only on the $j$'s. Also recall that the recoupling coefficients are proportional to the 6-j symbols of the group,
\begin{equation}
\hat{F}^{j_Dj_E}_{j_Aj_Bj_Cj_{ABC}} = \alpha \left\{\begin{array}{ccc} j_A&j_B&j_D\\j_C&j_{ABC}&j_E \end{array}\right\},
\label{eq:sixj}
\end{equation}
where 
\begin{equation}
\alpha \equiv (-1)^{(j_A+j_B+j_C+j_{ABC})}\sqrt{(2j_D+1)(2j_E+1)}.
\end{equation}

\subsection{Tensor product of three reducible representations}

Consider the action of SU(2) on the space $\mathbb{V}^{(ABC)}$, 
\begin{equation}
\mathbb{V}^{(ABC)} \cong \mathbb{V}^{(A)}\otimes \mathbb{V}^{(B)} \otimes \mathbb{V}^{(C)},\nonumber  
\end{equation}
where $\mathbb{V}^{(A)}, \mathbb{V}^{(B)}$ and $\mathbb{V}^{(C)}$ are reducible representations of SU(2). It induces a decomposition
\begin{equation}
\mathbb{V}^{(ABC)} \cong \bigoplus_{j_{ABC}} \left(\mathbb{D}^{(ABC)}_{j_{ABC}} \otimes \mathbb{V}^{(ABC)}_{j_{ABC}}\right),
\label{eq:decoVABC}
\end{equation}
where $j_{ABC}$ takes all values that are compatible with any $j_A, j_B$ and $j_C$.

\begin{figure}[t]
\begin{center}
  \includegraphics[width=12cm]{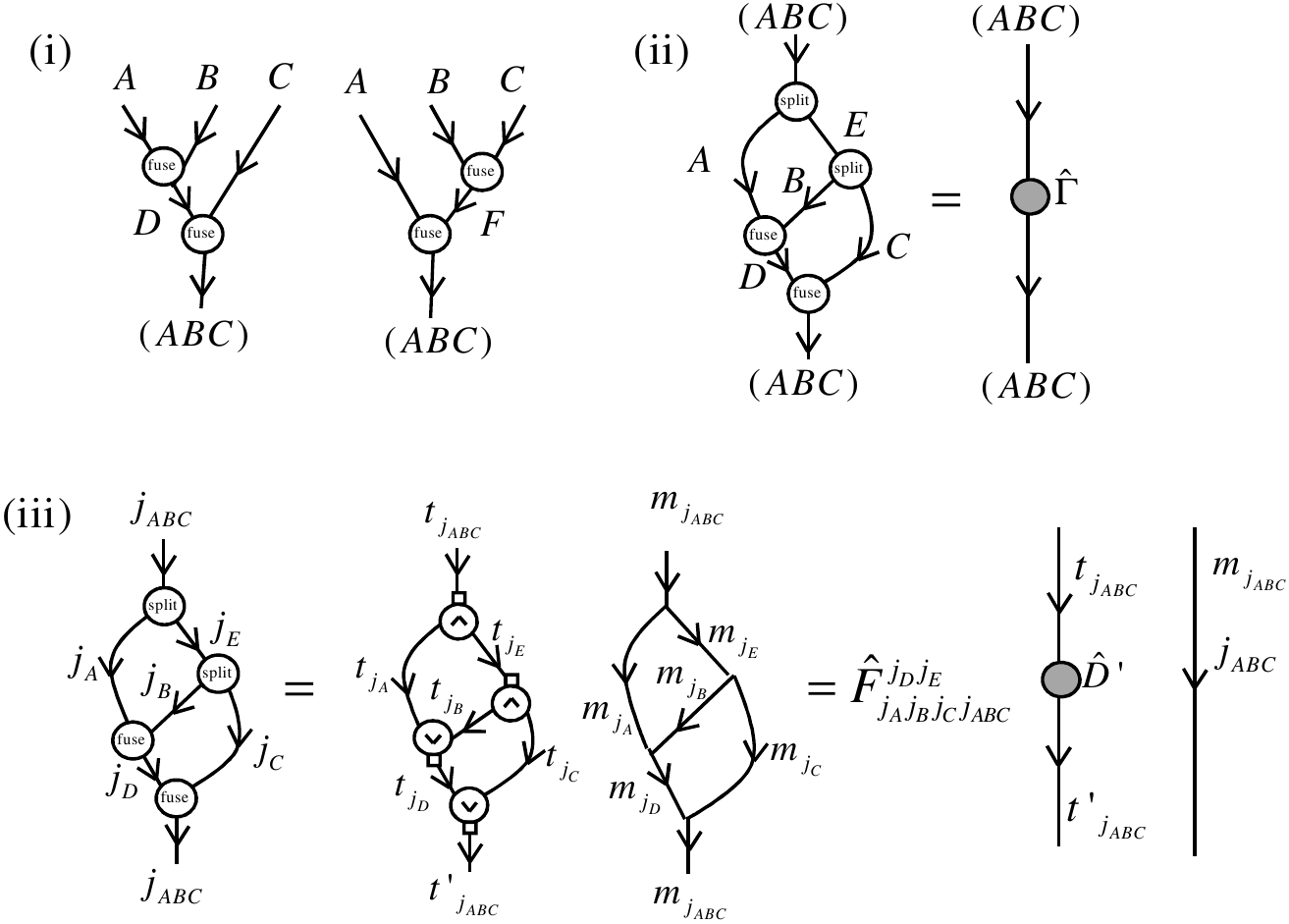}
\end{center} 
\caption{(i) The change of basis from the product basis to two different coupled bases in the fusion of three reducible representations is given in terms of two fusing tensors. (ii) The two coupled basis are related by means of the matrix $\hat{\Gamma}$ [Eq.~(\ref{eq:gammasplit})] that is obtained by contracting a tensor network made of tensors $\fuser$ and tensors $\spliter$. (iii) The components of $\hat{\Gamma}$ are given in terms of recoupling coefficients. This can be seen by performing the contraction piecewise. For fixed values of spins $j_A, j_B, j_C, j_{ABC}, j_D$ and $j_E$ the tensor network decomposes into a tensor network made of $X$ tensors and a spin network. The former is contracted to obtain a matrix $\hat{D}'$ whereas the latter can be replaced by the Identity $\hat{I}_{2j_{ABC}+1}$ and a recoupling coefficient [Fig.~\ref{fig:fmove}(ii)].
\label{fig:fmove1}}
\end{figure}

Extending the argument for irreps, we can relate the coupled basis of $\mathbb{V}^{(ABC)}$ to the product basis by first considering the sequence (\ref{eq:order1}) of tensor products and using two  $\fuser$ tensors
\begin{equation}
\fuse{A}{B}{D},~~~\fuse{D}{C}{(ABC)}
\end{equation}
to relate at each step the coupled basis with the product basis. Alternatively, we can consider the sequence (\ref{eq:order2}) of tensor products and use the different set of fusing tensors 
\begin{equation}
\fuse{B}{C}{E},~~~\fuse{A}{E}{(ABC)}
\end{equation}
to relate the product basis to the coupled basis at each step. The respective change of basis transformation for the two cases is depicted in Fig.~\ref{fig:fmove1}(i). 

The two coupled bases, so obtained, are related by means of a matrix $\hat{\Gamma}$ that decomposes, according to Schur's Lemma [Eq.~(\ref{eq:Schur})] as
\begin{equation}
\boxed{\hat{\Gamma} = \bigoplus_{j_{ABC}} (\hat{D}_{j_{ABC}} \otimes \hat{I}_{j_{ABC}}),} \label{eq:gammasplit}
\end{equation}
where the components of $\hat{D}_{j_{ABC}}$ can be expressed in terms of recoupling coefficients. This decomposition can be derived as follows.
%%%%%%%%%%%%%%%%%%%%%%%%%%%%%%%%%%%%%%%%%%%%%%%%%%%%%%%%%%%%%%%%%%%%%%%%%%%%%%%%%%%%%%%%%%%%%%%%%%%%%%%%%%%%%%%%%%%%%%%%%%
%%%%%%%%%%%%%%%%%%%%%%%%%%%%%%%%%%%%%%%%%%%%%%%%%%%%%%%%%%%%%%%%%%%%%%%%%%%%%%%%%%%%%%%%%%%%%%%%%%%%%%%%%%%%%%%%%%%%%%%%%%

The matrix $\hat{\Gamma}$ is obtained by contracting the tensor network made of tensors $\fuser$ and tensors $\spliter$ that is shown in Fig.~\ref{fig:fmove1}(ii). This contraction can be performed piecewise [Fig.~\ref{fig:fmove1}(iii)]. For fixed values of $j$'s on all links the tensor network factorizes into two pieces since each constituent tensor $\fuser$ and tensor $\spliter$ factorizes into a $X$ and a $C$ tensor. The tensor network made of $C$ tensors equates [Fig~\ref{fig:fmove}(ii)] the Identity times the recoupling coefficient $\hat{F}^{j_E}_{j_A j_B j_C j_{ABC}}$. The matrix $\hat{D}_{j_{ABC}}$ in Eq.~(\ref{eq:gammasplit}) is then defined as
\begin{equation}
\hat{D}_{j_{ABC}} \equiv \sum_{j_A j_B j_C j_D j_E} \hat{F}^{j_E}_{j_A j_B j_C j_{ABC}} \hat{D'}_{j_A j_B j_C j_{ABC}}^{j_E}, \label{eq:D}
\end{equation}
where $\hat{D'}_{j_A j_B j_C j_{ABC}}^{j_E}$ denotes the matrix that is obtained by contracting together the $X$ tensors. Here the sum is over all values of $j_A, j_B, j_C, j_D,$ and $j_E$ that are compatible with a given value of $j_{ABC}$.  

\subsection{Tensor product of $L$ irreps \label{sec:symmetry:tpL}}

In a similar way, we can consider the tensor product of four irreps; different choices of a coupled basis, corresponding to different fusion trees, are related by the $9-j$ symbols and so on. 

More generally, let us consider the tensor product of $L$ representations, Eq.~(\ref{eq:latticespace}), where each space $\mathbb{V}^{(l)}$ ($l=1,2,\ldots, L$) transforms as an irrep $j_{l}$. A coupled basis can be labeled by a fusion tree $\tree$ and the set of intermediate irreps $j_{e_1}, j_{e_2}, \ldots, j_{e_Z}$ that are assigned to the internal links of $\tree$. We denote by 
\begin{equation}
\ket{jm_{j}; \tree; j_{e_1} \ldots j_{e_{Z}}}~\in~\mathbb{V} \label{eq:couplebasis}
\end{equation}
such a basis. By attaching the appropriate Clebsch-Gordan tensor $\cfuser$ to each node of $\tree$ and contracting the resulting tree tensor network we can obtain tensors
\begin{equation}
\hat{Q}^{j_{e_1} \ldots j_{e_{Z}}}_{j_{1} \ldots j_{L}}(\tree), \label{eq:Q}
\end{equation}
that mediate the change from the product basis to this coupled basis.
%%%%%%%%%%%%%%%%%%%%%%%%%%%%%%%%%%%%%%%%%%%%%%%%%%%%%%%%%%%%%%%%%%%%%%%%%%%%%%%%%%%%%%%%%%%%%%%%%%%%%%%%%%%%%%%%%%%%%%%%%%
\begin{figure}[t]
\begin{center}
  \includegraphics[width=14cm]{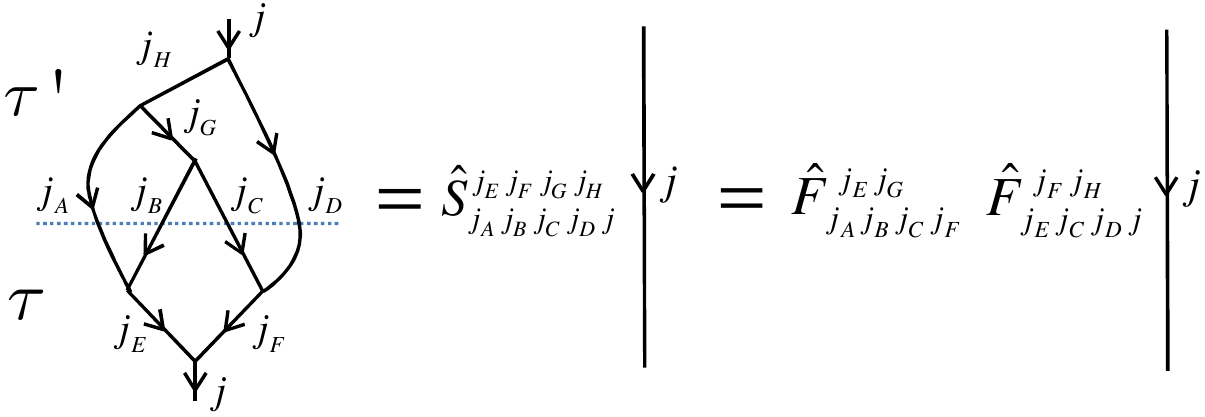}
\end{center}  
\caption{The spin network that relates two coupled bases labeled by fusion trees $\tree$ and $\tree'$. The spin network is proportional to the Identity, the proportionality factor is the coefficient $\hat{S}_{j_A j_B j_C j_D j}^{j_E j_F j_G j_H}$, which can be shown to be the product of two recoupling coefficients.
\label{fig:spin}}
\end{figure}
%%%%%%%%%%%%%%%%%%%%%%%%%%%%%%%%%%%%%%%%%%%%%%%%%%%%%%%%%%%%%%%%%%%%%%%%%%%%%%%%%%%%%%%%%%%%%%%%%%%%%%%%%%%%%%%%%%%%%%%%%%

Another coupled basis $\ket{jm_{j}; \tree'; j_{f_1} \ldots j_{f_{z}}}$ corresponding to a different fusion tree $\tree'$ is related to the basis (\ref{eq:couplebasis}) by the transformation
\begin{equation}
\boxed{\hat{Q}^{j_{f_1} \ldots j_{f_{Z}}}_{j_{1} \ldots j_{L}}(\tree') = \sum_{j_{f_1} \ldots j_{f_{Z}}} \hat{S}_{j_1 \ldots j_L}^{j_{e_1} \ldots j_{e_{Z}} j_{f_1} \ldots j_{f_{Z}}}(\tree, \tree') \hat{Q}^{j_{e_1} \ldots j_{e_{Z}}}_{j_{1} \ldots j_{L}}(\tree),} \label{eq:genrecoup}
\end{equation}
where the coefficients $\hat{S}_{j_1 \ldots j_k}^{j_{e_1} \ldots j_{e_z} j_{f_1} \ldots j_{f_z}}(\tree, \tree')$ can be expressed in terms of the recoupling coefficients [Eq.~(\ref{eq:fmove})]. 

As an example, consider two different ways of coupling four spins $j_A, j_B, j_C$ and $j_D$ according to the fusion trees $\tree$ and $\tree'$ that are shown in Fig.~\ref{fig:fusion}(ii). The two coupled bases are related by coefficients $\hat{S}_{j_A j_B j_C j_D j}^{j_{E} j_{F} j_{G} j_{H}}(\tree, \tree')$ that are defined according to the equality depicted in Fig.~\ref{fig:spin}. Note that the tensor network made of Clebsch-Gordan tensors, shown in Fig.~\ref{fig:spin}, is an instance of a \textit{spin network}. In this case, the spin network has two open links and can therefore be regarded as an SU(2)-invariant operator. The equality in the figure then simply depicts that the spin network is proportional to the Identity. The numerical value of the coefficient $\hat{S}_{j_A j_B j_C j_D j}^{j_{E} j_{F} j_{G} j_{H}}(\tree, \tree')$ can be calculated without contracting the spin network, but by instead following a procedure called \textit{evaluating} a spin network. Section~\ref{sec:sn} illustrates with simple examples the procedure to evaluate a spin network corresponding to a generic coefficient $\hat{S}_{j_1 \ldots j_k j}^{j_{e_1} \ldots j_{e_z} j_{f_1} \ldots j_{f_z}}(\tree, \tree')$. For instance, it is shown that $\hat{S}_{j_A j_B j_C j_D j}^{j_{E} j_{F} j_{G} j_{H}}(\tree, \tree')$ can be expressed in terms of two recoupling coefficients,
\begin{equation}
\hat{S}_{j_A j_B j_C j_D j}^{j_{E} j_{F} j_{G} j_{H}}(\tree, \tree') = \hat{F}_{j_A j_B j_C j_{F}}^{j_{E} j_{G}} \hat{F}_{j_{E} j_C j_D j}^{j_{F} j_{H}}.
\end{equation}

\subsection{Tensor product of $L$ reducible representations}

Finally, consider the tensor product of $L$ reducible representations. A coupled basis, labeled by a given fusion tree, is related to the product basis by means of a transformation that is obtained by attaching a tensor $\fuser$ to each node of the fusion tree, and contracting the resulting tree tensor network. 

\begin{figure}[t]
\begin{center}
  \includegraphics[width=10cm]{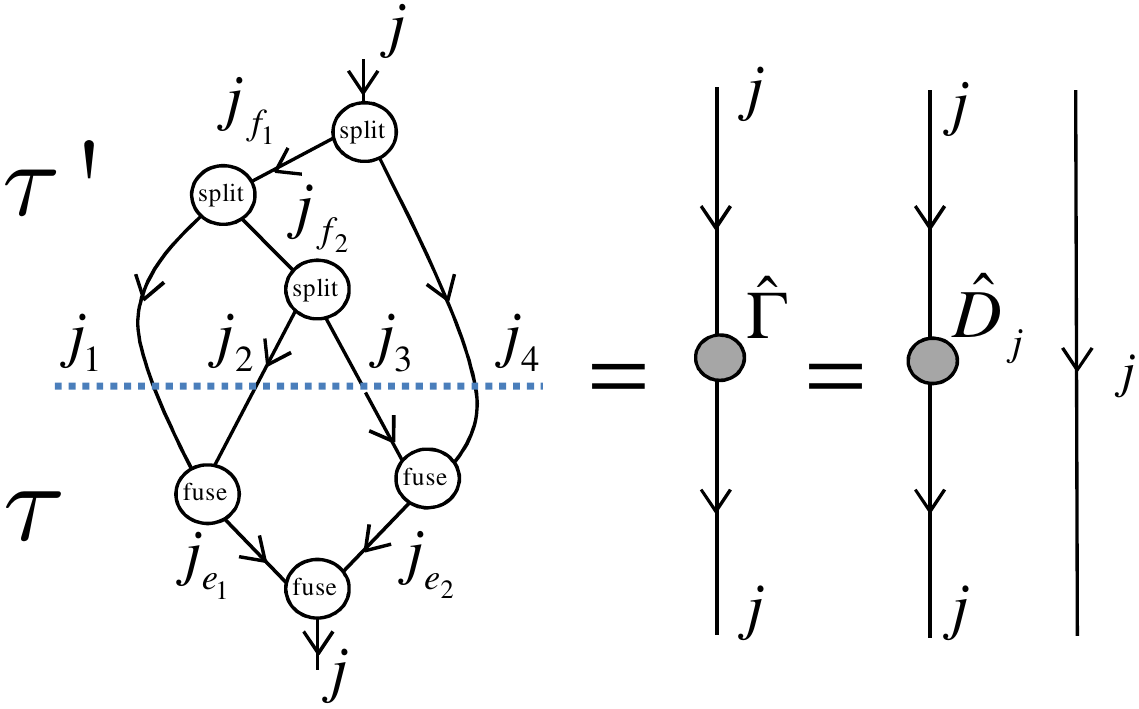}
\end{center}  
\caption{The transformation $\hat{\Gamma}$ that relates two different coupled bases, labeled by fusion trees $\tree$ and $\tree'$, when fusing four reducible representations. Matrix $\hat{\Gamma}$ decomposes into a degeneracy matrix $\hat{D}_j$ and the Identity.
\label{fig:spin1}}
\end{figure}

Two different choices of a coupled basis, corresponding to two different fusion trees $\tree$ and $\tree'$, are related by a matrix $\hat{\Gamma}(\tree, \tree')$. The matrix $\hat{\Gamma}(\tree, \tree')$ is a generalization of the matrix with the same name that appears Eq.~(\ref{eq:gammasplit}). This matrix is obtained by contracting a tensor network [e.g. Fig.~\ref{fig:spin1}] made of tensors $\fuser$ and tensors $\spliter$.

%%%%%%%%%%%%%%%%%%%%%%%%%%%%%%%%%%%%%%%%%%%%%%%%%%%%%%%%%%%%%%%%%%%%%%%%%%%%%%%%%%%%%%%%%%%%%%%%%%%%%%%%%%%%%%%%%%%%%%%%%%
%%%%%%%%%%%%%%%%%%%%%%%%%%%%%%%%%%%%%%%%%%%%%%%%%%%%%%%%%%%%%%%%%%%%%%%%%%%%%%%%%%%%%%%%%%%%%%%%%%%%%%%%%%%%%%%%%%%%%%%%%%

For a fixed value of the total spin $j$, matrix $\hat{\Gamma}(\tree, \tree')$ decomposes in terms of degeneracy matrices $\hat{D}_j$. The components of the latter can be expressed in terms of recoupling coefficients by generalizing Eq.~(\ref{eq:D}). We obtain,
\begin{equation}
\hat{D}_j = \sum \hat{S}_{j_1 \ldots j_k}^{j_{e_1} \ldots j_{e_{z}, j} j'_{e_1} \ldots j'_{e_{z}}} \hat{D}'^{j_{e_1}, \ldots, j_{e_z}}_{j_1, \ldots, j_k, j}, \label{eq:genD}
\end{equation}
where the sum runs over all spin labels but excluding $j$.

\section{Block structure of SU(2)-invariant tensors\label{sec:symTensor}}

In this section we consider tensors that are invariant under the action of the symmetry. We explain how such tensors decompose into a compact canonical form which exploits their symmetry. The canonical form can be understood as a block structure in the tensor components. In Sec.~\ref{sec:blockmoves} we then adapt the set $\mathcal{P}$ of primitive tensor network manipulations to work in this form. With the formalism of SU(2)-invariant tensors at hand we then consider tensor network decompositions made of SU(2)-invariant tensors in Sec.~\ref{sec:symTN}.

\subsection{SU(2)-invariant tensors}

%\subsection{SU(2)-invariant tensors \label{ssec:su2invtens}}
Consider a rank-$k$  tensor $\hat{T}$ with indices $\{i_1, i_2, \ldots, i_k\}$ and directions $\vec{D}$. Each index $i_l$ is associated with a vector space $\mathbb{V}^{(l)}$ on which SU(2) acts by means of transformations $\hat{W}_{\textbf{r}}^{(l)}$.

Also consider the action of SU(2) on the space $\mathbb{V}~\equiv~\bigotimes_{l=1}^{k}~\mathbb{V}^{(l)}$ given by
\begin{equation}
	\hat{Y}^{(1)}_{\textbf{r}}\otimes \hat{Y}^{(2)}_{\textbf{r}}\otimes \ldots \otimes \hat{Y}^{(k)}_{\textbf{r}},
\label{eq:Xtrans}
\end{equation}
where 
\begin{equation}
\hat{Y}^{(l)}_{\textbf{r}} = \left\{ 
	\begin{array}{cc} \hat{W}^{(l)~*}_{\textbf{r}}& ~~~\mbox{ if } \vec{D}(l) = \mbox{ `in' } ,\\ 
	 									\hat{W}^{(l)}_{\textbf{r}}& ~~~~\mbox{ if } \vec{D}(l) = \mbox{ `out' }.
	\end{array} \right.
\label{eq:y}
\end{equation}
($\hat{W}_{\textbf{r}}^{(l)~*}$ denotes the complex conjugate of $\hat{W}_{\textbf{r}}^{(l)}$.) That is, $\hat{Y}^{(l)}_{\textbf{r}}$ acts differently depending on whether index $i_l$ is an incoming or outgoing index. We then say that tensor $\hat{T}$ is SU(2) \textit{invariant} if it is invariant under the transformation of Eq.~(\ref{eq:Xtrans}). In components we have
\begin{equation}
\boxed{
\begin{split}
	\sum_{i_1, i_2, \ldots, i_k} 	\left(\hat{Y}^{(1)}_{\textbf{r}}\right)_{i_1'i_1} \left(\hat{Y}^{(2)}_{\textbf{r}} \right)_{i_2' i_2} \ldots \left( \hat{Y}^{(k)}_{\textbf{r}} \right)_{i_k' i_k} \hat{T}_{i_1 i_2 \ldots i_k}=\hat{T}_{i_1' i_2' \ldots i_k'}&,
	\label{eq:Tinv}
	\end{split}
	}
\end{equation}
for all $\textbf{r} \in \mathbb{R}^3$. 

In the remainder of this section we explore the consequences of the constraints in Eq.~(\ref{eq:Tinv}). The main result is as follows. By writing each index $i_l$ of the tensor in a spin basis, $i_{l} = (j_l, t_{j_l}, m_{j_l})$, the tensor is revealed to have a block structure, namely, the non-trivial components are organized into blocks that are supported on orthogonal subspaces. For a given value of spin $j_l$, the index $i_l$ splits into a \textit{degeneracy index} $(j_l, t_{j_l})$ and a \textit{spin index} $(j_l, m_{j_l})$. An SU(2)-invariant tensor $\hat{T}$ decomposes into a set of \textit{degeneracy tensors}, denoted by $\hat{P}$ and carrying all the degeneracy indices, and a set of \textit{structural} tensors, denoted $\hat{Q}$, carrying all the spin indices. The degeneracy tensors contain all the degrees of freedom and correspond to the `blocks' alluded above. On the other hand, the structural tensors are completely determined by the symmetry since they can be factorized into a trivalent tree tensor network made of Clebsch-Gordan coefficients. Examples of structural tensors include Eq.~(\ref{eq:Q}), however, a structural tensor may not generally decompose according to a fusion tree. We refer to the decomposition $(\hat{P}, \hat{Q})$ as the \textit{canonical decomposition} or the \textit{canonical form} of tensor $\hat{T}$. The main benefit of the canonical form lies in the fact that $\hat{T}$ can be specified compactly by means of only the degeneracy tensors. 

In the ensuing discussion we describe the canonical decomposition of SU(2)-invariant tensors on a case by case basis. We explicitly describe the canonical form of SU(2)-invariant tensors with one to three indices. The canonical form in these cases is unique up to overall numerical factors. On the other hand, an SU(2)-invariant tensor with four or more indices can be decomposed in several equivalent ways. We illustrate this with examples without resorting to a complete theoretical characterization of the canonical form in all cases. A more rigorous characterization is developed Chapter 6 where we consider a special canonical form of SU(2)-invariant tensors, namely, tree decompositions. A tree decomposition corresponds to decomposing \textit{both} the degeneracy tensors and the structural tensors according to a \textit{fusion} tree. We find this decomposition more convenient from an implementation point of view. In Chapter 6, we also describe how to construct a tree decomposition for any SU(2)-invariant tensor, how two different tree decompositions of the same tensor are related to one another, and how primitive tensor manipulations are adapted to tree decompositions.

%%%%%%%%%%%%%%%%%%%%%%%%%%%%%%%%%%%%%%%%%%%%%%%%%%%%%%%%%%%%%%%%%%%%%%%%%%%%%%%%%%%%%%%%%%%%%%%%
\begin{figure}[t]
\begin{center}
  \includegraphics[width=8cm]{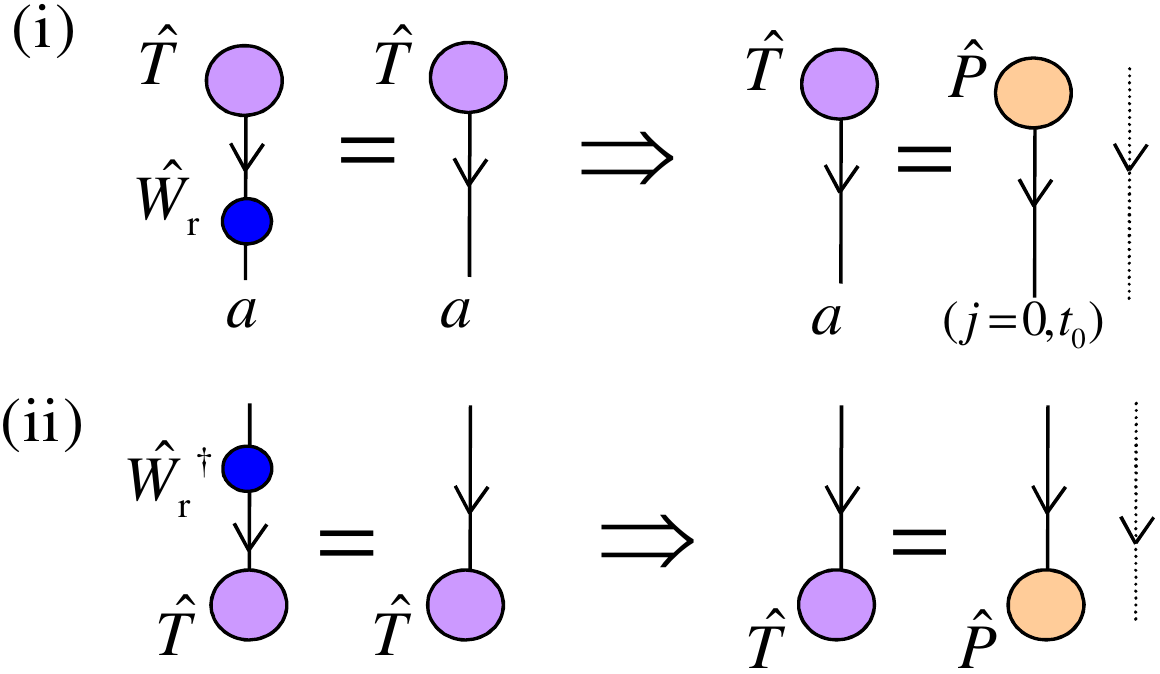}
\end{center}  
\caption{Implication of the symmetry constraints fulfilled by a rank-$1$ SU(2)-invariant tensor with (i) an outgoing index, and (ii) an incoming index. The only allowed spin on the one index is $j=0$.\label{fig:one}}
\end{figure}
%%%%%%%%%%%%%%%%%%%%%%%%%%%%%%%%%%%%%%%%%%%%%%%%%%%%%%%%%%%%%%%%%%%%%%%%%%%%%%%%%%%%%%%%%%%%%%%%
\subsection{One index\label{sec:symTensor:inv:one}}

An SU(2)-invariant tensor $\hat{T}$ with an outgoing index $a$ fulfills the constraint [Fig.~\ref{fig:one}(i)]
\begin{equation}
(\Psi)_{a'} = \sum_{a} (\hat{W}_{\textbf{r}})_{a'a}(\Psi)_{a},\label{eq:inv11}
\end{equation}
where $\hat{W}_{\textbf{r}}$ is the representation of SU(2) on the vector space associated to index $a$.

Let us now write index $a$ in the spin basis $a~=~(j,~t_{j},~m_{j})~=~(0,t_0,0)$. Then we have
\begin{equation}
\hat{T}_a = (\hat{P})_{t_0},
\label{eq:canonone}
\end{equation}
where $(\hat{P})_{t_0}$, shorthand for $(\hat{T}_{j_a=0})_{t_0, m_{0}=0}$, encodes the non-trivial components of $\hat{T}$. Since the only relevant irrep on the one index is $j=0$ the structural tensors are trivial. Therefore, tensor $\hat{T}$ can be stored compactly as $\hat{P}$.

On the other hand, an SU(2)-invariant tensor $\hat{T}$ with an incoming index $a$ fulfills
\begin{equation}
(\Psi)_{a'} = \sum_{a} (\hat{W}^*_{\textbf{r}})_{a'a}(\Psi)_{a},\label{eq:inv12}
\end{equation}
or equivalently
\begin{equation}
(\Psi)_{a'} = \sum_{a} (\Psi)_{a}(\hat{W}^{\dagger}_{\textbf{r}})_{a'a},\label{eq:inv12}
\end{equation}
where $\hat{W}_{\textbf{r}}^*$ and $\hat{W}_{\textbf{r}}^{\dagger}$ are the complex conjugate and adjoint of $\hat{W}_{\textbf{r}}$ respectively. The canonical form of $\hat{T}$ is the same as that stated as Eq.~(\ref{eq:canonone}).

\subsection{Two indices\label{sec:symTensor:inv:two}}

\begin{figure}[t]
\begin{center}
  \includegraphics[width=10cm]{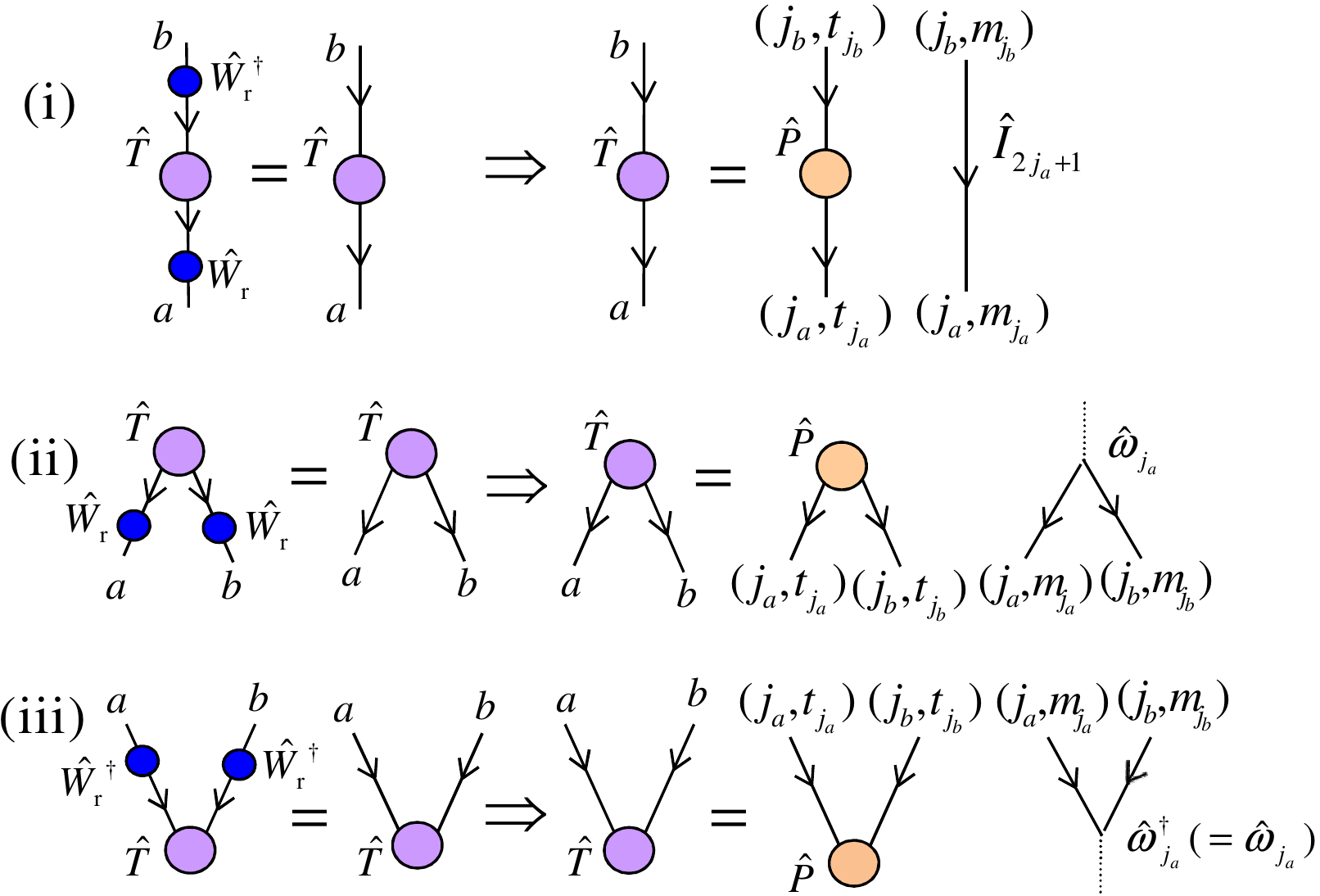}
\end{center}  
\caption{Implication of the symmetry constraints fulfilled by rank-$2$ SU(2)-invariant tensors as resulting in the decomposition of the tensors into  degeneracy tensors $\hat{P}$ and structural tensors. 
\label{fig:two}}
\end{figure}

An SU(2)-invariant matrix $\hat{T}$, possibly rectangular, with indices $a$ and $b$ fulfills [Fig.~\ref{fig:two}(i)]
\begin{align}
\hat{T}_{a'b'} &= \sum_{ab}\left(\hat{W}^{(A)*}_{\textbf{r}}\right)_{a'a}\left(\hat{W}_{\textbf{r}}^{(B)}\right)_{b'b}\hat{T}_{ab},\nonumber \\
&=\sum_{ab}\left(\hat{W}_{\textbf{r}}^{(B)}\right)_{b'b}\hat{T}_{ab}\left(\hat{W}^{(A)\dagger}_{\textbf{r}}\right)_{aa'},
\label{eq:twoinv}
\end{align}
where $\hat{W}_{\textbf{r}}^{(A)}$ and $\hat{W}_{\textbf{r}}^{(B)}$ are the representations of SU(2) on the vector space associated to index $a=(j_a,m_{j_a},t_{j_a})$ and $b=(j_b,m_{j_b},t_{j_b})$ respectively. Schur's Lemma establishes that the matrix $\hat{T}$ decomposes as
\begin{align}
(\hat{T})_{ab} = (\hat{P}_{j_aj_b})_{t_{j_a} t_{j_b}} \delta_{j_aj_b} \delta_{m_{j_a} m_{j_b}},
\label{eq:canon:two1}
\end{align}
which can also be written in a block-diagonal form,
\begin{align}
\hat{T} &= \bigoplus_j \hat{T}_j, \nonumber \\
&=\bigoplus_j (\hat{P}_j \otimes \hat{I}_j).\label{eq:canon:two2}
\end{align}
Here the sum is over all values $j$ of spin $j_a$ that are equal to a value of spin $j_b$.

A rank-$2$ SU(2)-invariant tensor $\hat{T}$ with both incoming indices $a$ and $b$ is associated with fusing spins $j_a$ and $j_b$ into a total spin 0. It fulfills [Fig.~\ref{fig:two}(ii)] 
\begin{align}
\hat{T}_{a'b'} &= \sum_{ab}\left(\hat{W}^{(A)*}_{\textbf{r}}\right)_{a'a}\left(\hat{W}_{\textbf{r}}^{(B)*}\right)_{b'b}\hat{T}_{ab}, \nonumber \\
&= \sum_{ab}\hat{T}_{ab}\left(\hat{W}^{(A)\dagger}_{\textbf{r}}\right)_{aa'}\left(\hat{W}_{\textbf{r}}^{(B)\dagger}\right)_{bb'}
\label{eq:twoinv}
\end{align}
and decomposes as
\begin{align}
(\hat{T})_{ab} = (\hat{P}_{j_aj_b})_{t_{j_a} t_{j_b}} \cfuse{j_am_{j_a}}{j_b m_{j_b}}{00}.
\label{eq:canon:two2}
\end{align}
%%%%%%%%%%%%%%%%%%%%%%%%%%%%%%%%%%%%%%%%%%%%%%%%%%%%%%%%%%%%%%%%%%%%%%%%%%%%%%%%%%%%%%%%%%%%%%%%
%%%%%%%%%%%%%%%%%%%%%%%%%%%%%%%%%%%%%%%%%%%%%%%%%%%%%%%%%%%%%%%%%%%%%%%%%%%%%%%%%%%%%%%%%%%%%%%%
Similarly, a rank-$2$ SU(2)-invariant tensor with both outgoing indices $a$ and $b$ fulfills [Fig.~\ref{fig:two}(iii)] 
\begin{align}
\hat{T}_{a'b'} = \sum_{ab}\left(\hat{W}^{(A)}_{\textbf{r}}\right)_{a'a}\left(\hat{W}_{\textbf{r}}^{(B)}\right)_{b'b}\hat{T}_{ab},
\label{eq:twoinv}
\end{align}
and decomposes as
\begin{align}
(\hat{T})_{ab} = (\hat{P}_{j_aj_b})_{t_{j_a} t_{j_b}} \csplitt{00}{j_am_{j_a}}{j_b m_{j_b}}.
\label{eq:canon:two3}
\end{align}

Both incoming (or both outgoing) spins $j_a$ and $j_b$ are compatible with the total spin 0 only for values $j$ of spin $j_a$ such that $j_a~=~j_b~=~j$ and for values $m$ of $m_a$ such that $m_{j_a}~=~-m_{j_b}~=~m$. Therefore, we can recast the canonical decompositions of Eqs.~(\ref{eq:canon:two2})-(\ref{eq:canon:two3}) in a block-diagonal form,
\begin{align}
\hat{T} &= \bigoplus_j \hat{T}_j, \nonumber \\
&=\bigoplus_{j} \left(\hat{P}_{j} \otimes \hat{\omega}_{j}\right),
\label{eq:canonblock2}
\end{align}
where $\hat{\omega}_j$ is a $(2j+1) \times (2j+1)$ reverse diagonal matrix with diagonal components
\begin{equation}
(\hat{\omega}_j)_{m,-m} \equiv \cfuse{jm}{j~-m}{00} = \csplitt{00}{jm}{j~-m} = \frac{(-1)^{j-m}}{\sqrt{2j+1}}. \label{eq:singletomega}
\end{equation}

To summarize, the canonical form of a rank-$2$ SU(2)-invariant tensor reads
\begin{equation}
\hat{T} = \bigoplus_{j} \left(\hat{P}_{j} \otimes \hat{Q}_{j}\right).
\label{eq:canonblock2}
\end{equation}
Here $\hat{P}_{j}$ contains the degrees of freedom of $\hat{T}$ that are not fixed by the symmetry, namely, $\hat{P}_{j}$ transforms trivially under the action of the SU(2), Eq.~(\ref{eq:decow}). On the other hand $\hat{Q}_j$ is determined by the symmetry according to the directions $\vec{D}$ of indices $a$ and $b$,
\begin{align}
\hat{Q}_j &= \hat{I}_{j}~~~~~~\mbox{ if } \vec{D} = \{\mbox{`in', `out'}\} \mbox{ or } \{\mbox{`out', `in'}\},\label{eq:blkmat}\\ 
&= \hat{\omega}_{j}~~~~~\mbox{ if } \vec{D} = \{\mbox{`in', `in'}\} \mbox{ or } \{\mbox{`out', `out'}\},
\end{align}

Thus, a rank-$2$ SU(2)-invariant tensor $\hat{T}$ can be stored compactly as
\begin{equation}
\{\{a=(j_a,t_{j_a}, m_{j_a}), b=(j_b,t_{j_b}, m_{j_b})\}, \vec{D}, \{\hat{P}_j\}\}.
\end{equation}

\textbf{Example 10:} Consider a rank-$2$ SU(2)-invariant tensor $\hat{T}$ with both outgoing indices and with each index associated to the vector space $\mathbb{V}$ of Example 8, $\mathbb{V}~\equiv~\mathbb{V}_0~\oplus~3\mathbb{V}_1~\oplus~\mathbb{V}_2$. Tensor $\hat{T}$ has the canonical form
\begin{equation}
\hat{T} \equiv (\hat{P}_0 \otimes \hat{\omega}_0) \oplus (\hat{P}_1 \otimes \hat{\omega}_1) \oplus (\hat{P}_2 \otimes \hat{\omega}_2), \label{eq:eg10}
\end{equation} 
where
\begin{align}\label{eq:eg101}
\hat{\omega}_0 &\equiv 1, \nonumber \\
\hat{\omega}_1 &\equiv \begin{pmatrix} 0 & 0 & \frac{1}{\sqrt{3}} \\ 0 & -\frac{1}{\sqrt{3}} & 0 \\ \frac{1}{\sqrt{3}} & 0 & 0\end{pmatrix}, \nonumber \\
\hat{\omega}_2 &\equiv \begin{pmatrix} 0 & 0 & 0 & 0 & \frac{1}{\sqrt{5}} \\ 0 & 0 & 0 & -\frac{1}{\sqrt{5}} & 0 \\ 0 & 0 & \frac{1}{\sqrt{5}} & 0 & 0 \\ 0 & -\frac{1}{\sqrt{5}} & 0 & 0 & 0 \\ \frac{1}{\sqrt{5}} & 0 & 0 & 0 & 0\end{pmatrix}.
\end{align}

The total number of complex coefficients contained in tensor $\hat{T}$ is $|\hat{T}| = 15 \times 15 = 225$. However, the tensor can be stored compactly as
\begin{equation}
\{\{a, b\}, \{\mbox{`out', `out'}\}, \{\hat{P}_0, \hat{P}_1, \hat{P}_2\}\}, \nonumber
\end{equation}
where the total number of complex coefficients that are contained in tensors $\hat{P}_0, \hat{P}_1$ and $\hat{P}_2$ is
\begin{equation} 
|\hat{P_0}|+|\hat{P_1}|+|\hat{P_2}| = 1\times 1 + 3\times 3 + 1\times 1 = 11.
\end{equation}
Therefore, by exploiting the symmetry the number of coefficients that need to be stored is twenty times smaller.\markend

\subsection{Three indices\label{sec:symTensor:inv:three}}

\begin{figure}[t]
\begin{center}
  \includegraphics[width=10cm]{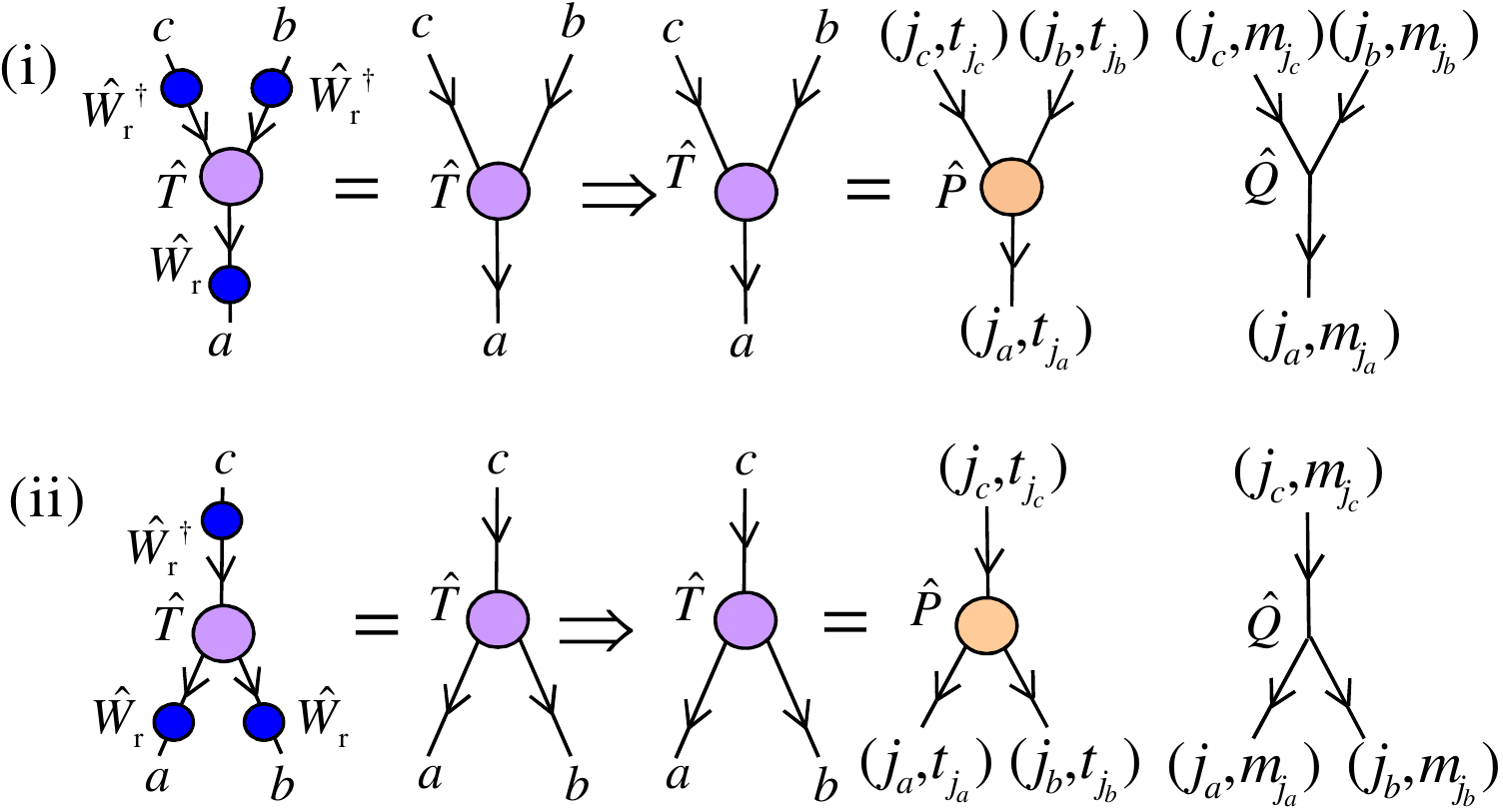}
\end{center}  
\caption{Examples of the constraints fulfilled by rank-$3$ SU(2)-invariant tensors and their implication as resulting in the decomposition of the tensors into degeneracy tensors $\hat{P}$ and a Clebsch-Gordan tensors. 
\label{fig:three}}
\end{figure}

Consider a rank-$3$ SU(2)-invariant tensor $\hat{T}$ with incoming indices $a$ and $b$ and outgoing index $c$. It fulfills [Fig.~\ref{fig:three}(i)]
\begin{align}
\hat{T}_{a'b'c'} &=\sum_{abc}\left(\hat{W}^{(A)*}_{\textbf{r}}\right)_{a'a}\left(\hat{W}_{\textbf{r}}^{(B)*}\right)_{b'b}\left(\hat{W}_{\textbf{r}}^{(C)}\right)_{c'c}\hat{T}_{abc},\nonumber\\
&=\sum_{abc}\left(\hat{W}_{\textbf{r}}^{(C)}\right)_{c'c}\hat{T}_{abc}\left(\hat{W}^{(A)\dagger}_{\textbf{r}}\right)_{aa'}\left(\hat{W}^{(B)\dagger}_{\textbf{r}}\right)_{bb'},
\label{eq:invthree2}
\end{align}
where $\hat{W}_{\textbf{r}}^{(A)}, \hat{W}_{\textbf{r}}^{(B)}$ and $\hat{W}_{\textbf{r}}^{(C)}$ are the representations of SU(2) on indices $a=(j_a,m_{j_a},t_{j_a}), b=(j_b,m_{j_b},t_{j_b})$ and $c=(j_c,m_{j_c},t_{j_c})$ respectively. The \textit{Wigner-Eckart theorem} establishes that $\hat{T}$ decomposes as
\begin{align}
(\hat{T})_{abc} = (\hat{P}_{j_aj_bj_c})_{t_{j_a} t_{j_b} t_{j_c}} \cfusespin{a}{b}{c}.
\label{eq:three22}
\end{align}
That is, for compatible values of the spins $j_a, j_b$ and $j_c$, tensor $\hat{T}$ factorizes into tensor $\hat{P}_{j_aj_bj_c}$ containing degrees of freedom and a Clebsch-Gordan tensor that mediates the fusion of spins $j_a$ and $j_b$ into spin $j_c$. 

%%%%%%%%%%%%%%%%%%%%%%%%%%%%%%%%%%%%%%%%%%%%%%%%%%%%%%%%%%%%%%%%%%%%%%%%%%%%%%%%%%%%%%%%%%%%%%%%%%%%%%%%%%%%%%%%%%%%%%%%%%
%%%%%%%%%%%%%%%%%%%%%%%%%%%%%%%%%%%%%%%%%%%%%%%%%%%%%%%%%%%%%%%%%%%%%%%%%%%%%%%%%%%%%%%%%%%%%%%%%%%%%%%%%%%%%%%%%%%%%%%%%%

An SU(2)-invariant tensor $\hat{T}$ with another combination of incoming and outgoing indices has a canonical decomposition that differs in the Clebsch-Gordan coefficients. For example, if $\hat{T}$ is an SU(2)-invariant tensor with incoming indices $b$ and outgoing indices $a$ and $c$ then it fulfills
\begin{align}
\hat{T}_{a'b'c'} &=\sum_{abc}\left(\hat{W}^{(A)}_{\textbf{r}}\right)_{a'a}\left(\hat{W}_{\textbf{r}}^{(B)*}\right)_{b'b}\left(\hat{W}_{\textbf{r}}^{(C)}\right)_{c'c}\hat{T}_{abc},\nonumber\\
&=\sum_{abc}\left(\hat{W}_{\textbf{r}}^{(A)}\right)_{a'a}\left(\hat{W}^{(C)}_{\textbf{r}}\right)_{cc'}\hat{T}_{abc}\left(\hat{W}^{(B)\dagger}_{\textbf{r}}\right)_{bb'},
\label{eq:invthree2}
\end{align}
and decomposes as
\begin{align}
(\hat{T})_{abc} = (\hat{P}_{j_aj_bj_c})_{t_{j_a} t_{j_b} t_{j_c}} \csplitspin{b}{a}{c}.\label{eq:three2}
\end{align}
More generally, a rank-$3$ SU(2)-invariant tensor with any combination of incoming and outgoing indices decomposes as
\begin{align} 
&(\hat{T})_{abc} = (\hat{P}_{j_aj_bj_c})_{t_{j_a} t_{j_b} t_{j_c}} (\hat{Q}_{j_aj_bj_c})_{m_{j_a} m_{j_b}m_{j_c}}. \nonumber \\
&\!\!\!\!\!\!\!\!\!\!\!\!\!\!\!\!\!\!\text{(Wigner Eckart Theorem)}\label{eq:wigner}
\end{align}
The block structure can be made more explicit by recasting Eq.~(\ref{eq:wigner}) as
\begin{align}
\hat{T} &\equiv \bigoplus_{j_aj_bj_c} \hat{T}_{j_aj_bj_c}, \nonumber \\
&\equiv \bigoplus_{j_aj_bj_c} \left(\hat{P}_{j_aj_bj_c} \otimes \hat{Q}_{j_aj_bj_c}\right),\label{eq:threetensor}
\end{align} 
where we use the direct sum symbol $\bigoplus$ to denote that the different tensors $\hat{T}_{j_aj_bj_c}$ are supported on orthonormal subspaces of the tensor product of the spaces associated with indices $a,b$ and $c$, and where the direct sum runs over all compatible values of $j_a, j_b$ and $j_c$. The components $(\hat{Q}_{j_aj_bj_c})_{m_{j_a}m_{j_b}m_{j_c}}$ are determined by the directions $\vec{D}$ of the indices,
\begin{align}
	  \cfusespin{a}{b}{c}&~~~~\mbox{if }\vec{D} = \{\mbox{`in', `in', `out'}\},\\ 
	 	\cfusespin{a}{c}{b}&~~~~\mbox{if }\vec{D} = \{\mbox{`in', `out', `in'}\},\\ 
	 	\cfusespin{a}{c}{b}&~~~~\mbox{if }\vec{D} = \{\mbox{`out', `in', `in'}\}, \\
	 	\csplitspin{b}{a}{b}&~~~~\mbox{if }\vec{D} = \{\mbox{`out', `in', `out'}\},\\ 
	 	\csplitspin{c}{a}{b}&~~~~\mbox{if }\vec{D} = \{\mbox{`out', `out', `in'}\},\\ 
	 	\csplitspin{a}{b}{c}&~~~~\mbox{if }\vec{D} = \{\mbox{`in', `out', `out'}\},\\
	 	\beta\csplitspin{a}{b}{c}&~~~~\mbox{if }\vec{D} = \{\mbox{`out', `out', `out'}\},\label{eq:case1}\\
	 	\beta\cfusespin{a}{b}{c}&~~~~\mbox{if }\vec{D} = \{\mbox{`in', `in', `in'}\},
\label{eq:threeall}
\end{align}
where $\beta = (-1)^{j_a-j_b+m_c}\sqrt{2j_c+1}$. 

To summarize, a rank-$3$ SU(2)-invariant tensor $\hat{T}$ can be stored in the most compact way as
\begin{equation}
\{\{a, b, c\}, \vec{D}, \{\hat{P}_{j_aj_bj_c}\}\},
\end{equation}
where the indices $a, b$ and $c$ are specified in the spin basis,
\begin{equation}
a = (j_a,t_{j_a},m_{j_a}),~~b = (j_b,t_{j_b},m_{j_b}),~~c = (j_c,t_{j_c},m_{j_c}).
\end{equation}

\textbf{Example 11:} Consider a rank-$3$ SU(2)-invariant tensor $\hat{T}$ such that each index, $a, b$ and $c$, is associated to the vector space $\mathbb{V}$ of Example 8. 
Tensor $\hat{T}$ can be stored by storing the degeneracy tensors, 
\begin{align}
&\hat{P}_{0,0,0},~~\hat{P}_{0,1,1},~~\hat{P}_{0,2,2},~~\hat{P}_{1,0,1},~~\hat{P}_{1,1,0},~~\hat{P}_{1,1,1},~~\hat{P}_{1,1,2}, \nonumber \\
&\hat{P}_{1,2,1},~~\hat{P}_{1,2,2},~~\hat{P}_{2,0,2},~~\hat{P}_{2,1,1},~~\hat{P}_{2,2,0},~~ \hat{P}_{2,2,1},~~\hat{P}_{2,2,2}, \nonumber
\end{align}
corresponding to all compatible values of $j_a, j_b$ and $j_c$.

The total number of complex coefficients that are contained in the degeneracy tensors is $\displaystyle \sum_{j_a j_b j_c}|\hat{P}_{j_aj_bj_c}| = 45$, whereas $|\hat{T}| = |a|\times|b|\times|c| = 15^3 = 3375$ components; the reduction in the number of coefficients is seventy-five times, much greater than that computed for rank-$2$ tensors in Example 10. In general, the sparsity of SU(2)-invariant tensors increases with increasing number of indices.\markend

\subsection{Four indices\label{sec:symTensor:inv:four}}

A rank-$4$ SU(2)-invariant tensor may be decomposed in several ways in terms of degeneracy tensors and structural tensors in correspondence with the existence of different fusion trees for four spins. 

Consider a rank-$4$ SU(2)-invariant tensor $\hat{T}$ with incoming indices $a=(j_a,t_{j_a},m_{j_a}), b=(j_b,t_{j_b},m_{j_b})$ and $c=(j_c,t_{j_c},m_{j_c})$ and outgoing index $d=(j_d,t_{j_d},m_{j_d})$. It fulfills
\begin{align}
\hat{T}_{a'b'c'd'}
%=\sum_{abcd}\left(\hat{W}^{(A)}_{\textbf{r}}\right)^*_{a'a}\left(\hat{W}_{\textbf{r}}^{(B)}\right)^*_{b'b}\left(\hat{W}_{\textbf{r}}^{(C)}\right)^*_{c'c}& \nonumber \\
%\left(\hat{W}_{\textbf{r}}^{(D)}\right)_{d'd}\hat{T}_{abcd}&, \nonumber \\
=\sum_{abcd}\left(\hat{W}_{\textbf{r}}^{(D)}\right)_{d'd}\hat{T}_{abcd} \left(\hat{W}^{(A)}_{\textbf{r}}\right)^{\dagger}_{aa'}
\left(\hat{W}_{\textbf{r}}^{(B)}\right)^{\dagger}_{bb'}
\left(\hat{W}_{\textbf{r}}^{(C)}\right)^{\dagger}_{cc'}
\label{eq:invfour}
\end{align}
where $\hat{W}^{(A)}_{\textbf{r}}, \hat{W}^{(B)}_{\textbf{r}}, \hat{W}^{(C)}_{\textbf{r}}$ and $\hat{W}^{(D)}_{\textbf{r}}$ are the representations of SU(2) on the indices $a, b, c$ and $d$ respectively.
%%%%%%%%%%%%%%%%%%%%%%%%%%%%%%%%%%%%%%%%%%%%%%%%%%%%%%%%%%%%%%%%%%%%%%%%%%%%%%%%%%%%%%%%%%%%%%%%%%%%%%%%%%%%%%%%%%%%%%%%%%
\begin{figure}[t]
\begin{center}
  \includegraphics[width=10cm]{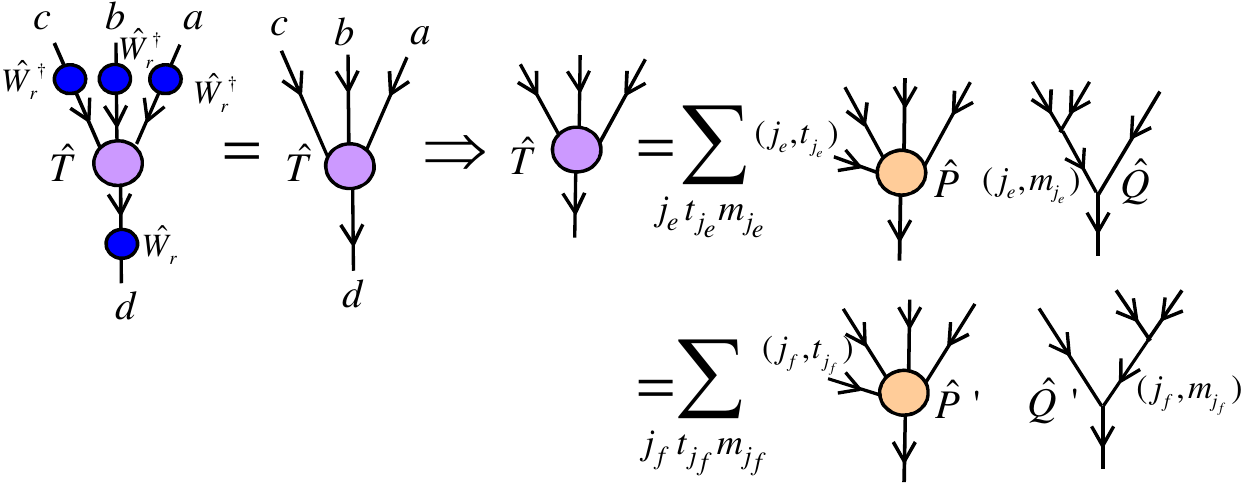}
\end{center}  
\caption{Two equivalent canonical decompositions of a rank-$4$ SU(2)-invariant tensor corresponding to two different ways of fusing the three incoming indices into the outgoing index.
\label{fig:four}}
\end{figure}
%%%%%%%%%%%%%%%%%%%%%%%%%%%%%%%%%%%%%%%%%%%%%%%%%%%%%%%%%%%%%%%%%%%%%%%%%%%%%%%%%%%%%%%%%%%%%%%%%%%%%%%%%%%%%%%%%%%%%%%%%%
Tensor $\hat{T}$ decomposes as
\begin{align}
(T)_{abcd} = \sum_{j_e}&(\hat{P}^{j_e}_{j_aj_bj_cj_d})_{t_{j_a}t_{j_b}t_{j_c}t_{j_d}} \cdot(\hat{Q}^{j_e}_{j_a j_b j_c j_d})_{m_{j_a} m_{j_b} m_{j_c} m_{j_d}},
\label{eq:deco41}
\end{align}
where the sum is over all values of the intermediate spin $j_e$.\\
The coefficients $(\hat{Q}^{j_e}_{j_a j_b j_c j_d})_{m_{j_a} m_{j_b} m_{j_c} m_{j_d}}$  [Eq.~(\ref{eq:cob1})] mediate the fusion of the spins $j_a, j_b$ and $j_c$ into a total spin $j_d$ according to a fusion tree, for example, first fusing $j_a$ and $j_b$ and then fusing the resulting spin with $j_c$.

Alternatively, tensor $\hat{T}$ can be decomposed as
\begin{align}
(T)_{abcd} = \sum_{j_f}&(\hat{P}'^{j_f}_{j_aj_bj_cj_d})_{t_{j_a}t_{j_b}t_{j_c}t_{j_d}} \cdot(\hat{Q}'^{j_f}_{j_a j_b j_c j_d})_{m_{j_a} m_{j_b} m_{j_c} m_{j_d}},
\label{eq:deco42}
\end{align} 
in terms of different structural coefficients $(\hat{Q}'^{j_f}_{j_a j_b j_c j_d})_{m_{j_a} m_{j_b} m_{j_c} m_{j_d}}$ [Eq.~(\ref{eq:cob2})] that are associated with fusing the spins according to a different fusion tree, namely, fusing spin $j_a$ with the spin obtained by first fusing $j_b$ and $j_c$. 

Since Eqs.~(\ref{eq:deco41}) and (\ref{eq:deco42}) represent the same tensor $\hat{T}$, the tensors $\hat{P}$ and $\hat{P}'$ are related by 
\begin{equation}
\hat{P}'^{j_f}_{j_aj_bj_cj_d} = \sum_{j_e}\hat{F}^{j_ej_f}_{j_aj_bj_cj_d} \hat{P}^{j_e}_{j_aj_bj_cj_d},
\label{eq:fmove2}
\end{equation}
where $\hat{F}^{j_ej_f}_{j_aj_bj_cj_d}$ are the recoupling coefficients [Eq.~\ref{eq:fmove}].

\subsection{$k$ indices\label{sec:symTensor:inv:k}}

Finally, consider a rank-$k$ SU(2)-invariant tensor $\hat{T}$ with all \textit{outgoing} indices and which fulfills Eq.~(\ref{eq:Tinv}). By writing each index in a spin basis, $i_l = (j_l, t_{j_l}, m_{j_l})$, tensor $\hat{T}$ can be decomposed as
\begin{align}
(\hat{T})_{i_1 \ldots i_k} \equiv \!\!\!\!\!\sum_{j_{e_1} \ldots j_{e_l}}\left(\hat{P}^{j_{e_1}\ldots j_{e_l}}_{j_{1} \ldots j_{k}}\right)_{t_{j_1} \ldots t_{j_k}}\!\!\!\!\!\cdot\left(\hat{Q}^{j_{e_1} \ldots j_{e_l}}_{j_{1} \ldots j_{k}}\right)_{m_{j_1}\ldots m_{j_k}}.
\label{eq:Tcanon}
\end{align}
Here tensor $\hat{Q}^{j_{e_1}\ldots j_{e_l}}_{j_{1} \ldots j_{k}}$ [Eq.~(\ref{eq:Q})] is a transformation characterized by a fusion tree $\tree$ whose internal links are decorated by the spins $\{j_{e_1},\ldots, j_{e_l}\}$. Another canonical form of the tensor $\hat{T}$,
\begin{align}
(\hat{T})_{i_1 \ldots i_k} \equiv \!\!\!\!\!\sum_{j'_{e_1} \ldots j'_{e_l}}\!\!\!\left(\hat{P}'^{j'_{e_1}\ldots j'_{e_l}}_{j_{1} \ldots j_{k}}\right)_{t_{j_1} \ldots t_{j_k}}\!\!\!\!\!\cdot\left(\hat{Q}'^{j'_{e_1} \ldots j'_{e_l}}_{j_{1} \ldots j_{k}}\right)_{m_{j_1}\ldots m_{j_k}},
\label{eq:Tcanon1}
\end{align}
comprises of different degeneracy tensors $\hat{P}'^{j'_{e_1}\ldots j'_{e_l}}_{j_{1} \ldots j_{k}}$ and different structural tensors $\hat{Q}'^{j'_{e_1} \ldots j'_{e_l}}_{j_{1} \ldots j_{k}}$ where the latter is a transformation characterized by another fusion tree $\tree'$. 

The two canonical forms, Eq.~(\ref{eq:Tcanon}) and (\ref{eq:Tcanon1}), are related as
\begin{equation}
\hat{P}'^{j'_{e_1} \ldots j'_{e_l}}_{j_{1} \ldots j_{k}} = \sum_{j_{e_1} \ldots j_{e_l}} \hat{S}^{j_{e_1} \ldots j_{e_l}j'_{e_1} \ldots j'_{e_l}}_{j_{1} \ldots j_{k}}(\tree, \tree') \hat{P}^{j_{e_1} \ldots j_{e_l}}_{j_{1} \ldots j_{k}},\label{eq:canonmap}
\end{equation}
where the coefficients $\hat{S}^{j_{e_1} \ldots j_{e_l}j'_{e_1} \ldots j'_{e_l}}_{j_{1} \ldots j_{k}}(\tree, \tree')$ are those which appear in Eq.~(\ref{eq:genrecoup}). 

Thus tensor $\hat{T}$ can be compactly stored as
\begin{equation}
\{\{i_1, i_2, \ldots, i_k\}, \tree, \vec{D}, \{\hat{P}^{j_{e_1} \ldots j_{e_l}}_{j_{1} \ldots j_{k}}\}\}. \label{eq:datak}
\end{equation}
For an arbitrary combination of incoming and outgoing indices of the tensor, the canonical decomposition is characterized by intermediate spins $j_{e_1} \ldots j_{e_l}$ that are assigned to the links of a trivalent tree that is more general than the fusion tree. Furthermore, two canonical decompositions are related by means of more generic spin networks than those considered (See Section~\ref{sec:sn}) for evaluating the coefficients $\hat{S}^{j_{e_1} \ldots j_{e_l}j'_{e_1} \ldots j'_{e_l}}_{j_{1} \ldots j_{k}}(\tree, \tree')$. A rigorous result is presented in Chapter 6 where we describe the generic transformation that relates two tree decompositions of an SU(2)-invariant tensor.

\section{Manipulations of SU(2)-invariant tensors \label{sec:blockmoves}}

In this section we consider manipulations of SU(2)-invariant tensors that belong to the set $\mathcal{P}$ [Sec.~\ref{sec:tensor:TN}] of primitives: reversal of indices, permutation of indices, reshaping of indices and matrix operations (matrix multiplication and matrix factorizations). We will adapt these manipulations to the presence of the symmetry by implementing them in such a way that the canonical form is maintained. 

Our approach will be to describe the basic transformations that are instrumental in implementing the symmetric version of these manipulations and demonstrate their use with simple examples. A more rigorous treatment of adapting the primitive tensor manipulations for SU(2)-invariant tensors is presented in Chapter 6. The basic transformations are symmetry preserving and can be described by means of specical SU(2)-invariant tensors. Consequently, a symmetric manipulation decomposes into the manipulation of the degeneracy tensors and the manipulation of the structural tensors. Computational cost is incurred only by the manipulation of degeneracy tensors. On the other hand, the manipulation of structural tensors can be performed \textit{algebraically} by applying relevant properties of Clebsch-Gordan coefficients. This fact is responsible for obtaining computational speedup from exploiting the symmetry.

%%%%%%%%%%%%%%%%%%%%%%%%%%%%%%%%%%%%%%%%%%%%%%%%%%%%%%%%%%%%%%%%%%%%%%%%%%%%%%%%%%%%%%%%%%%%%%%%
\begin{figure}[t]
\begin{center}
  \includegraphics[width=10cm]{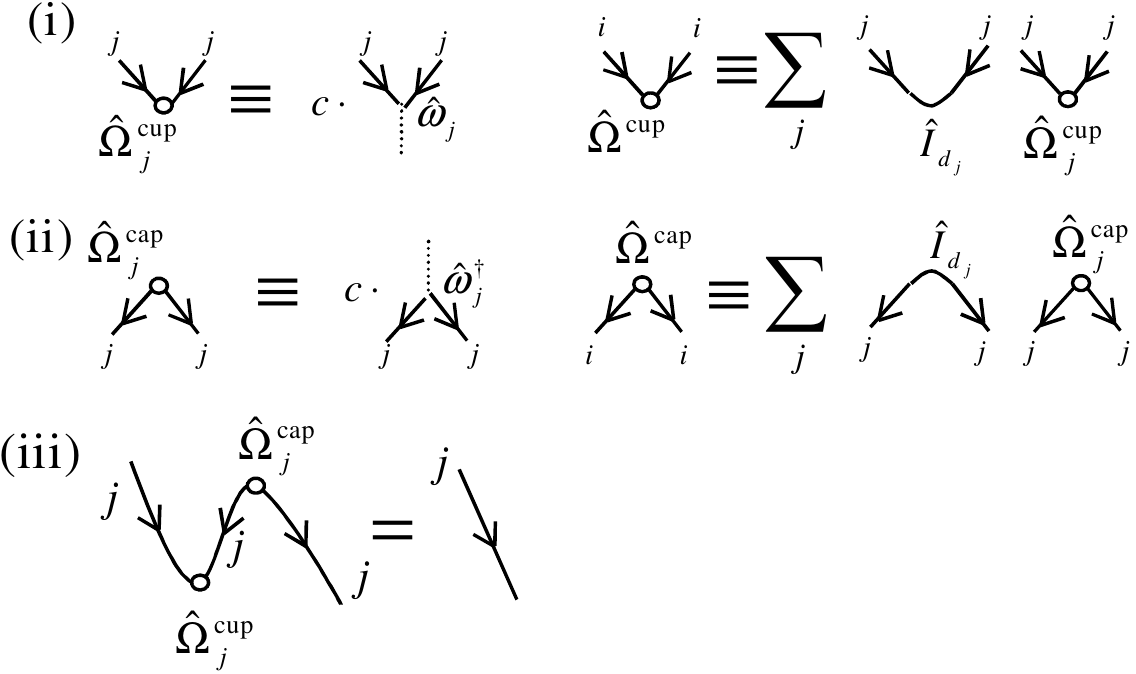}
\end{center}  
\caption{(i) Graphical representation of the cup tensor. For simplicity, we choose the same graphical representation for a cup tensor applied to an index that carries a single spin $j$ as well as for one that is applied on an index $i$ that carries several spins $j$ (possibly with degeneracy $d_j$). ($c = 2j+1$ is a normalization constant.) (ii) The analogous graphical representation of the cap tensor. (iii) Multiplying together a cup tensor and a cap tensor yields the Identity.}
\label{fig:cupcap}
\end{figure}
%%%%%%%%%%%%%%%%%%%%%%%%%%%%%%%%%%%%%%%%%%%%%%%%%%%%%%%%%%%%%%%%%%%%%%%%%%%%%%%%%%%%%%%%%%%%%%%%

\subsection{Reversal of indices \label{sec:blockmoves:symbend}}

An index $i=(j, t_j, m_j)$ of an SU(2)-invariant tensor can be reversed by means of the `cup' and `cap' transformations. The cup transformation is given by a rank-$2$ SU(2)-invariant tensor  $\mycup$ with both \textit{incoming} indices. It can be used to reverse an \textit{outgoing} index. Analogously, the cap transformation is given by a rank-$2$ SU(2)-invariant tensor $\mycap$ with both \textit{outgoing} indices and can be used to reverse an \textit{incoming} index. 

In the canonical form, the cup and cap tensors read as block-diagonal matrices,
\begin{align}
\mycup &\equiv \bigoplus_j (\hat{I}_{d_j} \otimes \mycup_j),\label{eq:cup} \\
\mycap &\equiv \bigoplus_j (\hat{I}_{d_j} \otimes \mycap_j),\label{eq:cap}
\end{align}
where $d_j = |t_j|$ and where
\begin{align}
\mycup_j &\equiv \sqrt{2j+1}\hat{\omega}_j, \label{eq:cupj} \\
\mycap_j &\equiv (-1)^{2j} \sqrt{2j+1}\hat{\omega}_j. \label{eq:capj}
\end{align}
Here $\hat{\omega}_j$ is the tensor defined in Eq.~(\ref{eq:singletomega}). By definition, the cup transformation inverts the action of the cap transformation and vice-versa,
\begin{align}
\mycup_j~\mycap_j = \mycap_j~\mycup_j = \hat{I}_{2j+1}. \label{eq:wiggle}
\end{align}

%%%%%%%%%%%%%%%%%%%%%%%%%%%%%%%%%%%%%%%%%%%%%%%%%%%%%%%%%%%%%%%%%%%%%%%%%%%%%%%%%%%%%%%%%%%%%%%
\begin{figure}[t]
\begin{center}
  \includegraphics[width=16cm]{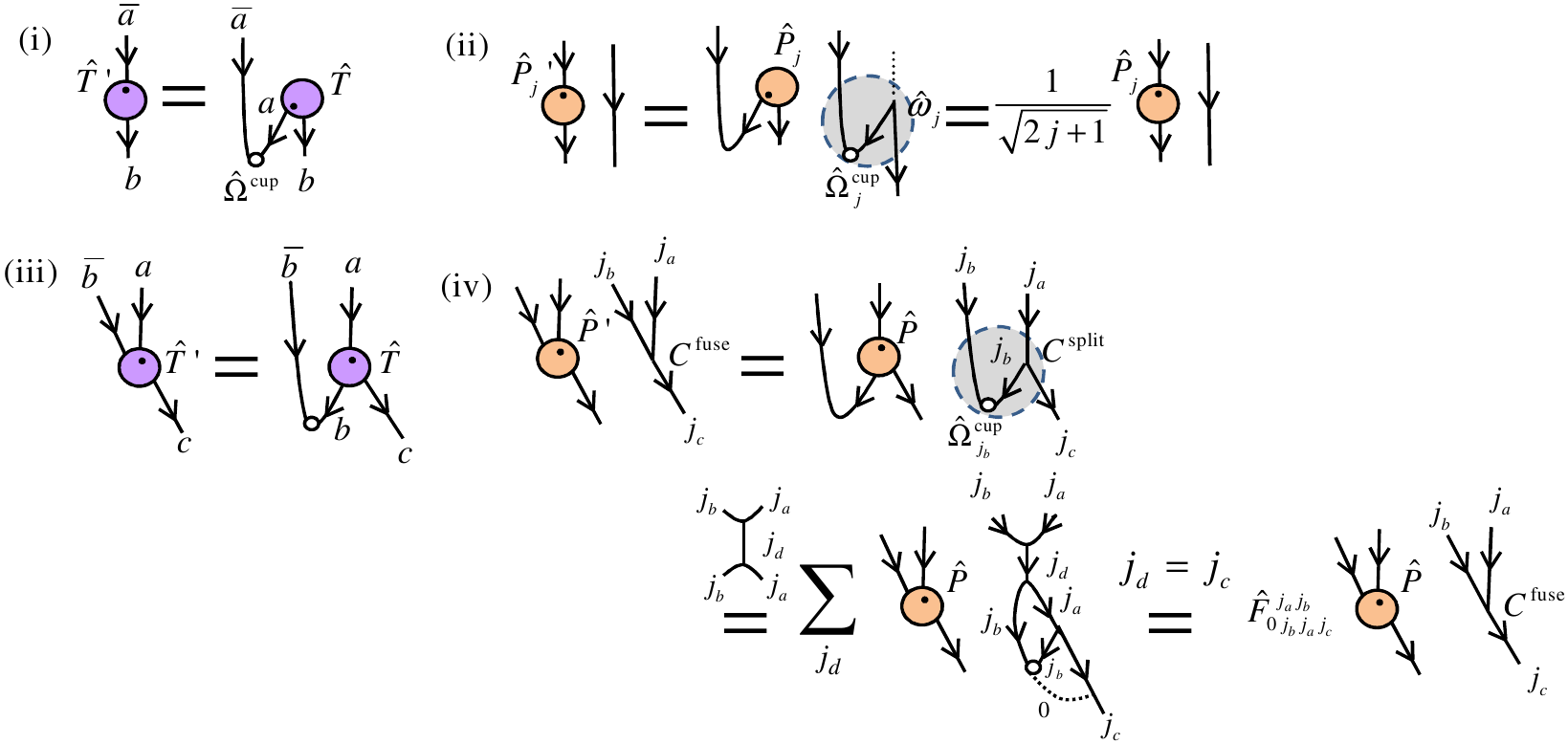}
\end{center}  
\caption{Reversing an index of an SU(2)-invariant tensor by means of a cup tensor. (i) Reversing an index of an SU(2)-invariant tensor $\hat{T}$ with two outgoing indices to obtain a matrix $\hat{T}'$. (ii) The reversal in $(i)$ as performed on the canonical form of $\hat{T}$. The degeneracy index can be reversed without affecting the components of the degeneracy tensor $\hat{P}$. The reversal of the spin index equates to replacing the shaded region with a Clebsch-Gordan tensor and a numerical factor $\frac{1}{2j+1}$, which is absorbed into $\hat{P}$ to obtain $\hat{P}'$. (iii) Reversing an index of a rank-$3$ SU(2)-invariant tensor. (iv) The reversal in $(iii)$ as performed on the canonical form of $\hat{T}$. The reversal of the spin index equates to replacing the shaded region with a Clebsch-Gordan tensor and a recoupling coefficient $\hat{F}^{j_aj_b}_{0j_bj_aj_c}$. The recoupling coefficient appears as a result of introducing a resolution of Identity [Fig.~\ref{fig:cg}(iii)] on the spins $j_b$ and $j_a$ and then simplifying the resulting diagram by applying the equality in Fig.~\ref{fig:fmove}(ii). 
\label{fig:bend}}
\end{figure}
%%%%%%%%%%%%%%%%%%%%%%%%%%%%%%%%%%%%%%%%%%%%%%%%%%%%%%%%%%%%%%%%%%%%%%%%%%%%%%%%%%%%%%%%%%%%%%%%

Reversal of index $i$ of tensor $\hat{T}$ can be decomposed into the reversal of the degeneracy index $(j, t_j)$ of the degeneracy tensors and reversal of the spin index $(j, m_j)$ of the structural tensors. Reversal of the degeneracy index is trivial since the cup and cap transformations act as the Identity $\hat{I}_{d_j}$ on it whereas the reversal of the spin index is mediated by transformations $\mycup_j$ and $\mycap_j$. 

Figure~\ref{fig:cupcap}(i)-(ii) introduces a graphical representation of the cup and cap tensors. The cup tensor is depicted as a small circle with two incoming lines (forming a `cup') whereas a cap tensor is depicted as a small circle with two outgoing lines (forming a `cap')  \citep{Baez, Biamonte10}.

Next, we illustrate how an outgoing index of a tensor can be reversed by means of the cup transformation. A cap transformation can be used to reverse an incoming index in an analogous way.

\textbf{Example 12:} Consider a rank-$2$ SU(2)-invariant tensor $\hat{T}$ with outgoing indices\\$a~=~(j_a, t_{j_a}, m_{j_a})$ and $b~=~(j_b, t_{j_b}, m_{j_b})$ and which is given in the canonical form,
\begin{equation}
\{\{a, b\}, \{\mbox{`out', `out'}\}, \{\hat{P}_{j}\}\},\label{eq:rev1}
\end{equation}
where $j$ assumes all values of $j_a$ that are equal to a value of $j_b$. Consider reversing index $a$ of $\hat{T}$ as shown in Fig.~\ref{fig:bend}(i). The resulting tensor (or matrix) $\hat{T}'$ is obtained by multiplying tensor $\hat{T}$ with a cup by contracting $a$. We follow the convention that multiplying with a cup corresponds to bending index $a$ upwards from the \textit{left} in the graphical representation. The same index can be bent upwards from the right by multiplying with the \textit{transpose} of the cup.

The resulting matrix $\hat{T}'$ has the canonical form
\begin{equation}
\{\{a, b\}, \{\mbox{`in', `out'}\}, \{\hat{P}'_{j}\}\},
\end{equation}
where
\begin{equation}
\boxed{
\hat{P}'_{j} = \frac{\hat{P}_j}{\sqrt{2j+1}}.\label{eq:degrev}
}
\end{equation}
In order to explain this expression consider Fig.~\ref{fig:bend}(ii) where the reversal is depicted as it is performed on the canonical form of $\hat{T}$. Reversal of the spin index equates to replacing the shaded region by a straight line. This corresponds to applying the following algebraic identity
\begin{equation}
\mycup_j~\hat{\omega}_{j} = \frac{\hat{I}_{2j+1}}{\sqrt{2j+1}}.
\end{equation}
The factor $\frac{1}{\sqrt{2j+1}}$ is absorbed into the degeneracy tensor $\hat{P}_j$, Eq.~(\ref{eq:degrev}), to obtain the final canonical form.\markend

\textbf{Example 13:} Consider a rank-$3$ SU(2)-invariant tensor $\hat{T}$ which is given in the canonical form,
\begin{equation}
\{\{a, b, c\}, \{\mbox{`in', `out', `out'}\}, \{\hat{P}_{j_aj_bj_c}\}\}. \label{eq:rev2}
\end{equation}
Consider reversing index $b$ of tensor $\hat{T}$ as shown in Fig.~\ref{fig:bend}(iii) by multiplying $\hat{T}$ with a cup such that $b$ is contracted. The canonical form of $\hat{T}'$ reads
\begin{equation}
\{\{a, b, c\}, \{\mbox{`in', `in', `out'}\}, \hat{P}'_{j_aj_bj_c}\},\label{eq:rev22}
\end{equation}
where
\begin{equation}
\boxed{
\hat{P}'_{j_aj_bj_c} = \hat{F}^{j_a j_b}_{0 j_b j_a j_c} \hat{P}_{j_aj_bj_c}.
}
\end{equation}
The recoupling coefficient $\hat{F}^{j_a j_b}_{0 j_b j_a j_c}$ appears due to the reversal of the spin index $(j_b, m_{j_b})$, as shown in Fig.~\ref{fig:bend}(iv). The Clebsch-Gordan tensor and the cup within the shaded region are replaced with another Clebsch-Gordan tensor and a recoupling coefficient. This is achieved by applying a resolution of Identity on spins $j_a$ and $j_b$ and simplifying the resulting diagram by applying the equality shown in Fig.~\ref{fig:fmove}(ii).\markend

The procedure of reversing the spin index illustrated in Example 13 can be applied to reverse a spin index of a generic rank-$k$ SU(2)-invariant tensor. Recall that a structural tensor is maintained as a trivalent tree of Clebsch-Gordan tensors. Reversal of the spin index corresponds to multiplying a cup with a Clebsch-Gordan tensor within this tree. Then, as in Example 13, we proceed by replacing the Clebsch-Gordan tensor and the cup by another Clebsch-Gordan tensor and a recoupling coefficient. The recoupling coefficient is absorbed into the degeneracy tensor to obtain the canonical form of the resulting tensor. In this way we can reverse an index of a generic rank-$k$ SU(2)-invariant tensor.

%%%%%%%%%%%%%%%%%%%%%%%%%%%%%%%%%%%%%%%%%%%%%%%%%%%%%%%%%%%%%%%%%%%%%%%%%%%%%%%%%%%%%%%%%%%%%%%%
\begin{figure}[t]
\begin{center}
  \includegraphics[width=12cm]{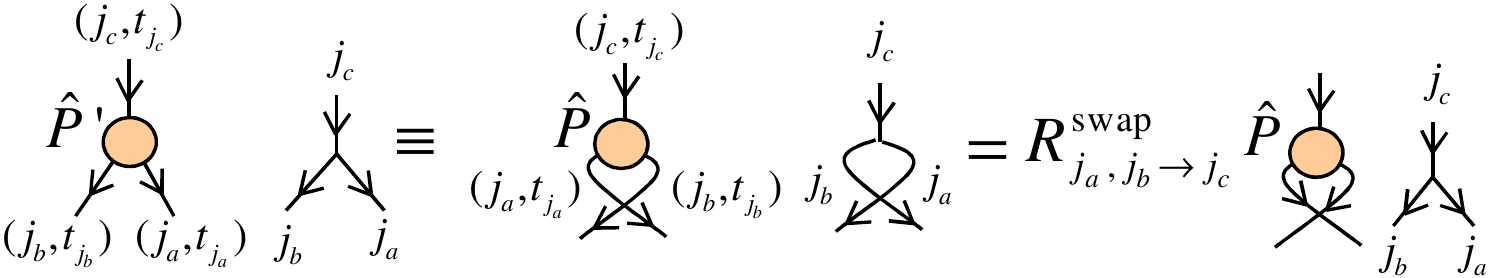}
\end{center} 
\caption{Permutation of indices [Fig.~\ref{fig:tensorman}(ii)] as performed on the canonical form of a rank-$3$ SU(2)-invariant tensor. The permutation decomposes into permutation of the degeneracy indices and permutation of the spin indices. The latter equates to replacing the Clebsch-Gordan tensor and a `cross' with a Clebsch-Gordan tensor and a numerical factor $\braid{a}{b}{c}$.\label{fig:permutecanon}} 
\end{figure}
%%%%%%%%%%%%%%%%%%%%%%%%%%%%%%%%%%%%%%%%%%%%%%%%%%%%%%%%%%%%%%%%%%%%%%%%%%%%%%%%%%%%%%%%%%%%%%%%
\subsection{Permutation of indices\label{sec:blockmoves:permute}}
Let us focus on the swap of two adjacent indices of an SU(2)-invariant tensor. As mentioned in Sec.~\ref{sec:tensor:manipulations} an arbitrary permutation of indices can be applied as a sequence of a number of such swaps.

Consider the swap e.g. Eq.~(\ref{eq:permute}) of two adjacent indices of a rank-$3$ SU(2)-invariant tensor $\hat{T}$ that is given in the canonical form,
\begin{equation}
\{\{a,b,c\}, \{\mbox{`in', `in', `out'}\}, \{\hat{P}_{j_aj_bj_c}\}\}. \nonumber
\end{equation}
Then tensor $\hat{T}'$ that is obtained as result of swapping indices $a$ and $b$ has the canonical form
\begin{equation}
	\{\{b,a,c\}, \{\mbox{`in', `in', `out'}\}, \{\hat{P}'_{j_b j_a j_c}\}\}, \nonumber
\end{equation}
where
\begin{equation}
\boxed{
\hat{P}_{j_b j_a j_c}' = \braid{a}{b}{c} \hat{P}_{j_a j_b j_c}.\label{eq:perm11}
}
\end{equation}
Here $\braider$ is a rank-$3$ SU(2)-invariant tensor with components $\braid{a}{b}{c}$,
\begin{equation}
\braid{a}{b}{c} \equiv (-1)^{j_a+j_b-j_c}, \label{eq:braid3}
\end{equation}
which mediates the swap of the spin indices $(j_a, m_{j_a})$ and $(j_b, j_{m_b})$ that fuse into index $(j_c, j_{m_c})$, see Fig.~\ref{fig:permutecanon}. That is,
\begin{align}
\cfusespin{b}{a}{c} = \braid{a}{b}{c} \cfusespin{a}{b}{c}.
\end{align}
(The same tensor $\braid{a}{b}{c}$ also relates, in a similar way, tensor $\csplitspin{c}{b}{a}$ and tensor $\csplitspin{c}{a}{b}$.) 

When swapping two adjacent indices of a generic rank-$k$ tensor the degeneracy tensors $\hat{P}$ and $\hat{P}'$ are also related directly by the swap tensor $\braider$, such as in Eq.~(\ref{eq:perm11}), if we work in a canonical form in which the indices that are swapped belong to the \textit{same} node of the trivalent tree that characterizes the canonical form. 

%%%%%%%%%%%%%%%%%%%%%%%%%%%%%%%%%%%%%%%%%%%%%%%%%%%%%%%%%%%%%%%%%%%%%%%%%%%%%%%%%%%%%%%%%%%%%%%%
\begin{figure}[t]
\begin{center}
  \includegraphics[width=5cm]{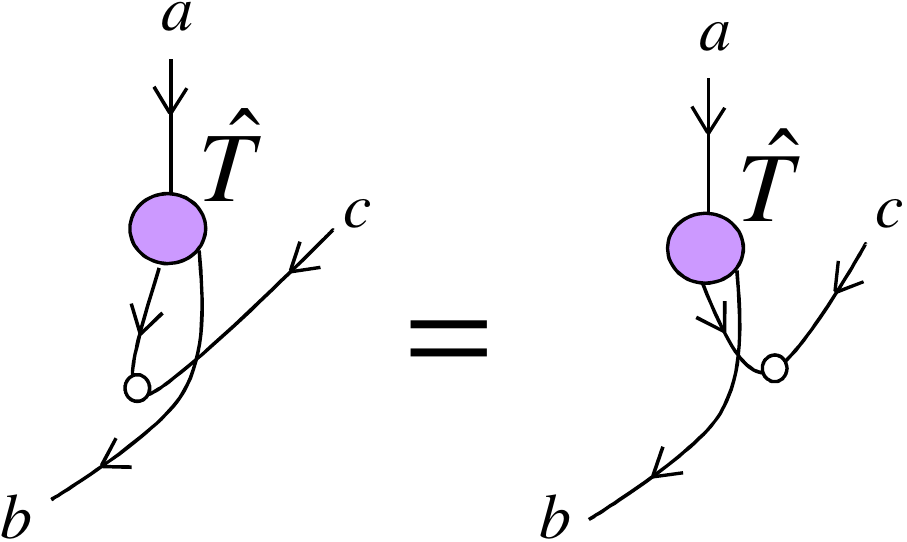}
\end{center}  
\caption{Illustration of the property that reversal of indices ``commutes '' with a permutation of them.\label{fig:bendpermute}} 
\end{figure}
%%%%%%%%%%%%%%%%%%%%%%%%%%%%%%%%%%%%%%%%%%%%%%%%%%%%%%%%%%%%%%%%%%%%%%%%%%%%%%%%%%%%%%%%%%%%%%%%

Notice how the canonical form of an SU(2)-invariant tensor facilitates a computational speedup for permutation of indices: permuting indices of the tensor is reduced to permuting indices of the much smaller degeneracy tensors. Figure~\ref{fig:permutereshapecompare} illustrates the computational speedup corresponding to a permutation of indices performed using our reference implementation MATLAB. In this implementation permutation of several indices is performed without necessarily breaking the permutation into swaps, see Sec.~\ref{sec:symTN:permute} in Chapter 6.

One can also consider manipulations that involve both reversing indices and permuting them. In this context it is useful to note that these manipulations ``commute'' with one another, as illustrated in Fig.~\ref{fig:bendpermute}. 

\subsection{Reshape of indices\label{sec:blockmoves:reshape}}

The transformation that implements the reshape of indices of an SU(2)-invariant tensor depends on the directions of the indices. We analyze three distinct cases. First, we consider fusion of two outgoing indices into an outgoing index and splitting of an outgoing index into two outgoing indices. Second, we consider the analogous reshape of incoming indices. And third, we consider the fusion of an incoming index with an outgoing index.

\begin{figure}[t]
\begin{center}
  \includegraphics[width=12cm]{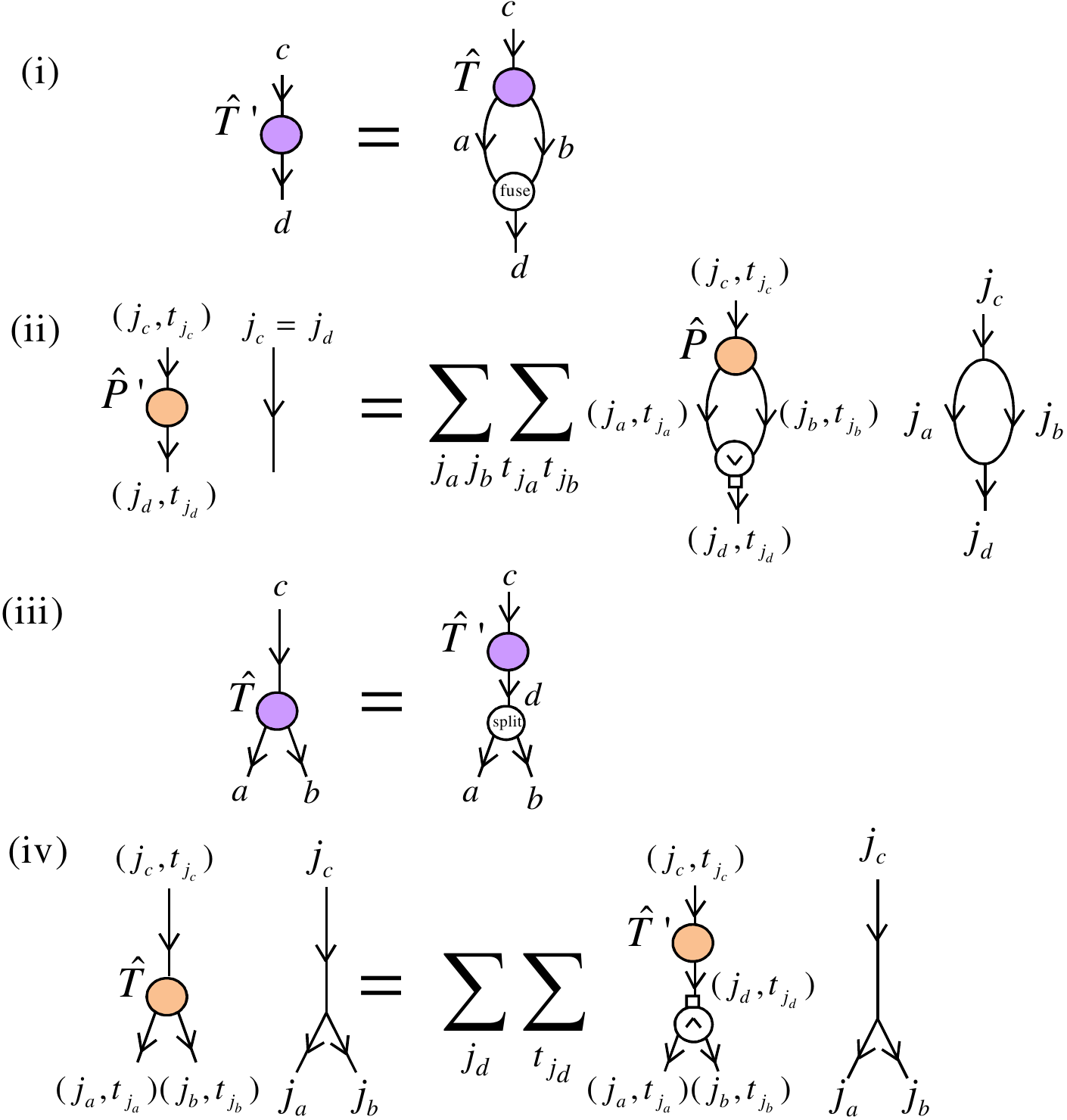}
\end{center}  
\caption{(i) Fusion of two outgoing indices of an SU(2)-invariant tensor by means of the fusing tensor $\fuser$. (ii) Fusion in the canonical form of the tensor decomposes into the fusion of the degeneracy indices using tensor $\tfuser$, and the fusion of the spin indices using tensor $\cfuser$. The latter can be performed for free since the loop can directly be replaced with a straight line [Fig.~\ref{fig:cg}(ii)]. (iii) Splitting of an outgoing index into two indices by means of the splitting tensor $\spliter$. (iv) Splitting the index in the canonical form of the tensor. \label{fig:reshapecanon}}
\end{figure}

Let us consider the fusion e.g. Eq.~(\ref{eq:fuse}) of two outgoing indices of an SU(2)-invariant tensor $\hat{T}$. In order to obtain the reshaped tensor $\hat{T}'$ in a canonical form it is required that the fused index be maintained in the spin basis. However, the direct product of indices $d = a \times b$ may result in an index that does not label a spin basis. Therefore, we fuse indices by multiplying $\hat{T}$ with the fusing tensor $\fuse{a}{b}{d}$ such that indices $a$ and $b$ are contracted [Fig.~(\ref{fig:reshapecanon}(i))], 
\begin{equation}
\hat{T}'_{dc} \equiv \sum_{ab} \hat{T}_{abc} \fuse{a}{b}{d},
\label{eq:su2fuseEx}
\end{equation}
or in the canonical form [Fig.~\ref{fig:reshapecanon}(ii)] ,
\begin{equation}
\boxed{
\hat{P}'_{j_dj_c} = \sum_{j_aj_b}\sum_{t_{j_a}t_{j_b}} \tfusespin{a}{b}{d} \hat{P}_{j_aj_bj_c}.\label{eq:su2fusecanon}
}
\end{equation}
Notice that the fusion of the spin indices, here, is straightforward. We proceed by multiplying with tensor $\cfuser$ and replacing the resulting `loop' in the figure with a straight line [Fig.~\ref{fig:cg}(ii)]. The fusion of two adjacent indices of a generic rank-$k$ SU(2)-invariant tensor follows a straightforward generalization of Eq.~(\ref{eq:su2fusecanon}). By working in a canonical form that is characterized by a trivalent tree in which the two indices belong to the same node, the fusion of the spin indices involves a simple loop elimination, similar to the one illustrated in Fig.~\ref{fig:reshapecanon}(ii).

%%%%%%%%%%%%%%%%%%%%%%%%%%%%%%%%%%%%%%%%%%%%%%%%%%%%%%%%%%%%%%%%%%%%%%%%%%%%%%%%%%%%%%%%%%%%%%%%
%%%%%%%%%%%%%%%%%%%%%%%%%%%%%%%%%%%%%%%%%%%%%%%%%%%%%%%%%%%%%%%%%%%%%%%%%%%%%%%%%%%%%%%%%%%%%%%%

The original tensor $\hat{T}$ can be recovered from $\hat{T}'$ by splitting the index $d$ back into indices $a$ and $b$. This is achieved by multiplying tensor $\hat{T}'$ with the splitting tensor $\splitt{d}{a}{b}$ such that $d$ is contracted [Fig.~\ref{fig:reshapecanon}(iii)],
\begin{equation}
\hat{T}_{abc} \equiv \sum_{d} \hat{T}'_{dc}  \splitt{d}{a}{b},\label{eq:su2splitEx}
\end{equation}
or in the canonical form [Fig.~\ref{fig:reshapecanon}(iv)],
\begin{equation}
\boxed{
\hat{P}'_{j_dj_c} = \sum_{j_aj_b}\sum_{t_{j_a}t_{j_b}} \tfusespin{a}{b}{d} \hat{P}_{j_aj_bj_c}.\label{eq:su2fusecanon}
}
\end{equation}

Notice that the sum in Eq.~(\ref{eq:su2fusecanon}) implies that a reshaped tensor $\hat{P}'_{j_aj_d}$ involves a linear combination of several tensors $\hat{P}_{j_aj_bj_c}$. Thus, performing the fusion in the canonical form requires more work than reshaping regular indices which is a simple rearrangement of the tensor components. As a result, fusing indices of SU(2)-invariant tensors can be more expensive than fusing indices of regular tensors, as illustrated in Fig.~\ref{fig:permutereshapecompare} for a reshaping done in MATLAB.

\begin{figure}[t]
\begin{center}
  \includegraphics[width=10cm]{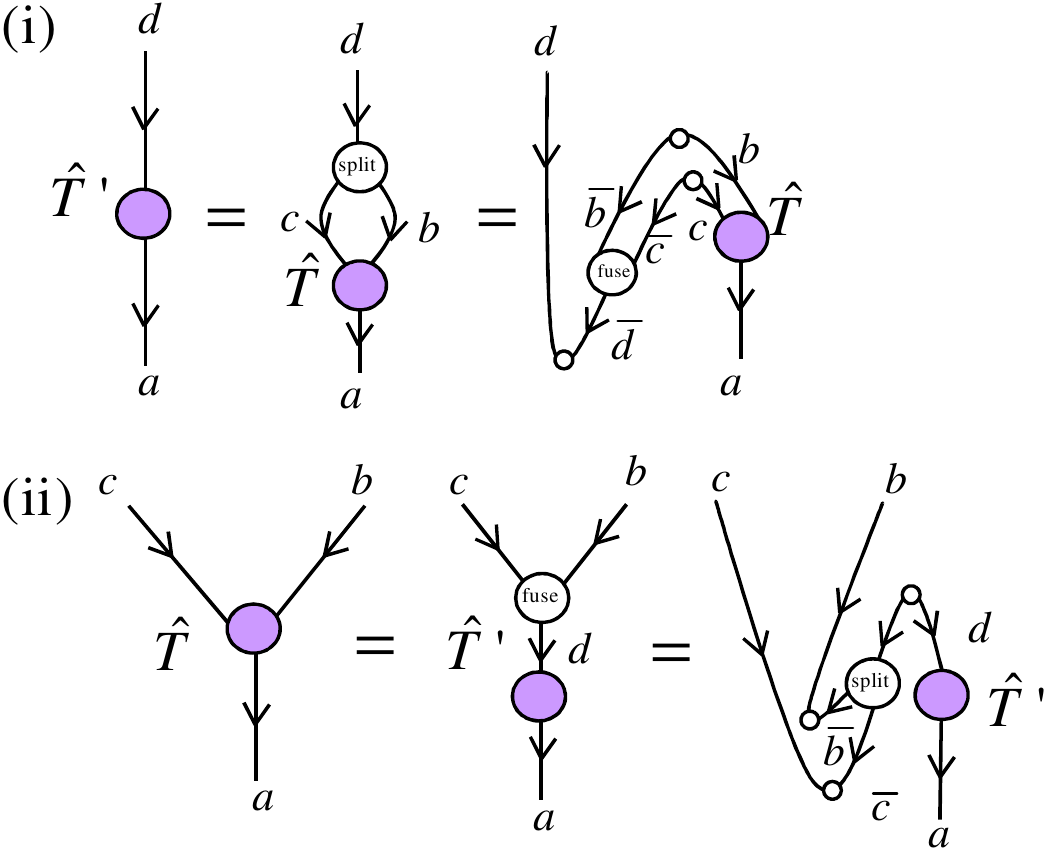}
\end{center}  
\caption{(i) Fusion of two incoming indices into an incoming index, and (ii) splitting an incoming index into two incoming indices can be performed in one of two way.\label{fig:reshapein}} 
\end{figure}

\begin{figure}[t]
\begin{center}
  \includegraphics[width=10cm]{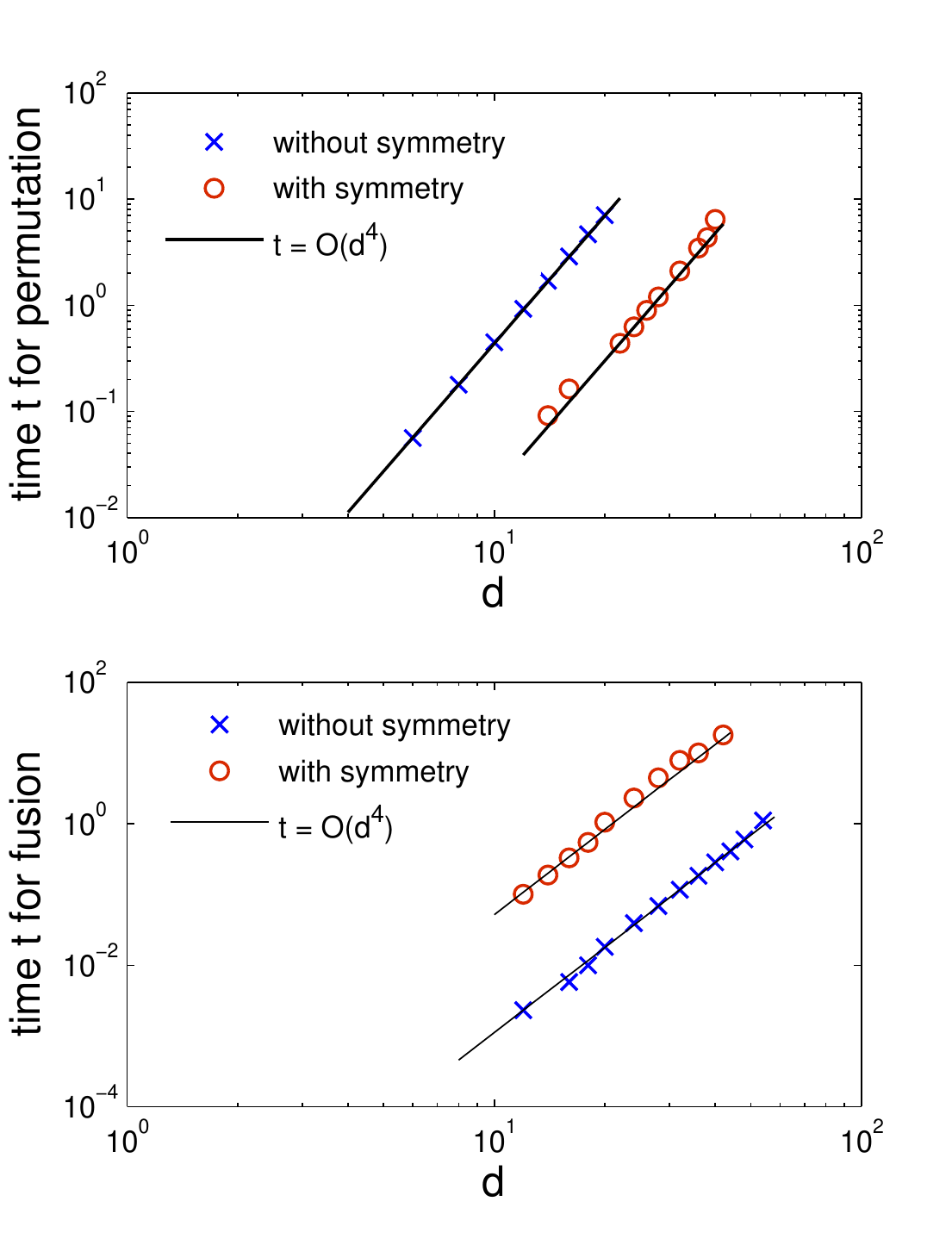}
\end{center}  
\caption{Computation times (in seconds) required to permute indices of a rank-four tensor $\hat{T}$, as a function of the size of the indices. All four indices of $\hat{T}$ have the same size $9d$, and therefore the tensor contains $|\hat{T}| = 9^4d^4$ coefficients. The figures compare the time required to perform these operations using a regular tensor and an SU(2)-invariant tensor, where in the second case each index contains three different values of spin $j=0,1,2$, each with degeneracy $d$, and the canonical form of Eq.~(\ref{eq:deco41}) is used. The upper figure shows the time required to permute two indices: For large $d$, exploiting the symmetry of an SU(2)-invariant tensor by using the canonical form results in shorter computation times. The lower figure shows the time required to fuse two adjacent indices. In this case, maintaining the canonical form requires more computation time. Notice that in both figures the asymptotic cost scales as $O(d^4)$, or the size of $\hat{T}$, since this is the number of coefficients which need to be rearranged. We note that the fixed-cost overheads associated with symmetric manipulations could potentially vary substantially with choice of programming language, compiler, and machine architecture. The results given here show the performance of our MATLAB implementation of SU(2) symmetry.}
\label{fig:permutereshapecompare}
\end{figure}

Next, let us consider fusing two incoming indices into a single incoming index. This can be done in one of two equivalent ways. The first involves multiplying the tensor with a \textit{splitting tensor} by contracting the two incoming indices. Equivalently, if one prefers to use the \textit{fusing tensor}, one can reverse the two indices, multiply with the fusing tensor, and finally reverse the fused index. The two approaches are depicted in Fig.~\ref{fig:reshapein}(i). The fused index can be split back into the original indices by reverting the fusion. In the first approach this is done by multiplying with a fusing tensor while in the second approach this is done by multiplying with a splitting tensor and then reversing the two indices [Fig.~\ref{fig:reshapein}(ii)].

Finally, consider the fusion of an incoming index with an outgoing index to produce, say, an outgoing index. This can be achieved by reversing the incoming index and then fusing the indices by means of a fusing tensor. The fused index should be split in a consistent manner by reverting this fusion procedure.
%%%%%%%%%%%%%%%%%%%%%%%%%%%%%%%%%%%%%%%%%%%%%%%%%%%%%%%%%%%%%%%%%%%%%%%%%%%%%%%%%%%%%%%%%%%%%%%%
%%%%%%%%%%%%%%%%%%%%%%%%%%%%%%%%%%%%%%%%%%%%%%%%%%%%%%%%%%%%%%%%%%%%%%%%%%%%%%%%%%%%%%%%%%%%%%%%

%%%%%%%%%%%%%%%%%%%%%%%%%%%%%%%%%%%%%%%%%%%%%%%%%%%%%%%%%%%%%%%%%%%%%%%%%%%%%%%%%%%%%%%%%%%%%%%%
%%%%%%%%%%%%%%%%%%%%%%%%%%%%%%%%%%%%%%%%%%%%%%%%%%%%%%%%%%%%%%%%%%%%%%%%%%%%%%%%%%%%%%%%%%%%%%%%%

\subsection{Multiplication of two matrices\label{sec:symTN:multiply}}

Let $\hat{M}$ and $\hat{N}$ be two SU(2)-invariant matrices given in the canonical form
\begin{align}
\hat{M} = \bigoplus_j (\hat{M}_j \otimes \hat{I}_{2j+1}), ~~~ \hat{N} = \bigoplus_j (\hat{N}_j \otimes \hat{I}_{2j+1}). \label{eq:matmult11}
\end{align}
Then the SU(2)-invariant matrix $\hat{T} = \hat{M} \hat{N}$ obtained by multiplying together matrices $\hat{M}$ and $\hat{N}$ has the canonical form
\begin{equation}
	\hat{T} = \bigoplus_j (\hat{T}_j \otimes \hat{I}_{2j+1}),\label{eq:matmult4}
\end{equation}
where $\hat{T}_j$ is obtained by multiplying matrices $\hat{M}_j$ and $\hat{N}_j$,
\begin{equation}
\hat{T}_j = \hat{M}_j \hat{N}_j.\label{eq:blockmult}
\end{equation}
Clearly, computational gain is obtained as a result of performing the multiplication $\hat{T} = \hat{R}\hat{S}$ block-wise. This is illustrated by the following example.

%%%%%%%%%%%%%%%%%%%%%%%%%%%%%%%%%%%%%%%%%%%%%%%%%%%%%%%%%%%%%%%%%%%%%%%%%%%%%%%%%%%%%%%%%%%%%%%%
\begin{figure}[t]
\begin{center}
  \includegraphics[width=10cm]{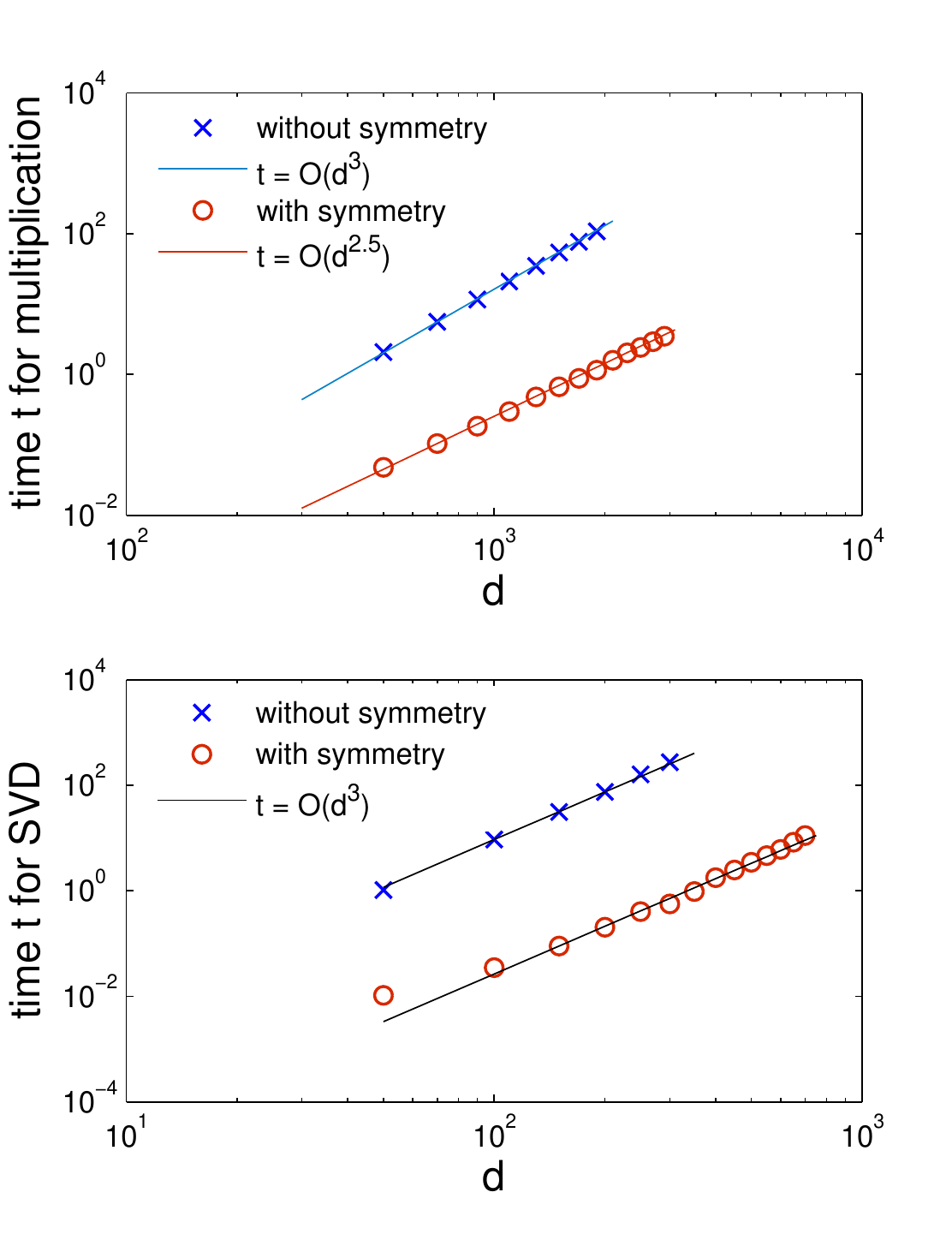}
\end{center}  
  \caption{Computation times (in seconds) required to multiply two matrices (upper panel) and to perform a singular value decomposition (lower panel), as a function of the size of the indices. Matrices of size $9d \times 9d$ are considered. The figures compare the time required to perform these operations using regular matrices and SU(2)-invariant matrices, where for the SU(2) matrices each index contains three different values of the spin $j=0,1,2$, each with degeneracy $d$, and the canonical form of Eq.~(\ref{eq:canonblock2}) is used. That is, each matrix decomposes into three blocks of size $d \times d$. For large $d$, exploiting the block diagonal form of SU(2)-invariant matrices results in shorter computation time for both multiplication and singular value decomposition. The asymptotic cost scales with $d$ as $O(d^3)$, while the size of the matrices grows as $O(d^2)$.(For matrix multiplication, a tighter bound of $O(d^{2.5})$ for the scaling of computational cost with $d$ is seen in this example.) We note that the fixed-cost overheads associated with symmetric manipulations could potentially vary substantially with choice of programming language, compiler, and machine architecture. The results given here show the performance of our MATLAB implementation of SU(2) symmetry.
  \label{fig:multsvdcompare}} 
\end{figure}
%%%%%%%%%%%%%%%%%%%%%%%%%%%%%%%%%%%%%%%%%%%%%%%%%%%%%%%%%%%%%%%%%%%%%%%%%%%%%%%%%%%%%%%%%%%%%%%%

\textbf{Example 13 : (Computational gain from blockwise multiplication)} Consider vector space $\mathbb{V}$ that decomposes as $\mathbb{V} \cong d_j\mathbb{V}_j$ where $j$ assumes values $1,\cdots,q$ and let $d_j=d, \forall j$. The dimension of the space $\mathbb{V}$ is $dp$ where $p=\sum_{j=1}^{q}(2j+1)=q^2+q$. 

Consider an SU(2)-invariant matrix $\hat{T}: \mathbb{V} \rightarrow \mathbb{V}$. Since there are $q$ blocks $\hat{T}_j$ and each block has size $d\times d$, the SU(2)-invariant matrix $\hat{T}$ contains $qd^2$ coefficients. For comparison, a regular matrix of the same size contains $d^2p^2$ coefficients, a number greater by a factor of $O(q^3)$.

Let us now consider multiplying two such matrices. We use an algorithm that requires $O(l^3)$ computational time to multiply two matrices of size $l\times l$. The cost of performing $q$ multiplications of $d\times d$ blocks in Eq.~\ref{eq:blockmult} scales as $O(qd^3)$. In contrast the cost of multiplying two regular matrices of the same size scales as $O(d^3p^3)$, requiring $O(q^5)$ times more computational time. Figure \ref{fig:multsvdcompare} shows a comparison of the computation times when multiplying two matrices for both SU(2)-invariant and regular matrices. \markend

\subsection{Factorization of a matrix\label{sec:symTN:factorize}}

The factorization of an SU(2)-invariant matrix $\hat{T}$ can also benefit from the block-diagonal structure. Consider, for instance, the singular value decomposition (SVD), $\hat{T} = \hat{U}\hat{S}\hat{V}$, where $\hat{U}$ and $\hat{V}$ are unitary matrices and $\hat{S}$ is a diagonal matrix with non-negative components. If $\hat{T}$ has the canonical form
\begin{equation}
\hat{T} = \bigoplus_j (\hat{T}_j \otimes \hat{I}_{2j+1}),
\end{equation}
we can obtain the SU(2)-invariant matrices
\begin{align}
	\hat{U} = \bigoplus_j (\hat{U}_j \otimes \hat{I}_{2j+1}),\nonumber \\
	~\hat{S} = \bigoplus_j (\hat{S}_j \otimes \hat{I}_{2j+1}),\nonumber \\
	~\hat{V} = \bigoplus_j (\hat{V}_j \otimes \hat{I}_{2j+1}),\nonumber
	\label{eq:svd0}
\end{align}
by performing SVD of each degeneracy matrix $\hat{T}_j$ independently,
\begin{equation}
	\hat{T}_j = \hat{U}_j \hat{S}_j \hat{V}_j.
	\label{eq:svd1}
\end{equation}
A different factorization of $\hat{T}$, such as spectral decomposition or polar decomposition, can be obtained by the analogous factorization of the blocks $\hat{T}_j$. 
The computational savings are analogous to those described in Example 13 for the multiplication of matrices. Figure \ref{fig:multsvdcompare} shows a comparison of computation times required to perform a singular value decomposition on SU(2)-invariant and regular matrices using MATLAB.

\section{Supplement: Examples of evaluating a spin network \label{sec:sn}}

%%%%%%%%%%%%%%%%%%%%%%%%%%%%%%%%%%%%%%%%%%%%%%%%%%%%%%%%%%%%%%%%%%%%%%%%%%%%%%%%%%%%%%%%%%%%%%%%
\begin{figure}[t]
\begin{center}
  \includegraphics[width=16cm]{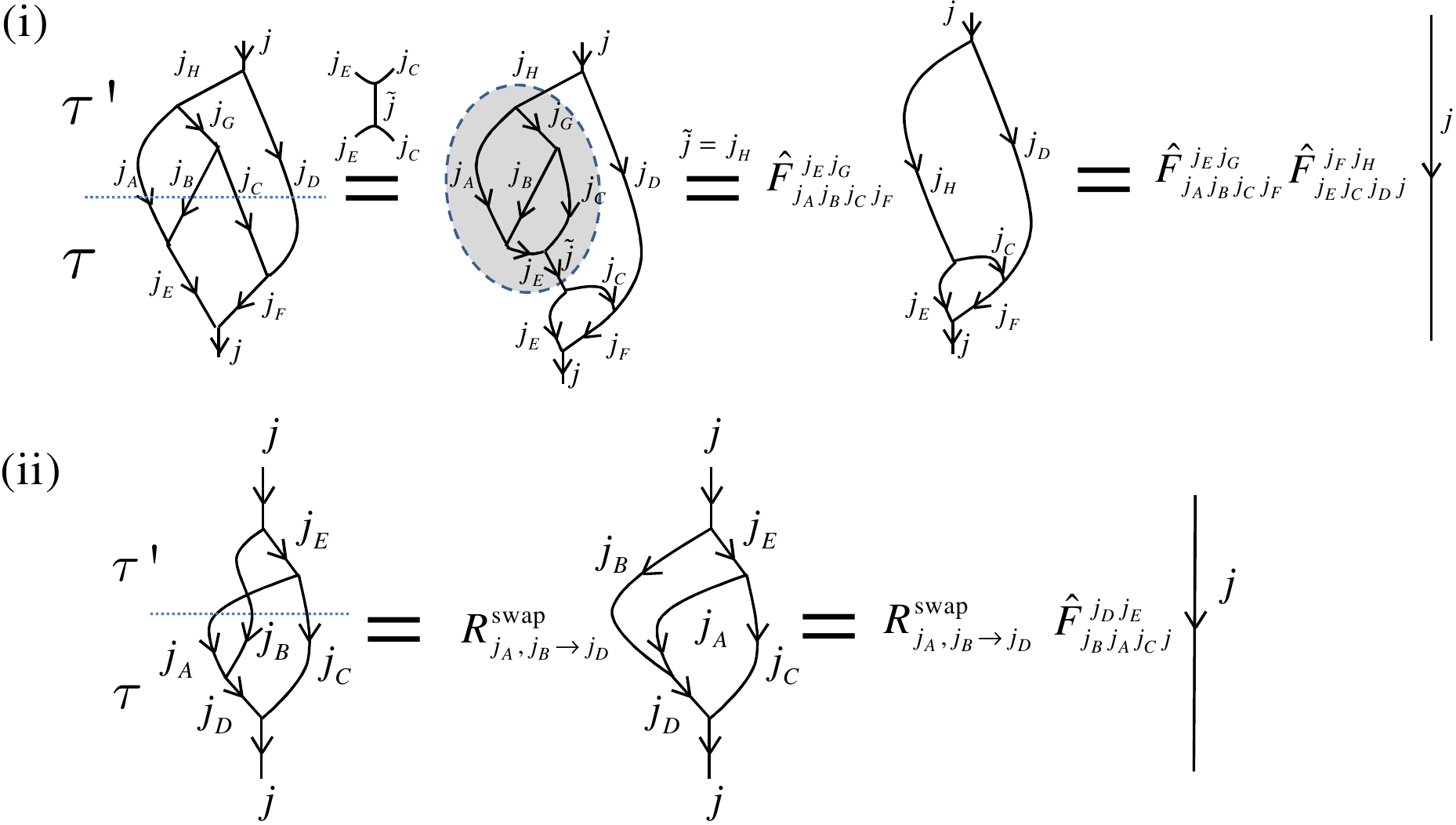}
\end{center}  
\caption{Illustration of evaluating a spin network.}
\label{fig:evalspin1}
\end{figure}
%%%%%%%%%%%%%%%%%%%%%%%%%%%%%%%%%%%%%%%%%%%%%%%%%%%%%%%%%%%%%%%%%%%%%%%%%%%%%%%%%%%%%%%%%%%%%%%%
Let us consider a spin network $\mathcal{S}(\tree, \tree')$ that is constructed by means of two fusion trees $\tree$ and $\tree'$ in the following way. First obtain a tree tensor network $\mathcal{T}$ by attaching a tensor $\cfuser$ to each node the fusion tree $\tree$. A tensor $\cfuser$ mediates the fusion of the incoming spins into the outgoing spin. Next, obtain the \textit{splitting tree} that is dual to $\tree'$. A splitting tree is obtained by reversing the direction of all links of a fusion tree. In the graphical representation this corresponds to a horizontal reflection of the fusion tree. Then obtain a tree tensor network $\mathcal{T}'$ by attaching to each node of the splitting tree a tensor $\cspliter$ that mediates the splitting of the incoming spin into outgoing spins. The spin network $\mathcal{S}(\tree, \tree')$ is obtained by connecting the open links of the two tree tensor networks: $\mathcal{T}$ and $\mathcal{T}'$.

Since the spin network $\mathcal{S}(\tree, \tree')$ has two open links it can be contracted to obtain an SU(2)-invariant matrix, which according to Schur's lemma is proportional to the Identity. An important property of $\mathcal{S}(\tree, \tree')$ is that this proportionality factor can be \textit{evaluated} algebraically without contracting the spin network. This property can be exploited to suppress the potentially high cost of contracting spin networks in numerical simulations.

The spin network $\mathcal{S}(\tree, \tree')$ can be evaluated in terms of the values of basic spin networks that are shown in Fig.~\ref{fig:cg}(ii) and Fig.~\ref{fig:fmove}(ii). The first step of the evaluation procedure generally involves expressing the spin network as a composition of these basic spin networks. This can be achieved by applying, possibly several times, a resolution of Identity, Fig.~\ref{fig:cg}(i), on appropriate links of the spin network. Then one proceeds by recursively applying the equalities in Fig.~\ref{fig:cg}(ii) and Fig.~\ref{fig:fmove}(ii) to regions of the spin network, eventually replacing the spin network with a straight line and an overall numerical factor.  Figure~\ref{fig:evalspin1}(i) illustrates these steps for the simple case of evaluating the spin network of Fig.~\ref{fig:spin} in terms of two recoupling coefficients.

We can also consider spin networks that have intercrossing lines such as those which appear when applying permuting indices of an SU(2)-invariant tensor. Such a spin network can be evaluated in terms of recoupling coefficients \textit{and} swap factors, as illustrated in Fig.~\ref{fig:evalspin1}(ii).

%*************************************************************************************************

\chapter{Implementation of non-Abelian symmetries \Rmnum{2}}

In this chapter we will describe a specific scheme for implementing non-Abelian symmetries. Our implementation is based on a special canonical form, the \textit{tree decomposition}, of symmetric tensors. 

A tree decomposition of a symmetric tensor corresponds to representing and storing the tensor as a tree tensor network made of two parts: (i) a symmetric \textit{vector} and (ii) possibly several \textit{splitting tensors} $\spliter$. We describe how to implement the tensor manipulations within the set $\mathcal{P}$ of primitive operations based on tree decompositions. In a tree decomposition, reshape and permutation of indices take a very simple form. In order to obtain the vector of the output tensor, one needs to simply multiply the vector of the initial tree decomposition with a matrix $\hat{\Gamma}$ [Fig~\ref{fig:spin}(ii)] that depends only on the permutation or reshape.

This approach offers several advantages. Reshapes and permutations can be performed without breaking them into pairwise fusions and swaps, as was described in the previous chapter. More importantly, one can precompute (that is, compute before running the algorithm) the matrices $\hat{\Gamma}$ since these do not depend on the actual components of the tensor being manipulated. This is of special advantage in the case of iterative algorithms, where by pre-computing these matrices one also eliminates the cost of evaluating spin networks at runtime, thus substantially reducing the computational costs.
%
%Without loss of generality, in this implementation we will only consider tensors with outgoing indices. Recall that any tensor $\hat{T}$ with incoming and outgoing indices can be transformed into another tensor $\hat{T}'$ with only outgoing indices by means of cap transformations. Tensor $\hat{T}$ can then be stored as tensor $\hat{T}'$ and additional cup transformations that relate  $\hat{T}'$ with $\hat{T}$. 

\subsection{Tree decompositions of SU(2)-invariant tensors\label{sec:vectorform:tree}}

Consider a rank-$k$ SU(2)-invariant tensor $\hat{T}$ with indices $\{i_1, i_2, \ldots, i_k\}$ and directions $\vec{D}$. Let us apply the following transformations on tensor $\hat{T}$ to obtain a vector. First reverse all incoming indices of $\hat{T}$ to obtain another tensor $\hat{T}'$. Then fuse the indices of $\hat{T}'$ according to a given fusion tree $\tree$ to obtain an SU(2)-invariant vector $\hat{v}$. This gives rise to a decomposition of tensor $\hat{T}$ in terms of the vector $\hat{v}$, a set of splitting tensors that revert the fusion sequence $\tree$, and a set of cup tensors that reverse the split indices that are identified with the incoming indices of $\hat{T}$. We refer to such a decomposition as a \textit{tree decomposition} of $\hat{T}$ and denote it as $\mathcal{D}(\hat{T})$. It is completely specified by the following list of elements:
\begin{equation}
\mathcal{D}(\hat{T}) \equiv (\{i_1, i_2, \ldots, i_k\}, \vec{D}, \tree, \hat{v}).  \label{eq:datastruct}
\end{equation}
Here the fusion tree $\tree$ determines the splitting tensors that are part of the decomposition while the directions $\vec{D}$ indicate the presence or absence of a cup tensor on the open indices of the tree decomposition.

\begin{figure}[t]
\begin{center}
  \includegraphics[width=8cm]{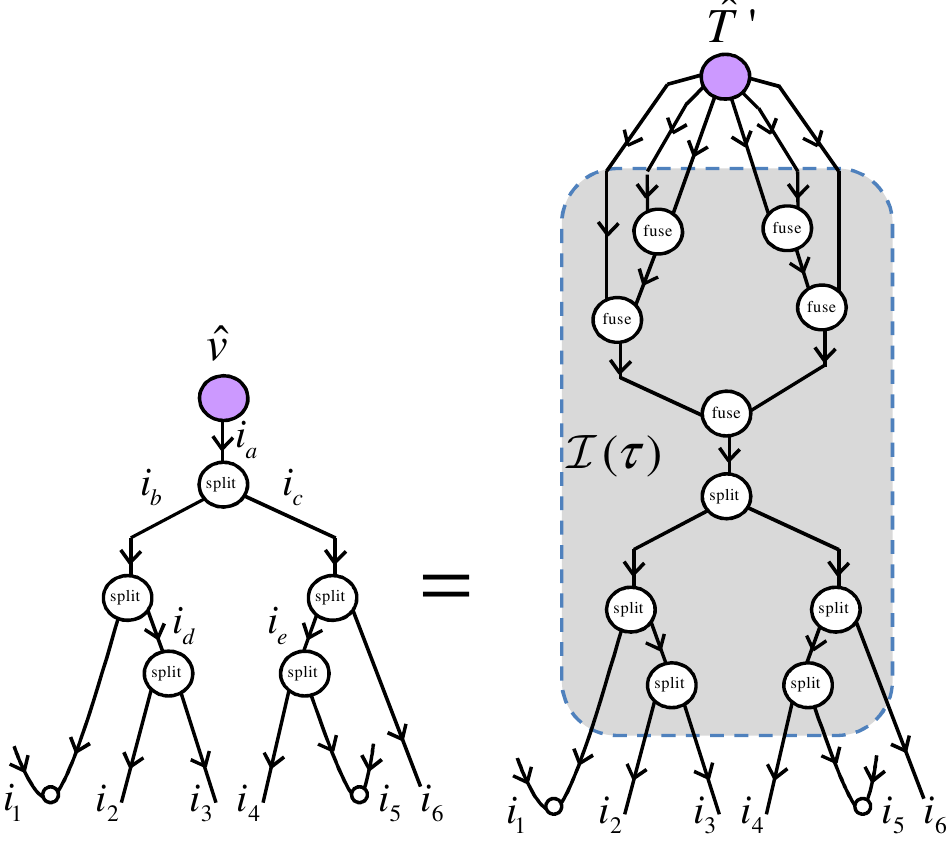}
\end{center}  
\caption{ A tree decomposition $\mathcal{D}(\hat{T})$ of a rank-$6$ SU(2)-invariant tensor $\hat{T}$ with indices $\{i_1, i_2, \ldots, i_6\}$. (All internal indices are summed over.) The tree decomposition consists of an SU(2)-invariant vector $\hat{v}$, a set of splitting tensors and cup tensors that are attached to the incoming indices of $\hat{T}$. The tree decomposition is obtained by applying a resolution of Identity $\mathcal{I}(\tree)$ on the tensor. The fusing and splitting tensors that constitute $\mathcal{I}(\tree)$ are connected together according to the fusion tree $\tree$. \label{fig:treeDeco1}}
\end{figure}

A tree decomposition of a rank-$6$ SU(2)-invariant tensor is shown in Fig.~\ref{fig:treeDeco1}. The tree decomposition in the diagram can be specified as,
\begin{align}
\mathcal{D}(\hat{T}) \equiv (\{i_1, i_2, i_3, i_4, i_5, i_6\}, 
\{\mbox{`in', `out', `out', `in-R', `out', `out'}\}, \tree, \hat{v}),\label{eq:tree6}
\end{align}
where
\begin{align}
\tree : \{i_4, i_5 \rightarrow  i_e;~~i_2, i_3 \rightarrow i_c;~~i_e, i_6 \rightarrow i_c;~~&i_1, i_d \rightarrow i_b;
i_b, i_c \rightarrow i_a\}. \nonumber
\end{align}
In the graphical representation of a tree decomposition $\mathcal{D}(\hat{T})$ the vector $\hat{v}$ appears at the top of the tree, the `body' of the tree comprises of splitting tensors that are connected according to the fusion tree $\tree$ and the indices $\{i_1, i_2, \ldots, i_6\}$ are associated, from left to right, to the open lines at the bottom of the tree. Some open indices are bent upwards by attaching cup tensors. A value $\vec{D}(l) = \mbox{`in'}$ indicates a cup tensor is attached to index $i_l$ while $\vec{D}(l) = \mbox{`out'}$ indicates its absence. We additionally denote by $\vec{D}(l) = \mbox{`in-R'}$ that the transpose of a cup tensor is attached to index $i_l$. In the graphical representation, the values $\vec{D}(l) = \mbox{`in'}$ or $\vec{D}(l) = \mbox{`in-R'}$ correspond to bending the index $i_l$ upwards from the left or from the right respectively. 
 
The vector $\hat{v}$ is obtained by applying an resolution of Identity, denoted $\mathcal{I}(\tree)$, on tensor $\hat{T}'$, as shown on the r.h.s. of Fig.~\ref{fig:treeDeco1}. The resolution of Identity $\mathcal{I}(\tree)$ is given by a tensor network made of a set of fusing tensors, that fuse the indices of $\hat{T}'$ according to the fusion tree $\tree$, and the corresponding set of splitting tensors that inverts the fusion. The vector $\hat{v}$ is obtained by contracting $\hat{T}'$ with the fusing tensors.

%%%%%%%%%%%%%%%%%%%%%%%%%%%%%%%%%%%%%%%%%%%%%%%%%%%%%%%%%%%%%%%%%%%%%%%%%%%%%%%%%%%%%%%%%%%%%%%%
%%%%%%%%%%%%%%%%%%%%%%%%%%%%%%%%%%%%%%%%%%%%%%%%%%%%%%%%%%%%%%%%%%%%%%%%%%%%%%%%%%%%%%%%%%%%%%%%

We emphasize that the cup tensors are stored \textit{as part} of the tree decomposition without consuming them into the tree. This is done to simplify reshape and permutation of indices of a tree decomposition since these operations can be performed without noticing the cup tensors. For instance, in order to permute the open indices of a tree decomposition one may proceed by detaching any cup tensors from the tree, permuting the indices and re-attaching the cup tensors to the updated tree, a direct application of the commutation property depicted in Fig.~\ref{fig:bendpermute}. On the other hand, manipulations that involve summing over an index that is attached to a cup tensor are an exception. For example, when multiplying two SU(2)-invariant matrices, each given as a tree decomposition, the cup tensor has to be properly considered to obtain the resultant matrix. This is discussed in Sec.~\ref{sec:symTN:matrixops}. 

%%%%%%%%%%%%%%%%%%%%%%%%%%%%%%%%%%%%%%%%%%%%%%%%%%%%%%%%%%%%%%%%%%%%%%%%%%%%%%%%%%%%%%%%%%%%%
\begin{figure}[t]
\begin{center}
  \includegraphics[width=8cm]{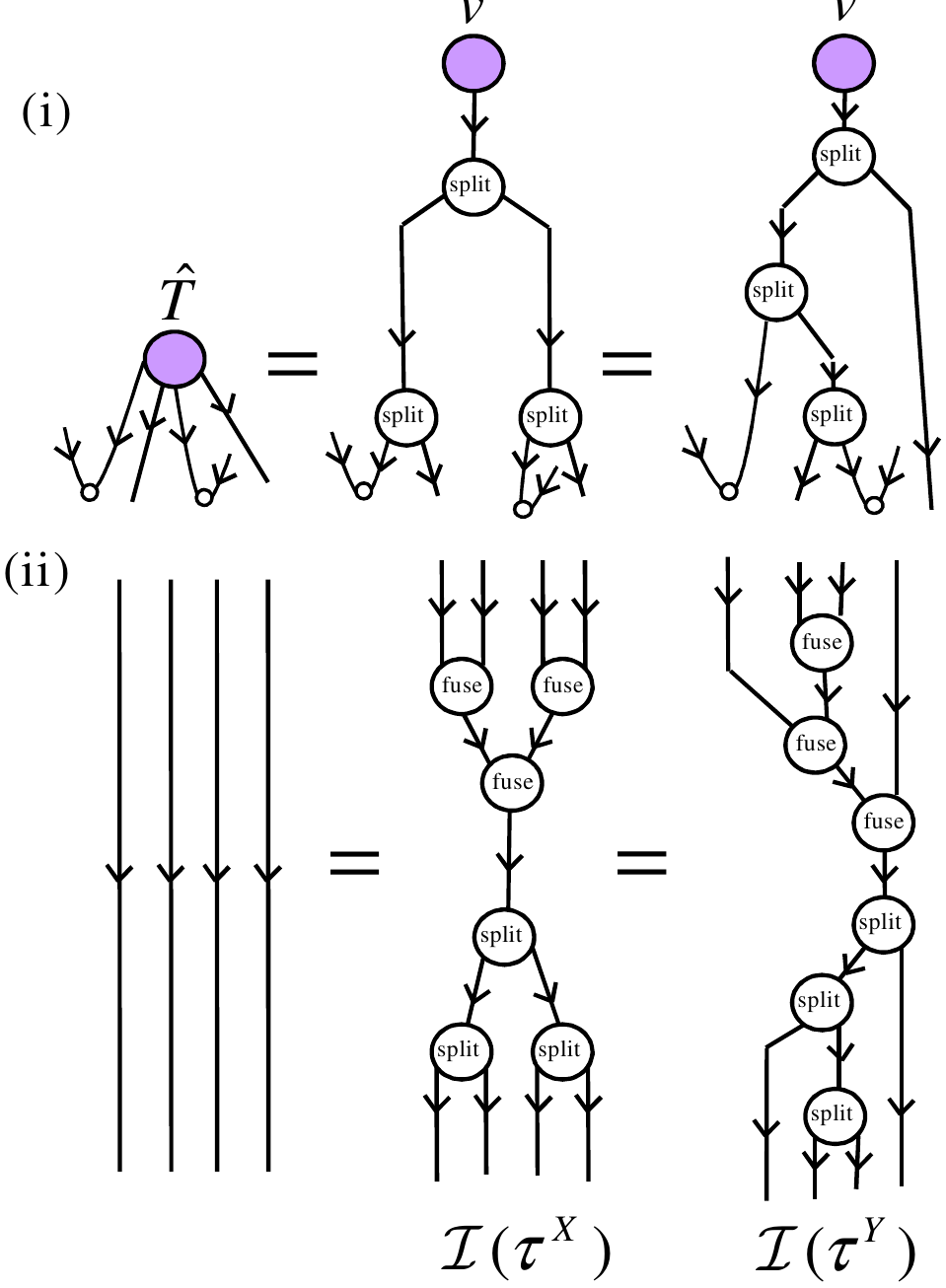}
\end{center}
\caption{ (i) Two different, but equivalent, tree decompositions $\mathcal{D}^{X}(\hat{T})$ and $\mathcal{D}^{Y}(\hat{T})$ of a rank-$4$ SU(2)-invariant tensor $\hat{T}$. The two decompositions are characterized by different fusion trees $\tree^{X}$ and $\tree^{Y}$. (ii) Tree decompositions $\mathcal{D}^{X}(\hat{T})$ and $\mathcal{D}^{Y}(\hat{T})$ are obtained by applying the resolutions of Identity $\mathcal{I}(\tree^{X})$ and $\mathcal{I}(\tree^{Y})$ on $\hat{T}$.
\label{fig:treeDeco4}}
\end{figure}
%%%%%%%%%%%%%%%%%%%%%%%%%%%%%%%%%%%%%%%%%%%%%%%%%%%%%%%%%%%%%%%%%%%%%%%%%%%%%%%%%%%%%%%%%%%%%%%%
\subsection{Mapping between tree decompositions\label{sec:symTN:tree:gamma}}

The same tensor $\hat{T}$ may be expressed in different tree decompositions corresponding to different choices of the fusion tree. Two different fusion trees $\tree^{X}$ and $\tree^{Y}$ lead to two different tree decompositions $\mathcal{D}^{X}(\hat{T})$ and $\mathcal{D}^{Y}(\hat{T})$ of the same tensor $\hat{T}$. As an example we show two different but equivalent tree decompositions of a rank-$4$ tensor in Fig. \ref{fig:treeDeco4}(i). The two decompositions 
\begin{equation}
(\{i_1, i_2, i_3, i_4\}, \vec{D}, \tree^{X}, \hat{v}^{X}) \mbox{ and } (\{i_1, i_2, i_3, i_4\}, \vec{D}, \tree^{Y}, \hat{v}^{Y}), \nonumber
\end{equation}
are obtained by applying on the tensor the resolutions of Identity $\mathcal{I}(\tree^{X})$ and $\mathcal{I}(\tree^{Y})$ respectively that are separately depicted in Fig.~\ref{fig:treeDeco4}(ii).

\begin{figure}[t]
\begin{center}
  \includegraphics[width=8cm]{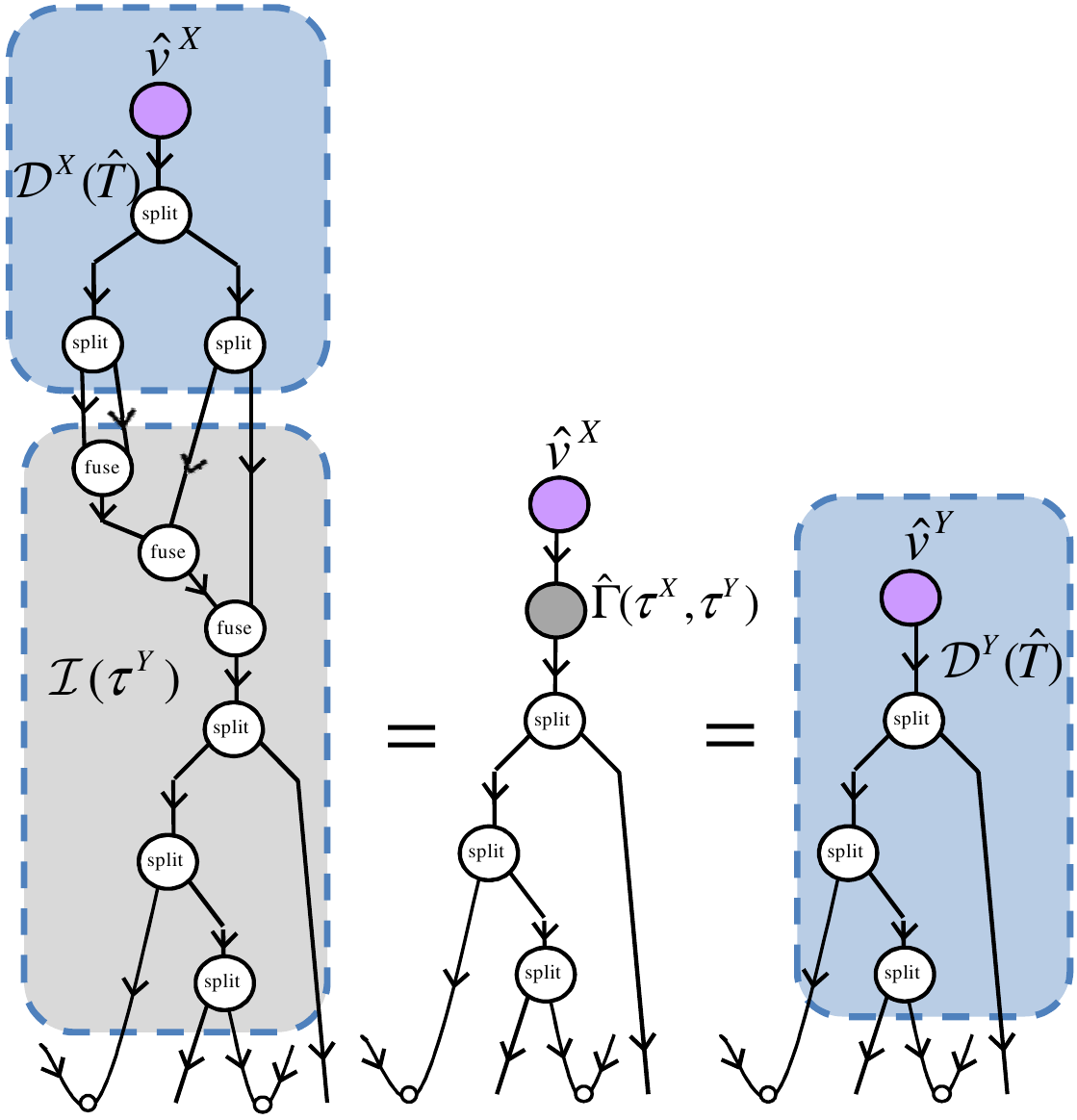}
\end{center}  
\caption{  Mapping a tree decomposition $\mathcal{D}^{X}(\hat{T})$ of an SU(2)-invariant tensor $\hat{T}$ to another tree decomposition $\mathcal{D}^{Y}(\hat{T})$ by applying a resolution of Identity $\mathcal{I}(\tree^Y)$ to $\mathcal{D}^{X}(\hat{T})$. We obtain an intermediate matrix $\hat{\Gamma}(\tree^X, \tree^Y)$ by contracting together the splitting tensors in $\mathcal{D}^{X}(\hat{T})$ and the fusing tensors in $\mathcal{I}(\tree^Y)$ and then multiply it with vector $\hat{v}^X$ to obtain vector $\hat{v}^Y$. \label{fig:MtoM}} 
\end{figure}

Suppose now that we have a tensor $\hat T$ in a tree decomposition $\mathcal{D}^{X}(\hat{T})$ and we wish to transform it into another tree decomposition $\mathcal{D}^{Y}(\hat{T})$. We find it convenient to obtain the vector $\hat{v}^{Y} \in \mathcal{D}^{Y}(\hat{T})$ in steps, as shown in Fig.~\ref{fig:MtoM}. First detach all cup tensors from $\mathcal{D}^{X}(\hat{T})$ and apply the resolution of Identity $\mathcal{I}(\tree^{Y})$ on the open indices of the tree. Then contract the splitting tensors in $\mathcal{D}^{X}(\hat{T})$ and the fusing tensors in $\mathcal{D}^{Y}(\hat{T})$ to obtain a matrix $\hat{\Gamma}(\tree^{X}, \tree^{Y})$. The new vector $\hat{v}^{Y}$ can be obtained by multiplying $\hat{v}^{X}$ with the matrix $\hat{\Gamma}(\tree^{X}, \tree^{Y})$,
\begin{equation}
	\hat{v}^{Y} = \hat{\Gamma}(\tree^X, \tree^Y)\hat{v}^X.
	\label{eq:Gamma}
\end{equation}
Thus, the matrix $\hat{\Gamma}(\tree^X, \tree^Y)$ can be used to map from one tree decomposition of an SU(2)-invariant tensor into another tree decomposition of the same tensor. Recall that the components of $\hat{\Gamma}(\tree^X, \tree^Y)$ can be expressed in terms of recoupling coefficients [Fig.~\ref{fig:spin1}, Eq.~(\ref{eq:genD})].
%%%%%%%%%%%%%%%%%%%%%%%%%%%%%%%%%%%%%%%%%%%%%%%%%%%%%%%%%%%%%%%%%%%%%%%%%%%%%%%%%%%%%%%%%%%%%%%%
%%%%%%%%%%%%%%%%%%%%%%%%%%%%%%%%%%%%%%%%%%%%%%%%%%%%%%%%%%%%%%%%%%%%%%%%%%%%%%%%%%%%%%%%%%%%%%%%

To summarize the above procedure we define a template routine NEWTREE that takes as input a tree decomposition $\mathcal{D}^X(\hat{T})$ and a fusion tree $\tree^Y$ and returns the tree decomposition $\mathcal{D}^Y(\hat{T})$ of the tensor. The routine reads
\begin{align}
&\textbf{NEWTREE} \nonumber \\
&\;\;\;\;\;\;\;\;\;\;\;\textbf{Input:} \nonumber \\
&\;\;\;\;\;\;\;\;\;\;\;\;\;\;\;\;\mathcal{D}^X(\hat{T}) := (\{i_1, i_2, \ldots, i_k\}, \vec{D}, \tree^X, \hat{v}^X) \nonumber \\
&\;\;\;\;\;\;\;\;\;\;\;\;\;\;\;\;\tree^Y \nonumber \\
&\;\;\;\;\;\;\;\;\;\;\;\textbf{Output:} \nonumber \\
&\;\;\;\;\;\;\;\;\;\;\;\;\;\;\;\;\mathcal{D}^Y(\hat{T}) := (\{i_1, i_2, \ldots, i_k\}, \vec{D}, \tree^Y, \hat{v}^Y) \nonumber \\
&\mbox{--------} \nonumber \\
&\mbox{Compute } \hat{\Gamma}(\tree^X, \tree^Y) \nonumber \\
&\hat{v}^Y = \hat{\Gamma}(\tree^X, \tree^Y) \hat{v}^X \nonumber \\
&\mathcal{D}^Y(\hat{T}) := (\{i_1, i_2, \ldots, i_k\}, \vec{D}, \tree^Y, \hat{v}^Y) \nonumber \\
&\textbf{return}(\mathcal{D}^Y(\hat{T})). \label{algo:newtree}
\end{align}
(where $:=$ denotes `stored as'.)

Recall that the matrix $\hat{\Gamma}$ is sparse [Eq.~(\ref{eq:D})]. It can be shown that the matrix-vector multiplication, Eq.~(\ref{eq:Gamma}), can be performed with a cost that is $O(|\hat{v}|)$ by means of sparse multiplication.
%\subsection{Manipulations of SU(2)-invariant tensors based on tree decompositions\label{sec:su2ops}}

Next we describe how manipulations in the set $\mathcal{P}$ of primitive tensor manipulations [Sec.~\ref{sec:tensor:TN}] are performed on tree decompositions. Consider an SU(2)-invariant tensor $\hat{T}$ that has been given as a tree decomposition $\mathcal{D}(\hat{T})$,
\begin{equation}
\mathcal{D}(\hat{T}) \equiv (\{i_1, i_2, \ldots, i_k\}, \vec{D}, \tree, \hat{v}).\nonumber
\end{equation}
Let $\hat{T}'$ denote the SU(2)-invariant tensor that is obtained from tensor $\hat{T}$ as a result of a manipulation in $\mathcal{P}$. Also, let $\mathcal{D}(\hat{T}')$ denote a tree decomposition of $\hat{T}'$,
\begin{equation}
\mathcal{D}(\hat{T}') \equiv (\{i'_1, i'_2, \ldots, i'_m\}, \vec{D}', \tree', \hat{v}').\nonumber
\end{equation}
We will describe how the components of vector $\hat{v}'$ are determined systematically in terms of components of the vector $\hat{v}$.

\subsection{Reversal of indices\label{sec:symTN:reverse}}

Reversal of an index of a tree decomposition is trivial since the cup tensors are stored as part of the tree decomposition. It corresponds to attaching a cup (or its transpose) to the corresponding open index of the tree in case the index is outgoing or detaching the cup from the index in case it is incoming. This simple procedure is summarized in the following template routine which describes reversal of possibly several indices of tensor $\hat{T}$ according to new directions $\vec{D}'$ provided as input. No computation is involved in the procedure. Only information pertaining to the directions of indices is updated,
\begin{align}
&\textbf{REVERSE} \nonumber \\
&\;\;\;\;\;\;\;\;\;\;\;\textbf{Input:} \nonumber \\
&\;\;\;\;\;\;\;\;\;\;\;\;\;\;\;\;\mathcal{D}(\hat{T}) := (\{i_1, i_2, \ldots, i_k\}, \vec{D}, \tree, \hat{v}) \nonumber \\
&\;\;\;\;\;\;\;\;\;\;\;\;\;\;\;\;\vec{D}' \nonumber \\
&\;\;\;\;\;\;\;\;\;\;\;\textbf{Output:} \nonumber \\
&\;\;\;\;\;\;\;\;\;\;\;\;\;\;\;\;\mathcal{D}(\hat{T}') := (\{i_1, i_2, \ldots, i_k\}, \vec{D}', \tree, \hat{v}) \nonumber \\
&\mbox{-----} \nonumber \\
& \mathcal{D}(\hat{T}') := (\{i_1, i_2, \ldots, i_k\}, \vec{D}', \tree, \hat{v}) \nonumber \\
&\textbf{return}(\mathcal{D}(\hat{T}')) \label{algo:reverse}
\end{align}
%Note that the cup and cap tensors will be explicitly involved only when obtaining a block-diagonal matrix from a tree decomposition and vice-versa, see Sec.~\ref{sec:symTN:matrixops}. On the other hand, permutes and reshapes of the indices of an SU(2)-invariant tensor are performed in a way that does not involve the cup and cap tensors explicitly.

\subsection{Permutation of indices\label{sec:symTN:permute}}

%%%%%%%%%%%%%%%%%%%%%%%%%%%%%%%%%%%%%%%%%%%%%%%%%%%%%%%%%%%%%%%%%%%%%%%%%%%%%%%%%%%%%%%%%%%%%%%%
\begin{figure}[t]
\begin{center}
  \includegraphics[width=10cm]{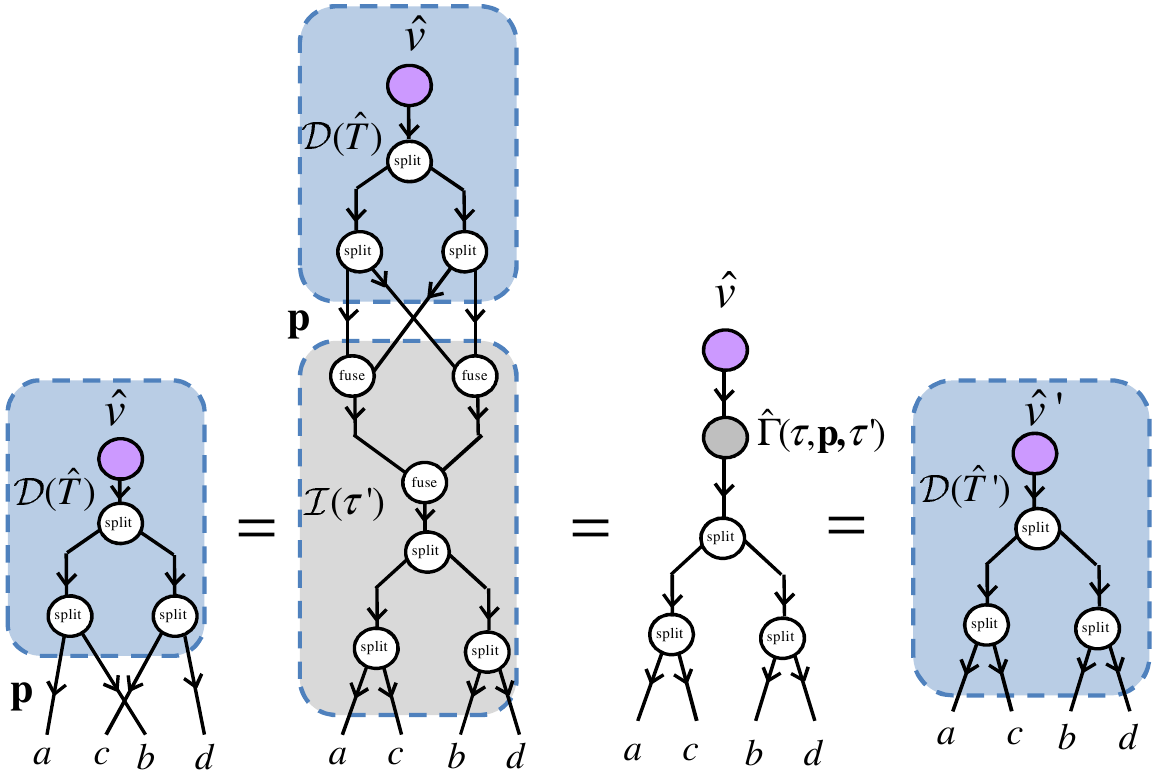}
\end{center}  
\caption{ Permuting indices of a rank-$4$ SU(2)-invariant tensor $\hat{T}$ given in a tree decomposition. The ``crossings'' in the diagram can be absorbed into the tree by applying a resolution of Identity $\mathcal{I}(\tree')$ (for a given fusion tree $\tree'$). An intermediate matrix $\hat{\Gamma}(\tree, \textbf{p}, \tree')$ is obtained by  contracting together the splitting tensors in $\mathcal{D}(\hat{T})$, the permutation \textbf{p} and the fusing tensors in $\mathcal{I}(\tree')$. The matrix is multiplied with the vector $\hat{v}$ to determine the updated vector $\hat{v}'$.}
\label{fig:permute1}
\end{figure}
%%%%%%%%%%%%%%%%%%%%%%%%%%%%%%%%%%%%%%%%%%%%%%%%%%%%%%%%%%%%%%%%%%%%%%%%%%%%%%%%%%%%%%%%%%%%%%%%

The procedure to permute indices of a tree decomposition is illustrated in Fig.~\ref{fig:permute1}(i). We consider a rank-$4$ SU(2)-invariant tensor with indices $\{a,b,c,d\}$ and apply a permutation \textbf{p},
\begin{equation}
\{a, c, b, d\} = \textbf{p}(\{a, b, c, d\}). \nonumber
\end{equation}
The permutation is depicted by intercrossing index $a$ and index $c$ of the tree. This crossing can be `absorbed' into the tree by applying the resolution of Identity $\mathcal{I}(\tree')$ on the tree. In order to determine the vector $\hat{v}'$ we contract the splitting tensors in $\mathcal{D}(\hat{T})$ and the fusing tensors in $\mathcal{I}(\tree')$ to obtain a matrix $\hat{\Gamma}(\tree, \textbf{p}, \tree')$, which is then multiplied with the initial vector $\hat{v}$,
\begin{equation}
\hat{v}' = \hat{\Gamma}(\tree, \textbf{p}, \tree') \hat{v}. \label{eq:permsu2}
\end{equation}
Clearly, the above procedure can be employed to apply any permutation \textbf{p} on the indices of a rank-$k$ SU(2)-invariant tensor. Matrix $\hat{\Gamma}(\tree, \textbf{p}, \tree')$ is a generalization of matrix $\hat{\Gamma}(\tree, \tree')$ [e.g. Fig.~\ref{fig:spin1}] in that it additionally includes a permutation of indices. The latter can then be seen as a special instance of $\hat{\Gamma}(\tree, \textbf{p}, \tree')$ with a trivial permutation of indices. The components of the degeneracy part $\hat{D}$ of matrix $\hat{\Gamma}(\tree, \textbf{p}, \tree')$ are given by Eq.~(\ref{eq:genD}) where coefficients $\hat{S}_{j_1 \ldots j_k}^{j_{e_1} \ldots j_{e_{z}, j} j'_{e_1} \ldots j'_{e_{z}}}$ can be expressed in terms of recoupling coefficients \textit{and} swap factors (see Sec.~\ref{sec:sn}). 
%Once again, obtaining the matrix $\hat{\Gamma}(\tree, \textbf{p}, \tree')$ as an intermediate step is convenient from an implementation point of view since $\hat{\Gamma}(\tree, \textbf{p}, \tree')$ is determined completely by the symmetry. Subsequently, for instance, the same matrix can be used to implement a permutation \textbf{p} of indices on multiple tensors that differ only in the components.

Finally, note that the cup tensors do not play any role in the permutation since reversal of indices commutes with a permutation of them [Fig.~\ref{fig:bendpermute}]. In practice, all cup tensors can be detached from the tree before applying the permutation and then re-attached to the updated tree.

The procedure to permute indices of a tree decomposition is summarized in the following template routine:
\begin{align}
&\textbf{PERMUTE} \nonumber \\
&\;\;\;\;\;\;\;\;\;\;\;\textbf{Input:} \nonumber \\
&\;\;\;\;\;\;\;\;\;\;\;\;\;\;\;\;\hat{T} := (\{i_1, i_2, \ldots, i_k\}, \vec{D}, \tree, \hat{v}) \nonumber \\
&\;\;\;\;\;\;\;\;\;\;\;\;\;\;\;\;\tree' \nonumber \\
&\;\;\;\;\;\;\;\;\;\;\;\;\;\;\;\;\textbf{p} \nonumber \\
&\;\;\;\;\;\;\;\;\;\;\;\textbf{Output:} \nonumber \\
&\;\;\;\;\;\;\;\;\;\;\;\;\;\;\;\;\mathcal{D}(\hat{T}') := (\{i'_1, i'_2, \ldots, i'_k\}, \vec{D}', \tree', \hat{v}') \nonumber \\
&\mbox{-----} \nonumber \\
&\mbox{Compute } \hat{\Gamma}(\tree, \textbf{p}, \tree') \nonumber \\
&\hat{v}' = \hat{\Gamma}(\tree, \textbf{p}, \tree') \hat{v} \nonumber \\
&\{i'_1, i'_2, \ldots, i'_k\} = \textbf{p}(\{i_1, i_2, \ldots, i_k\}) \nonumber \\
&\hat{T}' := (\{i'_1, i'_2, \ldots, i'_k\}, \vec{D}', \tree', \hat{v}') \nonumber \\
&\textbf{return}(\mathcal{D}(\hat{T}')) \label{algo:permute}
\end{align}

\subsection{Reshape of indices\label{sec:symTN:reshape}}

\begin{figure}[t]
\begin{center}
  \includegraphics[width=12cm]{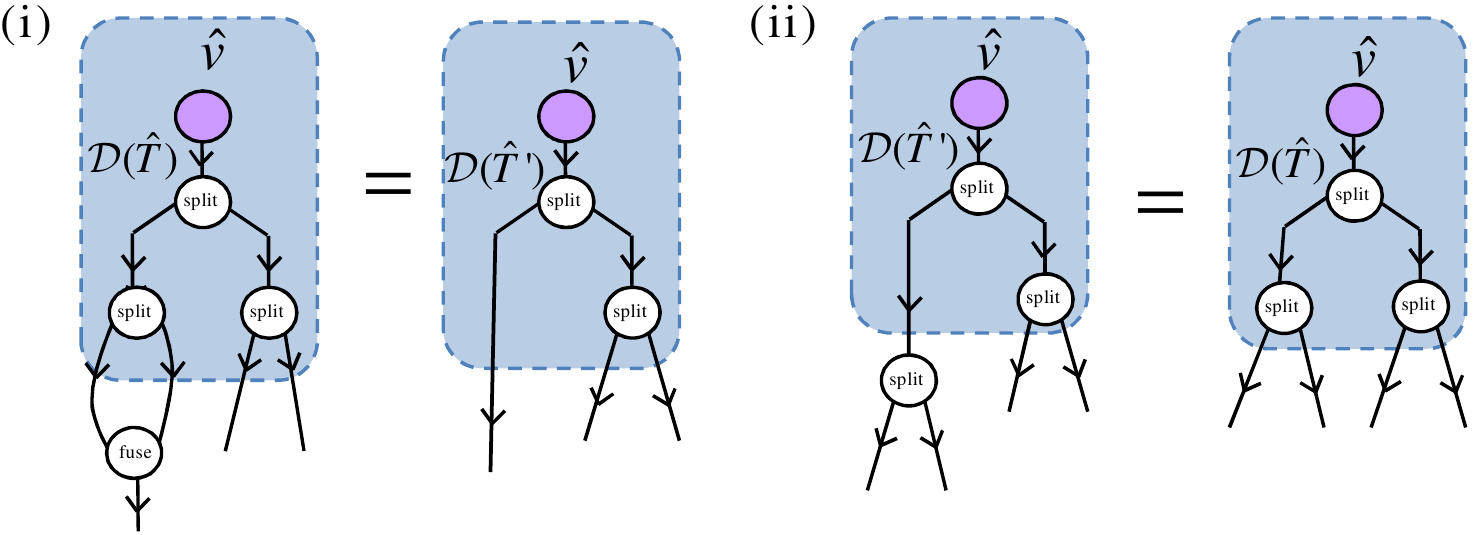}
\end{center}  
\caption{ (i) Fusion of two adjacent indices that belong to the same node of the tree decomposition proceeds by simply deleting the node to obtain the updated tree decomposition. (ii) Splitting an index corresponds to concatenating a new node to the tree decomposition.}
\label{fig:reshape}
\end{figure}

Consider fusing a pair of adjacent indices $i_l$ and $i_{l+1}$ of the tree decomposition $\mathcal{D}(\hat{T})$. Let us suppose that indices $i_l$ and $i_{l+1}$ do not carry cup tensors and also that they belong to the same node in $\mathcal{D}(\hat{T})$. Indices $i_l$ and $i_{l+1}$ can be fused into an index $i$ by applying the tensor $\fuse{i_l}{i_{l+1}}{i}$ and using the equality shown in Fig.~\ref{fig:su2fuse1}(i) to immediately obtain the tree decomposition $\mathcal{D}(\hat{T}')$. This is illustrated in Fig.~\ref{fig:reshape}(i). Note that the final vector $\hat{v}'$ is the same as the initial vector $\hat{v}$. The updated fusion tree $\tree'$ can be obtained from $\tree$ by deleting the node $\{i_l, i_{l+1} \rightarrow i\}$ from $\tree$. We denote this as,
\begin{equation}
\tree' = \tree - \{i_l, i_{l+1} \rightarrow i_m\}. \nonumber
\end{equation}

The original tree decomposition may be recovered from $\mathcal{D}(\hat{T}')$ by splitting index $i$ back into indices $i_l$ and $i_{l+1}$. This operation is again straightforward since it does not involve a computation of vector components, as illustrated in Fig.\ref{fig:reshape}(ii). The original fusion tree $\tree$ is recovered by concatenating a node to $\tree'$,
\begin{equation}
\tree \equiv \tree' \cup \{i\rightarrow i_l, i_{l+1}\}.
\end{equation}

Now let us consider fusing indices $i_l$ and $i_{l+1}$ that \textit{do not} belong to the same node of $\mathcal{D}(\hat{T})$. In this case one can first map $\mathcal{D}(\hat{T})$ into another tree decomposition $\tilde{\mathcal{D}}(\hat{T})$ in which indices $i_l$ and $i_{l+1}$ belong to the same node and then proceed with the fusion on the tree $\mathcal{D}(\hat{T})$ as described above. This can be done by applying the procedure NEWTREE (\ref{algo:newtree}) with inputs $\tilde{\mathcal{D}}(\hat{T})$ and the desired fusion tree.

Consider the template routine FUSE that fuses indices according to a set of \textit{disjoint} fusion trees $\tree_1, \tree_2, \ldots$ where each fusion tree specifies fusion of a subset of adjacent indices $\{i_m, i_{m+1}\}, \{i_n,i_{n+1},i_{n+2}\}, \ldots$:
\begin{align}
&\textbf{FUSE} \nonumber \\
&\;\;\;\;\;\;\;\;\;\;\;\textbf{Input args:} \nonumber \\
&\;\;\;\;\;\;\;\;\;\;\;\;\;\;\;\;\mathcal{D}(\hat{T}) := (\{i_1, i_2, \ldots, i_k\}, \vec{D}, \tree, \hat{v}) \nonumber \\
&\;\;\;\;\;\;\;\;\;\;\;\;\;\;\;\;\{\tree_1, \tree_2, \ldots\} \nonumber\\
&\;\;\;\;\;\;\;\;\;\;\;\;\;\;\;\;\mbox{Final indices: } \{i'_1, i'_2, \ldots, i'_{l}\} \nonumber \\
&\;\;\;\;\;\;\;\;\;\;\;\;\;\;\;\;\mbox{Fusion tree of final tensor: }\tree' \nonumber \\
&\;\;\;\;\;\;\;\;\;\;\;\textbf{Output args:} \nonumber \\
&\;\;\;\;\;\;\;\;\;\;\;\;\;\;\;\;\mathcal{D}(\hat{T}') := (\{i'_1, i'_2, \ldots, i'_{l}\}, \vec{D}, \tree', \hat{v}')\nonumber \\
&\mbox{--------} \nonumber\\
&\tree'' = \tree' \cup \tree_1 \cup \tree_2 \cup \ldots \nonumber \\
&\mathcal{D}''(\hat{T}) = \mbox{NEWTREE}(\mathcal{D}(\hat{T}), \tree'') \nonumber \\
&\hat{v}' = \hat{v}'' \nonumber \\
&\mathcal{D}(\hat{T}') := (\{i'_1, i'_2, \ldots, i'_{l}\}, \tree', \hat{v}') \nonumber \\
&\textbf{return}(\mathcal{D}(\hat{T}'))
\end{align}
Notice that here we essentially apply the total fusion at once by concatenating the input fusion trees $\tree_1, \tree_2, \ldots$ into a single tree and then applying the procedure NEWTREE. Consequently, the fusion is carried out by means of a single matrix-vector multiplication. Furthermore, the computational cost incurred by the procedure is dominated by the cost of this step. As mentioned previously, this cost is $O(|\hat{v}|)$.

Also consider the following routine to split indices $\{i'_1, i'_2, \ldots\}$ of a tree decomposition $\mathcal{D}(\hat{T}')$ (typically the output of FUSE) by reversing the fusion sequence encoded in fusion trees $\tree_1, \tree_2, \ldots$:
\begin{align}
&\textbf{SPLIT} \nonumber \\
&\;\;\;\;\;\;\;\;\;\;\;\textbf{Input args:} \nonumber \\
&\;\;\;\;\;\;\;\;\;\;\;\;\;\;\;\;\mathcal{D}(\hat{T}') := (\{i'_1, i'_2, \ldots, i'_l\}, \vec{D}, \tree', \hat{v}') \nonumber \\
&\;\;\;\;\;\;\;\;\;\;\;\;\;\;\;\;\mbox{Final indices: } \{i_1, i_2, \ldots, i_{k}\} \nonumber \\
&\;\;\;\;\;\;\;\;\;\;\;\;\;\;\;\;\tree_1, \tree_2, \ldots \nonumber \\
&\;\;\;\;\;\;\;\;\;\;\;\textbf{Output args:} \nonumber \\
&\;\;\;\;\;\;\;\;\;\;\;\;\;\;\;\;\mathcal{D}(\hat{T}) := (\{i_1, i_2, \ldots, i_{k}\}, \vec{D}, \tree, \hat{v})\nonumber\\
&\mbox{--------} \nonumber \\
&\tree = \tree' \cup \tree_1 \cup \tree_2 \cup \ldots \nonumber \\
&\hat{v} = \hat{v}' \nonumber \\
&\hat{T} := (\{i_1, i_2, \ldots, i_{k}\}, \tree, \hat{v}) \nonumber \\
&\textbf{return}(\hat{T})
\end{align}
Note that no computation of vector components is involved in this procedure. 

Let us now describe how to reshape indices that may carry cup tensors. First consider the fusion of indices each of which carries a cup tensor. We proceed by detaching the cup tensors from the indices, applying the procedure FUSE and finally attaching a cup tensor to each of the fused indices. Analogously, an index that carries a cup tensor may be split into two indices, by detaching the cup tensor, applying the procedure SPLIT and attaching a cup tensor to each of two indices so obtained.

Finally, consider the fusion of an index that carries a cup tensor with an index that does not carry a cup tensor. The fusion proceeds by detaching the cup tensor and then applying the procedure FUSE on the indices. The fusion is to be reversed in a consistent manner by first applying the procedure SPLIT on the fused index and then attaching a cup tensor to the originally incoming index.
%%%%%%%%%%%%%%%%%%%%%%%%%%%%%%%%%%%%%%%%%%%%%%%%%%%%%%%%%%%%%%%%%%%%%%%%%%%%%%%%%%%%%%%%%%%%%%%%%%%%%%%%%%%%%%%%%%%%%%%%%%%%%%%%%%%%%%%%%%%%%%%%%
%%%%%%%%%%%%%%%%%%%%%%%%%%%%%%%%%%%%%%%%%%%%%%%%%%%%%%%%%%%%%%%%%%%%%%%%%%%%%%%%%%%%%%%%%%%%%%%%

\subsection{Matrix multiplication and factorizations \label{sec:symTN:matrixops}}

Let us consider how matrix operations are performed on tree decompositions. Two SU(2)-invariant matrices, each given as a tree decomposition, may be multiplied together by first obtaining the matrices in a block-diagonal form (from the respective tree decompositions), performing a block-wise multiplication (Sec.~\ref{sec:symTN:multiply}) and recasting the resulting block-diagonal matrix into a tree decomposition. 

An SU(2)-invariant matrix may be factorized e.g. singular value decomposed in a similar way. One proceeds by obtaining the matrix in a block-diagonal form, performing block-wise factorization (Sec.~\ref{sec:symTN:factorize}), and recasting each of the factor block-diagonal matrices into a tree decomposition. 

In the remainder of the section we explain how a block-diagonal form is obtained from a tree decomposition and vice-versa.

%%%%%%%%%%%%%%%%%%%%%%%%%%%%%%%%%%%%%%%%%%%%%%%%%%%%%%%%%%%%%%%%%%%%%%%%%%%%%%%%%%%%%%%%%%%%%%%%
\begin{figure}[t]
\begin{center}
  \includegraphics[width=10cm]{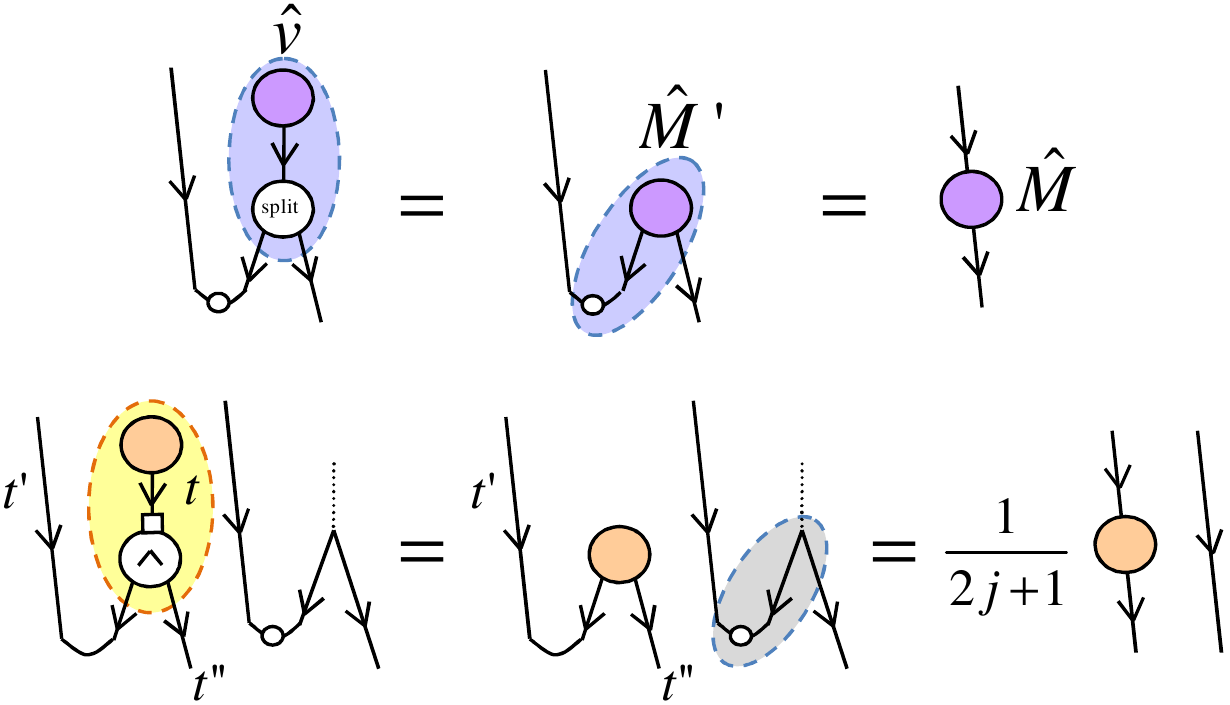}
\end{center}  
\caption{ The block-diagonal form of an SU(2)-invariant matrix is obtained from its tree decomposition by performing two multiplications. At each step the tensors within the shaded region are multiplied together. The same steps are also shown in the canonical form.}
\label{fig:recovermat}
\end{figure}
%%%%%%%%%%%%%%%%%%%%%%%%%%%%%%%%%%%%%%%%%%%%%%%%%%%%%%%%%%%%%%%%%%%%%%%%%%%%%%%%%%%%%%%%%%%%%%%%
\subsubsection{Block-diagonal matrix from the tree decomposition} 

Consider a tree decomposition $\mathcal{D}(\hat{T})$ of an SU(2)-invariant matrix $\hat{T}$. The decomposition $\mathcal{D}(\hat{T})$ comprises of a vector $\hat{v}$, a splitting tensor $\spliter$ and a cup tensor $\mycup$. We wish to obtain, from $\mathcal{D}(\hat{T})$, the corresponding block diagonal matrix,
\begin{equation}
\hat{T} = \bigoplus_j (\hat{T}_j \otimes \hat{I}_{2j+1}).\label{eq:blockform}
\end{equation}
This can be achieved by multiplying together the vector, the splitting tensor and the cup tensor. We perform this multiplication in two simple steps as shown in Fig.~\ref{fig:recovermat}. We first multiply $\hat{v}$ with $\spliter$ to obtain an intermediate SU(2)-invariant tensor $\hat{T}'$ that takes the canonical form,
\begin{equation}
\hat{T}'_j = \bigoplus_j (\hat{T}'_j \otimes \hat{\omega}_j),
\end{equation}
where the components $(\hat{T}'_j)_{t't''} $ of $\hat{T}'_j$ are given by
\begin{equation}
(\hat{T}'_j)_{t't''} = \sum_{t} \hat{v}_t~\tsplitt{0t}{jt'}{jt''}.\label{eq:absorbcup1}
\end{equation}
We then multiply (algebraically) tensor $\hat{T}'$ with the cup tensor to obtain the block-diagonal matrix $\hat{T}$ to obtain
\begin{equation}
\hat{T}_j = \frac{1}{2j+1} \hat{T}'_j.\label{eq:absorbcup2}
\end{equation}

\subsubsection{Tree decomposition from the block-diagonal matrix} 

The tree decomposition $\mathcal{D}(\hat{T})$ can be obtained from the block-diagonal form (\ref{eq:blockform}) in a straightforward manner by reverting the previous procedure. We first multiply $\hat{T}$ with a cap tensor to obtain tensor $\hat{T}'$. Once again, the outcome of this multiplication follows algebraically. We obtain
\begin{equation}
(\hat{T}'_j)_{t't''} = (2j+1)(\hat{T}_j)_{t't''}.\label{eq:conv2tree1}
\end{equation}
We then fuse the indices of $\hat{T}'$ to obtain a vector $\hat{v}$,
\begin{equation}
\hat{v}_t = \sum_j \sum_{t't''} (\hat{T}'_j)_{t't''} \tfuse{jt'}{jt''}{0t}.\label{eq:conv2tree2}
\end{equation}
The tree decomposition $\mathcal{D}(\hat{T})$ comprises of vector $\hat{v}$, the splitting tensor $\tspliter$ that reverts the fusion in Eq.~(\ref{eq:conv2tree2}) and a cup tensor. 

%%%%%%%%%%%%%%%%%%%%%%%%%%%%%%%%%%%%%%%%%%%%%%%%%%%%%%%%%%%%%%%%%%%%%%%%%%%%%%%%%%%%%%%%%%%%%%%%
\begin{figure}[t]
\begin{center}
  \includegraphics[width=12cm]{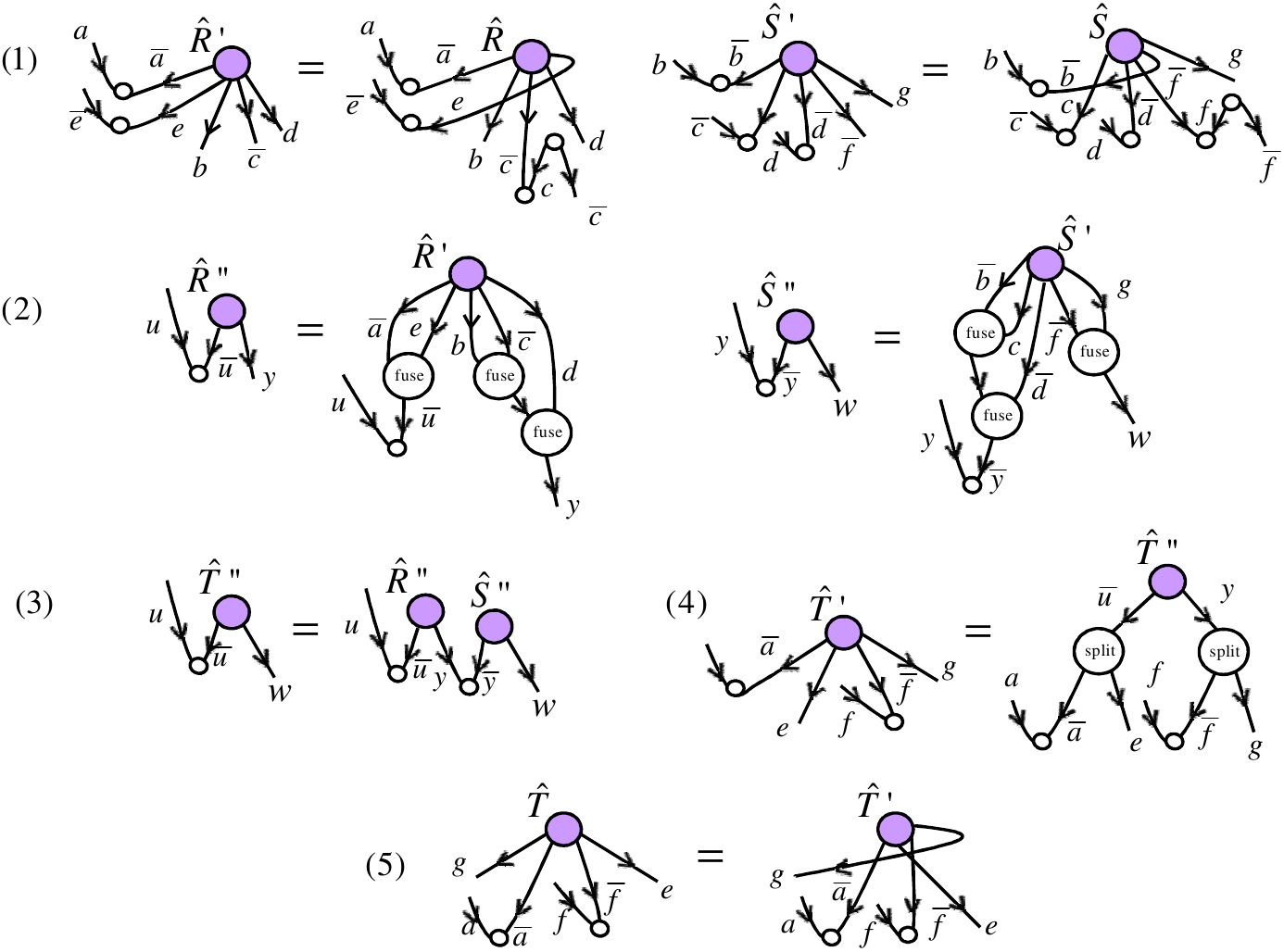}
\end{center}  
\caption{  The five steps of tensor multiplication [Fig.~\ref{fig:multiply2}] as adapted to the presence of the symmetry.  For simplicity, the tree decompositions are not explicitly shown; the circle representing a tensor masks the tree decomposition of the tensor.}
\label{fig:tensormult}
\end{figure}
%%%%%%%%%%%%%%%%%%%%%%%%%%%%%%%%%%%%%%%%%%%%%%%%%%%%%%%%%%%%%%%%%%%%%%%%%%%%%%%%%%%%%%%%%%%%%%%%
\subsection{Multiplication of two tensors\label{sec:symTN:multTens}}

We can now consider the multiplication of two SU(2)-invariant tensors by breaking it into a sequence of five elementary steps consisting of reversals, permutes, reshapes and matrix multiplication, as was exemplified in Sec.~\ref{sec:tensor:multiply} and Fig.~\ref{fig:multiply2}. Here the elementary steps are performed on the tree decompositions of tensors. Figure~\ref{fig:tensormult} illustrates the five steps of the tensor multiplication, Eq.~(\ref{eq:tensormult}), as adapted to tree decompositions. For simplicity, we have not shown the tree decompositions in the figure and the circle that depicts a tensor can be imagined to mask the tree decomposition of the tensor.

%%%%%%%%%%%%%%%%%%%%%%%%%%%%%%%%%%%%%%%%%%%%%%%%%%%%%%%%%%%%%%%%%%%%%%%%%%%%%%%%%%%%%%%%%%%%%%%%
\begin{figure}[t]
\begin{center}
  \includegraphics[width=10cm]{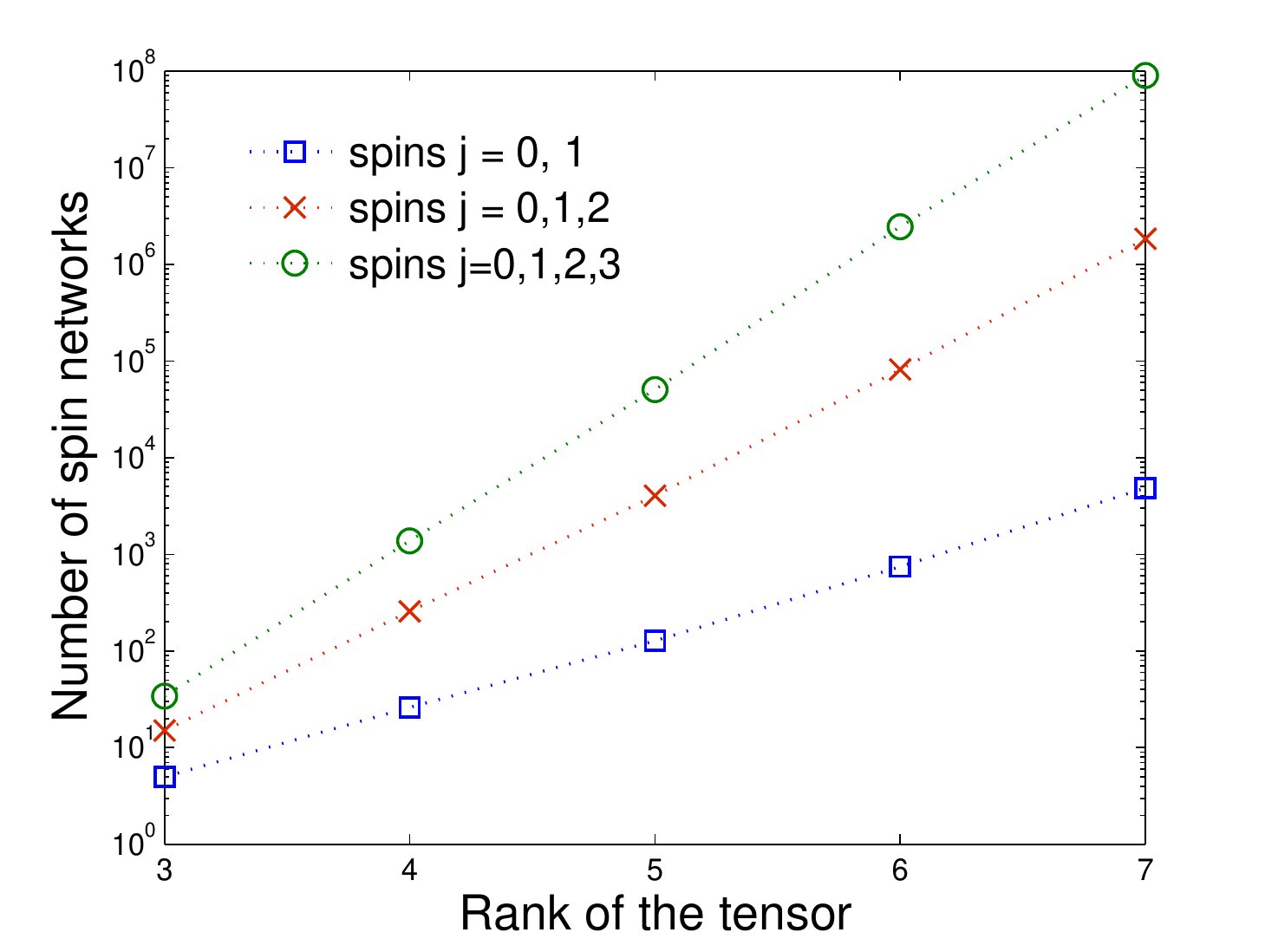}
\end{center}  
\caption{ The number of spin networks to be evaluated when permuting or reshaping an SU(2)-invariant tensor, as a function of the number of indices of the tensor and the number of different spins $j$ assigned to each index of the tensor. The plot shows the increase in the number of spin networks that are evaluated when reshaping tensors with increasing number of indices. It also illustrates the corresponding increase that results from increasing the number of different values of spin $j$ that are assigned to the indices of the tensor. (Note that the number of spin networks does not depend on the degeneracy dimension of a spin $j$.) Consequently, the cost of reshaping tensors with a large number of indices may potentially become significant.}
\label{fig:sn1}
\end{figure}
%%%%%%%%%%%%%%%%%%%%%%%%%%%%%%%%%%%%%%%%%%%%%%%%%%%%%%%%%%%%%%%%%%%%%%%%%
\subsection{Discussion on computational performance\label{sec:precompute}}

The core of obtaining computational gain from exploiting the symmetry lies in block-wise matrix operations [Fig.~\ref{fig:multsvdcompare}] while permutation and reshape of indices are applied mainly to obtain block-diagonal matrices from tensors. As has been illustrated in Fig.~\ref{fig:permutereshapecompare} the cost of reshaping, for instance, SU(2)-invariant tensors can be significantly larger than that incurred in reshaping regular tensors, and in some case can lead to a severe degradation of the overall gain obtained by exploiting the symmetry. Let us analyze the cost associated to reshape and permutation of indices of an SU(2)-invariant tensor.

We have described that by working on tree decompositions reshape and permutation of indices equates to multiplying a matrix $\hat{\Gamma}$ with a vector.
The computation of $\hat{\Gamma}$ may be costly since it generally involve evaluating many spin networks, see Fig.~\ref{fig:sn1}. Consequently, the cost of reshaping or permuting tensors with a large number of indices may become significant. This is more so the case for \textit{iterative} algorithms where a fixed set of manipulations repeat many times. For example, one may optimize tensors iteratively in a variational algorithm such that the components of the tensor are updated in the current iteration and used as an input to the subsequent iteration. Note that each iteration involves evaluating a large number of spin networks, albeit the \textit{same} spin networks are evaluated in each iteration. This fact can be exploited to \textit{pre-compute} the transformations $\hat{\Gamma}$ once, say in the first iteration of the algorithm, and storing them in memory for \textit{reuse} in subsequent iterations. By precomputation of these matrices the cost of evaluating many spin networks is suppressed from the runtime costs. 

In our MATLAB implementation the use of such a precomputation scheme resulted in a significant speed-up of simulations at the cost of storing additional amounts of precomputed data. In the passing we also remark that since all computations have been reduced to matrix operations, computational performance can also be potentially enhanced by parallelizing and vectorizing the underlying matrix operations.

\section{Tensor networks with SU(2) symmetry: A practical demonstration\label{sec:symTN}}

Consider a lattice $\mathcal{L}$ made of $L$ sites where each site $l$ is described by a vector space $\mathbb{V}^{(l)}$ that transforms as a finite dimensional representation of SU(2). The vector space $\mathbb{V}^{(\mathcal{L})}$ of the lattice is given as
\begin{equation}
\mathbb{V}^{(\mathcal{L})} \equiv \bigotimes_l \mathbb{V}^{(l)}.
\end{equation}
Consider a state $\ket{\Psi} \in \mathbb{V}^{(\mathcal{L})}$ that is invariant, Eq.~\ref{eq:latticeinv}, under the action of SU(2) on the vector space $\mathbb{V}^{(\mathcal{L})}$. \textit{We describe $\ket{\Psi}$ by means of a tensor network made of SU(2)-invariant tensors.}

\begin{figure}[t]
\begin{center}
  \includegraphics[width=10cm]{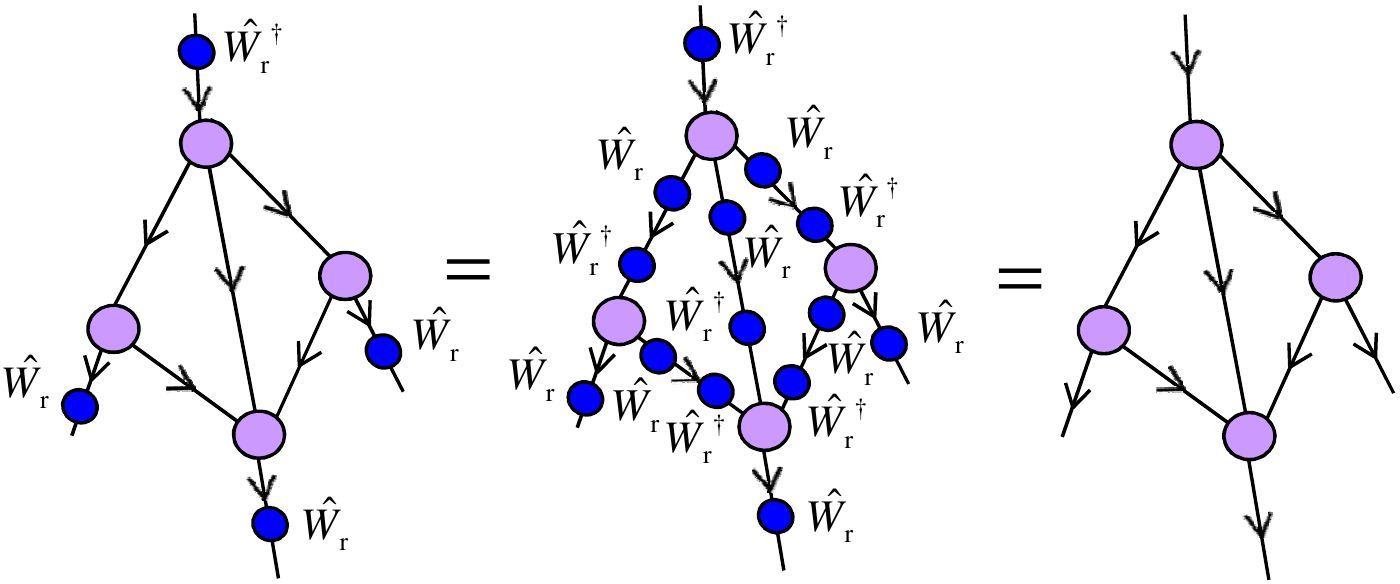}
\end{center}
\caption{ A tensor network $\mathcal{N}$ made of SU(2)-invariant tensors represents an SU(2)-invariant tensor $\hat{T}$. This is seen by means of two equalities. The first equality is obtained by inserting resolutions of the Identity $\hat{I} = \hat{W}_{\textbf{r}} \hat{W}^{\dagger}_{\textbf{r}}$ on each index connecting two tensors in $\mathcal{N}$. The second equality follows from the fact that each tensor in $\mathcal{N}$ is SU(2)-invariant. }
\label{fig:symTN}
\end{figure}

It readily follows that the tensor obtained by contracting such a tensor network is SU(2)-invariant, as illustrated in Fig. \ref{fig:symTN}. 

%%%%%%%%%%%%%%%%%%%%%%%%%%%%%%%%%%%%%%%%%%%%%%%%%%%%%%%%%%%%%%%%%%%%%%%%%%%%%%%%%%%%%%%%%%%%%%%%
%%%%%%%%%%%%%%%%%%%%%%%%%%%%%%%%%%%%%%%%%%%%%%%%%%%%%%%%%%%%%%%%%%%%%%%%%%%%%%%%%%%%%%%%%%%%%%%%

On the one hand, by storing each constituent tensor of the tensor network in a canonical form we can ensure a compact tensor network description of $\ket{\Psi}$. On the other, computational speedup can be obtained by exploiting the sparse canonical form of the tensors when performing manipulations of individual tensors in a tensor network algorithm.

In the remainder of the section we illustrate the implementation of SU(2) symmetry in tensor network algorithms with practical examples. We do so in the context of the Multi-scale Entanglement Renormalization Ansatz, or MERA, and present numerical results from our reference implementation of SU(2) symmetry in MATLAB.

\subsection{Multi-scale entanglement renormalization ansatz\label{sec:MERA:ansatz}}

\begin{figure}[t]
\begin{center}
  \includegraphics[width=10cm]{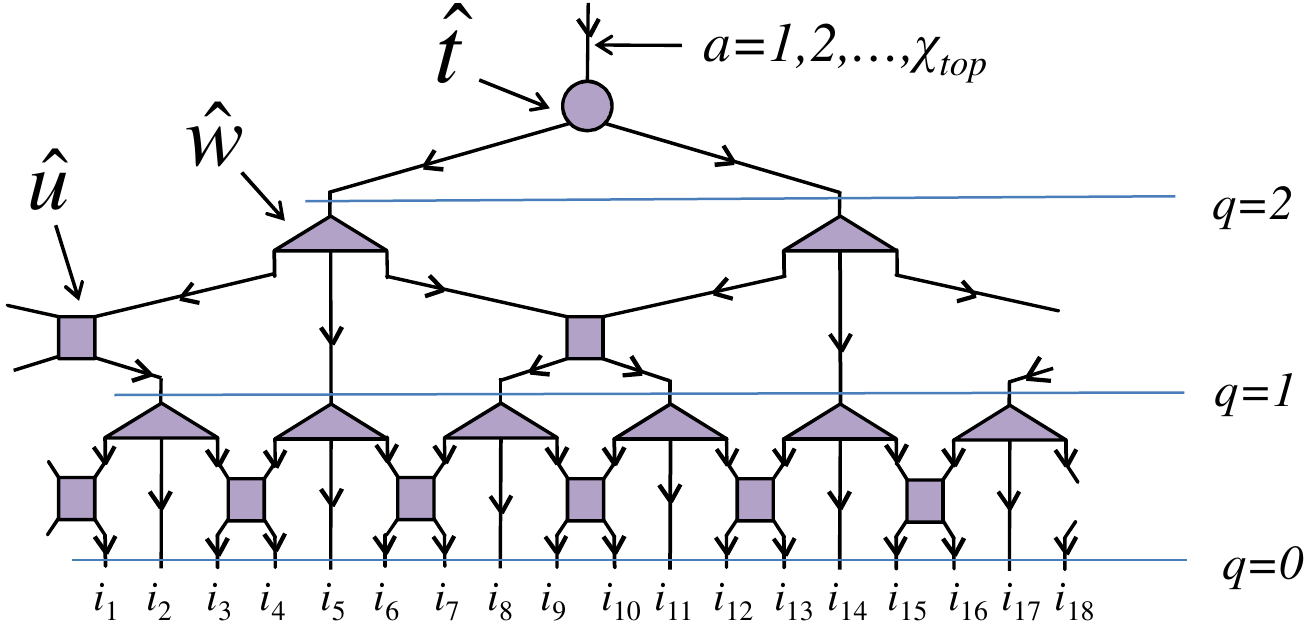}
\end{center}  
\caption{ MERA for a system of $L=2\times 3^{2}= 18$ sites, made of two layers of disentanglers $\hat{u}$ and isometries $\hat{w}$, and a top tensor $\hat{t}$.}
\label{fig:mpsmera}
\end{figure}

Figure \ref{fig:mpsmera} shows a MERA that represent states $\ket{\Psi} \in \mathbb{V}^{(\mathcal{L})}$ of a lattice $\mathcal{L}$ made of $L=18$ sites. Recall that the MERA is made of layers of isometric tensors, known as disentanglers $\hat{u}$ and isometries $\hat{w}$, that implement a coarse-graining transformation. In this particular scheme, isometries map three sites into one and the coarse-graining transformation reduces the $L=18$ sites of $\mathcal{L}$ into two sites using two layers of tensors. A collection of states on these two sites is then encoded in a top tensor $\hat{t}$, whose upper index $a=1,2,\cdots, \chi_{\tiny \mbox{top}}$ is used to label $\chi_{\tiny \mbox{top}}$ states $\ket{\Psi_a} \in \mathbb{V}^{(\mathcal{L})}$. This particular arrangement of tensors corresponds to the 3:1 MERA described in  \citep{Evenbly09}. We will consider a MERA analogous to that of Fig.~\ref{fig:mpsmera} but with $Q$ layers of disentanglers and isometries, which we will use to describe states on a lattice $\mathcal{L}$ made of $2\times 3^{Q}$ sites. 

%%%%%%%%%%%%%%%%%%%%%%%%%%%%%%%%%%%%%%%%%%%%%%%%%%%%%%%%%%%%%%%%%%%%%%%%%%%%%%%%%%%%%%%%%%%%%%%%
%%%%%%%%%%%%%%%%%%%%%%%%%%%%%%%%%%%%%%%%%%%%%%%%%%%%%%%%%%%%%%%%%%%%%%%%%%%%%%%%%%%%%%%%%%%%%%%%

We will use the MERA as a variational ansatz for ground states and excited states of quantum spin models described by a local Hamiltonian $\hat{H}$. In order to find an approximation to the ground state of $\hat{H}$, we set $\chi_{\tiny\mbox{top}}=1$ and optimize the tensors in the MERA so as to minimize the expectation value 
\begin{equation}
	\bra{\Psi} \hat{H} \ket{\Psi}
\end{equation}
where $\ket{\Psi}\in \mathbb{V}^{(\mathcal{L})}$ is the pure state represented by the MERA. In order to find an approximation to the $\chi_{\tiny\mbox{top}}>1$ eigenstates of $\hat{H}$ with lowest energies, we optimize the tensors in the MERA so as to minimize the expectation value
\begin{equation}
	\sum_{a=1}^{\chi_{\tiny\mbox{top}}}\bra{\Psi_a} \hat{H} \ket{\Psi_a},~~~~\braket{\Psi_a}{\Psi_{a'}} = \delta_{aa'}.
\end{equation}
The optimization is carried out using the MERA algorithm described in \citep{Evenbly09}, which requires contracting tensor networks (by sequentially multiplying pairs of tensors) and performing singular value decompositions.

\subsection{MERA with SU(2) symmetry\label{sec:MERA:ansatz}}

An SU(2)-invariant version of the MERA, or SU(2) MERA for short, is obtained by simply considering SU(2)-invariant versions of all of the isometric tensors, namely the disentanglers $\hat{u}$, isometries $\hat{w}$, and the top tensor $\hat{t}$. This requires assigning a spin operator to each index of the MERA. We can characterize the spin operator by two vectors, $\vec{j}$ and $\vec{d}$: a list of the different values the spin takes and the degeneracy associated with each such spin, respectively. For instance, an index characterized by $\vec{j} = \{0, 1\}$ and $\vec{d} = \{2, 1\}$ is associated to a vector space $\mathbb{V}$ that decomposes as $\mathbb{V} \cong d_0\mathbb{V}_0 \oplus d_1 \mathbb{V}_1$ with $d_0 = 2$ and $d_1 = 1$.

Let us explain how a spin operator is assigned to each link of the MERA. Each open index of the first layer of disentanglers corresponds to one site of $\mathcal{L}$. The spin operator on any such index is therefore given by the quantum spin model under consideration. For example, a lattice with a spin-$\frac{1}{2}$ associated to each site corresponds to assigning spin-$\frac{1}{2}$ operators [Eq.~(\ref{eq:eg2c1})] to each of the open indices. 

For the open index of the tensor $\hat{t}$ at the very top the MERA, the assignment of spins will depend on spin sector $J$ that one is interested in. For instance, in order to find an approximation to the ground state and first seven excited states of the quantum spin model within the spin sector $J$, we choose $\vec{j} = \{J\}$ and $\vec{d} = \{8\}$. 

For each of the remaining indices of the MERA, the assignment of the pair $(\vec{j}, \vec{d})$ needs careful consideration and a final choice may only be possible after numerically testing several options and selecting the one which produces the lowest expectation value of the energy.

For demonstrative purposes, we will use the SU(2) MERA as a variational ansatz to obtain the ground state and excited states of the spin-$\frac{1}{2}$ antiferromagnetic quantum Heisenberg chain that is given by,
\begin{align}
\hat{H} = \sum_{s=1}^L \hat{h}^{(s, s+1)}, \label{eq:heisenberg}
\end{align}
where 
\begin{align}
\hat{h}^{(s, s+1)} &= 4\left(\hat{J}_x^{(s)}\hat{J}_x^{(s+1)} + \hat{J}_y^{(s)}\hat{J}_y^{(s+1)} + \hat{J}_z^{(s)}\hat{J}_z^{(s+1)}\right),
\end{align}
$\hat{J}_{x}, \hat{J}_y$ and $\hat{J}_z$ are the spin-$\frac{1}{2}$ operators [Eq.~(\ref{eq:eg2c1})]. The model has a global SU(2) symmetry, since the Hamiltonian commutes with the spin operators acting on the lattice $\mathcal{L}$. This follows from the fact that each local term $\hat{h}^{(s, s+1)}$ in the Hamiltonian commutes with the two site spin operators,
\begin{equation}
[\hat{h}^{(s, s+1)}, \hat{J}^{(s)}_{\alpha} + \hat{J}^{(s+1)}_{\alpha}] = 0,~~~\alpha={x,y,x}.
\end{equation}
Each spin-$\frac{1}{2}$ degree of freedom of the Heisenberg chain is described by a vector space $\mathbb{V} \cong \mathbb{V}_{\half}$ that is spanned by two orthonormal states [Eq.~(\ref{eq:basiseg1})],
\begin{equation}
\ket{j=\half, m=-\half} \mbox{ and } \ket{j=\half, m=\half}. \nonumber
\end{equation}
For computational convenience, we will consider a lattice $\mathcal{L}$ where each site contains two spins. Therefore each site of $\mathcal{L}$ is described by a space $\mathbb{V} \cong \mathbb{V}_0 \oplus \mathbb{V}_1$, where $d_0=1$ and $d_1=1$, also discussed in Example 6. This corresponds to the assignment $\vec{j} = \{0, 1\}$ and $\vec{d} = \{1, 1\}$ at the open legs at the bottom of the MERA. Thus, a lattice $\mathcal{L}$ made of $L$ sites corresponds to a chain of $2L$ spins. 

\begin{figure}[t]
\begin{center}
  \includegraphics[width=12cm]{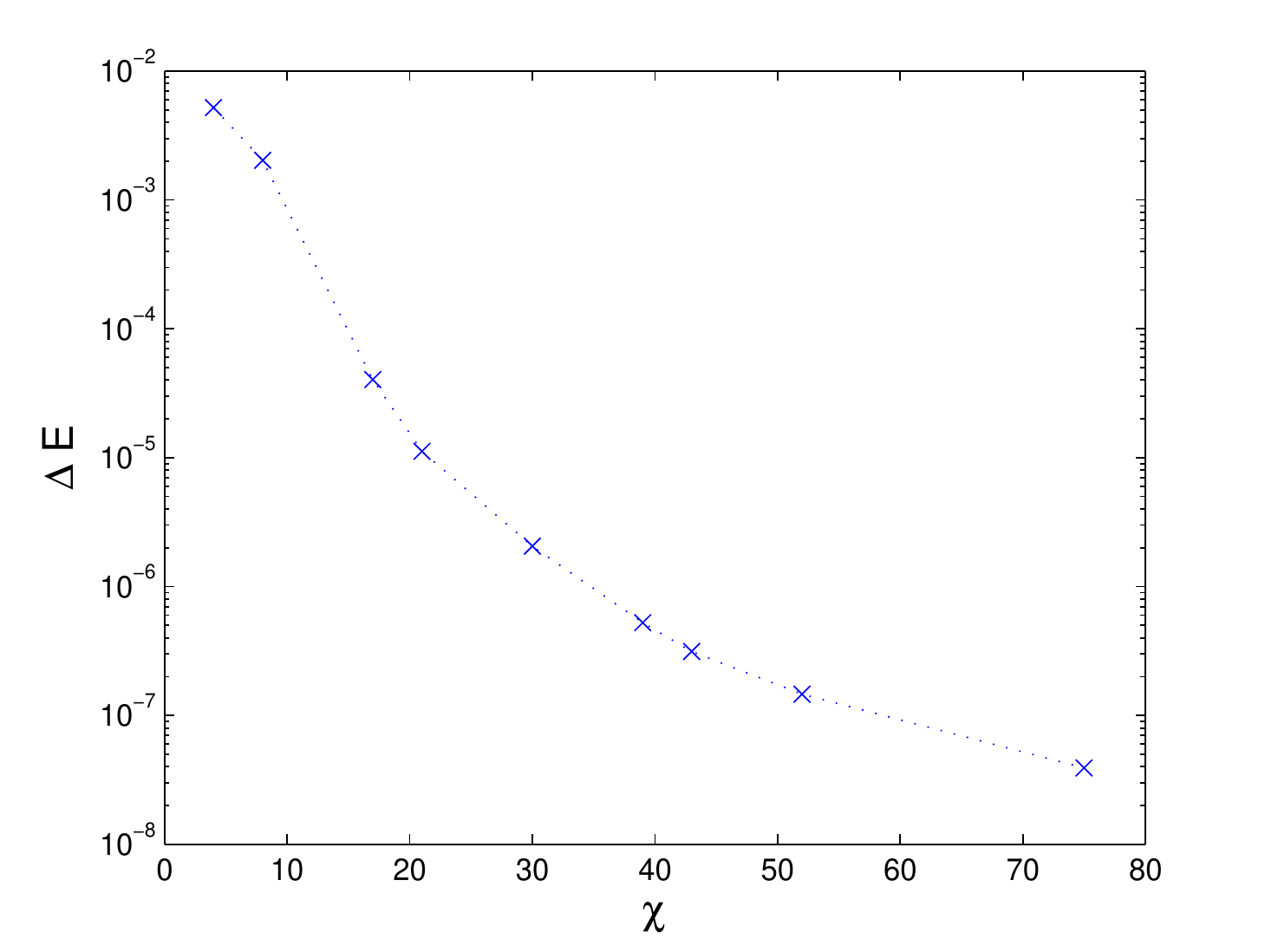}
\end{center}  
\caption{Error in ground state energy $\Delta E$ as a function of $\chi$ for the Heisenberg model with $2L=108$ spins and periodic boundary conditions, in the singlet sector, $J=0$. The error is calculated with respect to the exact solutions, and is seen to decay polynomially with $\chi$ for the particular choice of spins listed in Table \ref{table:degdist}. \label{fig:gserror}}
\end{figure}

Table \ref{table:degdist} lists some of the spin and degeneracy dimensions assignment (for the internal links of the MERA) that we have used in the numerical computations for $L=54$ (or 108 spins). For a given value of $\vec{j}$ and $\vec{d}$ the corresponding dimension $\chi$ can be obtained as,
\begin{equation}
\chi = \sum_{j \in \vec{j}} (2j+1)\times d_j.
\end{equation}
Figure~\ref{fig:gserror} shows the error in the ground state energy of the Heisenberg chain as a function of the bond dimension $\chi$, for the assignments of $\vec{j}$ and $\vec{d}$ that are listed in Table \ref{table:degdist}. For the choice of spin assignments listed in the table the error is seen to decay polynomially with $\chi$, indicating increasingly accurate approximations to the ground state. 
%%%%%%%%%%%%%%%%%%%%%%%%%%%%%%%%%%%%%%%%%%%%%%%%%%%%%%%%%%%%%%%%%%%%%%%%%%%%%%%%%%%%%%%%%%%%%%%%
%%%%%%%%%%%%%%%%%%%%%%%%%%%%%%%%%%%%%%%%%%%%%%%%%%%%%%%%%%%%%%%%%%%%%%%%%%%%%%%%%%%%%%%%%%%%%%%%

\begin{table}
%%%%%%%%%%%%%%%%%%%%%%%%%%%%%%%%%%%%%%%%%%%%%%%%%%%%%%%%%%%%%%%%%%%%%%%%%%%%%%%%%%%%%%%%%%%%%%%%
\centering % used for centering table
\begin{tabular}{c| c| c} % centered columns (4 columns)
\hline\hline %inserts double horizontal lines
Total bond dimension, $\chi$ &  Spins $\vec{j}$ & Degeneracies $\vec{d}$ \\ [0.5ex] % inserts table
\hline
4 & $\{0,1\}$ & $\{1,1\}$ \\ 
8 & $\left\{0,1\right\}$ & $\left\{2,2\right\}$  \\
17 & $\left\{0,1,2\right\}$ & $\left\{3,3,1\right\}$  \\
21 & $\left\{0,1,2\right\}$ & $\left\{4,4,1\right\}$  \\
30 & $\left\{0,1,2\right\}$ & $\left\{5,5,2\right\}$  \\
39 & $\left\{0, 1, 2\right\}$ & $\left\{6,6,3\right\}$  \\ % inserting body of the table
43 & $\left\{0, 1, 2\right\}$ & $\left\{7,7,3\right\}$  \\
52 & $\left\{0, 1, 2\right\}$ & $\left\{8,8,4\right\}$  \\
75 & $\left\{0, 1, 2, 3\right\}$ & $\left\{9,9,5,2\right\}$  \\
[1ex]% [1ex] adds vertical space
%heading
\hline % inserts single horizontal line
\hline %inserts single line
\end{tabular}
\caption{
%[table:degdist]
Example of spin assignment in an SU(2) MERA for the anti-ferromagnetic spin chain with $L = 54$ sites (or $108$ spins).\label{table:degdist} % is used to refer this table in the text
}
\end{table}

\subsection{Advantages of exploiting the symmetry}

We now discuss some of the advantages of using the SU(2) MERA.

\subsubsection{Selection of spin sector}

An important advantage of the SU(2) MERA is that it exactly preserves the SU(2) symmetry. In other words, the states resulting from a numerical optimization are exact eigenvectors of the total spin operator $\textbf{J}^2 : \mathbb{V}^{(\mathcal{L})} \rightarrow \mathbb{V}^{(\mathcal{L})}$. In addition, the total spin $J$ can be pre-selected at the onset of optimization by specifying it in the open index of the top tensor $\hat{t}$. 

\begin{figure}[t]
\begin{center}
  \includegraphics[width=12cm]{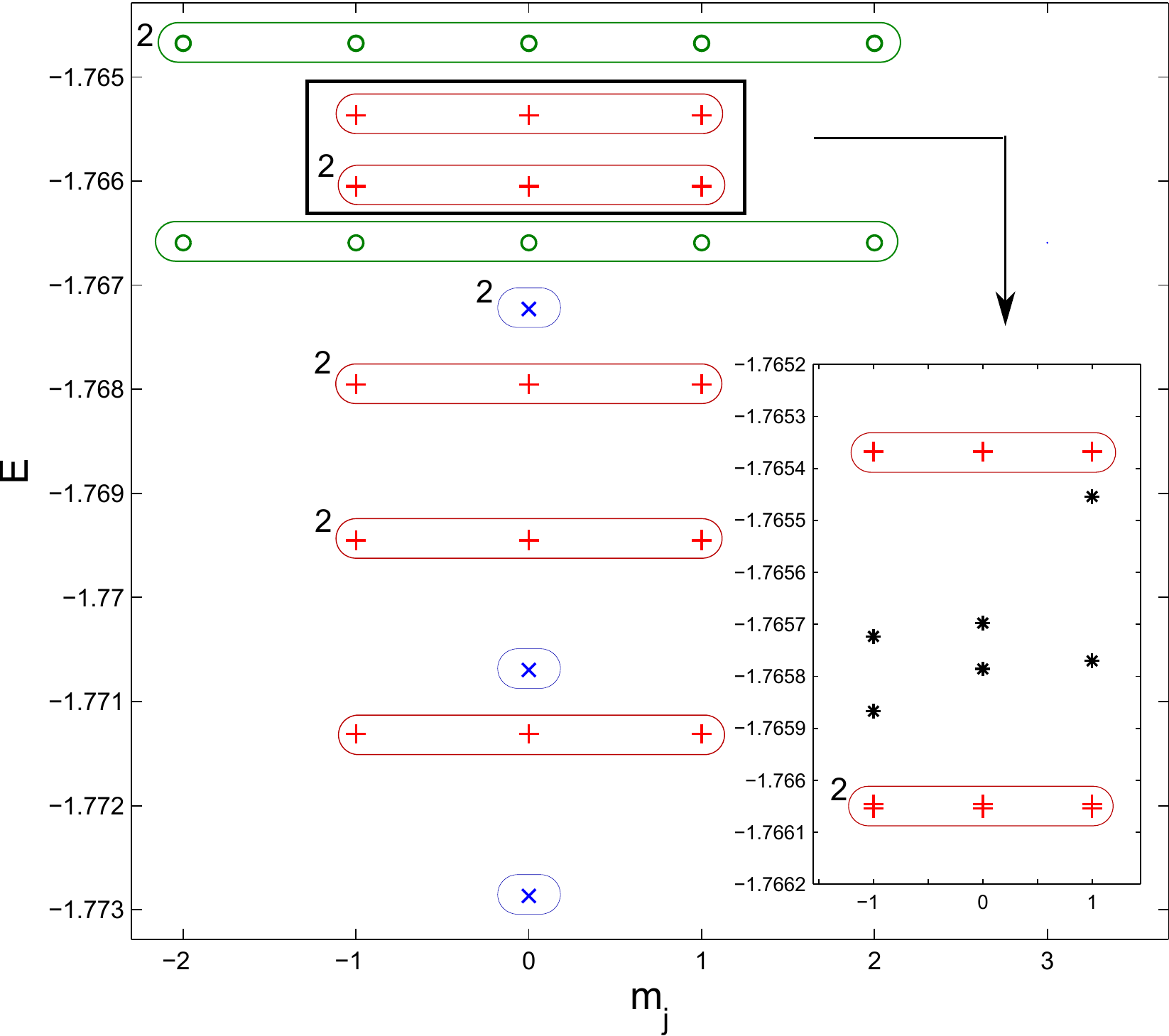}
\end{center}  
\caption{  Low energy spectrum of $\hat{H}$ with $L=54$ sites (=108 spins). Depicted states have spin $J$ of zero ($\times$, blue loops), one (+, red loops), or two ($\circ$, green loop). The superscript $^2$ close to the boundary of a loop indicates that the loop encloses two-fold degenerate states e.g., the second, third and fourth spin-1 triplets are twofold degenerate. The inset shows a zoom in of the region enclosed within the box. It compares the energies of the two-fold degenerate spin-one states within the box with those obtained using the regular MERA (black asterix points). Since the symmetry is not protected, the states obtained with the regular MERA corresponding to different $m_J$ do not have the same energies.
\label{fig:spec}}
\end{figure}

Figure~\ref{fig:spec} shows the low energy spectrum of the Heisenberg model $\hat{H}$ for a periodic system of $L=54$ sites (or $108$ spins), including the ground state and several excited states in the spin sectors $J=0, 1, 2$. The states have been organized according to spin projection $m_J$. We see that states with different spin projections $m_J$ (for a given $J$) are obtained to be exactly degenerate, as implied by the symmetry. 

Similar computations can be performed with the regular MERA. However, the regular MERA cannot guarantee that the states obtained in this way are exact eigenvectors of $\textbf{J}^2$. Instead the resulting states are likely to have total spin fluctuations. This is shown in inset of Fig.~\ref{fig:spec}, which corresponds to the zoom in of the region in the plot that is enclosed within the box. The inset shows (black asterix points) the corresponding energies obtained for the enclosed two-fold degenerate $J=1$ states using the regular MERA. We see that the states corresponding to different values of $m_J$ are obtained with different energies.

%%%%%%%%%%%%%%%%%%%%%%%%%%%%%%%%%%%%%%%%%%%%%%%%%%%%%%%%%%%%%%%%%%%%%%%%%%%%%%%%%%%%%%%%%%%%%%%%
%%%%%%%%%%%%%%%%%%%%%%%%%%%%%%%%%%%%%%%%%%%%%%%%%%%%%%%%%%%%%%%%%%%%%%%%%%%%%%%%%%%%%%%%%%%%%%%%

Also note that by using the SU(2) MERA, the three sectors $J=0,1$ and $2$ can be addressed with independent computations. This implies, for instance, that finding the gap between the first singlet ($J=0$) and the first $J=2$ state, can be addressed with two independent computations by respectively setting $(J=0, \chi_{\tiny\mbox{top}}=1)$ and $(J=2, \chi_{\tiny\mbox{top}}=1)$ on the open index of the top tensor $\hat{t}$. However, in order to capture the first $J=2$ state using the regular MERA, we would need to consider at least $\chi_{\tiny\mbox{top}} = 20$ (at a larger computational cost and possibly lower accuracy), since this state has only the $20$\textsuperscript{th} lowest energy overall.

%%%%%%%%%%%%%%%%%%%%%%%%%%%%%%%%%%%%%%%%%%%%%%%%%%%%%%%%%%%%%%%%%%%%%%%%%%%%%%%%%%%%%%%%%%%%%%%%
\begin{figure}[t]
\begin{center}
  \includegraphics[width=12cm]{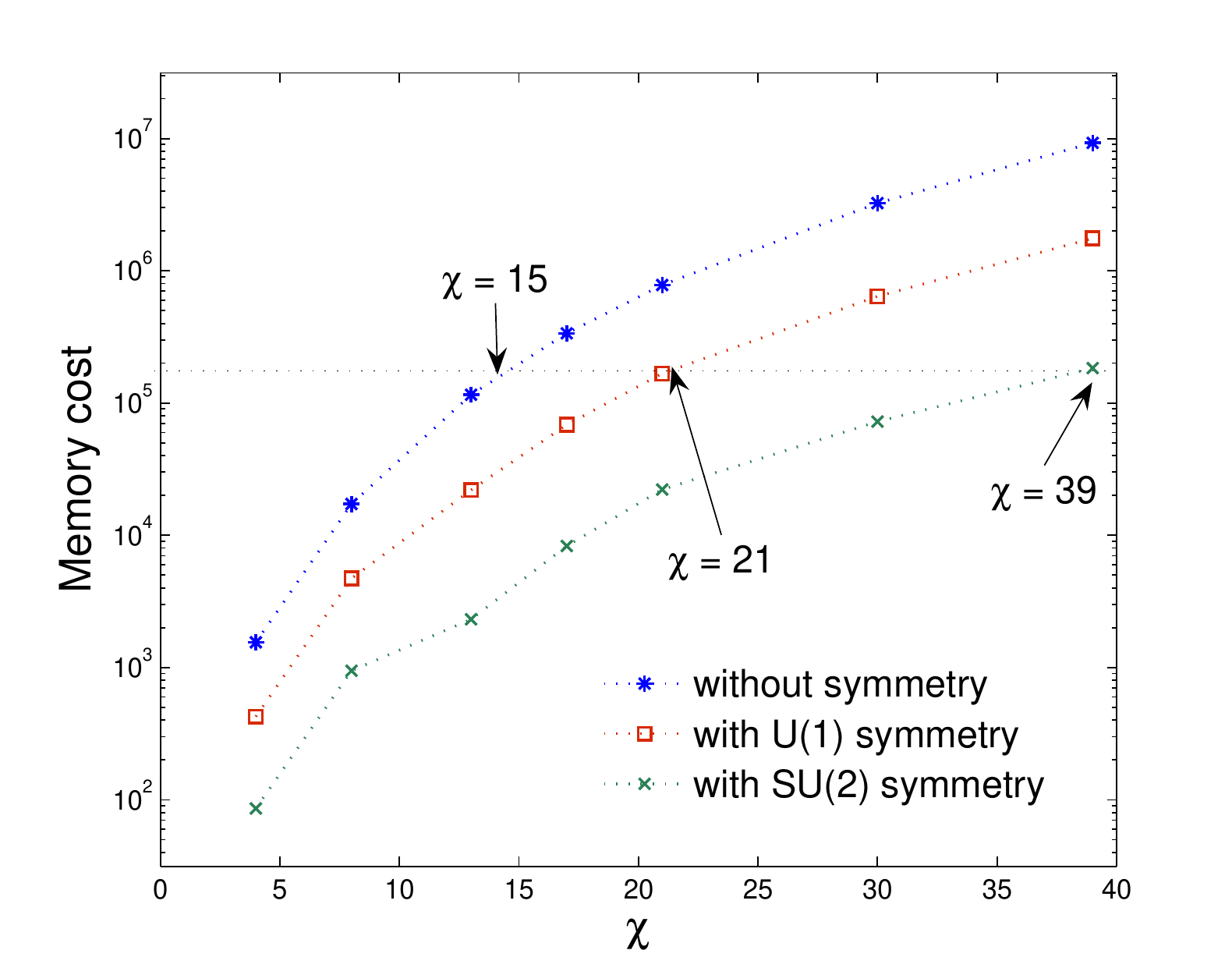}
\end{center}  
\caption{  Memory cost (in number of components) for storing the MERA as a function of the bond dimension $\chi$. The horizontal line on this graph shows that this reduction in memory cost equates to the ability to store MERAs with a higher bond dimension $\chi$: For the same amount of memory required to store a MERA with bond dimension $\chi=15$, one may choose instead to store a U(1)-symmetric MERA with $\chi=26$ or an SU(2)-symmetric MERA with $\chi=39$. \label{fig:memcompare}}
\end{figure}
%%%%%%%%%%%%%%%%%%%%%%%%%%%%%%%%%%%%%%%%%%%%%%%%%%%%%%%%%%%%%%%%%%%%%%%%%%%%%%%%%%%%%%%%%%%%%%%%
%%%%%%%%%%%%%%%%%%%%%%%%%%%%%%%%%%%%%%%%%%%%%%%%%%%%%%%%%%%%%%%%%%%%%%%%%%%%%%%%%%%%%%%%%%%%%%%%
\begin{figure}[t]
\begin{center}
  \includegraphics[width=12cm]{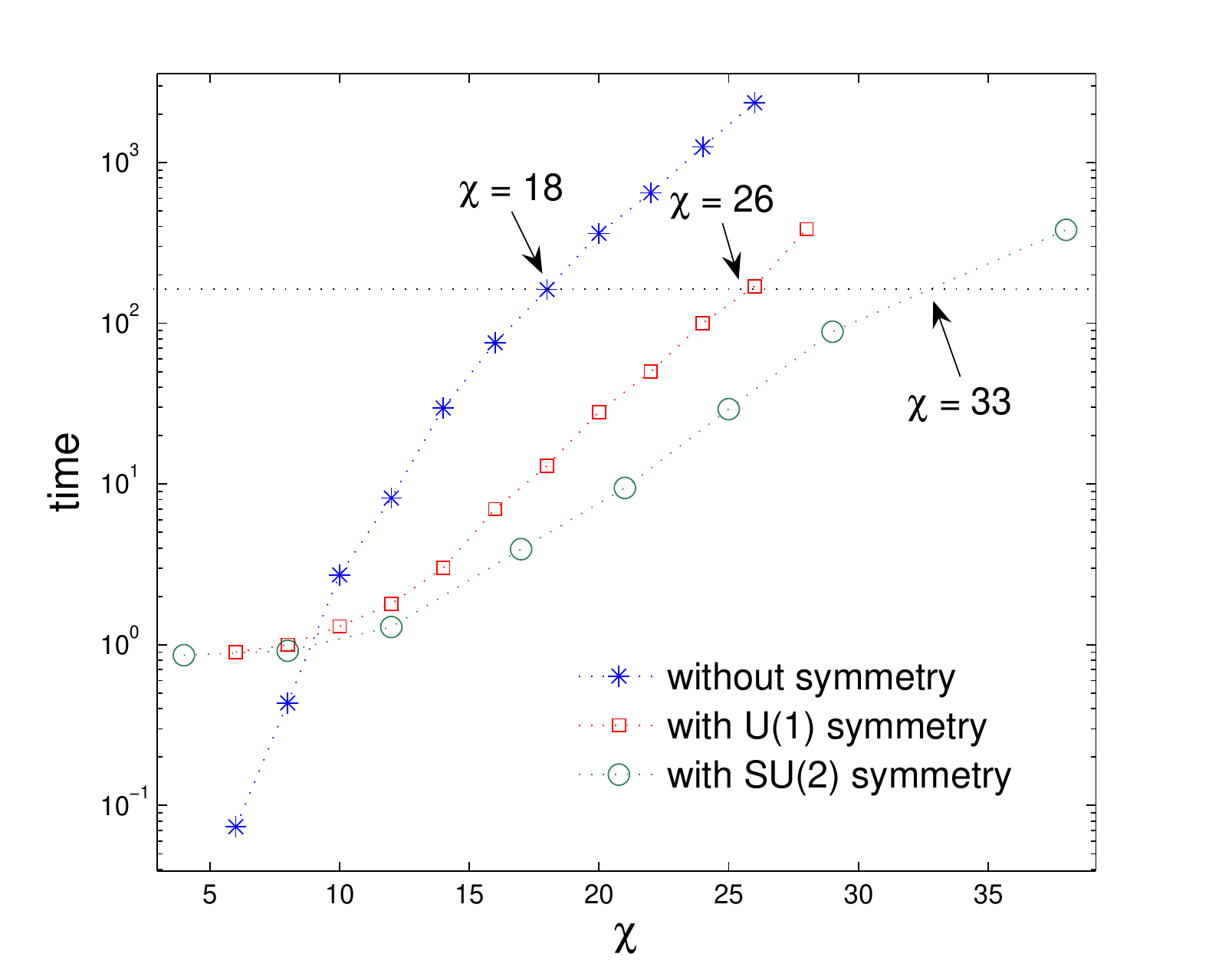}
\end{center}  
\caption{ Computation time (in seconds) for one iteration of the MERA energy minimization algorithm, as a function of the bond dimension $\chi$. For sufficiently large $\chi$, exploiting the SU(2) symmetry leads to reductions in computation time. The horizontal line on this graph shows that this reduction in computation time equates to the ability to evaluate MERAs with a higher bond dimension $\chi$: For the same cost per iteration incurred when optimizing a regular MERA in MATLAB with bond dimension $\chi=18$, one may choose instead to optimize a U(1)-symmetric MERA with $\chi=26$ or an SU(2)-symmetric MERA $\chi=33$.\label{fig:meracompare}}
\end{figure}
%%%%%%%%%%%%%%%%%%%%%%%%%%%%%%%%%%%%%%%%%%%%%%%%%%%%%%%%%%%%%%%%%%%%%%%%%%%%%%%%%%%%%%%%%%%%%%%%
\subsubsection{Reduction in memory and computational costs}

The use of SU(2)-invariant tensors in the MERA also results in a reduction of computational costs. We compared the memory and computational costs associated with using the regular MERA and the SU(2) MERA. We also found it instructive to compare the analogous costs associated with a MERA that is made of tensors that remain invariant under only a subgroup U(1) of the symmetry group. This entails introducing the spin projection operators $\hat{J}_z$ on the links of the MERA and imposing the invariance of constituent tensors under the action of these operators. For such a U(1)-MERA, imposing such constraints corresponds to conservation of the total spin projection $m_J$, while the total spin may fluctuate. (The explicit construction of the U(1)-MERA was discussed in Chapter 4).

Figure \ref{fig:memcompare} shows a comparison of the total number of complex coefficients that are required to be stored for $L=54$ sites (corresponding to 108 spins) in the three cases: regular MERA, U(1) MERA and the SU(2) MERA. U(1)-invariant tensors (see Chapter 4) have a block structure in the eigenbasis of $\hat{J}_z$ operators on each index of the tensor, and therefore they incur a smaller memory cost in comparison to regular tensors. For example, it can be seen that for the same memory required to store a regular MERA with $\chi=15$, one can instead consider storing a U(1)-MERA with $\chi=21$. On the other hand, SU(2)-invariant tensors are substantially more sparse. When written in the canonical form, SU(2)-invariant tensors are not only block-sparse but each block, in turn, decomposes into a degeneracy part and a structural part such that the structural part need not be stored in memory. With the same amount of memory that is required to store, for example, a $\chi=15$ regular MERA, one can already store a $\chi=39$ SU(2) MERA.

In Fig.~\ref{fig:meracompare} we show an analogous comparison of the computational performance in the three cases. We plot the computational time required for one iteration of the energy minimization algorithm of \citep{Evenbly09} (during which all tensors in the MERA are updated once), as a function of the total bond dimension $\chi$ for the cases of regular MERA, U(1) MERA  and SU(2) MERA. We see that for sufficiently large $\chi$, using SU(2)-invariant tensors leads to a shorter time per iteration of the optimization algorithm. In the case of symmetric tensors we considered pre-computation of repeated operations, see Sec.~\ref{sec:precompute}.

%*************************************************************************************************

\chapter{Conclusions and Outlook\label{sec:conclusion}}

In this thesis we have described how to incorporate global internal symmetries into tensor network states and algorithms.

On the theoretical side we developed a framework to characterize and manipulate symmetric tensors. Any given tensor network can be adapted to the presence of a symmetry by imposing the constituent tensors to be symmetric. Symmetric tensors are very sparse objects. Their judicious use and careful manipulation can lead to an enormous computational gain in numerical simulations. This has been extensively demonstrated in this thesis by means of our reference MATLAB implementation.
%
%We described how a symmetric tensor decomposed into a \textit{canonical decomposition} as a set of degeneracy tensors and a set of structural tensors. The tensor can be compactly stored in terms of the degeneracy tensor. On the other hand, the structural tensors need not be explicitly stored since they are completely determined by the symmetry. Moreover, one can manipulate them algebraically by using properties of the Clebsch-Gordan coefficients. Performing manipulations on the canonical decomposition of the tensor, allows for the symmetry to be both exactly preserved and exploited for computational gain in numerical simulations. 

On the implementation side, we have described a practical scheme for protecting and exploiting the symmetry in numerical simulations. We proposed the use of \textit{tree decompositions} of a symmetric tensor. Several advantages of this scheme were discussed, not excluding the overall simplicity and elegance of the method. We hope to have provided a suitable implementation framework for researchers who are familiar with the theoretical aspects of incorporating the symmetries into tensor network algorithm but nonetheless find the practical implementation challenging.

In implementing symmetries we have gone beyond the case of MPS, which being a trivalent tensor network is simpler to handle. We described the construction of the U(1) and SU(2) symmetric versions of the MERA. Our Abelian implementation led to computational gains measuring up to an increase of ten to twenty times. The analogous gain from the non-Abelian implementation was much larger, measuring up to an increase of forty to fifty times. These gains may be used either to reduce overall computation time or to permit substantial increases in the MERA bond dimension $\chi$, and consequently in the accuracy of the results obtained. Therefore the exploitation of symmetries, especially non-Abelian symmetries, can be an invaluable tool for numerically challenging systems. This is more so the case in two dimensional lattice models where simulation costs are much more severe. An example of a potential application is to a system of interacting fermions that appears in the context of high temperature conductivity. Here even though symmetries are present in the model, they have not been throughly exploited in the context of tensor network algorithms.

Although we have given special attention to specific symmetry groups, U(1) and SU(2), the formalism presented in this thesis may equally well be applied to any reducible compact non-Abelian group that is multiplicity free. In particular, one can consider composite symmetries such as SU(2)$\times$U(1), corresponding to spin isotropy and particle number conservation and SU(2)$\times$SU(2) corresponding to conservation of spin and isospin etc. Such a symmetry is characterized by a set of charges $(a_1, a_2, a_3,\ldots)$; when fusing two such sets of charges $(a_1,a_2,a_3,\ldots)$ and $(a'_1,a'_2,a'_3,\ldots)$, each charge $a_i$ is combined with its counterpart $a'_i$ according to the relevant fusion rule. Once again, this behaviour may be encoded into a single fusing tensor $\fuser$.

Our implementation scheme can also be readily extended to incorporate more general symmetry constraints such as those associated with conservation of total fermionic and anyonic charge. One proceeds by defining the following tensors for the relevant charges,
\begin{itemize}
	\item the fusing tensor $\fuser$ that encodes the fusion of two charges,
	\item the recoupling coefficients $\hat{F}$ that relate various ways of fusing three charges, and
	\item the swap tensor $\braider$ that encodes the swap behaviour of two charges.
\end{itemize}
Note that within our specific implementation framework, one may instead just define the linear maps $\hat{\Gamma}$ that mediate tensor manipulations for the respective charges.

As an example, consider fermionic constraints where the relevant charge, $p$, is the parity of fermion particle number. Charge $p$ takes two values, $p=0$ and $p=1$ corresponding to even or odd number of fermions. The fusing tensor $\fuser$ encodes the fusion rules that specify how charges $p$ and $p'$ fuse together to obtain a charge $p''$. These correspond to the fusion rules for the group $Z_2$, given as,
\begin{align}
(p=0) \times (p'=0) &\rightarrow (p''=0), \nonumber \\
(p=0)\times (p'=1) &\rightarrow (p''=1), \nonumber \\
(p=1) \times (p'=0) &\rightarrow (p''=1), \nonumber \\
(p=1) \times (p'=1) &\rightarrow (p''=0). \nonumber
\end{align}
The recoupling coefficients $\hat{F}_{p_a p_b p_c p_d}^{p_e p_f}$, associated with the fusion of three charges $p_a, p_b$ and $p_c$ are simple in this case owing to the Abelian fusion rules. They take a value $\hat{F}_{p_a p_b p_c p_d}^{p_e p_f} = 1$ for all values of intermediate charges $p_e$ and $p_f$ that appear when fusing the three charges one way or the other.
%
%They are given as,
%\begin{equation}
%\hat{F}_{p_a p_b p_c p_d}^{p_e p_f} = \left\{ 
%	\begin{array}{cc} 1&\mbox{ if } p_e = p_f,\\ 
%	 									0&\mbox{otherwise },
%	\end{array} \right.~~~~~\forall p_a, p_b, p_c, p_d, p_e, p_f \in \{0, 1\},
%\end{equation}
%where $p_d$ is the total charge obtained as a result of fusing the three charges. Charges $p_e$ and $p_f$ are the intermediate charges that appear when fusing the three charges one way or the other.

The final ingredient is the tensor $\braider$, which in this case is defined as,
\begin{align}
\braider_{p=0, p'=0 \rightarrow p''=0} &= 1,~~ \braider_{p=0, p'=1 \rightarrow p''=1} = 1, \nonumber \\
\braider_{p=1, p'=0 \rightarrow p''=1} &= 1,~~ \braider_{p=1, p'=1 \rightarrow p''=0} = -1.\nonumber
\end{align}
 
In a similar way, one can encode the corresponding fusion rules for anyonic charges into the fusing tensor $\fuser$. In the case of anyons, the recoupling coefficients $\hat{F}$ are obtained as solutions to the \textit{pentagon equations} whereas the tensors $\braider$ are replaced with the anyonic braid operators that are obtained as solutions to the \textit{hexagon equations}, see \citep{Trebst08, Feiguin07, Pfeifer10}. Thus, having defined these tensors for the relevant charges, the formalism and the implementation framework presented in this thesis can be readily adapted to incorporate the constraints corresponding to the presence of fermionic or anyonic charges.
%
%\begin{table}
%%%%%%%%%%%%%%%%%%%%%%%%%%%%%%%%%%%%%%%%%%%%%%%%%%%%%%%%%%%%%%%%%%%%%%%%%%%%%%%%%%%%%%%%%%%%%%%%%
%\centering % used for centering table
%\begin{tabular}{|c|c|c|c|} % centered columns (4 columns)
%\hline %inserts double horizontal lines
%\textit{Conserved physical quantity} &  \textit{Charges}, $p$ & a & b \\ [0.5ex] % inserts table
%\hline\hline
%$\begin{array}{l}\mbox{Parity} \\ \mbox{(Bosons)} \end{array}$ & 0,1 & $p'' = (p+p')\%2$ & $\braid{p}{p'}{p''} = 1 \forall p, p'$\\ \hline
%$\begin{array}{l}\mbox{Parity} \\ \mbox{(Fermions)} \end{array}$ & 0,1 & $p'' = (p+p')\%2$ & $\braid{p}{p'}{p''} = (-1)^{p''} \forall p, p'$\\ \hline
%$\begin{array}{l}\mbox{Particle number,} \\ \mbox{Spin Projection} \end{array}$ & $0,1,\ldots$ & $p'' = p+p'$ & $\braid{p}{p'}{p''} = 1 \forall p, p'$\\ \hline 
%%heading
%%\hline % inserts single horizontal line
%\end{tabular}
%\caption{
%%[table:degdist]
%Popular tensor networks in one and two spatial dimensions.\label{table:tn}
%}
%\end{table}

In a different context, preservation of symmetry can be crucial even without demanding a computational gain. Recently, a novel classification of symmetric
phases in 1D gapped spin systems was undertaken in \citep{Chen11}. In the absence of any symmetry, in this classification all states are equivalent to trivial product states. However, by preserving certain symmetries many phases were reported to exist with a different \textit{symmetry protected topological order}. As an alternative, such a classification could potentially be addressed with the symmetric version of the MERA, since the MERA is adept at characterizing fixed points of the renormalization group flow which correspond to different phases. 

Symmetries are a fundamental aspect of nature. Nearly all physical phenomenon can be explained by the presence or the absence of a symmetry. In numerical methods, the preservation of symmetries may well be a necessary requirement for simulating certain aspects of the system. As a computational aid, symmetries will play a crucial role in pushing forward the frontiers of tensor network algorithms in the coming years.

% --- End of main body of thesis
% --- Generate bibliography etc.
% Note: Bibliography style can be set up in header.tex, or here.

% Specify bibliographystyle. Necessary when using bibtex.
% Should be natbib-compliant if using natbib.
%\bibliographystyle{alpha}
%\bibliographystyle{unsrt}
\bibliographystyle{plainnat_dotfill} % Natbib default styles are plainnat, abbrvnat, unsrtnat
  % The _dotfill versions look nice when using the pagebackref option with the hyperref
  % package: They put a line of dots between the citation and the backreferences.
  % They also include support for the eprint field in BibTeX entries of type "misc".

\bibliography{thesis} % Generate bibliography (using BibTeX) - list all .bib files here
% (can include more than one bibfile in \bibliography{}

%\cleardoublepage % Move to beginning of a new right-hand page
%\printindex % Generate index
%\printglossary % Generate glossary
%\bibitem{su2mps} S. Singh, H.-Q. Zhou and G. Vidal, New J. Phys. 12 (2010) 033029.

\end{document}